\documentclass[me,dissertation]{puthesis}

\usepackage[numbers]{natbib}
\usepackage{epstopdf, epsfig}
\usepackage[colorlinks]{hyperref}
\usepackage{cancel}
\usepackage{multirow} % tabular multirow
\usepackage{amsmath}
\usepackage{array}
\usepackage{amssymb}
\usepackage{natbib} 		% already included in JFM class
\usepackage{hyperref}
\usepackage{color}
\usepackage{pbox}
\usepackage{subcaption}
\usepackage{tikz}
\usetikzlibrary{shapes}
\usepackage{mathrsfs}
\usepackage[utf8]{inputenc}
\usepackage[T1]{fontenc}
\usepackage{caption}
\usepackage{setspace}

\graphicspath{{./figures/}}

\title{%
  Theoretical and Numerical Investigation \\
  of Nonlinear Thermoacoustic, Acoustic, \\ and Detonation Waves%
}

\author{Prateek Gupta}{Prateek Gupta}

\pudegree{Doctor of Philosophy}{PhD}{August}{2019}

\majorprof{Carlo Scalo}

\campus{West Lafayette}

\newcolumntype{M}[1]{>{\centering\arraybackslash}m{#1}}
\newcommand{\tikzcircle}[2][black,fill=black]{\tikz[baseline=-0.5ex]\draw[#1,radius=#2] (0,0) circle ;}
\usepackage[T1]{fontenc}
\newcommand{\markerone}{\raisebox{0.5pt}{\tikz{\node[draw,scale=0.4,circle,fill=black!20!black](){};}}}
\newcommand{\markertwo}{\raisebox{0.5pt}{\tikz{\node[draw,scale=0.4,circle,fill=black!20!gray](){};}}}
\newcommand{\markerthree}{\raisebox{0.5pt}{\tikz{\node[draw,scale=0.4,circle,fill=none](){};}}}
\newcommand{\markerfour}{\raisebox{0.5pt}{\tikz{\node[draw,scale=0.4,regular polygon, regular polygon sides=4,fill=black!20!black](){};}}}
\newcommand{\markerfive}{\raisebox{0.5pt}{\tikz{\node[draw,scale=0.4,regular polygon, regular polygon sides=4,fill=black!20!gray](){};}}}
\newcommand{\markersix}{\raisebox{0.5pt}{\tikz{\node[draw,scale=0.4,regular polygon, regular polygon sides=4,fill=none](){};}}}
\usepackage{footmisc}

\usepackage{scalerel,stackengine}
\stackMath
\newcommand\reallywidehat[1]{%
\savestack{\tmpbox}{\stretchto{%
  \scaleto{%
    \scalerel*[\widthof{\ensuremath{#1}}]{\kern-.6pt\bigwedge\kern-.6pt}%
    {\rule[-\textheight/2]{1ex}{\textheight}}%WIDTH-LIMITED BIG WEDGE
  }{\textheight}% 
}{0.5ex}}%
\stackon[1pt]{#1}{\tmpbox}%
}
\newcommand{\uFluct}{u'}
\newcommand{\pFluct}{p'}
\newcommand{\rhoFluct}{\rho '}
\newcommand{\sFluct}{s'}
\newcommand{\TFluct}{T'}

\begin{document}

% Start a new volume for your thesis.
% All theses must have at least one volume.
% If your thesis has multiple volumes put another "\volume"
% command between chapters below.
\volume

% Front matter:
%     dedication
%     acknowledgments
%     preface
%     table of contents
%     list of tables
%     list of figures
%     list of symbols
%     list of abbreviations
%     nomenclature
%     glossary
%     abstract
%     publication

%
%  revised  front.tex  2017-01-08  Mark Senn  http://engineering.purdue.edu/~mark
%  created  front.tex  2003-06-02  Mark Senn  http://engineering.purdue.edu/~mark
%
%  This is ``front matter'' for the thesis.
%
%  Regarding ``References'' below:
%      KEY    MEANING
%      PU     ``A Manual for the Preparation of Graduate Theses'',
%             The Graduate School, Purdue University, 1996.
%      PU8    ``A Manual for the Preparation of Graduate Theses'',
%             Eighth Revise Edition, Purdue University.
%      TCMOS  The Chicago Manual of Style, Edition 14.
%      WNNCD  Webster's Ninth New Collegiate Dictionary.
%
%  Lines marked with "%%" may need to be changed.
%

  % Statement of Thesis/Dissertation Approval Page
  % This page is REQUIRED.  The page should be numbered page ``ii''
  % and should NOT be listed in your TABLE OF CONTENTS.
  % References: PU8 ordinal pages 5 and 29.
  % The web page https://engineering.purdue.edu/AAE retrieved on
  % January 8, 2017 had "School of Aeronautics and Astronautics"---that
  % is used instead of "Department af Aeronautics and Astronautics"
  % below.
\begin{statement}
  \entry{Dr.~Carlo Scalo, Chair}{School of Mechanical Engineering}
  \entry{Dr.~Ivan Christov}{School of Mechanical Engineering}
  \entry{Dr.~Stuart Bolton}{School of Mechanical Engineering}
  \entry{Dr.~Sally Bane}{School of Aeronautics and Astronautics}
  \approvedby{Dr.~Jay P. Gore}{Associate Head for Graduate Studies}
\end{statement}

  % Dedication page is optional.
  % A name and often a message in tribute to a person or cause.
  % References: PU 15, WNNCD 332.
%\begin{dedication}
%  This is the dedication.
%\end{dedication}

  % Acknowledgements page is optional but most theses include
  % a brief statement of apreciation or recognition of special
  % assistance.
  % Reference: PU 16.
\begin{acknowledgments}
  I would like to express my sincere gratitude to my advisor Prof. Carlo Scalo for his help and support. His passion to research upon a fundamental idea all the way up to application and strong advocacy of scientific rigor have been constant motivations. I would also like to thank Prof. Ivan Christov, Prof. Sally Bane, and Prof. Stuart Bolton for serving on my defense committee and for their feedback on this dissertation.

  I wish to thank my colleagues at Compressible Flow and Acoustics Lab for creating a vibrant atmosphere in the workplace. A special thanks to my colleagues and friends, Zongxin Yu, Yongkai Chen, Emmanuel Gill Torres, Karl Jantze, Julien Brillon, and Joel Redmond for their company. A very special thanks to my friends at Purdue, including Kumar Akash, Kunal Pardikar, Akash Patil, Yeshaswi Menghmalani, Rohil Jain, Viplove Arora, Vishwanath Ganesan, and Vishal Anand. 
  
  I would also like to thank Prof. Guido Lodato at INSA de Rouen for his guidance regarding the SD3DvisP solver in my early years of PhD.
  
  Besides people directly involved in my PhD, I am extremely grateful to my bachelor thesis advisor, Prof. Supreet Singh Bahga at IIT Delhi, Mechanical Engineering Department, for his support during tough times. His guidance played a pivotal role in conditioning me into a patient and thorough researcher. 

  I am grateful to my parents, brother, sister-in-law, and the extended family for their continued support and unconditional love. This dissertation is a culmination of their support, time, and energy put into me. To them, I dedicate this dissertation. 

\end{acknowledgments}

  % The preface is optional.
  % References: PU 16, TCMOS 1.49, WNNCD 927.
%\begin{preface}
%  This is the preface.
%\end{preface}

  % The Table of Contents is required.
  % The Table of Contents will be automatically created for you
  % using information you supply in
  %     \chapter
  %     \section
  %     \subsection
  %     \subsubsection
  % commands.
  % Reference: PU 16.
\tableofcontents

  % If your thesis has tables, a list of tables is required.
  % The List of Tables will be automatically created for you using
  % information you supply in
  %     \begin{table} ... \end{table}
  % environments.
  % Reference: PU 16.
\listoftables

  % If your thesis has figures, a list of figures is required.
  % The List of Figures will be automatically created for you using
  % information you supply in
  %     \begin{figure} ... \end{figure}
  % environments.
  % Reference: PU 16.
\listoffigures

  % List of Symbols is optional.
  % Reference: PU 17.
  
%\begin{symbols}
%  $m$& mass\cr
%  $v$& velocity\cr
%\end{symbols}

  % List of Abbreviations is optional.
  % Reference: PU 17.
  
%\begin{abbreviations}
%  abbr& abbreviation\cr
%  bcf& billion cubic feet\cr
%  BMOC& big man on campus\cr
%\end{abbreviations}

  % Nomenclature is optional.
  % Reference: PU 17.
  
%\begin{nomenclature}
%  Alanine& 2-Aminopropanoic acid\cr
%  Valine& 2-Amino-3-methylbutanoic acid\cr
%\end{nomenclature}

  % Glossary is optional
  % Reference: PU 17.
  
%\begin{glossary}
%  chick& female, usually young\cr
%  dude& male, usually young\cr
%\end{glossary}

  % Abstract is required.
  % Note that the information for the first paragraph of the output
  % doesn't need to be input here...it is put in automatically from
  % information you supplied earlier using \title, \author, \degree,
  % and \majorprof.
  % Reference: PU 17.
\begin{abstract}
  Finite amplitude perturbations in compressible media are ubiquitous in scientific and engineering applications such as gas-turbine engines, rocket propulsion systems, combustion instabilities, inhomogeneous solids, and traffic flow prediction models, to name a few. Small amplitude waves in compressible fluids propagate as sound and are very well described by linear theory. On the other hand, the theory of nonlinear acoustics, concerning high-amplitude wave propagation (Mach$<$2) is relatively underdeveloped. Most of the theoretical development in nonlinear acoustics has focused on wave steepening and has been centered around the Burgers' equation, which can be extended to nonlinear acoustics only for purely one-way traveling waves. In this dissertation, theoretical and computational developments are discussed with the objective of advancing the multi-fidelity modeling of nonlinear acoustics, ranging from quasi one-dimensional high-amplitude waves to combustion-induced detonation waves. 
  
  We begin with the theoretical study of spectral energy cascade due to the propagation of high amplitude sound in the absence of thermal sources. To this end, a first-principles-based system of governing equations, correct up to second order in perturbation variables is derived. The exact energy corollary of such second-order system of equations is then formulated and used to elucidate the spectral energy dynamics of nonlinear acoustic waves. We then extend this analysis to thermoacoustically unstable waves -- i.e. amplified as a result of thermoacoustic instability. We drive such instability up until the generation of shock waves. We further study the nonlinear wave propagation in geometrically complex case of waves induced by the spark plasma between the electrodes. This case adds the geometrical complexity of a curved, three-dimensional shock, yielding vorticity production due to baroclinic torque. Finally, detonation waves are simulated by using a low-order approach, in a periodic setup subjected to high pressure inlet and exhaust of combustible gaseous mixture. An order adaptive fully compressible and unstructured Navier Stokes solver is currently under development to enable higher fidelity studies of both the spark plasma and detonation wave problem in the future.  
  
\end{abstract}

%
% Put chapter \include commands here.
%

% Introductions may precede the first chapters or major divisions of theses.
% Reference: TM2006, page 31.
% CHANGE NEXT LINE?

% No citations in abstract
% short captions in TOC
% in abstract (below Mach<2) ?????
% Navier-Stokes 
% K41 theory
% Second-order
% remove math env. from Kelvin
% <> signs in the caption
% align equations instead of the gather
% 4.30 needs to be aligned
% 4.39(a)
% place holder should be a cdot
% replace const.. with const.
% change the multiple paranthesis
% lower case of Kg
% heat addition in (6.4)
% 7.11 slash
% cdot vs dot before 7.11
% replace ref with eqref
% %%%%%%%%%%%%%%%%%%%%%%%%%%%%%%%%%%%
\chapter{Introduction}

% What is nonlinear wave propagation

% What is the scope of the current work 

% Introduce thermoacoustics briefly with amplification plots

% Introduce spectral energy cascade 

% Show generation of high amplitude waves as a result of spark plasma and detonation combustion.
Nonlinear wave processes are observed in a variety of engineering and physics applications such as acoustics~\cite{Hamilton_NLA_1998, Lighthill_JSV_1978, Whitham2011}, thermoacoustics~\cite{ScaloLH_JFM_2015, GuptaLS_JFM_2017, Culik_AGARD_2006, Mcmanus1993review, Schwinn_CandF_2018, PoinsotV_numComb_2011}, surface waves~\cite{AblowitzClarkson1991}, plasma-physics~\cite{Gurbatov2012}, and inhomogeneous solids~\cite{Nazarov2015}. In acoustics, nonlinear wave propagation (finite amplitude acoustics) has been an active area of research since more than five decades now and can be classified as both hyperbolic and dispersive, based on the medium of propagation~\cite{Whitham2011}. Planar nonlinear waves in compressible fluids primarily exhibit non-dispersive hyperbolic nature. Propagation of such waves causes two main nonlinearities known, acoustic streaming~\cite{Lighthill_JSV_1978, Gedeon_1997_Cryocoolers} and wave steepening~\cite{Hamilton_NLA_1998, Naugolnykh1998}. While acoustic streaming is a kinematic nonlinear effect, wave steepening is caused by the variation of speed of propagation (local speed of sound) due to variation in pressure. For very high pressure variations, nonlinear acoustic waves form shock waves, which can also cause combustion of the medium due to intense temperature variation, depending on the ignition properties of the medium. Moreover, non-planar nonlinear waves can cause intense hydrodynamic events such as vorticity generation~\cite{Truesdell_1952_JAS}, which can be eventually utilized for changing scalar and momentum mixing properties of flows.

The scope of the present work includes theoretical and numerical study of such nonlinear wave propagation effects, in particular, mathematically quantifying the planar nonlinear acoustic waves identifying the extent of second order nonlinear theory, effect of boundary heating and combustion heating on such nonlinear waves, followed by the hydrodynamic effects of such waves on flows.  We begin with the basic definition and mathematical identification of such nonlinearities. As the high amplitude planar waves propagate, multiple length scales are generated which modify the energy dissipation dynamics significantly. Moreover, such nonlinear waves can be generated from a variety of heat sources such as thermoacoustic instability, heat release in detonation waves, spark plasma discharges etc. We outline the analysis of such nonlinear wave-heat source interactions for compressible fluids. Throughout, theoretical developments are supported by high-fidelity canonical numerical simulations.

This dissertation is organized in five chapters, each summarized briefly below.

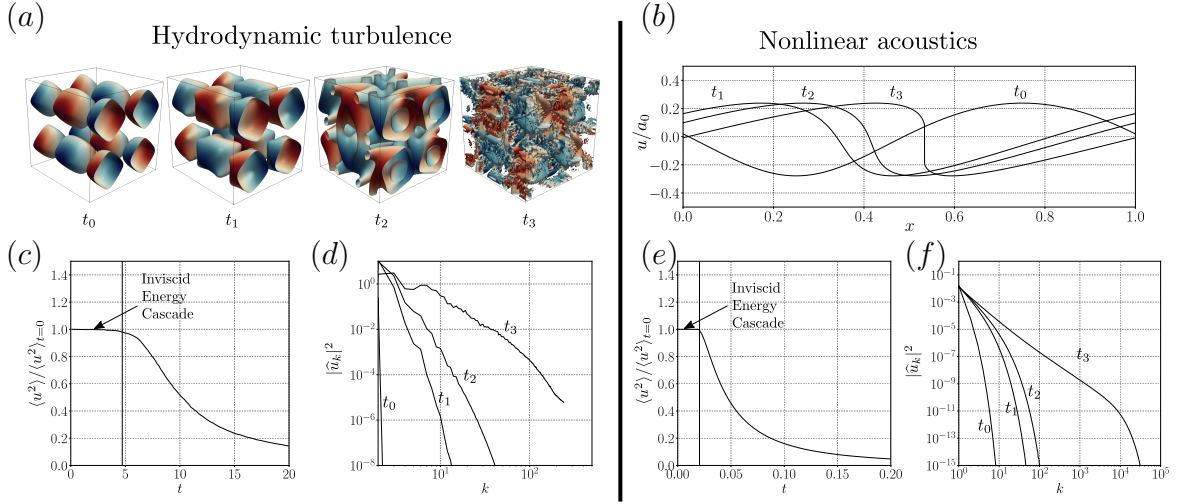
\begin{figure}[!t]
\centering
\includegraphics[width=\linewidth]{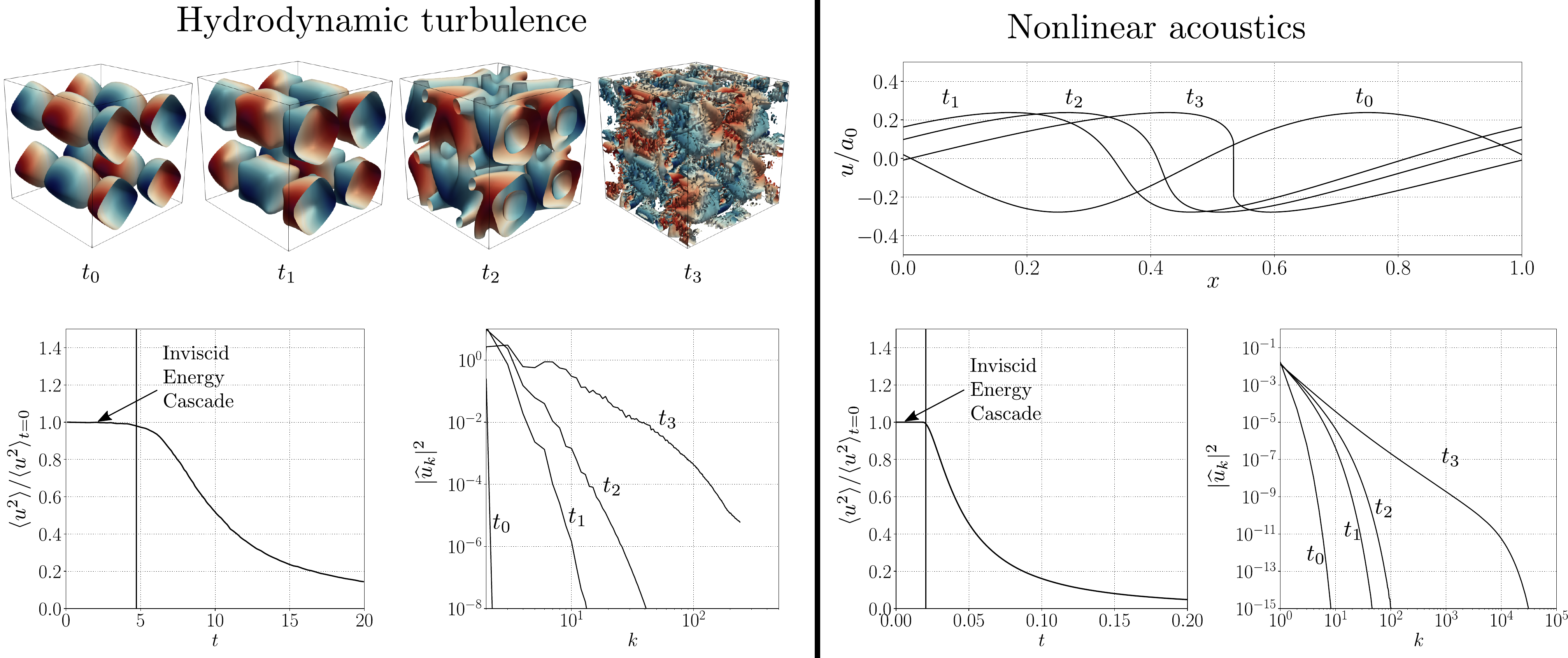}
\put(-440,180){$(a)$}
\put(-200,180){$(b)$}
\put(-440,90){$(c)$}
\put(-325,90){$(d)$}
\put(-200,90){$(e)$}
\put(-100,90){$(f)$}
\singlespace
\caption[Illustration of similar spectral energy dynamics in nonlinear acoustics and hydrodynamic turbulence through DNS of Navier-Stokes equations.]{ $Q$-criterion iso-surfaces colored with the local velocity magnitude obtained from a direct numerical simulation of a Taylor-Green vortex in a triply-periodic domain $\left[-\pi, \pi\right]^3$~\cite{Chapelier_jcp_2017}, exhibiting breakdown into hydrodynamic turbulence $(a)$, velocity perturbation field in a high amplitude nonlinear traveling acoustic wave (TW) $(b)$, evolution of normalized spatial average of $u^2$ $(c),(e)$, and velocity spectra $|\widehat{u}_k|^2$ $(d), (f)$ at times $t_0, t_1, t_2, t_3$. The spectral broadening occurs due to the nonlinear terms in the governing equations generating smaller length scales resulting in energy cascade from larger to smaller length scales.
}
\label{fig: Introfig1}
\end{figure}
\section*{Summary of Chapters 2 and 3: Decaying nonlinear acoustics }

Nonlinear acoustic wave steepening occurs due to gradients in the wave speed associated with thermodynamic nonlinearities, and it entails generation of smaller length scales (harmonics) via a nonlinear energy cascade, which can be realized mathematically via the multiplication of two truncated Fourier series, 
\begin{eqnarray}
\left(\sum^{n}_{k=-n}a_k e^{ikx}\right)\left(\sum^{m}_{l=-m}b_le^{ilx}\right) = 
={\sum_k a_k b_{-k}} +{\underset{k+l\neq 0}{\sum_k \sum_l} a_k b_l e^{i(k+l)x}}.
\label{eq: IntroTrig}
\end{eqnarray}

The left hand side of the Eq.~\eqref{eq: IntroTrig} represents any quadratic nonlinear term appearing in a governing equation. Continued nonlinear evolution results in further generation of smaller length scales, as depicted by the second term on the right hand side of Eq.~\eqref{eq: IntroTrig}, ultimately leading to spectral broadening, as shown in figure~\ref{fig: Introfig1}. In the specific case of steepening of nonlinear acoustic waves, the shock thickness denotes the smallest length scale generated in the flow, governed by the viscous dissipation. The latter causes saturation of energy spectra, hence establishing an energy flow among various scales.

In this work, we focus on the characterization of the spectral energy cascade in nonlinear acoustics. In particular, we study the spatio-temporal and spectro-temporal evolution of decaying finite amplitude planar nonlinear acoustic waves in three canonical configurations, traveling waves (TW), standing waves (SW), and randomly initialized Acoustic Wave Turbulence (AWT). Utilizing the second order nonlinear acoustics approximation, we derive analytical expressions for the spectral energy, energy transfer function, and dissipation. Analogous to the study of small scale generation in hydrodynamic turbulence, well quantified by the K41 theory~\cite{kolmogorov1941a, kolmogorov1941b, kolmogorov1941c}, we define relevant length scales associated with fully developed nonlinear acoustic waves elucidating the scaling features of the energy spectra.

\section*{Summary of Chapter 4: Thermoacoustic shock waves}
\begin{figure}[!b]
\centering
\includegraphics[width=0.9\textwidth]{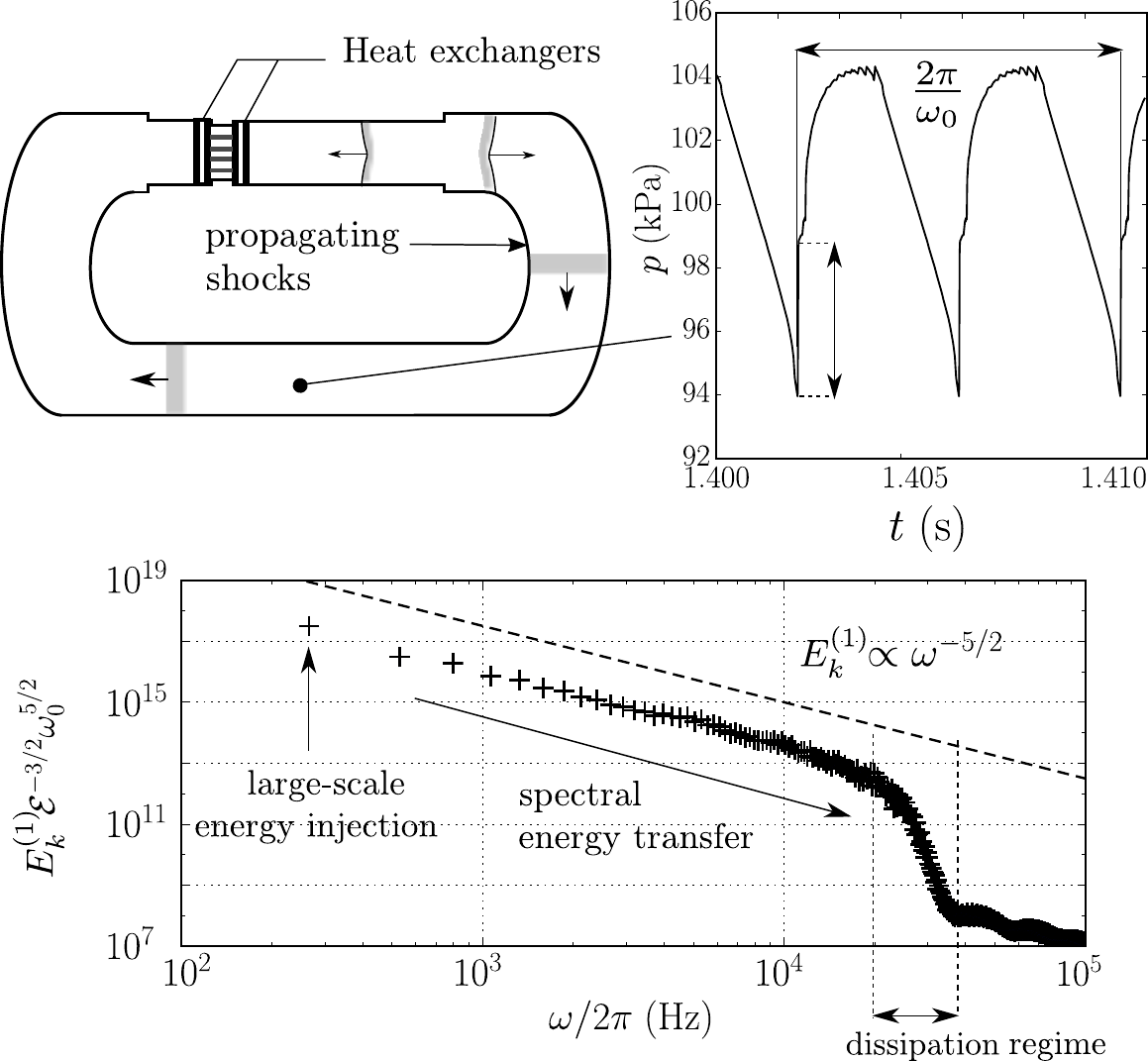}
\put(-400,360){$(a)$}
\put(-180,360){$(b)$}
\put(-400,170){$(c)$}
\caption[Physical setup of a looped thermoacoustically unstable resonator.]{Physical setup of a looped thermoacoustically unstable resonator $(a)$, pressure probed inside the resonator $(b)$, and the isentropic acoustic energy spectrum of the probed signal $(c)$ depicting spectral energy cascade.}
\label{fig: Introfig2}
\end{figure}

The acoustic energy cascade in the spectral space due to nonlinear acoustic wave propagation is analogous to the turbulent spectral energy cascade in which vortex stretching (hydrodynamic nonlinearity) causes breakdown of large length scale eddies into smaller eddies up to a point where the viscous dissipation dominates and prevents generation of further smaller scales~\cite{pope2000turbulent}. In equilibrium energy cascade, mean shear causes injection of turbulent kinetic energy into large eddies which breakdown into smaller eddies thus establishing an equilibrium energy spectrum~\cite{tennekes1972first}. Usually, hydrodynamic instabilities are associated as the mechanisms of energy injection into large length scales by mean shear, such as Tollmien-Schlichting instability in low speed hydrodynamic boundary layers~\cite{schlichting1960boundary}. Analogously, acoustic energy cascade due to nonlinear acoustic wave propagation can be sustained via acoustic instabilities. We show such existence of equilibrium spectral energy cascade in which energy in large length scales (harmonics) is injected by thermoacoustic instabilities resulting in formation of thermoacoustically sustained shock waves (see figure~\ref{fig: Introfig2}). 

Thermoacoustic amplification of waves in a compressible flow is the result of a fluid dynamic instability emerging from the favourable coupling between pressure and heat-release fluctuations. The wavelength of thermoacoustically unstable waves is set by the size of the enclosing resonant chamber, while the heat release, providing the energy source for the amplification, is confined to a compact region. The heat release rate is a function of local velocity and pressure fluctuations, affecting for example the instantaneous flame surface area in a combustion chamber or the rate of convective heat extraction from a hot wire-mesh screen in a Rijke tube. The resulting fluctuations in the heat release rate drive a cycle of compressions and dilatations, which act as a source for pressure fluctuations that, if within a quarter phase from the heat release itself, become thermoacoustically amplified, as identified by the Rayleigh's  criterion~\cite{Rayleigh_Nature_1878}), 
\begin{equation}
\int_{\Omega}p'q' d\Omega > 0.
\label{eq: RayleighCriterionIntro}
\end{equation}
Variables $p'$ and $q'$ denote the pressure and heat release rate perturbations in a confined domain $\Omega$. 

In this work, we show the application of the previously discussed theory on nonlinear acoustic energy cascade. We investigate the high amplitude (or macrosonic) limit of thermoacoustically driven nonlinear waves characterized by the formation of self-sustaining resonating shock waves and inter-scale energy transfer. A comprehensive nonlinear theoretical and high-fidelity modeling approach is adopted to accurately describe macrosconic thermoacoustic waves. To this end, a canonical travelling-wave looped resonator, inspired by Yazaki \emph{et al.}'s~\cite{Yazaki_PhysRevLet_1998} experimental setup but geometrically optimized via linear theory~\cite{Rott_ZAMP_1969,Rott_ZAMP_1973,LinSH_JFM_2016}, has been designed to maximize the growth rate of the quasi-travelling-wave second harmonic and thus achieve rapid shock wave formation. Yazaki \emph{et al.}'s~\cite{Yazaki_PhysRevLet_1998} looped configuration allows quasi-travelling-wave acoustic phasing which facilitates faster nonlinear energy cascade compared to standing wave resonators~\cite{BiwaEtAl_JASA_2014}. It is shown that the energy content in spectral domain resembles the equilibrium energy cascade observed in turbulence, similar to the spectral energy distribution of an ensemble of acoustic waves interacting nonlinearly among each other~\cite{Nazarenko_2011_WT, Zakharov_2012_WT}.

\section*{Summary of Chapter 5: Spark Plasma induced shock waves}
\begin{figure}[!t]
 \centerline{\includegraphics[width=\textwidth]{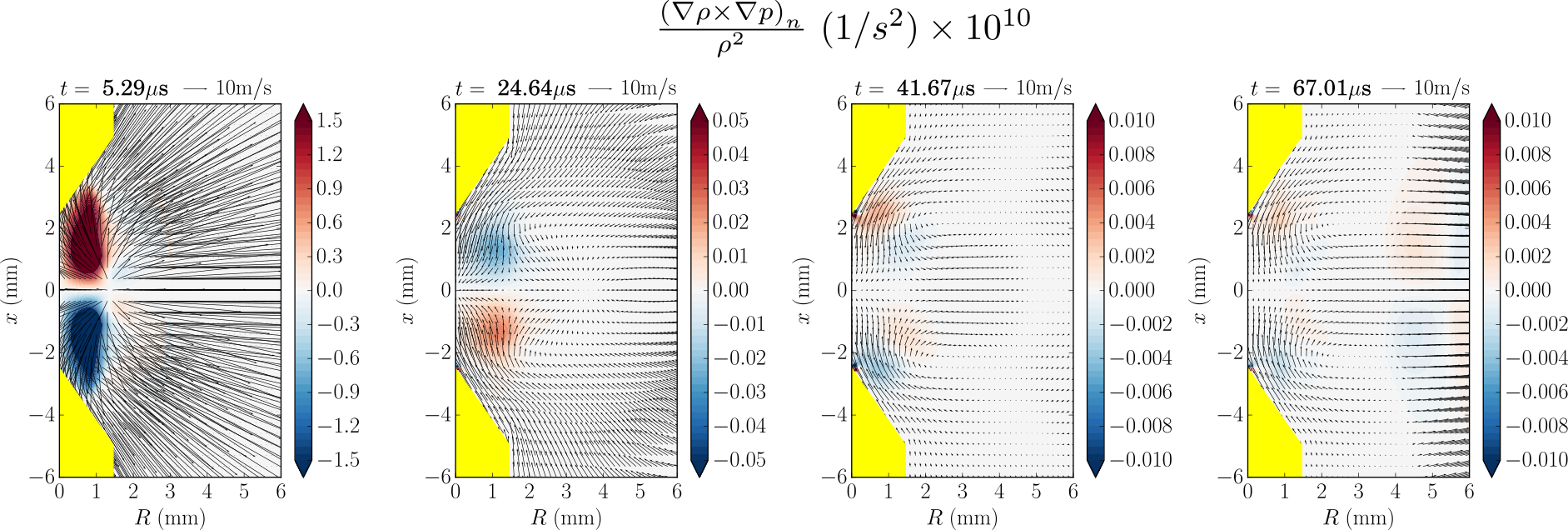}}
 \caption{Evolution of the normalized baroclinic torque  $\Gamma_n$ normal to the $x-y$ plane in the initial stages of the shock wave propagation for slip electrode walls.}
\label{fig: InitialStagesBaroclinic_Slip_intro}
\end{figure}

For planar weak shocks, the acoustic nonlinearities constitute the primary nonlinearities in the flow. However, for weak (or strong) shocks in 2D or 3D, more interesting features appear. One such interesting feature is the generation of vorticity behind a curved shock. 
As first derived by Truesdell~\cite{Truesdell_1952_JAS}, a steady 2D curved shock induces a vorticity jump across the shock directly proportional to the curvature of the shock. Later, this calculation was generalized and analyzed by Kevlahan~\cite{Kevlahan_1997_JFM} for an unsteady 2D shock, who showed the baroclinic moment applied by the curved shock also causes vorticity generation. 

In general, the baroclinic moment acting on the flow is given by, 
\begin{equation}
 \Gamma = \frac{\nabla p \times \nabla \rho}{\rho^2}.
\end{equation}

For an isentropic perturbation, the pressure gradient $\nabla p$ and density gradient $\nabla \rho$ are parallel. Consequently, isentropic waves can not apply baroclinic moments on the fluid. However, waves generated due to intense heat release may exhibit entropy gradients (even though perturbations are weak) not aligned with pressure perturbations. Such entropy variations result in baroclinic moments which cause vorticity changes in a flow. 

In this work, we analyze the flow field generated by the heat deposition between two electrodes due to dielectric breakdown. Such setups are utilized in controlling flow fields and combustion mixing due to coherent structures generated near the region of dielectric breakdown. We show that the baroclinic moment plays an important role in generating coherent structures near conical shaped electrodes. We model the spark plasma heat deposition be a spato-temporally varying heating kernel.

\section*{Summary of Chapter 6: High pressure detonation waves}

Combustion waves are primarily classified as deflagration or detonation waves, the latter being high amplitude pressure waves propagating at supersonic speeds. Heating caused by a propagating adiabatic shock-compression results in ignition of fuel-oxidizer mixture. The fluid containing reaction products expands behind the shock wave and depending on downstream boundary conditions, further accelerates the wave front sustaining its propagation \cite{JHlee_2008}. The classical Zel’dovich, von Neumann and D\"oring (ZND) model postulates equilibrium one-dimensional detonation waves. In experiments and detailed theoretical studies, unstable detonation wave propagation with complex reactive chemistry and compressible flow physics interactions are often observed~\cite{JHlee_2008, SHEPHERD2009}. Consequently, unsteady dynamics of unstable one-dimensional pulsating detonation waves have received wide-spread attention \cite{Erpenbeck_PoF_1964, Lee_JFM_1990, Sharpe_PRSA_1999}. 

\begin{figure}[!h]
\centering
\includegraphics[width=0.7\textwidth]{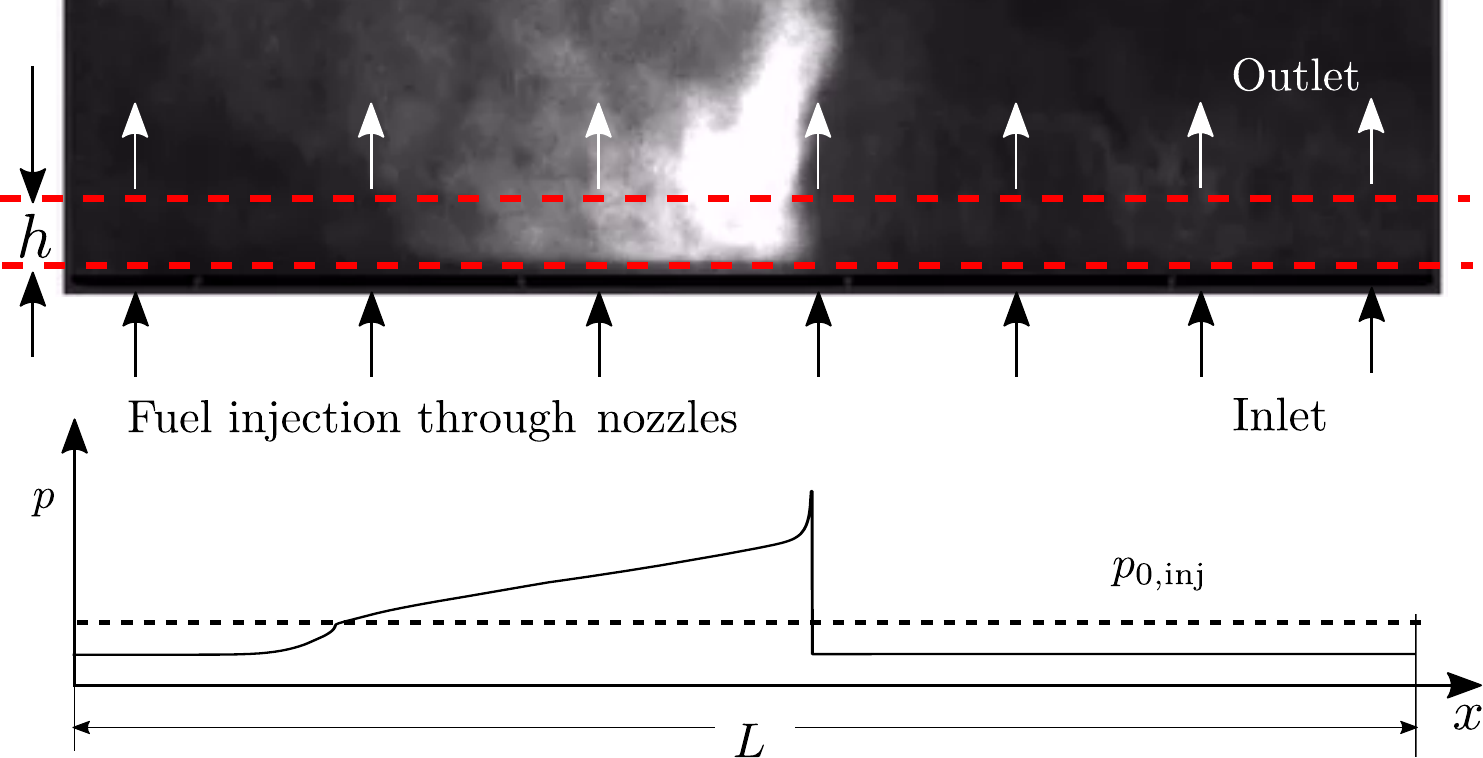}
\put(-320,140){$(a)$}
\put(-320,70){$(b)$}
  \caption{Schematic illustrating the one-dimensional model reduction based on the OH$^*$ chemiluminescence visualization of experimental study by Schwinn \emph{et al.}~\cite{Schwinn_CandF_2018} $(a)$ and sketch of the spatial profile of pressure corresponding to the visualized detonation wave $(b)$. Dashed line (- -) shows the stagnation injection pressure.}
\label{fig: ConceptProblem_intro}
\end{figure}
Detonation waves result in very high pressure gains, which exhibit higher thermodynamic efficiencies when utilized for mechanical work~\cite{Wolanski_RDE_2015,WintenbergerShepherd_JPP_2006}. Rotating Detonation Engines (RDEs) are propulsion devices, which utilize such pressure gains through continuously spinning detonation waves for generating thrust~\cite{Bykovskii_JPP_2006}. Fuel-oxidizer mixture is injected axially into an annular shaped device which is undergoing combustion due to the rotating detonation waves. High pressure combustion expels products axially from the opposite end thus generating thrust. However, the complex combustion wave propagation dynamics in such devices require further attention and careful analysis for effective design~\cite{Schwinn_CandF_2018}. 

In this work, we perform numerical investigations of such sustained detonation dynamics in a periodic domain, inspired by the recent experimental study of Schwinn \emph{et al.}~\cite{Schwinn_CandF_2018} in which a straight-line detonation chamber was analyzed experimentally exhibiting sustained resonance of detonation waves. We numerically investigate the effect of fuel-oxidizer injection rates on the dynamics of detonation waves and sustenance in a periodic one-dimensional domain. To this end, we model fuel-oxidizer injection rates utilizing a one-dimensional model reduction which represents solving the governing equations close to the injector plate in an experimental setup. Moreover, we adopt a device-scale dynamical system perspective in the current study to elucidate the wave propagation dynamics.

\section*{Summary of Chapter 7: Mesh adaptive Navier Stokes solver}

In the previous two topics investigated, an extremely wide spectrum of length scales is observed. In vortex breakdown behind a curved shock wave induced by spark-plasma heat deposition, vortex rings have been observed in experiments~\cite{BsinghEtAl_2019_AIAA}, which are generated by the curved shock wave. These vortex rings collide and break down generating very small scale structures in the flow. For detonation wave resonance in a periodic domain, the detonation waves propagate in a domain $~10^6$ times longer than the thickness~\cite{Schwer_AIAA_2010, Schwer_CandF_2011}. As a result, very high numerical resolution is required in only a very small region inside the domain which mandates the mesh adaptive approach in multiple dimensions for efficient computations. 

\begin{figure}[!t]
\centering
\includegraphics[width=\textwidth]{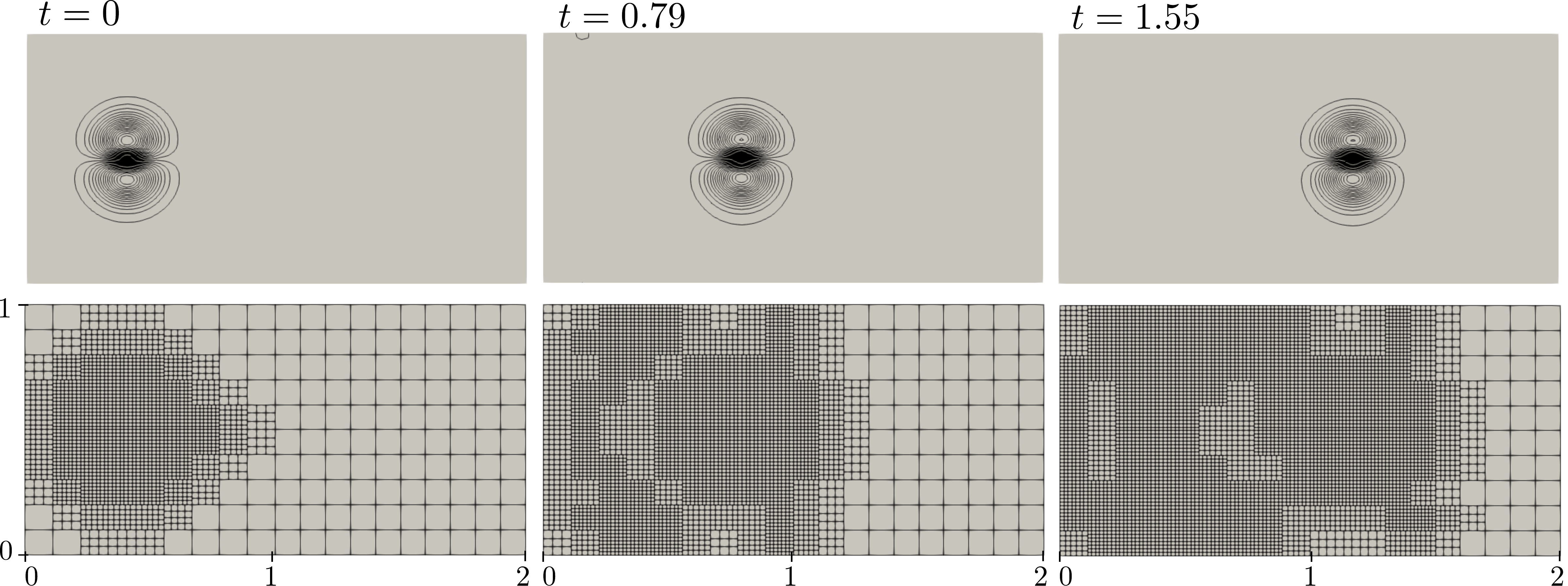}
  \caption{Advection of an isentropic vortex in a subsonic base flow (top) and adaptive order refinement of the underlying mesh (bottom).}
\label{fig: Vortex_INTRO}
\end{figure}

To this end, a high fidelity mesh adaptive Navier-Stokes solver is developed and presented in this work, as a tool for future development and study of the topics discussed above. The solver adapts the local order of interpolation and derivative calculation upon detecting shock waves and vorticity, thus enabling higher degrees of freedom locally for higher resolution. Test results on cases involving high gradients in flow fields due to shock waves and vorticity (see figure~\ref{fig: Vortex_INTRO}) are presented. 

% %%%%%%%%%%%%%%%%%%%%%%%%%%%%%%%%%%%

% Summary and/or conclusions are optional but often used.
% The summary and/or conclusions often are the last
% the last major division(s) of the text.
% Reference: TM2006 page 32.
% CHANGE NEXT LINE?
%\include{summary}
%%%%%%%%%%%%%%%%%%%%%%%%%%%%%%%%%%%%%%%%%
% \include{SecondOrderAcoustics}
\chapter{Planar second order nonlinear acoustics}
The contents of this chapter have been published in \emph{Physical Review E}~\cite{GuptaScalo_PRE_2018} (see Section II) and have been reported here in abridged form with minor modifications.

\section{Introduction} \label{sec: Intro2}
High amplitude planar nonlinear acoustic waves exhibit two main nonlinearities :- acoustic streaming~\cite{Lighthill_JSV_1978, Gedeon_1997_Cryocoolers} and wave steepening~\cite{Hamilton_NLA_1998, Naugolnykh1998}. Usually, problems in nondispersive nonlinear wave propagation are studied utilizing the model Burgers equation \cite{Whitham2011, Gurbatov_1981_JETP, Naugolnykh1998, Gurbatov2012}.
However, in this work, we focus on quantifying the spatio-temporal and spectro-temporal evolution of finite amplitude planar nonlinear acoustic waves exactly via second order nonlinear governing equations, as derived from the continuum gas dynamics (compressible Navier Stokes) equations in this chapter. In later chapters, we study the evolution of finite amplitude planar nonlinear acoustic waves in three canonical configurations, traveling waves (TW), standing waves (SW), and randomly initialized Acoustic Wave Turbulence (AWT) utilizing these equations.

%%%%%%%%%%%%%%%%%%%%%%%%%%%%%%%%%%%%%%%%%%%%%%%%%%%%%%%%%%%%%%%%%%%%%%%%%%%%%%%%%%%%%%%%%%%
\section{Governing equations and scaling analysis}
\label{sec: EntropyScaling}
{In this section, we derive the governing equations for nonlinear acoustics truncated up to second order (in the acoustic perturbation variables) for a single-component ideal gas. We begin with fully compressible one-dimensional Navier-Stokes equations for continuum gas dynamics and analysis of entropy scaling with pressure jumps in weak shocks formed due to the steepening of nonlinear acoustic waves~(Section~\ref{sec: 1DNS}). We then briefly discuss the variable decomposition and non-dimensionalization in 
Section~\ref{sec: 2ndOrderEntropy}, followed by the derivation of second order governing equations for nonlinear acoustics in Section~\ref{sec: 2ndOrderNLA}.}

\subsection{Fully compressible 1D Navier-Stokes and entropy scaling in weak shocks}
\label{sec: 1DNS}

One-dimensional governing equations of continuum gas dynamics (compressible Navier-Stokes) for an ideal gas are given by, 
\begin{align}
&\frac{\partial \rho^*}{\partial t^*} + \frac{\partial (\rho^* u^*) }{\partial x^*} = 0,\label{eq: NS_eqns1}
\\
&\frac{\partial}{\partial t^*}\left(\rho^* u^*\right)  + \frac{\partial}{\partial x^*}\left(\rho^* u^{*2}\right) = -\frac{\partial p^*}{\partial x^*} + \frac{\partial }{\partial x}\left(\left(\frac{4}{3}\mu^* + \mu^*_B\right)\frac{\partial u^*}{\partial x^*}\right),\label{eq: NS_eqns2}
\\
&\rho^* T^*\left(\frac{\partial  s^*}{\partial t^*} + u^*\frac{\partial s^*}{\partial x^*}\right) = \frac{\partial }{\partial x^*}\left( \frac{\mu^*C^*_p}{Pr} \frac{\partial T^*}{\partial x^*}\right) + \left(\frac{4}{3}\mu^*+\mu^*_B\right)\left(\frac{\partial u^*}{\partial x^*}\right)^2,
\label{eq: NS_eqns3}
\end{align}
which are closed by the ideal gas equation of state, 
\begin{align}
p^* = \rho^* R^* T^*,\label{eq: IdealGas}
\end{align}
where $p^*, u^*, \rho^*, T^*, s^*$, respectively, denote total pressure, velocity, density, temperature, and entropy of the fluid, $ x^*$ and $t^*$  denote space and time, and $\mu^*$ denotes dynamic viscosity. Results in this chapter and subsequent chapters have been obtained by DNS of Eqs.~\eqref{eq: NS_eqns1}-\eqref{eq: NS_eqns3}, to resolve all the length scales of planar nonlinear acoustic waves. For our simulations (see Section~\ref{sec: numerics_PRE}), {\color{black}{we choose the gas specific constants for air at standard temperature and pressure (STP),
\begin{equation}
R^* = 287.105~\frac{\mathrm{m}^2}{\mathrm{s}^2\cdot K},\quad\quad\mu^*_B=0,\quad\quad Pr=0.72.
\end{equation}
\begin{figure}[!t]
\centering
\includegraphics[width=0.9\linewidth]{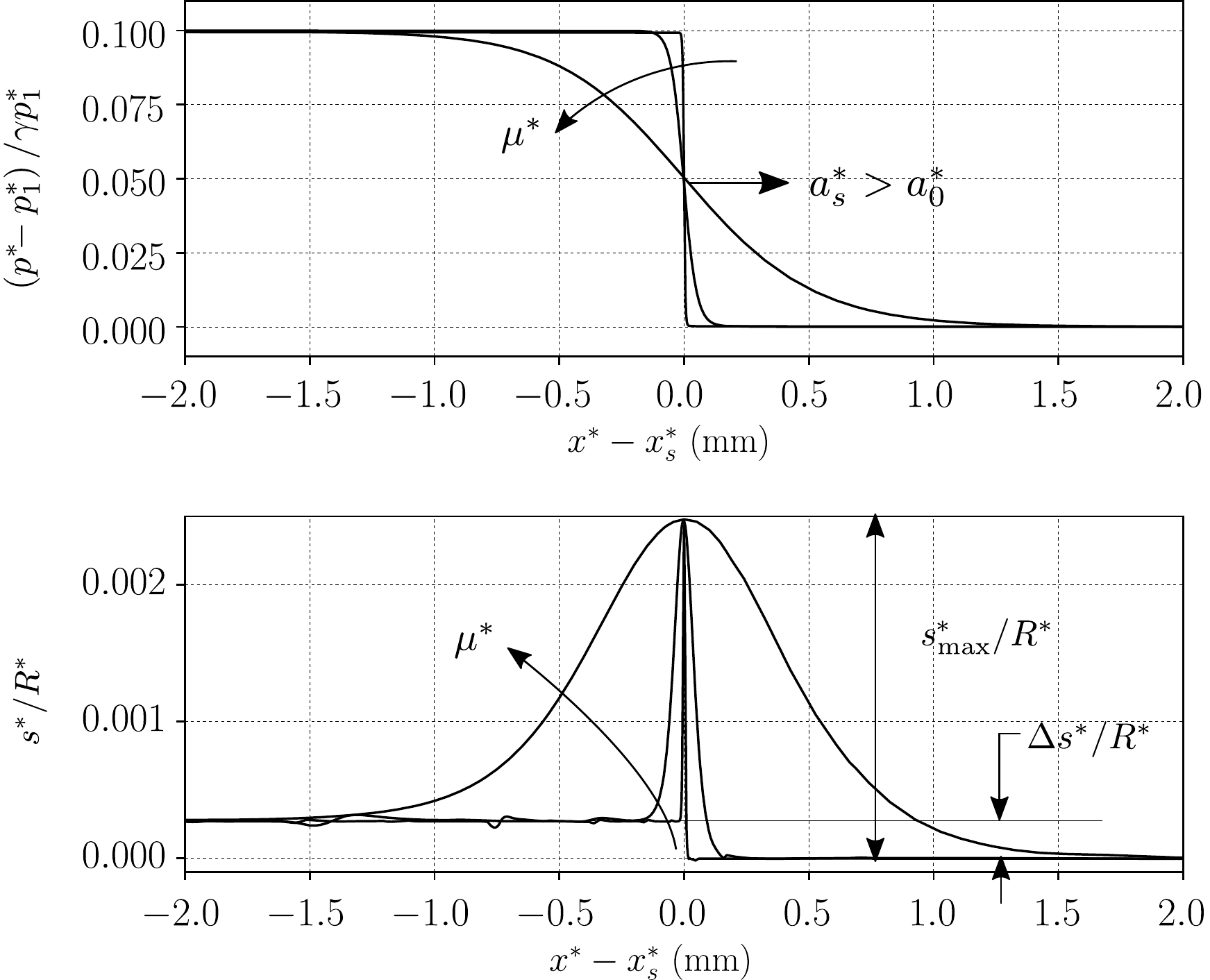}
\put(-390,330){$(a)$}
\put(-390,160){$(b)$}
\caption[Variation of entropy and pressure within a weak shock wave for varying viscosity values.]{Weak shock wave structure $(a)$ pressure $p^*$ and $(b)$ entropy $s^*$ propagating with a speed $a^*_s>a^*_0$ obtained from DNS (see Section~\ref{sec: numerics_PRE}). $\Delta s^*/R^*$ and $s^*_{\mathrm{max}}/R^*$ are the entropy jump and maximum entropy respectively. With increasing viscosity, the peak in entropy remains constant. The DNS data has been obtained for base state viscosity values given in Table~\ref{tab: test_cases}.}
\label{fig: EntropyJumps}
\end{figure}
\begin{figure}[!b]
\centering
\includegraphics[width=0.75\linewidth]{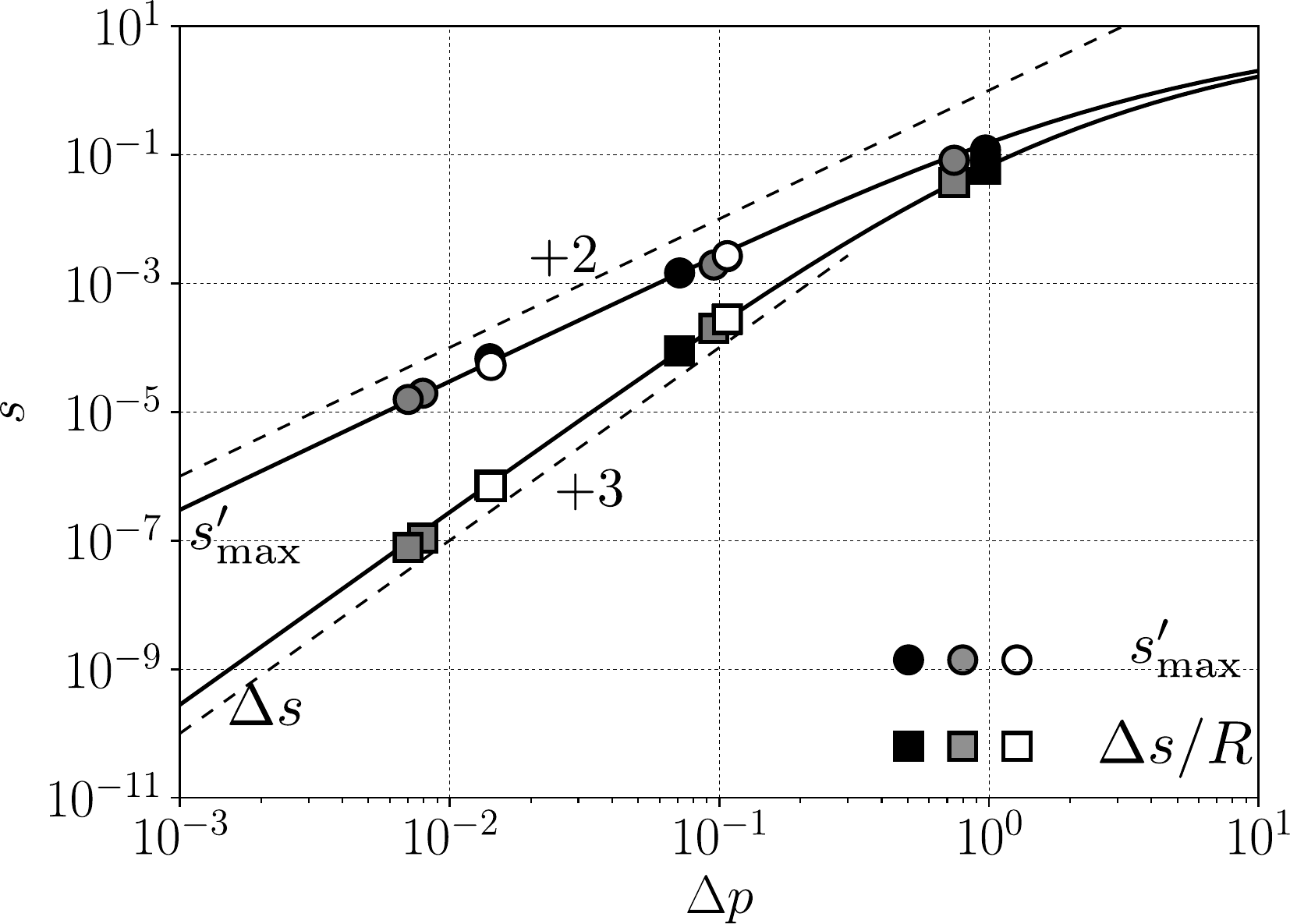}
\caption[Scaling of entropy jump across the shock wave and maximum entropy inside the shock wave with pressure jump.]{{\color{black}{Entropy jump $\Delta s^*=s^*_2-s^*_1$ and maximum entropy generated $s^*_{\mathrm{max}}$ versus pressure jump $\Delta p^*$ across a planar shock wave. In the labeled region ($\Delta p^*/\gamma p^*_1 < 1$, referred as `weak shocks' hereafter), the entropy jump $\Delta s^*$ scales as $\mathcal{O}\left(\Delta p^{*3}\right)$, whereas the maximum entropy generated $s^*_{\mathrm{max}}$ scales as $\mathcal{O}\left(\Delta p^{*2}\right)$, approximately. Markers denote DNS data (see Section~\ref{sec: numerics_PRE}), (\protect\markerone,\protect\markerfour) $\mu^*=7.5\times10^{-3}~$kg$\cdot$m$^{-1}\cdot$s$^{-1}$; (\protect\markertwo,\protect\markerfive), $\mu^*=7.5\times10^{-4}~$kg$\cdot$m$^{-1}\cdot$s$^{-1}$; (\protect\markerthree,\protect\markersix), $\mu^*=7.5\times10^{-5}~$kg$\cdot$m$^{-1}\cdot$s$^{-1}$ for varying values of $\Delta p^*$. Solid lines correspond to Eqs.~\eqref{eq: EntropyMach} and~\eqref{eq: EntropyMax}.}}}
\label{fig: EntropyScaling}
\end{figure}
Planar nonlinear acoustic waves steepen and form weak shocks. {For weak shocks, the smallest length scale (shock-thickness) is also significantly larger than the molecular length scales. Hence, in this work, we neglect the molecular vibrational effects in the single component ideal gas ($\mu^*_B=0$) , typically modeled via bulk viscosity effects~\cite{Cramer_PoF_2012}}. Across a freely propagating planar weak shock (figure~\ref{fig: EntropyJumps}), the entropy jump ($\Delta s^* = s^*_2 - s^*_1$) is given by the classical gas-dynamic relation~\cite{LiepmannRoshko},
\begin{align}
\frac{\Delta s^*}{R^*} &= \frac{1}{\gamma - 1}\ln\left(1 + \frac{2\gamma}{\gamma + 1}\left(M^2 -1 \right)\right) - \frac{\gamma}{\gamma - 1}\ln\left(\frac{\gamma + 1}{\gamma - 1 + 2/M^2}\right),
\label{eq: EntropyMach}
\end{align}
where $M$ is the Mach number, given by, 
\begin{equation}
 \frac{\Delta p^*}{\gamma p^*_1} = \frac{p^*_2 - p^*_1}{\gamma p^*_1} = \frac{2}{\gamma + 1}\left(M^2-1\right),
\label{eq: PressureMach}
\end{equation}
and $\Delta p^* = p^*_2 - p^*_1$ is the pressure jump with $p^*_1$ and $p^*_2$ being the pre-shock and post-shock pressures, respectively.}} Near the inflection point of the fluid velocity profile, the entropy reaches a local maximum $(s^* = s^*_\mathrm{max})$. According to Morduchow and Libby~\cite{Morris_JASc_1949}, maximum entropy $s^*_{\mathrm{max}}$ assuming $\mu^*_B = 0$ and $Pr = 3/4$, can be obtained as, 
\begin{equation}
\frac{s^*_{\mathrm{max}}}{R^*} = \frac{1}{\gamma - 1}\ln\left(1 + \frac{\gamma - 1}{2}M^2\left(1-\xi\right)\xi^{\frac{\gamma - 1}{2}}\right),
\label{eq: EntropyMax}
\end{equation}

where, 
\begin{equation}
\xi = \frac{\gamma - 1}{\gamma + 1} + \frac{2}{\gamma + 1}\frac{1}{M^2}.
\end{equation} 

For weak shock waves, ($\Delta p^*/\gamma p^*_1 <1$), the entropy jump $\Delta s^*$ and maximum entropy $s^*_{\mathrm{max}}$ scale with pressure jumps as (cf. figure~\ref{fig: EntropyScaling}),
\begin{equation}
{\Delta s^*} = \mathcal{O}\left(\Delta p^{*3}\right), \quad s^*_{\mathrm{max}} = \mathcal{O}\left({\Delta p}^{*2}\right),
\end{equation}
independent of $\mu^*$ (cf. Eqs.~\eqref{eq: EntropyMach} and~\eqref{eq: EntropyMax}). The overall entropy jump $\Delta s^*$ is due to irreversible thermoviscous losses occurring within the shocks. However, the overshoot in entropy ($s^*_{\mathrm{max}}>\Delta s^*$) is due to both reversible and irreversible processes, and is not in violation of the second law of thermodynamics~\cite{Morris_JASc_1949}. Moreover, in the range of pressure jumps considered in the DNS (see Section~\ref{sec: numerics_PRE}), the maximum Mach number of the shock is around $M\approx 1.4$, which is well within the limits of validity of the continuum approach~\cite{Bird_MolecularGasDynamicsBook_1994}. Hence, it is physically justified to draw conclusions regarding the smallest length scales through the governing equations based on continuum approach and assuming thermodynamic equilibrium.

%{\color{red} \#\#\#\#\#\#\#\#\#\# I would keep everything dimensional up to here. Our dimensionalization only makes sense from an acoustics perspective. This part on shocks only cares about conditions immediately before and after the shock. Even in Figure 3 where I would report $s^*/R^* and \Delta p^* / p_1^*$ since for the reader $ \Delta p$ is, in reality,  $\Delta p = \Delta p^*/p_0^*$ as he/she will 'discover later', causing confusion. So for this section there is no reason to bring up anything related to the base flow or acoustic base flow. }

\subsection{Perturbation variables and non-dimensionalization}
\label{sec: 2ndOrderEntropy}

{In this section, we utilize the previous consideration on the second order scaling of the maximum entropy $s^*_{\mathrm{max}}$ inside a weak shock wave to derive second order nonlinear acoustics equations. To this end, we decompose the variables in base state and perturbation fields and derive equations containing only linear and quadratic terms in perturbation fields. Denoting the base state with the superscript $( )_0$ and the perturbation fields with the superscript $({ })'$, we obtain, }
%{\color{red} \#\#\#\#\#\#\#\#\#\# We need a few sentences here somewhere clarifying that we are now switching to an acoustics perspective where the base conditions are important, whereas before we were focusing on the neighborhood of the shock and only upstream/downstream information is relevant and that data shown in Figure 3 is normalized with respect to conditions at the foot of the shock but from this point on we use base conditions, indicated with subscript 0}
\begin{subequations}
\begin{align}
& \rho^* = \rho^{*}_0 + {\rho^*}',\quad p^* = p^*_0 + {p^*}', \\
& u^* = {u^*}', \quad s^* = {s^*}', \quad T^* = T^*_0 + {T^*}',
 \label{eq: PertDecom}
 \end{align}
\end{subequations}
where no mean flow $u^*_0 = 0$ is considered and $s^*_0$ is arbitrarily set to zero.
We neglect the fluctuations in the dynamic viscosity as well, i.e.,
\begin{equation}
\mu^* = \mu^*_0.
\end{equation}
{\color{black}{While in classic gas dynamics, pre-shock values are used to normalize fluctuations or jumps across the shock (e.g. see Eq.~\eqref{eq: PressureMach}), hereafter we choose base state values to non-dimensionalize the nonlinear acoustics equations,
\begin{subequations}
\begin{align}
& \rho = \frac{{\rho^*}}{\rho^*_0} = 1 + \rhoFluct ,~~p = \frac{p^*}{\gamma p^*_0} = \frac{1}{\gamma} + \pFluct, \\
& u = \frac{u^*}{a^*_0} = \uFluct, ~~s = \frac{s^*}{R^*} = \sFluct,~~ T = \frac{T^*}{T^*_0} = 1 + \TFluct,\\
&x = \frac{x^*}{L^*}, ~~ t = \frac{a^*_0 t^*}{L^*}.
\end{align}
\label{eq: per_norm}
\end{subequations}
where $L^*$ is the length of the one-dimensional periodic domain.  {As also typically done in classical studies of homogeneous isotropic turbulence~\cite{Ishihara_ARFM_2009,Batchelor1953theory,pope2000turbulent,monin1971statistical,Hosokawa_PRE_2008, Burattini_PRE_2006}, periodic boundary conditions represent a common (yet not ideal) way to approximate infinite domains; as such, a spurious interaction between the flow physics that one wishes to isolate and the periodic box size may occur. For the TW and SW test cases analyzed herein, $L^*$ corresponds to the initial (and hence largest) reference length scale of the acoustic perturbation; in the AWT case, the value of $L^*$ should be chosen as much larger than the integral length scale $\ell$ or Taylor microscale $\lambda$ (see Section~\ref{sec: ScalesAcousticCascade}), which truly define the state of turbulence.}}

% In this study, we consider normalized pressure perturbations $p'<10^{-1}$ (i.e. $p'^*<10^{-1}\gamma p^*_0$}). \Carlo{Random placement for this last sentence -- also, as I look up to Figure 3 and I see the same number $10^{-1}$, but different normalization, that is, with respect to pre-shock values and makes me confused. However, given that $p^*_1/p^*_0<1$ you are fine? Clarify expanding a bit. But this consideration is either a paragraph on its own or should go somewhere else.} }

%%Equations~\eqref{eq: NS_eqns1}-\eqref{eq: IdealGas} can be combined into truncated nonlinear governing equations for acoustic perturbations ($\pFluct$ and $\uFluct$) upon utilizing the appropriate scaling of entropy perturbation $\sFluct$ with pressure perturbation $\pFluct$. 

{Due to thermodynamic nonlinearities, wave propagation velocity increases across a high-amplitude compression front, resulting in wave-steepening ~\cite{Whitham2011} and hence generation of small length scales associated with increasing temperature and velocity gradients responsible for thermoviscous dissipation. The increase in thermoviscous dissipation results in positive entropy perturbations peaking within the shock structure. For pressure jumps $\Delta p^*/\gamma p^*_1<1$, the maximum entropy scales approximately as $\mathcal{O}\left(\Delta p^{*2}\right)$ (cf. figure~\ref{fig: EntropyScaling}). Moreover, as we discuss in a later section (see Section~\ref{sec: EnergyCorollary}), the second order nonlinear acoustic equations impose a strict limit of $|p'|<1/\gamma$ ($\simeq 0.714$ for $\gamma = 0.72$) for base state normalized (Eq.~\eqref{eq: per_norm}) (not pre-shock state normalized (Eq.~\eqref{eq: PressureMach})) perturbations. Hence, in our simulations~(see Section~\ref{sec: numerics_PRE}), we consider a suitable range of $10^{-3}<p'<10^{-1}$, which satisfies the aforementioned constraints. Thus, the second order scaling of entropy holds in our simulations. 

Below, we utilize this entropy scaling to derive the correct second order nonlinear acoustics equations governing the spatio-temporal evolution of dimensionless perturbation variables $p'$ and $u'$, as defined in Eq.~\eqref{eq: per_norm}.}

\subsection{Second order nonlinear acoustics equations}
\label{sec: 2ndOrderNLA}
For a thermally perfect gas, the differential in dimensionless density $\rho$ can be related to differentials in dimensionless pressure $p$ and dimensionless entropy $s$ as,
\begin{align}
d\rho &= \left(\frac{\partial \rho}{\partial p}\right)_{s} dp + \left(\frac{\partial \rho}{\partial s}\right)_{p} ds,\nonumber\\
& = \frac{\rho}{\gamma p}dp - \frac{\rho(\gamma - 1)}{\gamma}ds.
\label{eq: ConstitutiveEquation}
\end{align}
Nondimensionalizing the continuity Eq.~\eqref{eq: NS_eqns1} and substituting Eq.~\eqref{eq: ConstitutiveEquation}, we obtain, 
\begin{align}
&\frac{\partial \rho}{\partial t} + \frac{\partial \rho u}{\partial x} = 0,\nonumber\\
&\frac{\partial p}{\partial t} + u\frac{\partial p}{\partial x} + \gamma p\frac{\partial u}{\partial x} = (\gamma -1)p\left(\frac{\partial s}{\partial t} + u\frac{\partial s}{\partial x}\right).
\label{eq: PressureStep1}
\end{align}
Substituting the dimensionless forms of Eqs.~\eqref{eq: NS_eqns3} and~\eqref{eq: IdealGas} and utilizing the decomposition given in Eqs.~\eqref{eq: per_norm}, we obtain the following truncated equation for pressure perturbation $\pFluct$,
\begin{align}
\frac{\partial \pFluct}{\partial t} + \frac{\partial \uFluct}{\partial x} + \gamma \pFluct \frac{\partial \uFluct}{\partial x} &+ \uFluct\frac{\partial \pFluct}{\partial x} = \nu_0\left(\frac{\gamma - 1}{Pr}\right)\frac{\partial^2 \pFluct}{\partial x^2} \nonumber \\
& + \mathcal{O}\left(\pFluct\sFluct, \sFluct^2, \pFluct^3, \left(\frac{\partial \uFluct}{\partial x}\right)^2\right).\label{eq: pressure}
\end{align}
Similarly, the truncated equation for velocity perturbation $\uFluct$ is obtained as,
\begin{align}
\frac{\partial \uFluct}{\partial t} + \frac{\partial \pFluct}{\partial x} + \frac{\partial}{\partial x}\left(\frac{\uFluct^2}{2} - \frac{\pFluct^2}{2}\right) &= \frac{4}{3}\nu_0\frac{\partial^2 \uFluct}{\partial x^2} \nonumber \\
&+\mathcal{O}\left(\rho'^2\pFluct,\rho'^3\pFluct \right). \label{eq: velocity}
\end{align}
In Eqs.~\eqref{eq: pressure} and~\eqref{eq: velocity}, $\nu_0$ is the dimensionless kinematic viscosity given by, 
\begin{equation}
\nu_0 = \frac{\mu^*_0}{\rho^*_0 a^*_0L^*},
\label{eq: DimensionlessVisc}
\end{equation}
and quantifies viscous dissipation of waves relative to propagation. Equations~\eqref{eq: pressure} and~\eqref{eq: velocity} constitute the nonlinear acoustics equations truncated up to second order, governing spatio-temporal evolution of finite amplitude acoustic perturbations $\pFluct$ and $\uFluct$. The entropy scaling ($s^*_{\mathrm{max}}=\mathcal{O}\left(\Delta p^{*2}\right)$) discussed previously results in the dissipation term on the right hand side of Eq.~\eqref{eq: pressure}. The left hand side of Eqs.~\eqref{eq: pressure} and~\eqref{eq: velocity} contains terms denoting linear and nonlinear isentropic acoustic wave propagation. {The Detailed derivation of Eqs.~\eqref{eq: pressure} and~\eqref{eq: velocity} is given in Appendix~\ref{sec: appedixA}, where we also show that the nonlinear terms on the left hand side (LHS) of Eqs.~\eqref{eq: pressure} and~\eqref{eq: velocity} are independent of the thermal equation of state. The functional form of the second order perturbation energy norm ($E^{(2)}$, Eq \eqref{eq: Energy_norm}) -- being exclusively dictated by such terms (see Section \ref{sec: EnergyCorollary}) -- is independent of the thermal equation of state of the gas. The results shown in this work focus on ideal-gas simulations merely for the sake of simplicity, with no loss of generality pertaining to inviscid nonlinear (up to second order) spectral energy transfer dynamics.}

  We note that, Eq.~\eqref{eq: pressure} consists of the velocity derivative term ($\gamma p'{\partial u'}/{\partial x}$), and is different from those obtained by Naugol'nykh and Rybak~\cite{Naugol1975spectrum}, which in dimensionless form read,
\begin{align}
&\frac{\partial \pFluct}{\partial t} - (\gamma -1)\pFluct\frac{\partial \pFluct}{\partial t} + \frac{\partial \uFluct}{\partial x} + \pFluct\frac{\partial \uFluct}{\partial x} = 0,\\
&\frac{\partial \uFluct}{\partial t} + \frac{\partial \pFluct}{\partial x} + \frac{\partial }{\partial x}\left(\frac{\uFluct^2}{2} - \frac{\pFluct^2}{2}\right) = 0.
\end{align} 
We adopt Eqs.~\eqref{eq: pressure} and~\eqref{eq: velocity} throughout the study since they represent the truncated governing equations exactly. Unlike Naugol'nykh and Rybak~\cite{Naugol1975spectrum}, we do not approximate the density $\rho$ using the Taylor series and only use the total differential form given in Eq.~\eqref{eq: ConstitutiveEquation}. 

Additionally, we note that Eqs.~\eqref{eq: pressure} and~\eqref{eq: velocity} can be combined into Westervelt's equation~\cite{Hamilton_NLA_1998} only if the Lagrangian defined as, 
\begin{equation}
 \mathcal{L} = \frac{\uFluct ^2}{2} - \frac{\pFluct ^2}{2},
\end{equation}
is zero, which holds only for linear pure traveling waves. The derivation of Burgers' equation in nonlinear acoustics follows from the Westervelt's equation~\cite{Hamilton_NLA_1998}. Hence, it is inadequate in modeling general nonlinear acoustics phenomena involving mixed phasing of nonlinear waves which occurs in the Standing Wave (SW) and Acoustic Wave Turbulence (AWT) cases analysed here.

% %%%%%%%%%%%%%%%%%%%%%%%%%%%%%%
% %%%%%%%%%%%%%%%%%%%%%%%%%%%%%%
% \include{InterScaleEnergyDynamics}
\chapter{Energy cascade and decay of nonlinear acoustic waves}
The contents of this chapter have been published as a regular article in \emph{Physical Review E}~\cite{GuptaScalo_PRE_2018} (see Sections III - VII) and have been reported here in abridged form with minor modifications.

\section{Introduction} \label{sec: Intro1}
In this chapter, we analyze the energy dynamics of decaying nonlinear acoustic waves utilizing three canonical configurations, traveling waves (TW), standing waves (SW), and randomly initialized Acoustic Wave Turbulence (AWT) utilizing the second order equations derived in the previous chapter and DNS of the fully compressible Navier-Stokes equations.
 The spectral energy and decay dynamics of 1D Burgers' turbulence have been studied extensively by Kida~\cite{Kida_1979_JFM}, Gurbatov et al.~\cite{Gurbatov_JFM_1997, Gurbatov1991}, Woyczynski~\cite{Woyczynski_Gottingen_2006}, Fournier and Frisch~\cite{Fournier1983Burgers}, and Burgers~\cite{Burgers1974nonlinear}. 
However, the equations of second order nonlinear acoustics can be reduced to Burgers' equations only with the restrictive assumption of pure traveling waves (TW). Consequently, we utilize the continuum gas dynamics governing equations to elucidate the spectral energy cascade and decay dynamics of nonlinear acoustics. The decay of nonlinear perturbations is governed by viscosity and thermal conductivity (thermoviscous diffusion). The decay exhibits a power law in time due to the gradual increase of the length scale over which the thermoviscous diffusion acts. Such decay dynamics occur due to the separation of energy containing and diffusive length scales and resemble those of decaying homogeneous isotropic turbulence (HIT), which also have been  studied extensively in the literature both theoretically  and experimentally \cite{batchelor1948decay, batchelor1949nature, hinze1975turbulence, Ishihara_ARFM_2009}.

We develop theoretical arguments based on the second order nonlinear acoustics equations derived in the previous chapter, and compare our results with DNS results of fully compressible Navier Stokes Eqs.~\eqref{eq: NS_eqns1}-\eqref{eq: NS_eqns3}. We analyze various length scales associated with the decaying nonlinear acoustic waves, as well as the spectra of waveforms. 

%%%%%%%%%%%%%%%%%%%%%%%%%%%%%%%%%%%%%%%%%%%%%%%%%%%%%%%%%%%%%%%%%%%%%%%%%%%%%%%%%%%%%%%%%%%

%%%%%%%%%%%%%%%%%%%%%%%%%%%%%%%%%%%%%%%%%%%%%%%%%%%%%%%%%%%%%%%%%%%%%%%%%%%%%%%%%%%%%%%%%%%
%\input{3_InviscidCascade.tex}
\section{Second order perturbation energy}
\label{sec: EnergyCorollary}
In this section we derive a new perturbation energy function for nonlinear acoustic waves utilizing Eqs.~\eqref{eq: pressure} and \eqref{eq: velocity}. To this end, we derive the perturbation energy conservation relation (energy corollary) for high amplitude acoustic perturbations. We show that the spatial average of the perturbation energy function satisfies the definition of the Lyapunov function for high amplitude acoustic perturbations and evolves monotonically in time (cf.~figure~\ref{fig: EnergyTime}). Utilizing the energy corollary, we derive spectral energy transport relations in further sections.

Multiplying Eqs.~\eqref{eq: pressure} and \eqref{eq: velocity} with $\pFluct$ and $\uFluct$, respectively, and adding, we obtain,
\begin{align}
\frac{\partial }{\partial t}\left(\frac{\pFluct ^2}{2} + \frac{\uFluct ^2}{2}\right) + \frac{\partial }{\partial x}\left(\uFluct\pFluct + \frac{\uFluct^3}{3}\right) + \gamma \pFluct ^2\frac{\partial \uFluct}{\partial x} = \nu_0\left(\frac{\gamma - 1}{Pr}\right)\pFluct\frac{\partial^2 \pFluct}{\partial x^2} +   \frac{4}{3}\nu_0 u'\frac{\partial^2 \uFluct}{\partial x^2}. \quad\quad
\label{eq: EnergyCorollary_Open}
\end{align}
Spatial averaging of Eq.~\eqref{eq: EnergyCorollary_Open} over a periodic domain $[0, L]$ yields, 
\begin{align}
\frac{d\left\langle E^{(1)}\right\rangle}{dt} = -\left\langle \gamma p'^2\frac{\partial u'}{\partial x}\right\rangle &- \nu_0\left(\frac{\gamma - 1}{Pr}\right)\left\langle\left(\frac{\partial p'}{\partial x}\right)^2\right\rangle-\frac{4}{3}\nu_0\left\langle\left(\frac{\partial u'}{\partial x}\right)^2\right\rangle,
\label{eq: LinearAcousticEnergy}
\end{align}
where $\left\langle . \right\rangle$ is the spatial averaging operator,
\begin{equation}
\left\langle . \right\rangle = \frac{1}{L}\int^L_0\left(.\right)dx,
\end{equation}
and 
\begin{equation}
E^{(1)} = \frac{\uFluct^2}{2} + \frac{\pFluct^2}{2},
\end{equation}
is the first order isentropic acoustic energy. Equation~\eqref{eq: LinearAcousticEnergy} suggests that, in a lossless medium ($\nu_0\rightarrow 0$), $\left\langle E^{(1)}\right\rangle$ would exhibit spurious non-monotonic behavior in time due to the first term on right hand side. Such non-monotonic behaviour is confirmed by the DNS results shown in figure~\ref{fig: EnergyTime}. Consequently, the linear acoustic energy norm $E^{(1)}$ does not quantify the perturbation energy correctly for high amplitude perturbations since the spatial average $\left\langle E^{(1)}\right\rangle$ supports spurious growth and decay in the absence of physical sources of energy.
\begin{figure}[!b]
\centering
\includegraphics[width=0.75\linewidth]{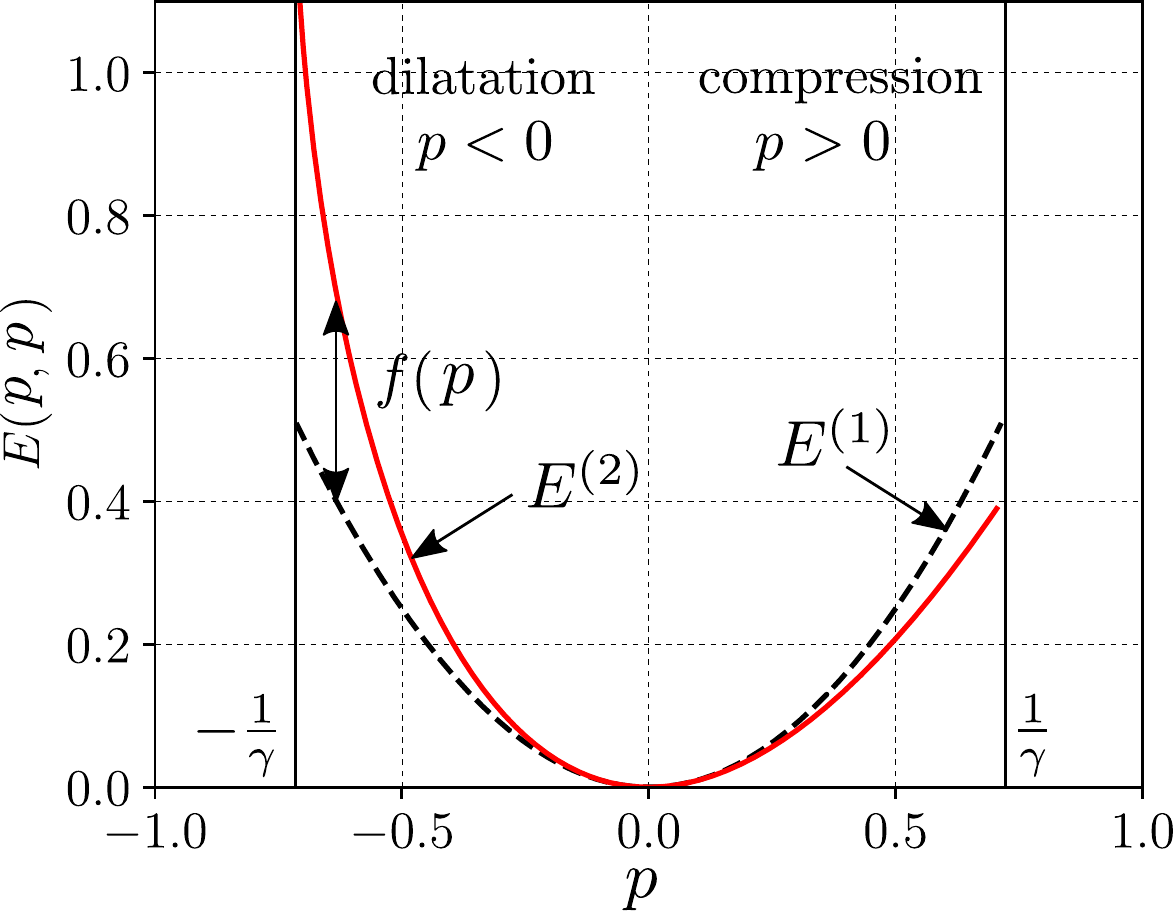}
\caption{Comparison of perturbation energy function for nonlinear acoustic waves $E^{(2)}$ with the linear acoustic energy $E^{(1)}$ in the case of $\pFluct=\uFluct$ (assumed for illustrative purpose). The correction $f(\pFluct)$ is independent of $\uFluct$.}
\label{fig: EnergyComp}
\end{figure}
The corrected perturbation energy function can be obtained upon recursively evaluating the velocity derivative term ($\gamma p'^2\partial u'/\partial x$ in Eq.~\eqref{eq: EnergyCorollary_Open}) utilizing Eq.~\eqref{eq: pressure} as, 
{\color{black}{
\begin{align}
\gamma p'^2\frac{\partial u'}{\partial x} = -\frac{\partial }{\partial t}\left(\frac{\gamma \pFluct^3}{3}\right) - \frac{\partial}{\partial x}\left(\frac{\gamma \uFluct \pFluct^3 }{3}\right) - &\gamma\left(\gamma - \frac{1}{3}\right)\pFluct^3\frac{\partial \uFluct}{\partial x} 
- \frac{\nu_0(\gamma - 1)}{Pr}\gamma \pFluct^2\frac{\partial \pFluct}{\partial x}.
\label{eq: Rec1}
\end{align}
Furthermore, the third term in above Eq.~\eqref{eq: Rec1} on the right can be further evaluated as,
\begin{align}
&\gamma\left(\gamma - \frac{1}{3}\right)\pFluct^3\frac{\partial \uFluct}{\partial x} = -\frac{\partial }{\partial t}\left(\frac{\gamma}{4}\left(\gamma - \frac{1}{3}\right)\pFluct^4\right) \nonumber \\
&- \frac{\partial}{\partial x}\left(\frac{\gamma}{4}\left(\gamma - \frac{1}{3}\right)\uFluct\pFluct^4\right) - \gamma\left(\gamma - \frac{1}{3}\right)\left(\gamma - \frac{1}{4}\right)\pFluct^4\frac{\partial \uFluct}{\partial x}-\frac{\nu_0(\gamma -1 )}{Pr}\gamma\left(\gamma - \frac{1}{3}\right)\pFluct^3\frac{\partial^2 \pFluct}{\partial x^2},
\label{eq: Rec2}
\end{align}
and so on}}. Continued substitution according to Eqs.~\eqref{eq: Rec1} and~\eqref{eq: Rec2} yields the closure of the system and the following energy corollary,
\begin{equation}
\frac{\partial E^{(2)}}{\partial t} + \frac{\partial I}{\partial x}= \nu_0\left(\frac{\gamma - 1}{Pr}\right) h(\pFluct)\frac{\partial^2 \pFluct}{\partial x^2} +  \frac{4}{3}\nu_0 \uFluct\frac{\partial^2 \uFluct}{\partial x^2} ,
\label{eq: energy_cons}
\end{equation}
where,  
\begin{equation}
I(\pFluct,\uFluct) = \pFluct \uFluct + \frac{\uFluct^3}{3} + \uFluct f(\pFluct),
\label{eq: Intensity}
\end{equation} 
is the intensity (energy flux) of the field, $h(\pFluct)$ is given by,
\begin{equation}
h(\pFluct) = \pFluct + \frac{\partial f(\pFluct)}{\partial \pFluct} = \frac{\partial E^{(2)}}{\partial \pFluct}.
\end{equation}
and $E^{(2)}$ is given by,
\begin{equation}
E^{(2)}(\pFluct,\uFluct) =  \frac{\uFluct ^2}{2} + \frac{\pFluct ^2}{2} + f(\pFluct) = E^{(1)} + f(\pFluct),
\label{eq: Energy_norm}
\end{equation}
and defines the second order perturbation energy for high amplitude acoustic perturbations. The energy corollary Eq.~\eqref{eq: energy_cons} is mathematically exact for the governing Eqs.~\eqref{eq: pressure} and~\eqref{eq: velocity}.

\begin{figure}[!b]
\centering
\includegraphics[width=\textwidth]{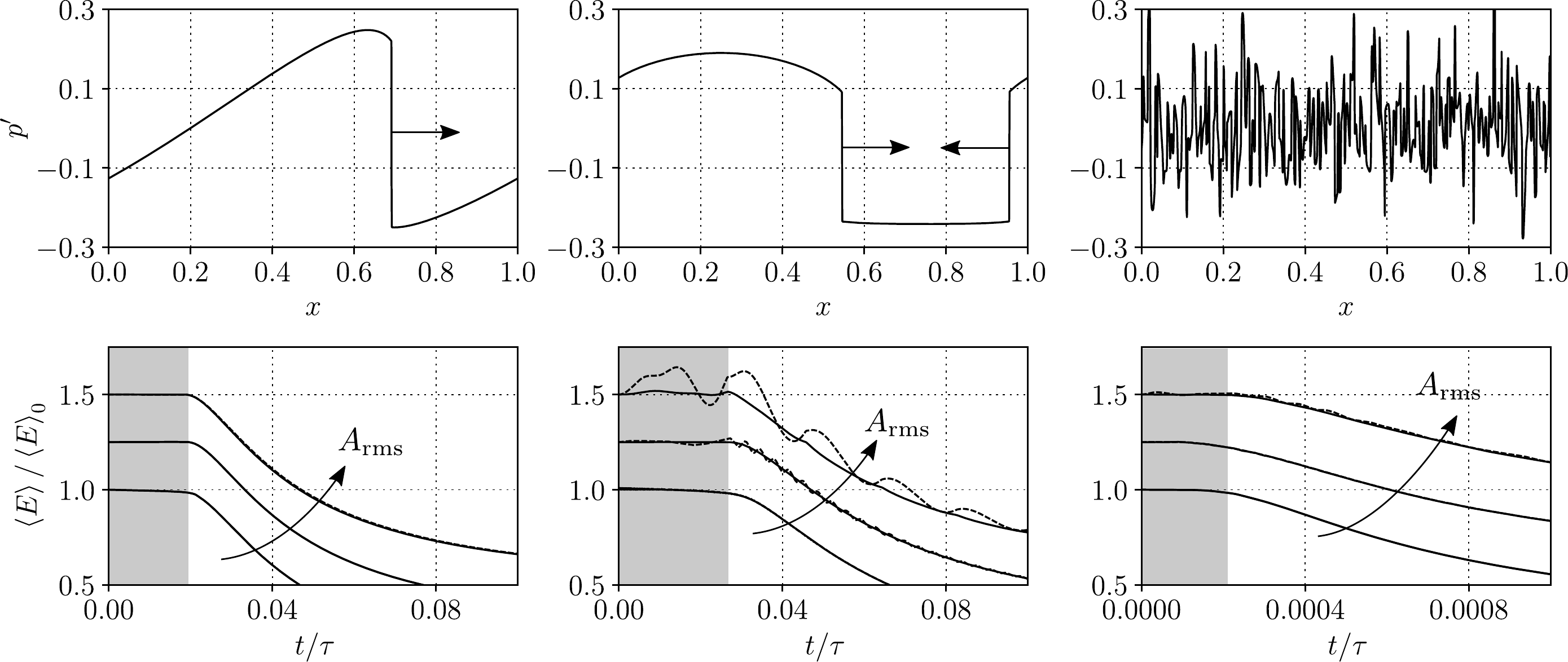}
\put(-425,190){$(a)$}
\put(-285,190){$(b)$}
\put(-145,190){$(c)$}
\caption[Spatial profile of finite amplitude waves for a traveling wave, a standing wave, and an acoustic wave turbulence case and temporal evolution of the average perturbation energy ($\left\langle E^{(2)} \right\rangle$ (--); $\left\langle E^{(1)} \right\rangle$ ($--$)) evaluated from the DNS data]{Spatial profile of finite amplitude waves (top) for TW($a$), SW($b$), and AWT($c$). Evolution of the average perturbation energy ($\left\langle E^{(2)} \right\rangle$ (--); $\left\langle E^{(1)} \right\rangle$ ($--$)) evaluated from the DNS data (bottom) scaled by the initial value against scaled time $t/\tau$ (cf. Eq.~\eqref{eq: SteepeningTime}) for increasing values of perturbation amplitude $A_{\mathrm{rms}}$ defined in Eq.~\eqref{eq: PcDef} at $\nu_0 = $ 1.836$\times$10$^{-7}$ (see Table~\ref{tab: test_cases}). The curves are shifted vertically by 0.25 for illustrative purpose only. With increasing perturbation amplitude $A_{\mathrm{rms}}$, the variation of linear acoustic energy norm < $E^{(1)}$ > becomes increasingly non-monotonic. The vertical dashed line (bottom) highlights the end of approximately inviscid spectral energy cascade regime. In this regime, the energy is primarily redistributed in the spectral space due to the nonlinear propagation ($\epsilon\simeq 0$).}
\label{fig: EnergyTime}
\end{figure}

The correction term $f(\pFluct)$ in $E^{(2)}$ appears due the thermodynamic nonlinearities and can be derived in the closed form as,
\begin{equation}
f(\pFluct) = \sum^{\infty}_{n=2}T_n =  \sum^{\infty}_{n=2}(-1)^{n+1}\frac{\gamma \pFluct^{n+1}}{n+1}\prod^{n}_{i=3}\left(\gamma - \frac{1}{i}\right),
\label{eq: EnergyCorrection_app}
\end{equation}
{where $T_2$ and $T_3$ can be identified in Eqs.~\eqref{eq: Rec1} and \eqref{eq: Rec2}, respectively.} Upon isolating the $n^{th}$ term of the above infinite series as,
\begin{equation}
T_n = (-1)^{n+1}\frac{\gamma \pFluct^{n+1}}{n+1}\underbrace{\left(\gamma - \frac{1}{3}\right)\left(\gamma - \frac{1}{4}\right)\cdots\left(\gamma - \frac{1}{n}\right)}_{n-2~\mathrm{terms}}.
\label{eq: nth-term}
\end{equation}
Multiplied fractions in the Eq.~\eqref{eq: nth-term} above yield the $n^{th}$ term as,
\begin{equation}
T_n = -\frac{2\gamma}{\left(\gamma-1\right)\left(2\gamma - 1\right)}\left(\gamma \pFluct\right)^{n+1}\binom{1/\gamma}{n+1}.
\end{equation}
Finally, the energy correction $f(\pFluct)$ can be recast as, 
\begin{align}
f(\pFluct) = \sum^{\infty}_{n=2}T_n = &-\frac{2\gamma}{\left(\gamma-1\right)\left(2\gamma - 1\right)}\Big(\left(1+\gamma \pFluct\right)^{1/\gamma} - 1 -  \pFluct+ \frac{\left(\gamma - 1\right)\pFluct^2}{2}\Big).
 \label{eq: EnergyCorrection}
\end{align}
The correction function $f(\pFluct)$ defined in Eq.~\eqref{eq: EnergyCorrection} accounts for second order isentropic nonlinearities and is not a function of entropy perturbation. Hence, $E^{(2)}$ accounts for the effect of high 
amplitude perturbations on perturbation energy isentropically. We note that this separates $E^{(2)}$ fundamentally from generalized linear perturbation energy norms, such as the ones derived by Chu~\cite{Chu1965EnergyNorm} for small amplitude non-isentropic perturbations, and by Meyers~\cite{myers1986exact} for acoustic wave propagation in a steady flow. {Moreover, as discussed in the previous section (and shown in Appendix~\ref{sec: appedixA}), since the isentropic nonlinearities on the LHS of Eqs.~\eqref{eq: pressure} and~\eqref{eq: velocity} are independent of the thermal equation of state, the functional form of $E^{(2)}$ and $I$ are also independent of the equation of state. However, the dissipation term on the right hand side of the energy corollary Eq.~\eqref{eq: energy_cons} may change with the thermal equation of state.}

The energy correction $f(\pFluct)$ {is infinite order in pressure perturbation $\pFluct$} and converges only for perturbation magnitude $|\pFluct|<1/\gamma$ thus naturally yielding the strict limit of validity of second order acoustic equations in modeling wave propagation and wave steepening. Figure~\ref{fig: EnergyComp} shows the newly derived second order perturbation energy $E^{(2)}$ compared against the isentropic acoustic energy $E^{(1)}$. Both $E^{(2)}$ and $E^{(1)}$ are non-negative in the range $|\pFluct|<1/\gamma$ ($\pFluct=\uFluct$ is assumed for illustrative purpose). Furthermore, $E^{(2)}$ is asymmetric in nature, with larger energy in dilatations compared to compressions of same magnitude, as shown in figure~\ref{fig: EnergyComp}. Such asymmetry signifies that the medium ({compressible ideal gas} in the present study) relaxes towards the base state faster for finite dilatations compared to compressions. 

For compact supported or spatially periodic perturbations, the energy conservation Eq.~\eqref{eq: energy_cons} shows that the spatially averaged energy $\left\langle E^{(2)}\right\rangle$ decays monotonically in time (in the absence of energy sources) accounting for the nonlinear interactions i.e.,
\begin{align}
\dot{\mathscr{V}}=\frac{d \left\langle E^{(2)}\right\rangle}{dt} &= -\nu_0\left(\frac{\gamma - 1}{{Pr}}\right)\left\langle \frac{\partial^2 E^{(2)}}{\partial \pFluct^2}\left(\frac{\partial \pFluct}{\partial x}\right)^2\right\rangle -  \frac{4}{3}\nu_0\left\langle\left(\frac{\partial \uFluct}{\partial x}\right)^2\right\rangle \nonumber \\
&= \left\langle \mathcal{D}\right\rangle= -\epsilon\leq 0,
\label{eq: LyapunovFunction}
\end{align}
where $\mathcal{D}$ is the perturbation energy dissipation and $\epsilon$ is the negative of its spatial average. The spatial average $\left\langle E^{(2)}\right\rangle$ is non-negative ($E^{(2)}\geq 0$), and Eq.~\eqref{eq: LyapunovFunction} and figure~\ref{fig: EnergyTime} confirm that $\left\langle E^{(2)}\right\rangle$ evolves monotonically in time in the absence of physical energy sources. Hence, the spatial average of the perturbation energy function $\left\langle E^{(2)}\right\rangle$ defines the Lyapunov function $\mathscr{V}$ of the nonlinear acoustic system governed by the set of second order governing Eqs.~\eqref{eq: pressure} and \eqref{eq: velocity} exactly. The spatial average $\left\langle E^{(2)}\right\rangle$ should be used for studying the stability of nonlinear acoustic systems~\cite{George2012, Strogatz2018NonlinearDynamics}, which, however, falls beyond the scope of this work.

Wave-front steepening entails cascade of perturbation energy into higher wavenumbers thus broadening the energy spectrum. A fully broadened spectrum of acoustic perturbations exhibits energy at very small length scales which causes high thermoviscous energy dissipation. We analyse the separation of length scales and energy decay caused by nonlinear wave steepening and thermoviscous energy dissipation in the following sections. To this end, we utilize the direct numerical integration of Navier-Stokes Eqs.~\eqref{eq: NS_eqns1}-\eqref{eq: IdealGas} resolving all the length scales (DNS) and the exact energy corollary~Eq.~\eqref{eq: energy_cons} for second order truncated Eqs.~\eqref{eq: pressure} and~\eqref{eq: velocity}.

%*******************************************************************************************%
%*******************************************************************************************%
%*******************************************************************************************%

\section{High Fidelity Simulations with Adaptive Mesh Refinement}
\label{sec: numerics_PRE}
We perform shock-resolved numerical simulations of 1D Navier-Stokes (DNS) Eqs.~\eqref{eq: NS_eqns1}-\eqref{eq: IdealGas} with Adaptive Mesh Refinement (AMR). We use the perturbation energy $E^{(2)}$ defined in Eq.~\eqref{eq: Energy_norm} to define the characteristic dimensionless perturbation amplitude $A_{\mathrm{rms}}$ as, 
\begin{equation}
 A_{\mathrm{rms}} = \sqrt{\left\langle E^{(2)}\right\rangle},
\label{eq: PcDef}
\end{equation}
which is varied in the range $10^{-3}-10^{-1}$. The dimensionless kinematic viscosity at base state $\nu_0$ is also varied from 1.836$\times$10$^{-5}$ to 1.836$\times$10$^{-7}$. The base state conditions in the numerical simulations correspond to STP, i.e. $p^*_0 = 101325~\mathrm{Pa}$ and $T^*_0 = 300~\mathrm{K}$.
\begin{table}[!t]
\centering
\caption{Simulation parameter space for TW, SW, and AWT cases listing base state dimensionless viscosity $\nu_0$ (cf. Eq.~\eqref{eq: DimensionlessVisc}), initial characteristic perturbation amplitude $A_{\mathrm{rms},0}$ (cf. Eq.~\eqref{eq: PcDef}), and dimensional characteristic perturbation in velocity $u^*_{\mathrm{rms}}$ and pressure $p^*_{\mathrm{rms}}$ fields (Eq.~\eqref{eq: AcousticReynolds}).}
\def\arraystretch{2}
\begin{tabular}{| c | c | c | c |}
\hline
\hline
$\nu_0$	&  1.836$\times$10$^{-5}$ &  1.836$\times$10$^{-6}$ &  1.836$\times$10$^{-7}$\\
\hline
$A_{\mathrm{rms},0}$ 	& $10^{-3}$ & $10^{-2}$ & $10^{-1}$ \\
$u^*_{\mathrm{rms}}~\left(\mathrm{m}/\mathrm{s}\right)$	&  0.347 &  3.472&  34.725\\
$p^*_{\mathrm{rms}}~\left(\mathrm{kPa}\right)$	&  0.142 &  1.419 &  14.185\\
\hline
\hline
\end{tabular}
 \label{tab: test_cases}
\end{table}

The goal of spanning $A_{\mathrm{rms}}$ and $\nu_0$ over three orders of magnitude is to achieve widest possible range of energy cascade rate and dissipation within computationally feasible times. Equation~\eqref{eq: PcDef} yields the definitions of the perturbation Reynolds number~$\mathrm{Re}_L$ , characteristic perturbation velocity field~$u^*_{\mathrm{rms}}$, and pressure field $p^*_{\mathrm{rms}}$ as, 
\begin{equation}
\mathrm{Re}_L = \frac{A_{\mathrm{rms}}a^*_0 L^*}{\nu_0^*},~\quad u^*_{\mathrm{rms}} = a^*_0A_{\mathrm{rms}},~\quad p^*_{\mathrm{rms}} = \rho^*_0{a_0^*}^2A_{\mathrm{rms}},
\label{eq: AcousticReynolds}
\end{equation}
where $\mathrm{Re}_L$ denotes ratio of diffusive to wave steepening time scale over the length $L$.
{In the simulations, we keep $\mathrm{Re}_L\gg 1$, which corresponds to very fast wave steepening rates compared to diffusion.} In further sections (see Section~\ref{sec: ScalesAcousticCascade}), we define the wave turbulence Reynolds number $\mathrm{Re}_{\ell}$ based on the integral length scale $\ell$. Below, we briefly discuss the numerical scheme utilized for shock-resolved simulations and outline the initialization of the three configurations (TW, SW, and AWT) for numerical simulations.

\subsection{Numerical approach}
\begin{figure}[!t]
\centering
\includegraphics[width=0.75\linewidth]{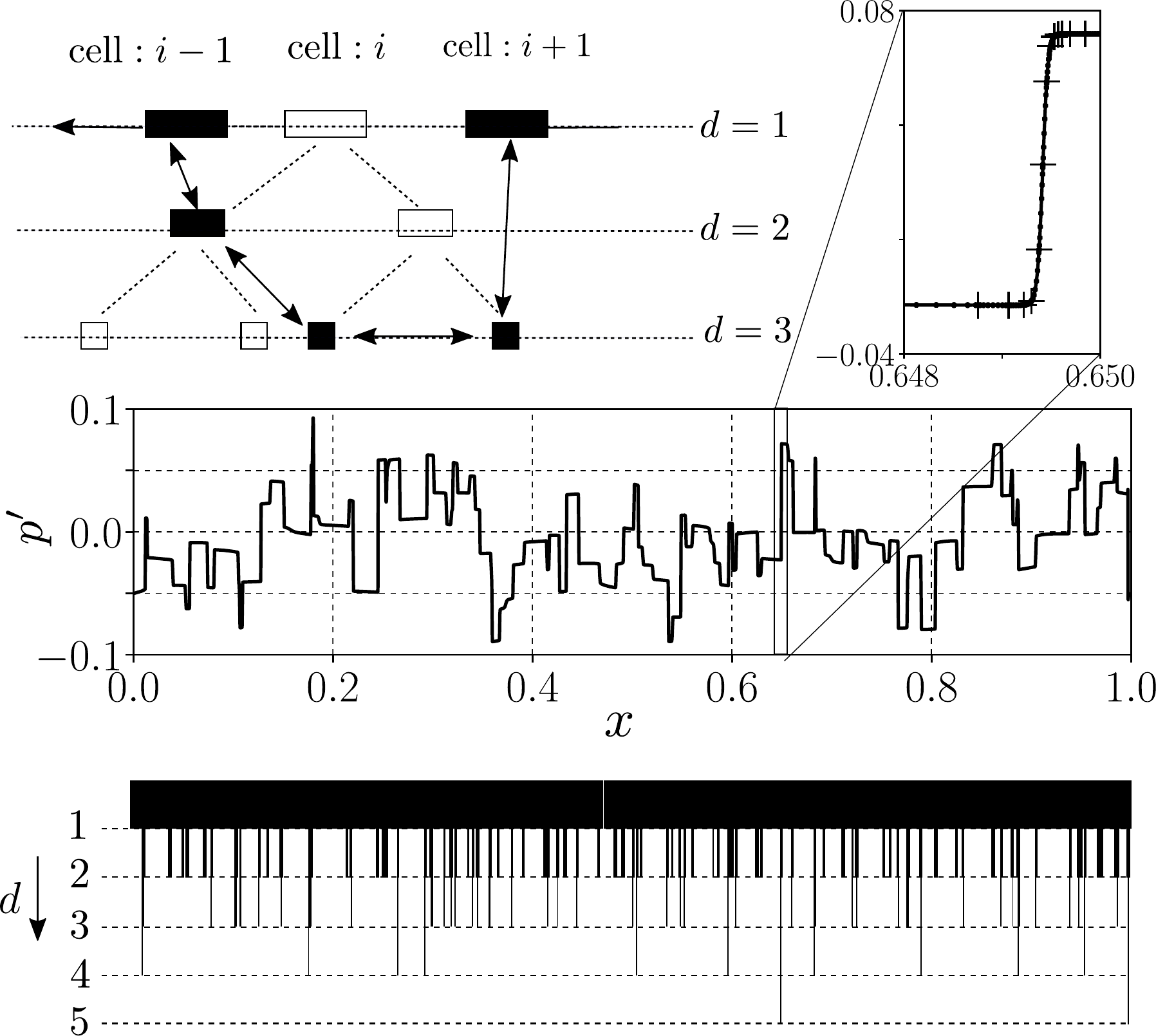}
\caption[Illustration of the binary tree implementation of Adaptive Mesh Refinement (AMR) technique.]{Illustration of the binary tree implementation of Adaptive Mesh Refinement (AMR) technique (top left). The mesh is refined based on the resolution error in pressure field in each cell acting as a node of a binary tree. The pressure field shown (middle) corresponds to the randomly initialized AWT case with $A_{\mathrm{rms}}=10^{-1}$, $\nu_0 =  1.836\times$10$^{-6}$ (Table~\ref{tab: test_cases}) at $t/\tau = 0.04$. The inset shows the resolved shock wave with $(+)$ denoting the cell interfaces. The mesh refinement levels (bottom) show the depth $d$ of the binary tree.}
\label{fig: AMR}
\end{figure}
We integrate the fully compressible 1D Navier-Stokes Eqs.~\eqref{eq: NS_eqns1}-\eqref{eq: NS_eqns3} in time utilizing the staggered spectral difference (SD) spatial discretization approach~\cite{kopriva1996conservative}. In the SD approach, the domain is discretized into 
cells. Within each cell, the orthogonal polynomial reconstruction of variables allows numerical differentiation with spectral accuracy. We refer the reader to the work by Kopriva and Kolias~\cite{kopriva1996conservative} for further details.

To accurately resolve spectral energy dynamics at all length scales, i.e. for resolved weak shock waves, we combine the SD approach with the adaptive mesh refinement (AMR) approach as first introduced by Mavriplis~\cite{mavriplis1994adaptive} for spectral methods. The SD-AMR approach eliminates the computational need of very fine grid everywhere for resolving the propagating shock waves. To this end, we expand the values of a generic variable $\phi$ local to the cell in the Legendre polynomial space as,
\begin{equation}
\phi = \sum^{N}_{i=1}\hat{\phi}_i \psi_i(x)
\end{equation}
where $\psi_i(x)$ is the Legendre polynomial of $(i-1)^{\mathrm{th}}$ degree. The polynomial coefficients $\hat{\phi}_i$ are utilized for estimating the local resolution error $\varepsilon$ defined as~\cite{mavriplis1994adaptive},
\begin{equation}
\varepsilon = \left(\frac{2 \hat{\phi}^2_N}{2N + 1} + \int^{\infty}_{N+1}\frac{2 f_{\varepsilon}^2(n)}{2n + 1} dn\right)^{1/2}, \quad f_{\varepsilon}(n) = ce^{-\sigma n},
\end{equation} 
where $f_{\varepsilon}$ is the exponential fit through the coefficients of the last four modes in the Legendre polynomial space.
When the estimated resolution error $\varepsilon$ exceeds a pre-defined tolerance, the cell divides into two subcells, which are connected utilizing a binary tree (shown in figure~\ref{fig: AMR}). The subcells merge together if the resolution error decreases below a pre-defined limit.

\subsection{Initial conditions}
\begin{table}[!b]
\centering
\setlength{\tabcolsep}{0.95em}
 \caption{Initial spectral compositions for traveling wave (TW), standing wave (SW), and acoustic wave turbulence (AWT). $\delta\left(\cdot\right)$ is the Dirac delta function.}
\def\arraystretch{1.2}
\begin{tabular}{| c | c | c | c |}
\hline
\hline
{$~~$} 	& TW & SW & {AWT}\\
\hline
$k_0$	&  1  &  1 &  1 \\
$k_E$	&  1  &  1 &  100 \\
$b_0(k)$	&  0  &  0 &  $e^{-(|k|-k_E)^2}$ \\
{$\widehat{E}_k$} & {${A}_{\mathrm{rms}}^2\delta(k_0)$} & {${A}_{\mathrm{rms}}^2\delta(k_0)$} & ${A}_{\mathrm{rms}}^2$\\
\hline
\hline
\end{tabular}\quad
 \label{tab: InitialSpectra}
\end{table}
We utilize the Riemann invariants for compressible Euler equations to initialize the propagating traveling and standing wave cases in the numerical simulations. The Riemann invariants in terms of perturbation variables assuming nonlinear isentropic changes are given by, 
\begin{align}
 R_{-} = \frac{2}{\gamma -1 }\left(\left(1 + {\rhoFluct}\right)^{\frac{\gamma - 1}{2}} - 1\right) - \uFluct,\label{eq: InvariantL} \\
 R_{+} = \frac{2}{\gamma -1 }\left(\left(1 + {\rhoFluct}\right)^{\frac{\gamma - 1}{2}} - 1\right) + \uFluct,
\label{eq: InvariantR}
\end{align}
where $R_{-}$ and $R_{+}$ are the left and right propagating invariants, respectively, and $\uFluct$ and ${\rhoFluct}$ are normalized velocity and density perturbations, as defined in Eq.~\eqref{eq: per_norm}. Initial conditions for TW and SW cases correspond to $R_{-} = 0$ and $R_{-} = R_{+}$ respectively. 

To initialize the broadband noise case, we first choose $\pFluct$ and $\uFluct$ pseudo-randomly from a uniform distribution for the whole set of discretization points in $x$. Low-pass filtering of $\pFluct$ and $\uFluct$ yields, 
\begin{align}
 \widetilde{\widehat{p}_k}(t&=0) = \widehat{p}_kb_0(k),\quad \widetilde{\widehat{u}_k}(t=0) = \widehat{u}_kb_0(k)\nonumber\\
  b_0(k) &= \begin{cases} 
      1 & k_0\leq|k|\leq k_E \\
     e^{-(|k|-k_E)^2} & |k|> k_E
   \end{cases}.
\label{eq: FilterPU1}
\end{align}
where ${\widehat{p}_k}$ and ${\widehat{p}_k}$ are the Fourier coefficients of pseudo-random fields $\pFluct$ and $\uFluct$, respectively. $\widetilde{\widehat{p}_k}$ and $\widetilde{\widehat{u}_k}$ are the low-pass filtered coefficients. The inverse Fourier transform of Eq.~\eqref{eq: FilterPU1} yields smooth initial conditions with the initial spectral energy ${\widehat{E}_k}$, as defined in Section~\ref{sec: ScalesAcousticCascade} (cf. Eq.~\eqref{eq: spectral_energy}). For TW and SW, only the single harmonic ($k=1$ in the current work) contains all of the initial energy. However, for AWT, $\widehat{E}_k$ is governed by the correlation function of velocity and pressure fields. In Table~\ref{tab: InitialSpectra}, we summarize the initial spectral energy for all three cases based on Eq.~\eqref{eq: FilterPU1}.\section{Scales of acoustic energy cascade and dissipation}
\label{sec: ScalesAcousticCascade}

In this section, we derive the analytical expressions of spectral energy, energy cascade flux, and spectral energy dissipation utilizing the exact energy corollary Eq.~\eqref{eq: energy_cons} (see Section~\ref{sec: EnergyCorollary}). We then identify the integral length scale $\ell$, the Taylor microscale $\lambda$, and the Kolmogorov length scale $\eta$ for TW, SW, and AWT cases in a periodic domain utilizing the DNS data (see figures~\ref{fig: LengthScalesSummary_1} and~\ref{fig: LengthScalesSummary_2} and Table~\ref{tab: LengthScalesTable}). Temporal evolution laws of these length scales yield energy decay laws, which are used for dimensionless spectral scaling relations (see Section~\ref{sec: SpectralScaling}).

\subsection{Spectral energy flux and dissipation rate for periodic perturbations}

The exact perturbation energy conservation equation is given by (cf. Eq.~\eqref{eq: energy_cons}),
\begin{equation}
\frac{\partial E^{(2)}}{\partial t} + \frac{\partial I}{\partial x} = \mathcal{D}
\end{equation}
Integrating over the periodic domain, the above energy corollary can be converted into the following statement of conservation of perturbation energy in the spectral space,
\begin{equation}
\frac{d}{d t} \sum_{|k'|\leq k} \widehat{E}_{k'} + \widehat{\Pi}_k = \sum_{|k'|\leq k}\widehat{\mathcal{D}}_{k'},
\label{eq: spectral_conservation}
\end{equation}

\begin{figure}[!t]
\centering
\includegraphics[width=0.75\linewidth]{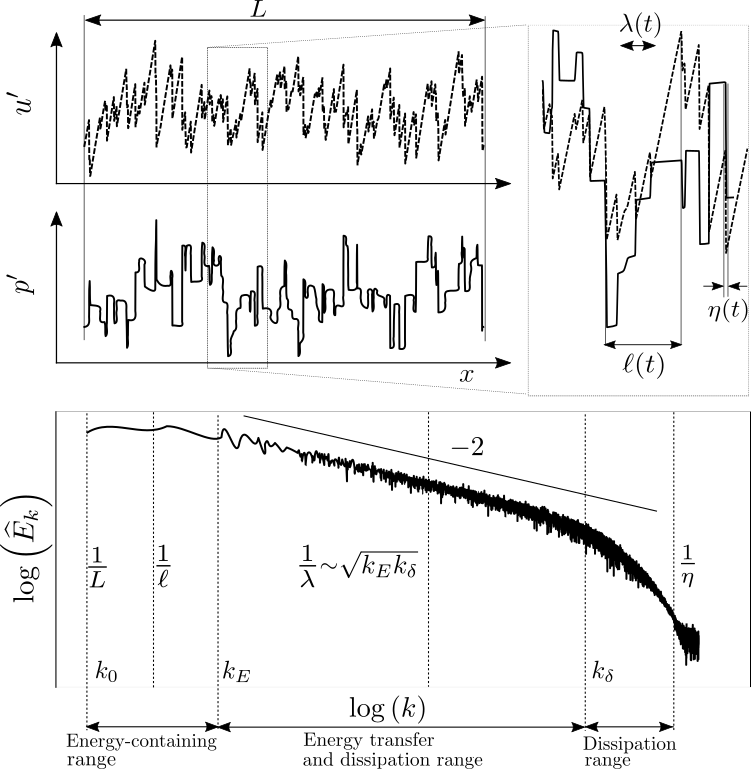}
\put(-330,320){$(a)$}
\put(-330,160){$(b)$}
\caption{Schematic illustrating the comparison of various length scales associated with spectral energy cascade in nonlinear acoustics in both spatial $(a)$ and spectral $(b)$ space. $(a)$ shows the perturbation velocity $\uFluct$ ($--$) and pressure $\pFluct$ (--) fields in AWT obtained from the DNS data for $\nu_0 = $1.836$\times$10$^{-7}$ and $A_{\mathrm{rms}}=10^{-1}$. $(b)$ shows the corresponding spectral energy $\widehat{E}_k$ in log-log space. The integral length scale $\ell$ corresponds to the characteristic distance between the shock waves traveling in the same direction. The Kolmogorov length scale $\eta$ corresponds to the shock wave thickness. The Taylor microscale $\lambda$ is the diffusive length scale and satisfies $\ell\gg\lambda\gg\eta$. $L$ corresponds to the length of the domain.}
\label{fig: LengthScalesSummary_1}
\end{figure}

where the first term corresponds to the temporal rate of change of cumulative spectral energy density,
\begin{equation}
\frac{d \widehat{E}_k}{dt} \approx \frac{d}{dt}\left(\frac{|\widehat{u}_k|^2}{2} + \frac{|\widehat{p}_k|^2}{2}\right) + \Re\left(\widehat{p}_{-k}\frac{d\widehat{g}_k}{dt}\right),
\label{eq: SpectralEnergyRate}
\end{equation}
and $\widehat{g}$ is the Fourier transform of $g(\pFluct)$ given by,
\begin{equation}
g(\pFluct) = \frac{\gamma}{\gamma-1}\left(\left(1+\gamma \pFluct\right)^{1/\gamma} - 1 -\pFluct \right).
\end{equation}
The spectral energy $\widehat{E}_k$ is given by, 
\begin{equation}
\widehat{E}_k = \frac{|\widehat{u}_k|^2}{2} + \frac{|\widehat{p}_k|^2}{2} + \Re \left(\widehat{p}_{-k}\left(\widehat{\frac{f(\pFluct)}{\pFluct}}\right)_k\right).
\label{eq: spectral_energy}
\end{equation}
\begin{figure}[!t]
\centering
\includegraphics[width=0.75\linewidth]{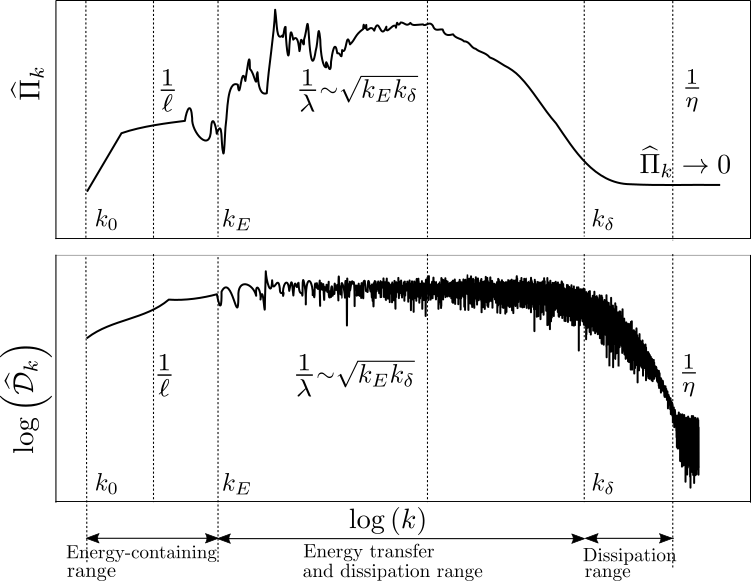}
\put(-330,260){$(a)$}
\put(-330,140){$(b)$}
\caption{ {Schematic illustrating the comparison of various length scales associated with spectral energy cascade in nonlinear acoustics in the spectral space, and corresponding variation of spectral energy flux $\hat{\Pi}_k$ ($a$) and spectral dissipation $\mathcal{D}_k$ $(b)$. The integral length scale $\ell$ corresponds to the characteristic distance between the shock waves traveling in the same direction. The Kolmogorov length scale $\eta$ corresponds to the shock wave thickness. The Taylor microscale $\lambda$ is the diffusive length scale and satisfies $\ell\gg\lambda\gg\eta$. $L$ corresponds to the length of the domain.}}
\label{fig: LengthScalesSummary_2}
\end{figure}

\begin{table}[!t]
\centering
 \caption{ Summary of the three length scales $\ell$, $\lambda$, and $\eta$, respective definitions, and the range of spectrum characterized by them. The integral length scale characterizes the energy containing range $(k_0, k_E)$. The Taylor microscale is the characteristic of the energy transfer and dissipation range $(k_E, k_{\delta})$. The Kolmogorov length scale corresponds to the highest wavenumber generated as a result of nonlinear acoustic energy cascade.}
\setlength{\tabcolsep}{0.7em}
\def\arraystretch{1.5}
\begin{tabular}{| c  c  c  c |}
\hline
\hline
\multirow{2}{*}{Length scale} 	& Integral & Taylor & Kolmogorov\\
                            	& length scale & Microscale & length scale\\
                                &$\ell$  & $\lambda$ & $\eta$\\
\hline
{Definition}	& $\sqrt{\frac{\sum_k {\widehat{E}_k}/{k^2}}{\sum_k \widehat{E}_k}}$  &  $\sqrt{\frac{2\delta\left\langle E^{(2)}\right\rangle }{\epsilon}} $ &  $\frac{\delta}{\sqrt{\left\langle E^{(2)}\right\rangle}}$ \\
Characteristic	&  \multirow{2}{*}{$(k_0,k_E)$} &  \multirow{2}{*}{$(k_E, k_{\delta})$} &  \multirow{2}{*}{$(k_\delta, \infty)$} \\
spectral range	&     &    &      \\
\hline
\hline
\end{tabular}\quad
 \label{tab: LengthScalesTable}
\end{table}
It is noteworthy that the correction in spectral energy does not follow directly from the nonlinear correction function $f(\pFluct)$ derived in the physical space. In Eq.~\eqref{eq: SpectralEnergyRate}, we have made the following approximation,
\begin{equation}
\frac{d}{dt}\left( \Re \left(\widehat{p}_{-k}\left(\widehat{\frac{f(\pFluct)}{\pFluct}}\right)_k\right)\right) \approx \Re\left(\widehat{p}_{-k}\frac{d\widehat{g}_k}{dt}\right)
\end{equation}
The second term $\widehat{\Pi}_k$ in Eq.~\eqref{eq: spectral_conservation} is the flux of spectral energy density from wavenumbers $|k'|\leq k$ to $|k'| > k$ and is given by, 
\begin{align}
\widehat{\Pi}_k = \sum_{|k'|\leq k}\Re\Big(\widehat{p}_{-k'}\left(\frac{\partial (\widehat{\uFluct g})}{\partial x}\right)_{k'} +  \widehat{p}_{-k'}\left(\widehat{\uFluct\frac{\partial \pFluct}{\partial x}}\right)_{k'} +\frac{1}{2}\widehat{u}_{-k'}\reallywidehat{\frac{\partial}{\partial x}\left(\uFluct ^2 - \pFluct ^2\right)_{k'}}\Big).
\label{eq: SpectralFlux}
\end{align}
Finally, the spectral dissipation $\widehat{\mathcal{D}}_k$ is given by, 
{\color{black}{
\begin{align}
\widehat{\mathcal{D}}_k = \nu_0\frac{\gamma - 1}{Pr}&\Re\left(\widehat{p}_{-k}\left(\reallywidehat{\left(1 + \frac{\partial g}{\partial \pFluct}\right)\left(\frac{\partial^2 \pFluct}{\partial x^2}\right)}\right)_{k}\right)- {\frac{16\pi^2}{3}\nu_0} k^2|\widehat{u}_k|^2.
\label{eq: SpectralDissipation}
\end{align}
}}

\begin{figure}[!t]
\centering
\includegraphics[width=\textwidth]{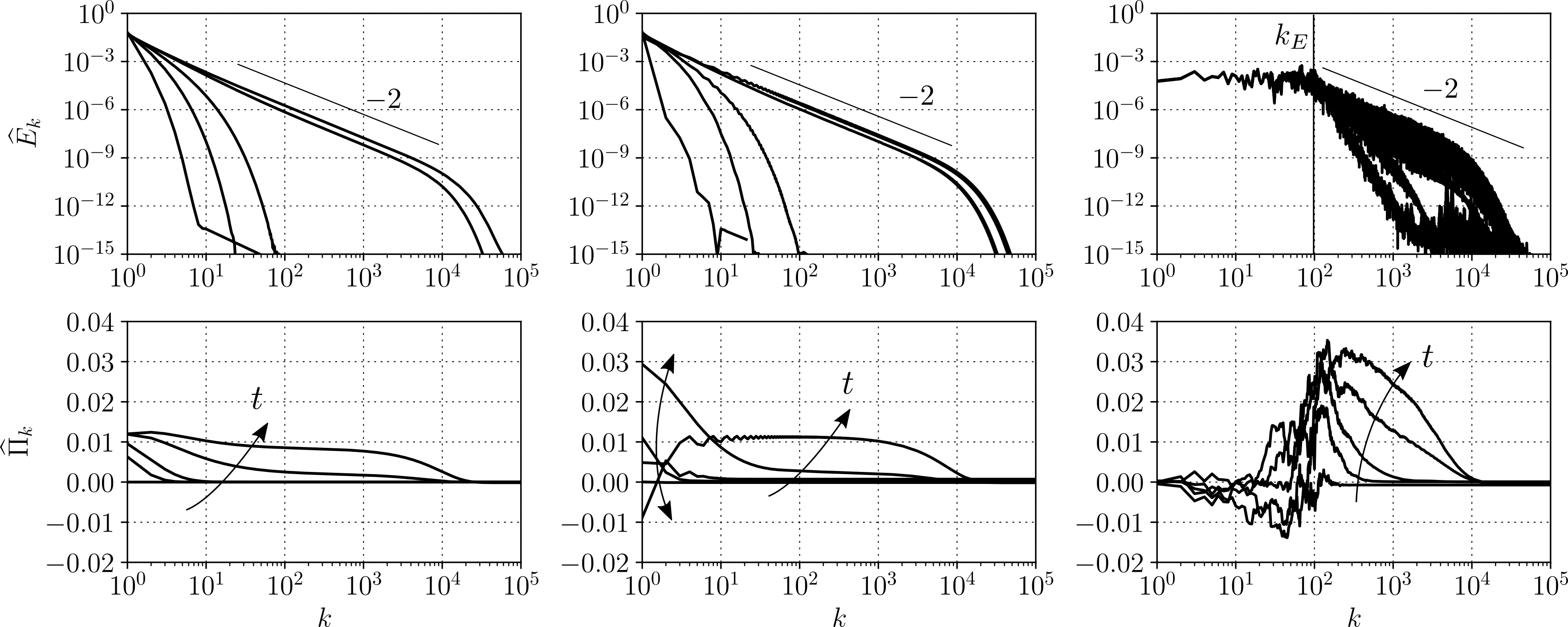}
\put(-425,180){$(a)$}
\put(-285,180){$(b)$}
\put(-145,180){$(c)$}
\caption{Spectro-temporal evolution of $\widehat{E}_k$ (top) and spectral flux $\widehat{\Pi}_k$ (bottom) for a TW ($a$), SW ($b$) and AWT ($c$). Spectral flux $\widehat{\Pi}_k$ for a traveling wave simply increases towards high wavenumbers. For a standing wave, $\widehat{\Pi}_k$ oscillates at low wavenumbers cyclically due to collisions of oppositely traveling shock waves while high wavenumber behaviour resembles that of a traveling shock. For AWT, the spectral broadening occurs for $k>k_E$ {with small fluctuations in time for $k<k_E$}.}
\label{fig: EnergyFLux}
\end{figure}

%\Carlo{Make this images pngs since they take a lot to load}

Detailed derivation of Eqs.~\eqref{eq: spectral_conservation}-\eqref{eq: SpectralDissipation} is given in appendix~\ref{sec: appedixB}. Figures~\ref{fig: LengthScalesSummary_1} and~\ref{fig: LengthScalesSummary_2} summarize the typical shape of the spectral energy $\widehat{E}_k$, spectral energy flux $\widehat{\Pi}_k$, and the spectral dissipation $\widehat{\mathcal{D}}_k$ along with the relative positions of the three relevant length scales, the integral length scale, $\ell$, the Taylor microscale, $\lambda$, and the Kolomogorov length scale $\eta$ in the spectral space. The spectro-temporal evolution of any configuration of nonlinear acoustic waves can be quantified utilizing these length scales and the respective evolution in time which is discussed in detail in the subsections below. Table~\ref{tab: LengthScalesTable} summarizes these length scales and the characteristic spectral range. {In further sections, we discuss all the spectral quantities as functions of absolute value of wavenumbers and drop the $|.|$ notation for convenience.}

The spectral energy flux $\widehat{\Pi}_k$, defined in Eq.~\eqref{eq: SpectralFlux}, is in terms of interactions of the Fourier coefficients of the pressure $\widehat{p}_k$ and velocity $\widehat{u}_k$ perturbations. For compact support or periodic perturbations, $\widehat{\Pi}_k$ approaches zero in the limit of very large wavenumbers $k\rightarrow \infty$, 
\begin{equation}
\lim_{k\rightarrow\infty} \widehat{\Pi}_{k} = \left\langle \frac{\partial I}{\partial x}\right\rangle = 0.
\end{equation} 
The last two terms in Eq.~\eqref{eq: SpectralFlux} result in $\widehat{\Pi}_{k}\rightarrow 0$ for large $k$ for general acoustic phasing. Hence, they are most relevant in SW and AWT cases. In a pure traveling wave (TW), $\uFluct = \pFluct$ at first order due to which the last two terms in Eq.~\eqref{eq: SpectralFlux} become negligible. Furthermore, the sequence of $\widehat{\Pi}_k$ also converges monotonically i.e.,
\begin{equation}
 \lim_{k\rightarrow\infty}\left(\widehat{\Pi}_{k-1} - \widehat{\Pi}_{k}\right) \rightarrow 0^{+},
 \label{eq: MonotonicConv}
\end{equation}
as shown in Figs.~\ref{fig: LengthScalesSummary_2}$(a)$ and~\ref{fig: EnergyFLux}. The flattening of the spectral energy flux $\widehat{\Pi}_k$ (Eq.~\eqref{eq: MonotonicConv}) begins at a specific wavenumber $k_\delta$ associated to the {Kolmogorov length scale}, $\eta$, as shown in figures~\ref{fig: LengthScalesSummary_1} and \ref{fig: LengthScalesSummary_2}. The spectral energy $\widehat{E}_k$ deviates off the $k^{-2}$ decay near the wavenumber $k_\delta$. Figure~\ref{fig: EnergyFLux} shows the spectro-temporal evolution of the spectral energy $\widehat{E}_k$ and the flux $\widehat{\Pi}_k$ for TW, SW, and AWT prior to formation of shock waves.

For TW, $\widehat{\Pi}_k$ increases in time due to spectral broadening. In SW, $\widehat{\Pi}_k$, while increasing, also oscillates at low wavenumbers due to the periodic collisions of oppositely propagating shocks. A combination of these processes takes place in a randomly initialized smooth finite amplitude perturbation, which at later times develops into AWT. At later times, nonlinear waves in all three configurations fully develop in to shock waves. Up to the shock formation, the spectral dynamics of all configurations simply involve increase of the spectral flux $\widehat{\Pi}_k$. The dimensionless shock formation time $\tau$ can be estimated as,
\begin{equation}
\tau = \frac{2}{(\gamma - 1)A_{\mathrm{rms},0}}.
\label{eq: SteepeningTime}
\end{equation}
Upon shock formation, the dynamic evolution of TW and SW remains phenomenologically identical. The isolated shocks propagate and the total perturbation energy of the system decays due to thermoviscous dissipation localized around the shock wave. {\color{black}{However, for AWT, along with collisions of oppositely propagating shocks, those propagating in the same direction coalesce due to differential propagating speeds. As we discuss below, this modifies the energy decay and spectral energy dynamics in AWT significantly compared to  TW and SW.}}

In the sub-sections below, we elucidate the energy dynamics before and after shock formation for TW, SW, and AWT. To this end, we define and discuss the relevant length scales as mentioned above, namely: the Taylor microscale $\lambda$, the integral length scale $\ell$, and the Kolmogorov length scale $\eta$. Particular focus is given to the AWT case due to modified dynamics caused by shock coalescence.

\begin{figure}[!t]
\centering
\includegraphics[width=\textwidth]{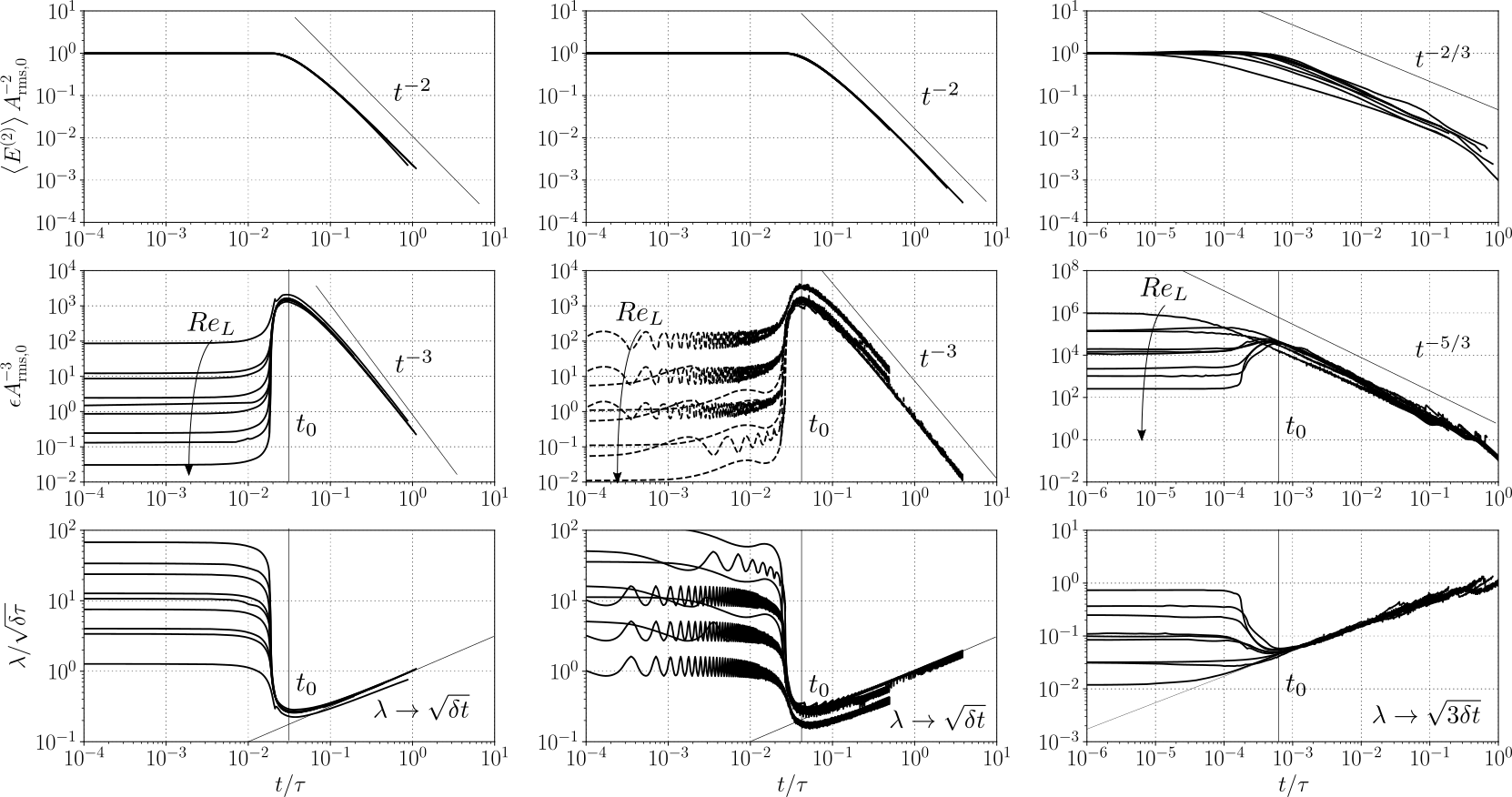}
\put(-427,235){$(a)$}
\put(-286,235){$(b)$}
\put(-145,235){$(c)$}
\caption{Temporal evolution of scaled total energy <$E^{(2)}$> ${A^{-2}_{\mathrm{rms},0}}$ (--) (top), dissipation rate $\epsilon {A^{-3}_{\mathrm{rms},0}}$ ($--$) (mid) and normalized Taylor microscale $\lambda/\sqrt{\delta \tau}$ (bottom) for TW ($a$), SW ($b$) and AWT ($c$) against the scaled time $t/\tau$ for varying perturbation Reynolds number $\mathrm{Re}_L$. The time $t_0$ signifies fully broadened spectrum of the perturbation field.}
\label{fig: EnergyDecayAll}
\end{figure}

\subsection{Taylor microscale}
In hydrodynamic turbulence, the Taylor microscale $\lambda$ separates the inviscid length scales from the viscous length scales~\cite{tennekes1972first,pope2000turbulent}. Due to the spectral energy cascade in planar nonlinear acoustics, we note that the spectral energy varies as $\widehat{E}_k\sim k^{-2}$ due to the formation of shocks and the spectral dissipation due to thermoviscous diffusion varies as $\widehat{\mathcal{D}}_k\sim k^2 \widehat{E}_k$. Consequently, the dissipation acts over most of the length scales with $k>k_E$ (figure~\ref{fig: LengthScalesSummary_2}$(b)$), unlike hydrodynamic turbulence where the viscous dissipation dominates only the smaller length scales~\cite{tennekes1972first, pope2000turbulent}. As shown in figure~\ref{fig: LengthScalesSummary_2}$(a)$, length scales in the range $(k_E, k_{\delta})$ exhibit both dissipation $\widehat{\mathcal{D}}_k$ and energy transfer $\widehat{\Pi}_k$. For $k>k_{\delta}$, $\widehat{\Pi}_k$ begins to converge monotonically to 0 and the interval $(k_{\delta}, 1/\eta)$ primarily exhibits dissipation $\widehat{\mathcal{D}}_k$ only. The Taylor microscale $\lambda$ quantifies the length scale associated to the whole dissipation range. 

Utilizing the definition of the total perturbation energy $\left\langle E^{(2)} \right\rangle$ and the dissipation rate $\epsilon$ (cf. Eq.~\eqref{eq: LyapunovFunction}), the microscale $\lambda$ can be defined as, 
\begin{equation}
\lambda(t) = \sqrt{\frac{2\delta\left\langle E^{(2)}\right\rangle }{\epsilon}},
\label{eq: Microscale}
\end{equation}
where $\delta$ is the thermoviscous diffusivity, given by, 
{\color{black}{
\begin{equation}
\delta = \nu_0\left(\frac{4}{3} + \frac{\gamma - 1}{\mathrm{Pr}}\right).
\label{eq: EffectiveDiff}
\end{equation}
}}
Equation~\eqref{eq: Microscale} indicates that the Taylor microscale can be identified as the geometrical centroid of full energy spectrum, i.e.
\begin{equation}
\lambda \sim \sqrt{\frac{\sum_k \widehat{E}_k}{\sum_k k^2 \widehat{E}_k}}.
\label{eq: lengthscales_centroids}
\end{equation}

As the smaller length scales (higher harmonics) are generated, the dissipation rate $\epsilon$ tends to increase reaching a maximum in time. The increase of dissipation rate $\epsilon$ implies decrease of the length scale $\lambda$ in time. Minima of $\lambda$ indicates the fully-broadened spectrum of energy limited by the thermoviscous diffusivity at very large wavenumbers. Further spatio-temporal evolution of the system is dominated by dissipation thus indicating the purely diffusive nature of the Taylor microscale, i.e.,
\begin{equation}
 \lambda\rightarrow C\sqrt{\delta t}.
\label{eq: TaylorMicroscaleAssy}
\end{equation}
The temporal evolution of $\lambda$ is qualitatively similar for TW, SW, and AWT, the constant $C$ in Eq.~\eqref{eq: TaylorMicroscaleAssy} differs for TW and SW compared with AWT due to the different spatial structure of perturbations. The time $t_0$ at which $\lambda$ reaches minimum signifies {fully developed nonlinear acoustic waves}. In case of AWT, it signifies fully developed acoustic wave turbulence. 

Figure~\ref{fig: EnergyDecayAll} shows the decay of scaled total perturbation energy $\left\langle E^{(2)}\right\rangle A_{\mathrm{rms},0}^{-2}$ and total dissipation rate $\epsilon A_{\mathrm{rms},0}^{-3}$ for the TW, SW, and AWT. We note that the total energy decays as a power law $t^{-2}$ for both TW and SW, whereas, for AWT, the initial decay law is $t^{-2/3}$. Asymptotic evolution (at large $t$) of the Taylor microscale follows from the decay laws as $\lambda = \sqrt{\delta t}$ and $\lambda = \sqrt{3\delta t}$, respectively. Since energy decay law of a single harmonic traveling and standing waves is rather trivial, we focus primarily on the AWT case for further discussion.

\subsection{Integral length scale}
\begin{figure}[!b]
\centering
\includegraphics[width=0.65\linewidth]{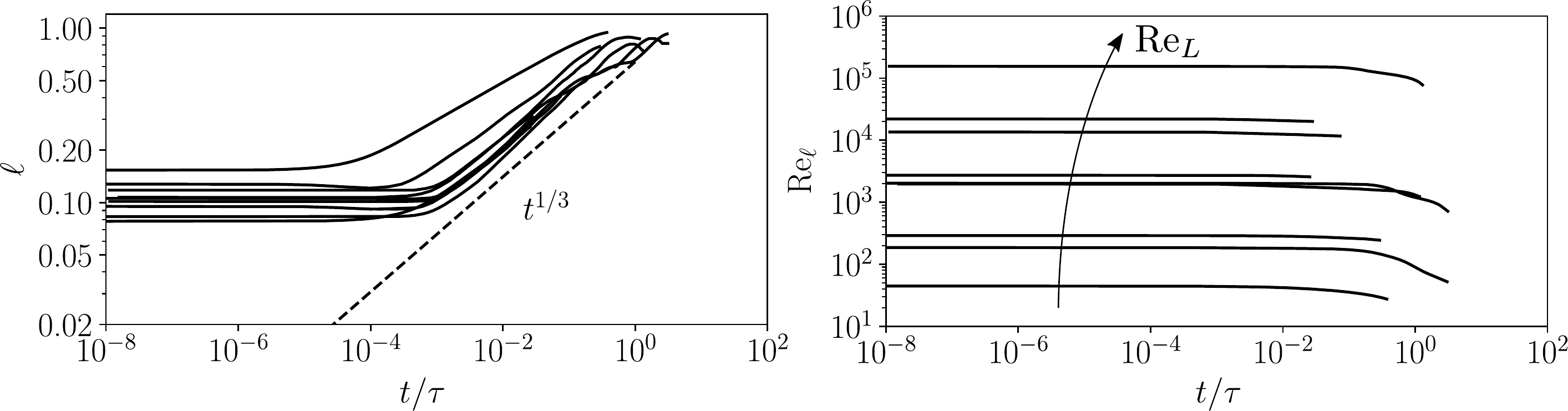}
\put(-285,280){$(a)$}
\put(-285,150){$(b)$}
\caption{Evolution of the integral length scale $\ell$ $(a)$ and the Reynolds number $\mathrm{Re}_{\ell}$ ($b$) defined in Eqs.~\eqref{eq: IntegralLengthScale} and~\eqref{eq: IntegralReynoldsNumber}, respectively, for all the cases of AWT considered. For small thermoviscous diffusivity, $\ell$ increases approximately as $t^{1/3}$ before saturating to the dimensionless domain length $L=1$ and $\mathrm{Re}_{\ell}$ remains approximately constant.}
\label{fig: IntegralLengthScaleEvolution}
\end{figure}
We identify the integral length scale $\ell$ as the characteristic length scale of the energy containing scales. In general, random {smooth} broadband noise (AWT) develops into an ensemble of shocks, propagating left and right in a one-dimensional system. For an ensemble of shock waves distributed spatially along a line, $\ell$ corresponds to the characteristic distance between consecutive shock waves traveling in the same direction, as shown schematically in figures~\ref{fig: LengthScalesSummary_1} and~\ref{fig: LengthScalesSummary_2}.
Formally, we define $\ell$ as, 
\begin{equation}
\ell = \sqrt{\frac{\sum_k \frac{\widehat{E}_k}{k^2}}{\sum_k \widehat{E}_k}},
\label{eq: IntegralLengthScale}
\end{equation}
which is identical to the integral length scale defined in Burgers' turbulence~\cite{Gurbatov_JFM_1997}. {\color{black}{Definition in Eq.~\eqref{eq: IntegralLengthScale} yields the centroid wavenumber of the initial energy spectrum (unlike the Taylor microscale, which corresponds to the full energy spectrum) and hence is characteristic of the large length scales of fully developed AWT.}}
To elucidate the evolution of the total perturbation energy $\left\langle E^{(2)}\right\rangle$ utilizing the integral length scale, we assume the following model spectral energy density $\widehat{E}_k$,
\begin{equation}
\widehat{E}_k = \begin{cases} 
      C_1 k^n & k_0\leq k \leq k_E \\
     C_2 k^{-2} & k_{\delta}> k > k_E
   \end{cases},
\label{eq: SpectralEnergyAssump}
\end{equation}
where $k^n$ corresponds to the shape of initialized energy spectrum in the range $(k_0,k_E)$ (figures~\ref{fig: LengthScalesSummary_1} and~\ref{fig: LengthScalesSummary_2}). In this work, we only utilize the white noise initialized AWT cases which correspond to $n=0$ (see Table~\ref{tab: InitialSpectra}). Moreover,
\begin{equation}
C_2 = C_1 k_E^{n+2}.
\end{equation}
The wave numbers $k_E$ and $k_\delta$ vary in time due to decaying energy. By definition, the mean of perturbations is zero. Hence, the smallest wavenumber containing energy $k_0$ (cf. figures~\ref{fig: LengthScalesSummary_1} and~\ref{fig: LengthScalesSummary_2}) is the reciprocal of the domain length $L$, i.e., 
\begin{equation}
k_0 = 1/L.
\end{equation}

We note that the above model spectral energy $\widehat{E}_k$ holds for two primary reasons. Firstly, the energy cascade results in the $k^{-2}$ decay of the spectral energy $\widehat{E}_k$ due to formation of shock waves~\cite{Gurbatov_JFM_1997}. In the limit of vanishing viscosity $\delta \rightarrow 0 $, such decay extends up to $k\rightarrow \infty$ in which case the developed shock waves render the system $C_0$ discontinuous. Secondly, the shape of the spectral energy $\widehat{E}_k$ for $k\rightarrow k_0$ corresponds to $k^{n}$, which is also the shape of initial energy spectral at time $t=0$. Such argument corresponds to the concept of \emph{permanence of large eddies} in hydrodynamic turbulence~\cite{Batchelor1953theory}, which in spectral space can be written as, 
\begin{equation}
\widehat{E}_k(t) \approx \widehat{E}_k(t=0),~~\mathrm{as}~~k\rightarrow k_0.
\label{eq: PLE}
\end{equation}
Gurbatov \emph{et al.}~\cite{Gurbatov_JFM_1997} utilized a similar argument in the context of Burgers' turbulence. Combining the Eqs.~\eqref{eq: IntegralLengthScale}-\eqref{eq: PLE}, the integral length scale $\ell$ is given by, 
\begin{align}
\ell &\approx 
\begin{cases} 
     \sqrt{\frac{n+1}{n-1}\left(\frac{k^{n-2}_E + k^{n-3}_Ek_0 + \cdots k^{n-2}_0}{k^{n}_E + k^{n-1}_Ek_0 + \cdots k^{n}_0}\right)} & n\neq 1 \\
     \sqrt{\frac{2\ln(k_E/k_0)}{k^2_E - k^2_0}} & n = 1.
   \end{cases},
\label{eq: IntegralLengthN}
\end{align}
where we have used the simplifying approximation of $k_\delta \gg k_E$. We note that the Eq.~\eqref{eq: IntegralLengthN} indicates the dependence of $\ell$ and consequently the energy decay law on $n$. In the present work, we perform numerical simulations for an uncorrelated white noise (filtered) which corresponds to $n=0$, and 
\begin{equation}
\ell \approx \frac{1}{\sqrt{k_0 k_E}}.
\label{eq: IntegralApprox}
\end{equation}
As a result of \emph{permanence of large eddies}, the decay of energy in the initial regime of AWT is associated only with the decreasing $k_E$ or increasing integral length scale $\ell$. Integrating Eq.~\eqref{eq: SpectralEnergyAssump} in the spectral space and differentiating in time yields (for $n=0$ in the current simulations), 
\begin{align}
\frac{d \left\langle E^{(2)} \right\rangle}{dt} &= C_1\left(2\frac{dk_E}{dt}\left(1 - \frac{k_E}{k_\delta}\right) + \left(\frac{k_E}{k_{\delta}}\right)^2\frac{dk_\delta}{dt}\right)  \\ &\approx -\frac{2C_1}{k_0 \ell^3}\frac{d\ell}{dt}.
\label{eq: EdotLdot}
\end{align}
Above relation shows that derivation of the energy decay power law amounts to finding the kinetic equations of the integral length scale $\ell$ and the limiting wavenumber $k_\delta$. For TW and SW, $\ell$ remains constant by definition. Consequently, the energy decay rate only depends on decrease of wavenumber $k_\delta$ and the coefficient $C_2$ due to the thermoviscous diffusion (cf. Eq.~\eqref{eq: dEdt1}
). However, for an ensemble of shock waves in AWT, $\ell$ increases monotonically in time, as shown in figure~\ref{fig: IntegralLengthScaleEvolution}$(a)$ due to the coalescence of shock waves propagating in the same direction. At large times, the domain consists of only two shock waves propagating in opposite directions. 

In the context of Burgers' turbulence in an infinite one-dimensional domain, Burgers~\cite{Burgers1974nonlinear} and Kida~\cite{Kida_1979_JFM} have derived the appropriate asymptotic evolution laws for the integral length scale $\ell$ based on the dimensional arguments. However, in the present work, the finiteness of the domain renders the asymptotic analysis infeasible. Our numerical results indicate that $\ell\sim t^{1/3}$ ($k_E\sim t^{-2/3}$) for randomly distributed shock waves at various $\mathrm{Re}_L$ values considered, as shown in figure~\ref{fig: IntegralLengthScaleEvolution}$(a)$. Equations~\eqref{eq: EdotLdot} and~\eqref{eq: IntegralApprox} show that such scaling is consistent with the observed energy decay law $\left\langle E^{(2)} \right\rangle \sim t^{-2/3}$ thus validating the result in Eq.~\eqref{eq: EdotLdot}. It is noteworthy that decay $k_E\sim t^{-2/3}$ is a result analogous to the one discussed in Burgers' turbulence~\cite{Burgers1974nonlinear,Kida_1979_JFM,Gurbatov_JFM_1997} considered in an infinite one-dimensional domain. {\color{black}{Due to infinitely long domain, the average distance between the shocks approaches $1/k_E$ (not $\ell$) simply due to larger number of shocks in the domain separated by the distance $1/k_E$ since $k_E$ corresponds to the largest wavenumber carrying initial energy, thus implying that mean distance between the shocks increases as $t^{2/3}$ as noted by Burgers~\cite{Burgers1974nonlinear}.}}

Based on the integral length scale, the Reynolds number $\mathrm{Re}_{\ell}$ can be defined as, 
\begin{equation}
\mathrm{Re}_{\ell} = \mathrm{Re}_{L}\ell,
\label{eq: IntegralReynoldsNumber}
\end{equation}
which captures the ratio of the diffusive time scale to the wave turbulence time. Upon formation of shock waves, the perturbation energy decays due to coalescence. Shock waves coalesce locally thus increasing the characteristic separation between the shock waves thus causing $\ell$ to increase. In this regime, the Reynolds number $\mathrm{Re}_{\ell}$ remains constant (figure~\ref{fig: IntegralLengthScaleEvolution}$(b)$) which denotes that the ratio of shock coalescence time scale $(\ell L^*)/(a^*_0A_{\mathrm{rms}})$ and the diffusive time scale $(\ell L^*)^2/\nu^*_0$ remains constant. As the wave turbulence decays further, $\ell \rightarrow L$ with continued decay of energy. Consequently, $\mathrm{Re}_{\ell}$ also begins to decay.

\subsection{Kolmogorov length scale}
For spectral energy $\widehat{E}_k\sim k^{-2}$ over the intermediate range of wavenumbers, $k\in (k_E, k_\delta)$
(cf. figures~\ref{fig: LengthScalesSummary_1} and~\ref{fig: LengthScalesSummary_2}), the Taylor microscale can be estimated as, 
\begin{equation}
\lambda\sim \frac{1}{\sqrt{k_E k_\delta}},
\label{eq: TaylorMicroscaleApp}
\end{equation}
utilizing the Eq.~\eqref{eq: lengthscales_centroids}. Equation~\eqref{eq: TaylorMicroscaleApp} shows that $\lambda$, despite being a dissipative scale, is not the smallest scale generated due to the energy cascade.
Analogous to the hydrodynamic turbulence, we define the Kolmogorov length scale $\eta$~\cite{tennekes1972first} as the smallest length scale generated as a result of the acoustic energy cascade. The length scale $\eta$ can be approximated by the balance of nonlinear steepening and energy dissipation, i.e.,
\begin{equation}
\frac{{A}^2_{\mathrm{rms}}}{\eta} \sim \delta\frac{A_{\mathrm{rms}}}{\eta^2}, \quad \eta \sim \frac{\delta}{{A}_{\mathrm{rms}}},
\label{eq: KolomogorovScale}
\end{equation}
where $A_{\mathrm{rms}}$ is defined in Eq.~\eqref{eq: PcDef}. Figures~\ref{fig: LengthScalesSummary_1} and~\ref{fig: LengthScalesSummary_2} illustrates the integral length scale $\ell$ and the Kolmogorov length scale $\eta$ in a typical AWT field. Visual inspection indicates $\ell \gg \eta$ which is as expected. We note that $\eta$ and $1/k_{\delta}$ evolve in time similarly, differing only by a constant value. For AWT, this is immediately understandable since, Eq.~\eqref{eq: TaylorMicroscaleApp} shows that $k_\delta \sim t^{-1/3}$ and Eq.~\eqref{eq: KolomogorovScale} shows that $\eta \sim t^{1/3}$ which implies $k_\delta \eta$ remains constant when the energy decays. 
\begin{figure}
\centering
\includegraphics[width=0.7\linewidth]{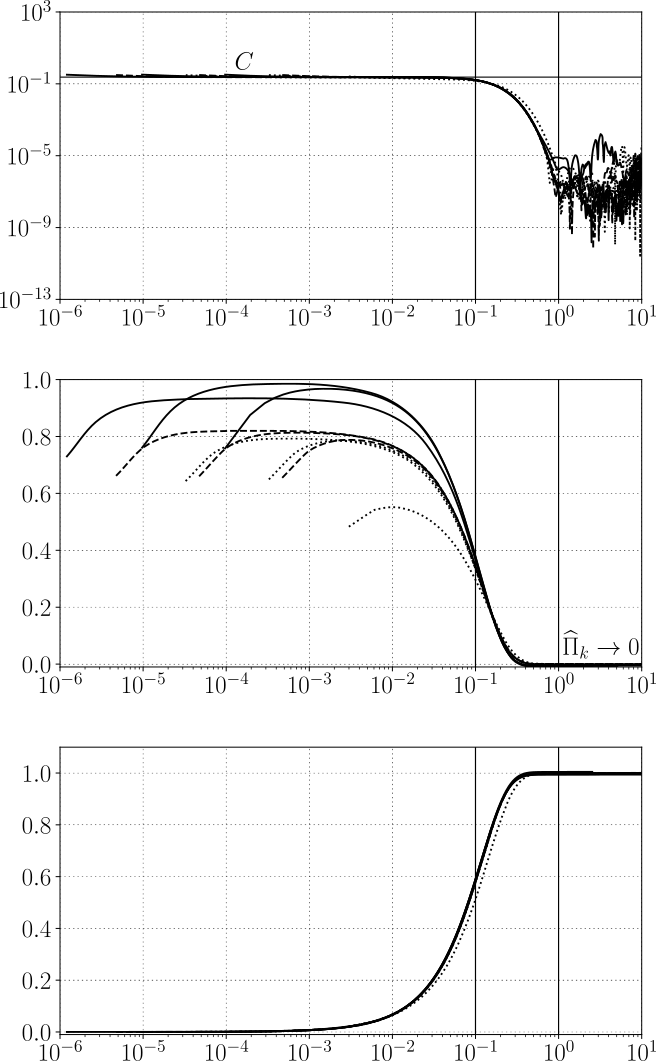}
\put(-290,470){$(a)$}
\put(-290,310){$(b)$}
\put(-290,150){$(c)$}
\caption{{\color{black}{Fully developed spectra of compensated energy $(a)$, spectral energy flux $(b)$, and cumulative dissipation $(c)$ for TW at time instant $t_0\approx0.03$. Harmonics with wavenumbers such that $k\eta < 1$ contain all the energy. The spectral energy flux vanishes at $k\eta \approx 1$ thus indicating numerical resolution of all the energy containing harmonics. The marked regime $0.1<k\eta<1$ signifies the dissipation range. The constant $C\approx 0.075$. (--) $A_{\mathrm{rms},0}=10^{-1}$; ($--$) $A_{\mathrm{rms},0}=10^{-2}$; ($\cdots$) $A_{\mathrm{rms},0}=10^{-3}$}}}
\label{fig: SpectralFlux_Traveling}
\end{figure}
For TW and SW, the spectral energy given by Eq.~\eqref{eq: SpectralEnergyAssump} corresponds to the degenerate case of $k_0 = k_E = 1$. For such a form of spectral energy, the energy evolution (cf. Eq.~\eqref{eq: EdotLdot}) changes to, 
\begin{equation}
 \frac{d\left\langle E^{(2)}\right\rangle}{dt} = \frac{1}{k_0}\frac{dC_2}{dt}\left(1 - \frac{k_0}{k_\delta}\right) + \frac{C_2}{k^2_\delta}\frac{dk_\delta}{dt}.
 \label{eq: dEdt1}
\end{equation}
As shown in figure~\ref{fig: EnergyDecayAll}, the Taylor microscale $\lambda \rightarrow \sqrt{\delta t}$. Consequently, for $k_E = k_0$ constant, Eq.~\eqref{eq: TaylorMicroscaleApp} shows that $k_\delta \sim t^{-1}$. Equation~\eqref{eq: dEdt1} shows that the decay of perturbation energy is due to decay in $C_2$ and $k_{\delta}$. Our numerical results (cf.~figure~\ref{fig: EnergyDecayAll}) show that for TW and SW, $\left\langle E^{(2)} \right\rangle \sim t^{-2}$ which suggests that $C_2\sim t^{-2}$ for $k_\delta\gg 1$ from Eq.~\eqref{eq: dEdt1}. Hence, the compensated energy spectrum $k^2\widehat{E}_k\sim t^{-2}$ for both TW and SW indicating that dissipation $\widehat{\mathcal{D}}_k$ remains active over all the length scales $k>k_0$ while the energy decays. 

Equation~\eqref{eq: KolomogorovScale} shows that the Reynolds number based on the Kolmogorov length scale or the shock thickness $Re_{\eta} = \eta \mathrm{Re}_L$ remains constant in time, 
{\color{black}{
\begin{equation}
 Re_{\eta} = \frac{\rho^*_0 a^*_0 L^* \eta A_{\mathrm{rms}}}{\mu^*_0} = \frac{4}{3} + \frac{\gamma - 1}{Pr}.
\end{equation}
}}
Above relation shows that $Re_{\eta} = \mathcal{O}\left(1\right)$ indicating that $\eta$ is the length scale at which diffusion dominates the nonlinear wave steepening.
\section{Scaling of spectral quantities}
\label{sec: SpectralScaling}

In this section, we discuss the variation and scaling of the energy $\widehat{E}_k$, the spectral energy flux $\widehat{\Pi}_k$, and the cumulative dissipation $\sum_{k'<k} \widehat{\mathcal{D}}_{k'}$ for high amplitude TW, SW, and AWT cases utilizing the length scale analysis presented in the previous sections. We show that the spectral energy $\widehat{E}_k$ and the cumulative dissipation $\sum_{k'<k} \widehat{\mathcal{D}}_{k'}$ for all the cases can be collapsed on to a common structure versus the reduced wavenumber $k\eta$ however, the flux $\widehat{\Pi}_k$ lacks such a universality. 

 As discussed in previous section (cf. Eq.~\eqref{eq: dEdt1}), the decay of total energy $\left\langle E^{(2)}\right\rangle$ and dissipation rate $\epsilon$ for TW is given by, 
\begin{equation}
\left\langle E^{(2)}\right\rangle \sim t^{-2},~\mathrm{and}~\epsilon\sim t^{-3},
\label{eq: EnergyTimeTraveling}
\end{equation}
which are well known results for the Burgers' equation as well~\cite{Bec_PhyRep_2007}.

While the results in Eq.~\eqref{eq: EnergyTimeTraveling} are well known, we note that such power law decay results in a universally constant structure of shock waves in the spectral space, as shown in Figs.~\ref{fig: SpectralFlux_Traveling} and~\ref{fig: SpectralFlux_Standing}. Utilizing the estimate of Kolmogorov length scale $\eta$ given in Eq.~\eqref{eq: KolomogorovScale}, the energy dissipation rate $\epsilon$ and the Kolmogorov length scale $\eta$ can be related as,
\begin{equation}
\epsilon \sim \frac{{A_{\mathrm{rms}}^3}}{\ell},~\mathrm{and}~ \eta \sim \frac{\delta}{\left(\epsilon \ell\right)^{1/3}}.
\end{equation}
{\color{black}{Hence, the energy spectrum $\widehat{E}_k$ can be written in the following collapsed form (figure~\ref{fig: SpectralFlux_Traveling}$a$).
\begin{equation}
\widehat{E}_kk^{2}\epsilon^{-2/3}\ell^{1/3} \sim C F(k\eta).
\label{eq: SpectralForm}
\end{equation}
In Eq.~\eqref{eq: SpectralForm}, the integral length scale $\ell$ is used for making the left hand expression dimensionless}}. For TW and SW, the integral length scale $\ell$ remains constant by definition ($\ell = L$). Hence, $C$ in Eq.~\eqref{eq: SpectralForm} is constant and can be attributed to the Kolmogorov's universal equilibrium theory for hydrodynamic turbulence. $F(.)$ is a function which decays as the reduced wavenumber $k\eta$ increases to 1. From the numerical simulations for cases listed in Table~\ref{tab: test_cases} we obtain,
\begin{equation}
C \approx 0.075.
\label{eq: ConstantValue}
\end{equation} 

\begin{figure}[!thbp]
\centering
\includegraphics[width=0.7\linewidth]{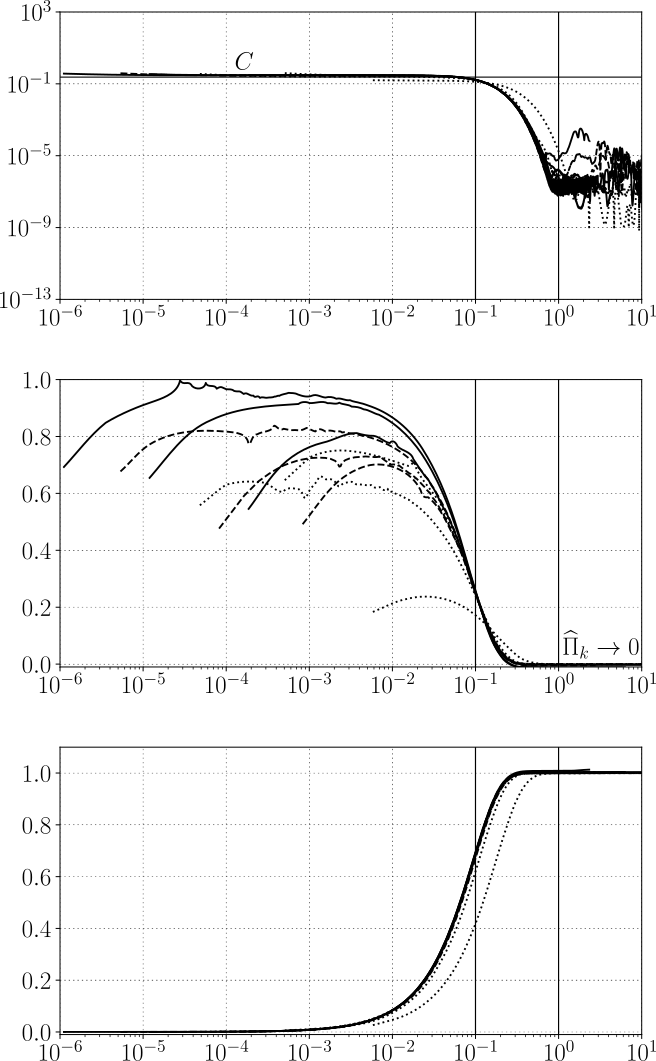}
\put(-290,470){$(a)$}
\put(-290,310){$(b)$}
\put(-290,150){$(c)$}
\caption{Fully developed spectra of compensated energy $(a)$, spectral energy flux $(b)$, and cumulative dissipation $(c)$ for SW averaged over one time cycle after $t_0\approx0.04$. Harmonics with wavenumbers $k\eta < 1$ contain all the energy. The spectral energy flux vanishes at $k\eta \approx 1$ thus indicating numerical resolution of all the energy containing harmonics. The constant $C\approx 0.075$. (--) $A_{\mathrm{rms},0}=10^{-1}$; ($--$) $A_{\mathrm{rms},0}=10^{-2}$; ($\cdots$) $A_{\mathrm{rms},0}=10^{-3}$}
\label{fig: SpectralFlux_Standing}
\end{figure}

Scaling of $\widehat{\Pi}_k$ with the energy dissipation rate $\epsilon$ shows the relative magnitude of spectral energy flux compared to the energy dissipation.{\color{black}{ For increasing Reynolds numbers $\mathrm{Re}_L$, we note that $\widehat{\Pi}_k/\epsilon$ increases but still remains less than 1 in the energy transfer and dissipation range, as shown in figure~\ref{fig: SpectralFlux_Traveling}$(b)$. This highlights the primary difference between energy spectra of nonlinear acoustic waves and hydrodynamic turbulence, in which, the energy transfer range does not exhibit viscous dissipation~\cite{pope2000turbulent}. However, in nonlinear acoustics, the dissipation occurs over all the smaller length scales which do not contain energy initially (figure~\ref{fig: LengthScalesSummary_2}$(b)$). Moreover, for $k\eta\approx 0.1$, the flux $\widehat{\Pi}_k$ rapidly approaches to zero. In the regime $k\eta>0.1$, scaled cumulative dissipation $\sum_{k'<k} \widehat{\mathcal{D}}_{k'}/\epsilon\rightarrow 1$ as $k\eta\rightarrow 1$.}}
\begin{figure}[!thbp]
\centering
\includegraphics[width=0.7\linewidth]{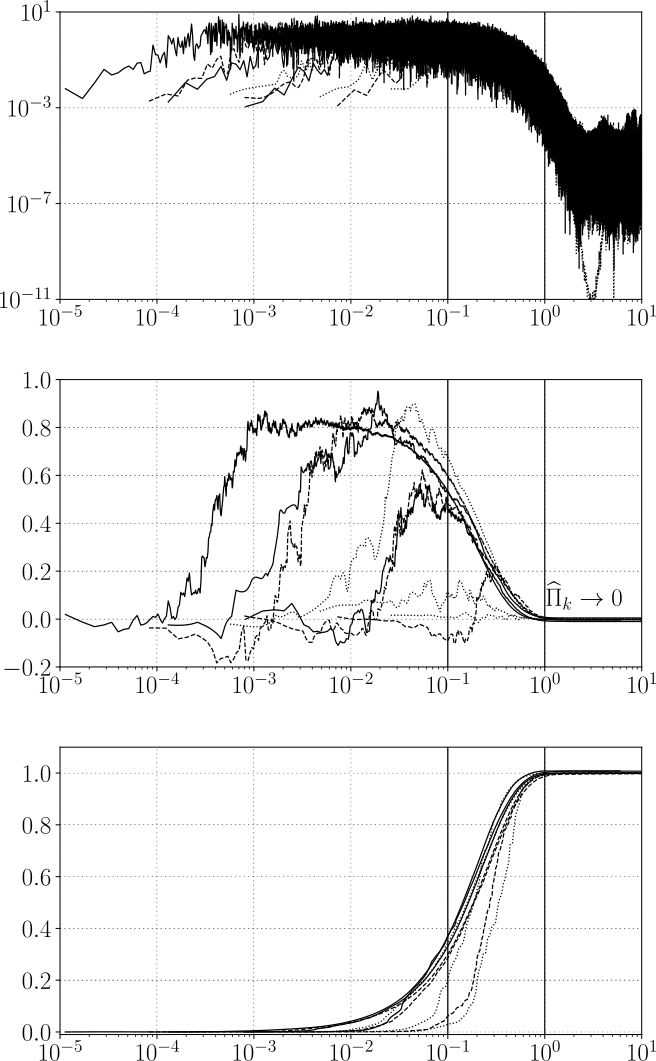}
\put(-290,470){$(a)$}
\put(-290,310){$(b)$}
\put(-290,150){$(c)$}
\caption{Fully developed spectra of compensated energy, $(a)$, spectral energy flux $(b)$, and cumulative dissipation $(c)$ against scaled wavenumber $k\eta$ for the randomly initialized broadband noise (AWT) cases with $A_{\mathrm{rms},0}$ and $\nu_0$ listed in Table~\ref{tab: test_cases} at dimensionless time $t/\tau = \tau_0\approx 6\times10^{-4}$. The marked regime $0.1<k\eta<1$ signifies the dissipation range. (--) $A_{\mathrm{rms},0}=10^{-1}$; ($--$) $A_{\mathrm{rms},0}=10^{-2}$; ($\cdots$) $A_{\mathrm{rms},0}=10^{-3}$}
\label{fig: SpectralEnergyTurbulence}
\end{figure}
Such functional forms of spectral energy, spectral energy flux, and cumulative dissipation can also be realized for the SW case. At later times, the nonlinear evolution results in two opposite traveling shock waves which collide with each other twice in one time period. Such collisions cause instantaneous peaks in the dissipation rate $\epsilon$ and corresponding oscillations in the Taylor microscale $\lambda$, as shown in figure~\ref{fig: EnergyDecayAll}. However, the total energy $\left\langle E^{(2)} \right\rangle$ decays monotonically by definition. In the spectral space, such collisions generate periodic oscillations in the spectral energy flux $\widehat{\Pi}_k$, as shown in figure~\ref{fig: EnergyFLux}. Averaging over one such time cycle yields the energy spectra forms similar to that for TW, as shown in figure~\ref{fig: SpectralFlux_Standing}. Such cycle averaging is allowed since the total energy $\left\langle E^{(2)}\right\rangle$ and the dissipation rate $\epsilon$ decay such that averaged behavior is identical to the one of traveling waves. Furthermore, the value of the constant $C$ is identical for SW.{\color{black}{ We further note that for the case with lowest Reynolds number $\mathrm{Re}_L$ ($\nu_0 = 1.836\times10^{-5}$ and $A_{0,\mathrm{rms}} = 10^{-3}$), the spectra exhibit energy for $k\eta > 1$ (figure~\ref{fig: SpectralFlux_Standing}$(a)$ since the Eq.~\eqref{eq: KolomogorovScale} underpredicts $\eta$. This suggests that the nonlinear spectral energy transfer is small compared to the spectral dissipation, as shown by figure~\ref{fig: SpectralFlux_Standing}$(b)$.}}

As discussed in previous sections, the decay phenomenology of AWT is different from that of TW and SW. Typical acoustic field $\uFluct(x,t), \pFluct(x,t)$ for a randomly initialized perturbation at a time after shock formation is shown in figure~\ref{fig: LengthScalesSummary_1}$(a)$. The velocity field corresponds to randomly positioned shocks connected with almost straight slant lines (expansion waves) and the pressure field with identical distribution of shocks but connected with horizontal lines. Shocks traveling in the same direction collide inelastically and coalesce, while those traveling in opposite directions pass through. As discussed previously, the integral length scale $\ell$ defines the average distance between the adjacent shock traveling in the same direction. Due to gradual coalescence of the shocks, $\ell$ increases in time. Moreover, as $t\rightarrow \infty$, it is obvious that two opposite traveling shocks remain in the domain and $\ell \rightarrow L$. We note that such behaviour is similar to the Burgers' turbulence~\cite{Kida_1979_JFM}. Figure~\ref{fig: SpectralEnergyTurbulence} shows the fully developed compensated spectra at scaled dimensionless time $t/\tau = t_0\approx 6\times10^{-4}$. For AWT, the compensated energy spectrum $\widehat{E}_kk^{2}\epsilon^{-2/3}\ell^{1/3}$ defined in Eq.~\eqref{eq: SpectralForm} does not remain constant in the energy transfer range of wavenumbers due to decay laws of energy and dissipation derived in the previous section.{\color{black}{ Moreover, for lowest $\mathrm{Re}_L$ case, the spectra exhibit energy for $k\eta>1$ (figure~\ref{fig: SpectralEnergyTurbulence}$(a)$) due to underprediction of $\eta$ obtained via balancing of nonlinear wave propagation and thermoviscous dissipation effects. The dissipation acts at large length scales also in the lowest $\mathrm{Re}_L$ case. Consequently, the spectral energy flux $\widehat{\Pi}_k$ is very small compared to dissipation $\epsilon$ and the length scale $\eta$ is primarily governed by diffusion only.}}

\section{Concluding remarks}
We have studied the spectral energy transport and decay of finite amplitude planar nonlinear acoustic perturbations governed by fully compressible 1D Navier-Stokes equations through shock-resolved direct numerical simulations (DNS) focusing on propagating single harmonic traveling wave (TW), standing wave (SW), and randomly initialized Acoustic Wave Turbulence (AWT). The maximum entropy perturbations scale as $\pFluct^2$ for normalized pressure perturbation $\pFluct \sim \mathcal{O}\left(10^{-3} - 10^{-1}\right)$. Consequently, the second order nonlinear acoustic equations are adequate to derive physical conclusions about spectral energy transfer in the system. Utilizing the second order equations, we derived the analytical expression for corrected energy corollary for finite amplitude acoustic perturbations yielding infinite order correction term in the perturbation energy density. We have shown that the spatial average of the corrected perturbation energy density can be classified as a Lyapunov function for the second order nonlinear acoustic system with strictly monotonic behaviour in time. 

Utilizing the corrected energy corollary, we derived the expressions for spectral energy, spectral energy flux, and spectral dissipation, analogous to the spectral energy equation studied in hydrodynamic turbulence. Utilizing the spectral expressions, we performed theoretical study of three possible length scales characterizing a general nonlinear acoustic system, namely, the integral length scale $\ell$, the Taylor microscale $\lambda$, and the Kolmogorov length scale $\eta$. 

In traveling waves (TW) and standing waves (SW), $\ell$ remains constant in the decaying regime. Spatial average of perturbation energy decays as $\left\langle E^{(2)}\right\rangle\sim t^{-2}$ and dissipation rate as $\epsilon \sim t^{-3}$ in time. The Kolmogorov scale increases linearly in time ($\eta\sim t$) in the decaying regime. Moreover, the spectral energy for both traveling and standing waves assumes the self-similar form: $\widehat{E}_k k^2 \epsilon^{-2/3}\ell^{1/3}\sim 0.075 f(k\eta)$. 

In acoustic wave turbulence (AWT), due to gradual increase of the integral length scale $\ell$ caused by the shock coalescence, the approximate decay laws are $\left\langle E^{(2)}\right\rangle \sim t^{-2/3}$ and $\epsilon \sim t^{-5/3}$, similar to the Burgers' turbulence~\cite{Burgers1974nonlinear}. While, various cases for AWT qualitatively collapse with the scaling $\widehat{E}_k k^2 \epsilon^{-2/3}\ell^{1/3}$, quantitative scaling can only be obtained utilizing a statistically stationary ensemble of shock waves combined with random forcing, which falls beyond the current scope. 
%%%%%%%%%%%%%%%%%%%%%%%%%%%%%%%%%%%%%%%%%%%%%%%%%%%%%%%%%%%%%%%%%%%%%%%%%%%%%%%%%%%%%%%%%%%\textsl{•}

% %%%%%%%%%%%%%%%%%%%%%%%%%%%%%%
\chapter{Thermoacoustic nonlinear energy cascade}
The contents of this chapter were previously published by Gupta, Lodato, Scalo in \emph{Journal of Fluid Mechanics}~\cite{GuptaLS_JFM_2017} and have been reported here in abridged form with minor modifications.

\section{Introduction} \label{sec: Intro2}
In previous chapter, we discussed the spectral energy cascade and decay in initialized nonlinear acoustic waves. In this chapter, we discuss thermoacoustically sustained spectral energy cascade in a canonical thermoacoustically unstable resonator exhibiting high amplitude (macrosonic) thermoacoustic waves at limit cycle. A comprehensive nonlinear theoretical and high-fidelity modeling approach is adopted to accurately describe macrosconic thermoacoustic waves. The travelling-wave looped resonator, inspired by Yazaki \emph{et al.}~\cite{Yazaki_PhysRevLet_1998}'s experimental setup but geometrically optimized via linear theory developed by Rott ~\cite{Rott_ZAMP_1969,Rott_ZAMP_1973,LinSH_JFM_2016}, has been designed to maximize the growth rate of the quasi-travelling-wave second harmonic and thus achieve rapid shock wave formation. Yazaki \emph{et al.}'s~\cite{Yazaki_PhysRevLet_1998} looped configuration allows quasi-travelling-wave acoustic phasing which facilitates faster nonlinear energy cascade compared to standing wave resonators~\cite{BiwaEtAl_JASA_2014}. It is shown that the energy content in spectral domain resembles the equilibrium energy cascade observed in turbulence, similar to the spectral energy distribution of an ensemble of acoustic waves interacting nonlinearly among each other~\cite{Nazarenko_2011_WT, Zakharov_2012_WT}. As demonstrated by the numerical simulation data and companion low-order nonlinear modeling, thermoacoustically sustained shock waves exhibit inter-scale energy transfer dynamics analogous to Kolmogorov's equilibrium hydrodynamic turbulent energy cascade~\cite{kolmogorov1941local}. Throughout, the results obtained via the proposed  nonlinear model are verified and compared with fully compressible high-fidelity Navier-Stokes simulations.

%% High-fidelity tools
The development of an accurate nonlinear thermoacoustic wave propagation theory and modeling framework warrants the support of high-fidelity numerical simulations. A high-order spectral difference numerical framework~\cite{kopriva1996conservative,kopriva1996conservative2,sun2007high,Jameson_JSC_2010} combined with an artificial Laplacian viscosity~\cite{persson:06} shock-capturing scheme has been adopted for the present study. Moreover, the computational setup has been reduced to a minimal-unit (or single-pore) configuration, as done by Rahman \emph{et al.}~\cite{RahmanEtAl_IntJHMT_2017}, to reduce the computational cost and ensure the maximum possible numerical resolution in the direction of shock propagation for a given number of discretization points, or degrees of (numerical) freedom. In spite of this choice, full resolution of the propagating shocks was still not attainable with the available resources.

%%%%%%%%%%%%%%%%%%%%%%%%%%%%%%%%%%%%%%%%%%%%%%%%%%%%%%%%%%%%%%%%%%%%%%%%%%%%%%%%%%%%%%%%%%%
\section{Problem Formulation} \label{sec: ComputationalSetup}

\subsection{Design of the minimal-unit model}
\label{sec: optimization}

% Brief description of the setup
The proposed computational setup (figure~\ref{fig:computational_setup}, top) is a straight, two-dimensional, axially periodic minimal-unit (or single-pore)  thermoacoustic device composed of four constant-area sections ($a$, $b$, $c$, and $d$). Such configuration represents an idealization of a looped thermoacoustic resonator (figure~\ref{fig:computational_setup}, bottom) similar---but not identical---to the one adopted by Yazaki \emph{et al.}~\cite{Yazaki_PhysRevLet_1998}. Adiabatic slip conditions are applied everywhere, with the exception of the thermoacoustic regenerator, or core (section $b$), where isothermal no-slip walls are used to impose a linear wall-temperature distribution $T_w(x)$ from the cold, $T_C$, to the hot side temperature, $T_H$, resulting in the base temperature distribution, $T_0(x)$. A body force is added to suppress Gedeon streaming (Section ~\ref{sec:gov_equations}), which would otherwise cause convective heat transport away from the hot end of section $b$ and require the introduction of a thermal buffer tube and a secondary ambient heat exchanger~\cite{PeneletGLB_PhysRevE_2005}, thus introducing further complications in the proposed canonical setup (discussed in Section ~\ref{sec:gov_equations}). The resulting relaxation of base state from $T_H$ to $T_C$ outside the regenerator is due to molecular diffusion. Consequently, a diffusive thermal layer develops which is very thin compared to the acoustic wave length and is neglected in the subsequent analysis.
The uniform base pressure and cold-side values of density and temperature are set to be equal to the reference thermodynamic quantities $P_0=P_{\mathrm{ref}}$, $\rho_C=\rho_{\mathrm{ref}}$, $T_C=T_{\mathrm{ref}}$ (table~\ref{tbl:base_state}), chosen for air (Section ~\ref{sec:gov_equations}). Stacking of any number of thus-conceived single-pore models in the $y$ direction, i.e. preserving the area ratio and increasing the number of pores in the regenerator, would yield the same numerical results. 

% Justification of the setup
The minimal-unit choice is dictated by the need to maximize the numerical resolution in the propagation direction of the captured shocks. Although full resolution of the propagating shocks is still not computationally feasible due to the very large length of the setup (of the order of $1$\,m) compared to the typical shock thickness scale (of the order of $100\,\mu$m). Performing a fully resolved three-dimensional simulation of an equivalent experimental setup would be (even more so) unfeasible: for instance, the setup studied by Yazaki \emph{et al.}~\cite{Yazaki_PhysRevLet_1998} consists of two heat exchangers and a ceramic catalyst with approximately 1000 pores. By design, the minimal-unit configuration neglects the thermoviscous losses outside the regenerator that attenuate the thermoacoustic amplification. The two-dimensional effects of the curvature of the resonator walls are also neglected. The former become less important as the ratio of the pore diameter to resonator diameter and number of pores increases (42 and 1000, respectively, in Yazaki's setup), and the latter are only relevant for very small ratios of the curvature radius to acoustic wavelength ($\sim 0.3$ in Yazaki's setup, assuming curvature radius of $L/6$ where $L$ is the total length). In conclusion, performing a full three-dimensional simulation taking into account approximately 1000 thermoacoustic pores, viscous losses in the resonator, and the curvature effects, would not lead to any significant additional insights into the physics of thermoacoustically generated shock waves, especially within the theoretical scope of the present study.

\begin{figure}[!t]
\centering
\includegraphics[width=0.85\textwidth]{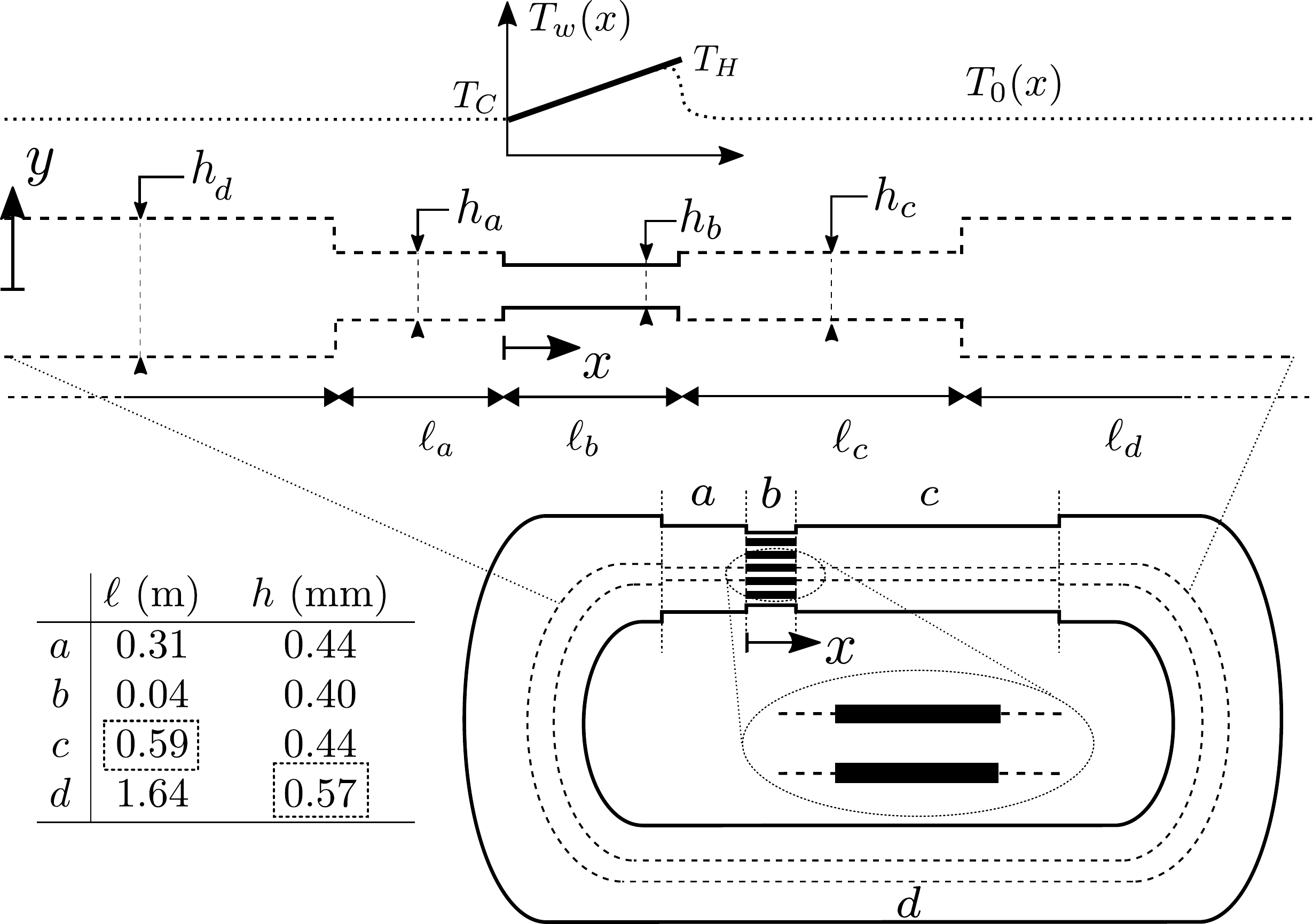}
  \caption{Two-dimensional axially periodic computational setup for minimal-unit simulations (top, not to scale), geometrical parameters (bottom left), and qualitative illustration of the equivalent experimental build-up of a variable-area looped thermoacoustic resonator (bottom right). Top figure: (-- --), adiabatic/sip conditions; (--), isothermal/no-slip conditions. The minimal unit is traced in the bottom right with dashed lines (-- --). The listed geometrical parameters provide sub-optimal growth rates across all the values of $T_H$, the boxed quantities have been determined via optimization (cf.~figure~\ref{fig:optimization_lst} and table~\ref{tbl:base_state}).}
\label{fig:computational_setup}
\end{figure}

\begin{table}[!b]
\centering
\caption{Thermodynamic parameters for base state. Base pressure and cold-side values of density and temperature are set to reference values: $P_0=P_{\mathrm{ref}}$, $\rho_{c}=\rho_{\mathrm{ref}}$, $T_C=T_{\mathrm{ref}}$.}
  \begin{tabular}{M{1.7cm}M{2.0cm}M{1.3cm}M{2.4cm}}
 \hline
$P_0$ & $\rho_C$ & $T_C$ & $T_H$ \\
 \multirow{2}{*}{101325\,Pa} & \multirow{2}{*}{1.176 kg/m$^3$} & \multirow{2}{*}{300\,K} & 400\,K, 450\,K, \\
&   &  & 500\,K, 550\,K \\
  \hline
\end{tabular}
 \label{tbl:base_state}
 \end{table}

% Now explain how you picked the specific geometry and what temperature settings you are going to run at.
The total length of the device is fixed to $\ell_a + \ell_b + \ell_c + \ell_d = 2.58~\mathrm{m}$, taken from the experimental setup of~\cite{Yazaki_PhysRevLet_1998}. The height of the regenerator has been chosen such that $h_b\sim2\delta_k$, where
\begin{equation}
\delta_k = \sqrt{\frac{2\nu}{\omega Pr}},
\end{equation}
$\nu$ is the kinematic viscosity, $Pr$ is the Prandtl number, and $\omega$ is the angular frequency of the unstable mode.
This results in a porosity $h_b/h_c=0.91$. The area ratio $h_d/h_c$ and the length $\ell_c$ have been chosen to yield sub-optimal values of growth rates across all hot-side temperature settings (figure~\ref{fig:optimization_lst}), assuring high enough thermoacoustic instability to achieve rapid shock wave formation. This optimization has been carried out with the system-wide numerical approach developed by Lin \emph{et al.}~\cite{LinSH_JFM_2016}. Further details of the linear stability analysis are given in~\cite{GuptaLS_JFM_2017}.

\subsection{Navier-Stokes calculations}

\subsubsection{Governing Equations} \label{sec:gov_equations}
Fully compressible Navier-Stokes simulations are carried out by solving the conservation laws for mass, momentum, and total energy in two dimensions, given by
\begin{subequations}
	\label{eq:navierstokes}
	\begin{align}
		\frac{\partial}{\partial t} \left(\rho\right) &+ \frac{\partial}{\partial x_j} \left(\rho u_j \right)  = 0,
		\label{subeq:ns1}
		\\
		\frac{\partial}{\partial t} \left(\rho u_i\right) &+ \frac{\partial}{\partial x_j} \left(\rho u_i u_j\right)  =  -\frac{\partial}{\partial x_i} p  +
		\frac{\partial}{\partial x_j} \tau_{ij} + \delta_{1i} f_{B},
		\label{subeq:ns2}
		\\
		\frac{\partial}{\partial t} \left(\rho \, E\right) &+ \frac{\partial}{\partial x_j} \left[ u_j \left(\rho \, E + p \right) \right] =
		\frac{\partial}{\partial x_j } \left(u_i \tau_{ij} - q_j\right),
		\label{subeq:ns3}
	\end{align}
\end{subequations}
respectively, where $x_i$($x_1$, $x_2$ or equivalently, $x$, $y$) are the axial and cross-sectional coordinates, $u_i$ are the velocity components in each of those directions, and $p$, $\rho$, $T$, and $E$ are the instantaneous pressure, density, temperature, and total energy per unit mass, respectively. 
%%%%%%%%%%%%%
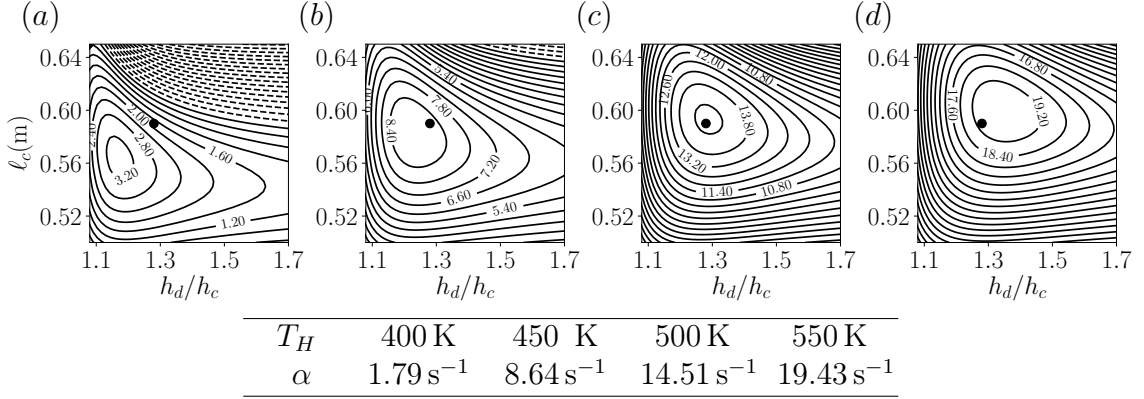
\begin{figure}[!b]
\centering
\includegraphics[width=\textwidth]{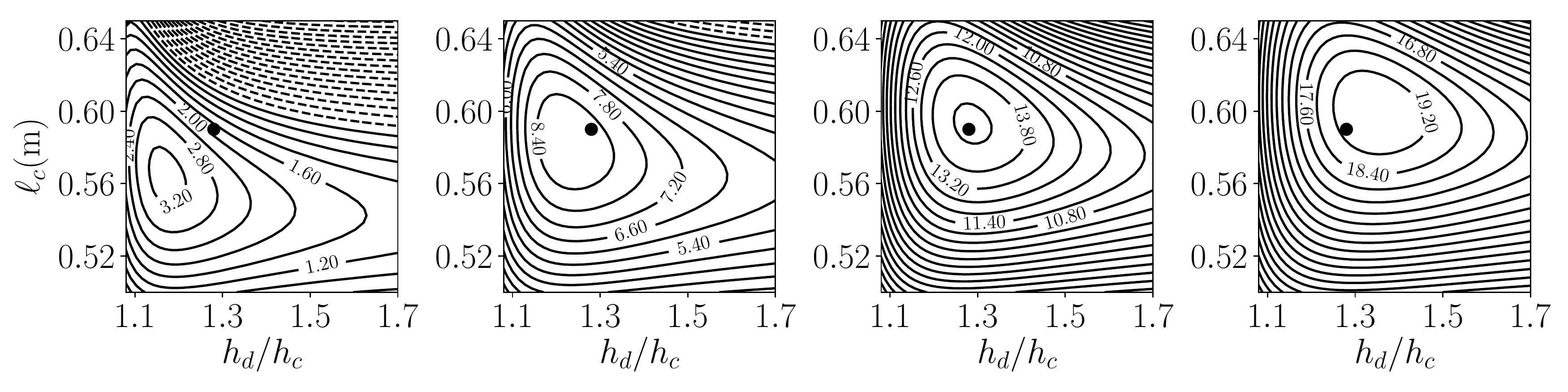}
\put(-425,110){$(a)$}
\put(-320, 110){$(b)$}
\put(-215,110){$(c)$}
\put(-110,110){$(d)$}\\
  \begin{tabular}{M{1cm}M{1.4cm}M{1.4cm}M{1.4cm}M{1.4cm}}
 \hline
$T_H$ & 400\,K & 450 \,K & 500\,K & 550\,K \\
\rule{0pt}{1ex}
$\alpha$ & 1.79\,$\mathrm{s}^{-1}$ & 8.64\,$\mathrm{s}^{-1}$ & 14.51\,$\mathrm{s}^{-1}$ & 19.43\,$\mathrm{s}^{-1}$\\
  \hline
\end{tabular}
  \caption{Iso-contours of thermoacoustic growth rate, $\alpha$, of the second harmonic versus resonator area ratio, $h_d/h_c$, and length $\ell_c$ (see figure~\ref{fig:computational_setup}) for hot-side temperatures $T_H=400~\mathrm{K}$ ($a$), $T_H=450~\mathrm{K}$ ($b$), $T_H=500~\mathrm{K}$ ($c$), and $T_H=550~\mathrm{K}$ ($d$). (--), $\alpha > 0$; (- -), $\alpha < 0$; (\tikzcircle{2pt}), sub-optimal values $\ell_c = 0.59$ m and $h_{d}/h_{c} = 1.28$ chosen for present investigation (figure~\ref{fig:computational_setup}). Growth rate values for each $T_H$ at sub-optimal geometry are listed in the table.}
\label{fig:optimization_lst}
\label{tbl:growth_rates}
\end{figure}
Due to the propagation of finite amplitude nonlinear acoustic waves and the periodic nature of the setup, Gedeon streaming~\cite{Gedeon_Cryo_1995} is expected. The mean flow caused by the Gedeon streaming results in the transport of heat away from the hot side of the regenerator into the whole device affecting the mean temperature distribution outside the regenerator. Such thermal leakage is usually mitigated by a secondary cold heat exchanger and a thermal buffer tube to achieve steady-state conditions. Due to the inhomogeneous mean temperature outside the regenerator, such a setup would exhibit a very large design parameter space. Moreover, the mean flow caused by the streaming would further disperse the acoustic waves, affecting the wave steepening and thus delaying the steady state further. Such effects fall beyond the scope of the present investigation. Thus, acoustic streaming in the current work is purposefully suppressed. To this end, a uniform mean pressure gradient $f_{B}$ in the axial direction is dynamically adjusted to relax the net axial mass flow rate to zero. The relevant expression for $f_{B}$ is: 
\begin{equation}
f_B = \frac{\alpha}{\Delta t}\left(\dot{m} - \dot{m}_0\right), 
\end{equation}
where $\alpha = 0.3$ is a relaxation coefficient, $\Delta t$ is the time step, $\dot{m}$ is the instantaneous volume averaged mass flow rate, and $\dot{m}_0 = 0$ is the target value. The viscous stress tensor, $\tau_{ij}$, and heat flux, $q_{i}$, are formulated based on the Stokes and the Fourier laws as
\begin{subequations}
\label{eq:heatfluxes}
  \begin{eqnarray}
		\tau_{ij} = 2 \mu \left[S_{ij} - \frac{1}{3} \frac{\partial u_k}{\partial x_k} \delta_{ij} \right],\quad q_j &=& -\frac{\mu\,C_p}{Pr} \frac{\partial T}{\partial x_j} ,
		\label{subeq:hf2}
  \end{eqnarray}
\end{subequations}
respectively, where $S_{ij}$ is the strain-rate tensor, given by $S_{ij}=\frac{1}{2} \left(\partial u_j/\partial x_i + \partial u_i /\partial x_j \right)$, $Pr$ is the Prandtl number, and $C_p$ is the specific heat capacity at constant pressure. The working fluid is air, assumed to be calorically and thermally perfect. The dynamic viscosity, $\mu(T)$, is varied with the temperature according to the Sutherland's law, $\mu(T) = \mu_\textrm{ref}(T/T_{S,\textrm{ref}})^{1.5}(T_{S,\textrm{ref}}+S)/(T+S)$ where $S = 120$ K is the Sutherland constant, $T_{S,\textrm{ref}} = 291.15$ K and $\mu_\textrm{ref} = 1.827\times10^{-5}$~\,kg$\cdot$m$^{-1}\cdot$s$^{-1}$. The values of the other unspecified fluid parameters, valid for air, are $Pr = 0.72$, the ratio of isobaric to isochoric specific heat capacities $\gamma=1.4$, the reference density $\rho_{\textrm{ref}} = 1.176$~\,kg$\cdot$m$^{-3}$, the pressure $p_{\textrm{ref}} = 101\,325$~\,Pa, the temperature $T_{\textrm{ref}}=300$~\,K, and the gas constant $R=p_{\textrm{ref}}/(\rho_{\textrm{ref}}\,T_{\textrm{ref}})$.
\begin{figure}
  \centerline{\includegraphics[width=1.0\textwidth]{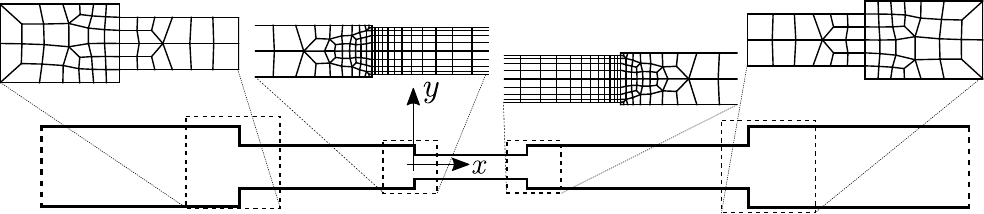}}
  \caption{Illustration of selection portions of the unstructured mesh near duct interfaces for Grid A with total number of elements $N_{\mathrm{el}}=2004$.}
\label{fig:mesh}
\end{figure}

%\begin{table}[!t]
%\centering
%\hspace*{-0.2cm}
%\begin{tabular}{c}
%\multicolumn{1}{c}{} \\
%\rule{0pt}{4.9ex} 
%\\
%$p=3$\\
%$p=5$ \\
%$p=7$ \\
%\end{tabular}
%\begin{tabular}{ccc}
%\multicolumn{3}{c}{Grid A ($N_{\mathrm{el}}=2004$) } \\
%\rule{0pt}{3.8ex}  
%$N_{\mathrm{dof}}$ & ${\Delta_h}_{\mathrm{(min)}}$ & ${\Delta_h}_{\mathrm{(max)}}$ \\
%\hline
%18036 & 7.33 $\mu$m & 0.63 mm\\
%50100 & 4.40 $\mu$m & 0.37 mm\\
%98196 & 3.14 $\mu$m & 0.26 mm\\
%\hline
%\end{tabular}\quad\vline\quad
%\begin{tabular}{ccc}
%\multicolumn{3}{c}{Grid B ($N_{\mathrm{el}}=7087$) } \\
%\rule{0pt}{3.8ex}  
%$N_{\mathrm{dof}}$ & ${\Delta_h}_{\mathrm{(min)}}$ & ${\Delta_h}_{\mathrm{(max)}}$ \\
%\hline
%63783 & 3.67 $\mu$m & 0.52 mm\\
%177175 & 2.20 $\mu$m & 0.31 mm\\
%347263 & 1.57 $\mu$m & 0.22 mm\\
%\hline
%\end{tabular}\quad
% \caption{Discretization order used per element, $p$, total number of elements, $N_{\mathrm{el}}$, degrees of freedom, $N_{\mathrm{dof}} = p^2 N_{\mathrm{el}}$, and minimum and maximum linear element size ${\Delta_h}_{\mathrm{(min)}}$, ${\Delta_h}_{\mathrm{(max)}}$ (cf.~(\ref{eq:linear_refinement_ratio})). Reported values are the ones used at the limit cycle. In all cases, simulations are carried out with $p=3$ and Grid A throughout the transient.}
% \label{tab:GridParams}
%\end{table}
The fully compressible high-fidelity Navier-Stokes calculations have been carried out with the discontinuous finite element \textsc{sd3DvisP} solver, an MPI parallelized Fortran 90 code employing the spectral difference local spatial reconstruction for hexahedral elements on unstructured grids~\cite{kopriva1996conservative,kopriva1996conservative2,sun2007high,Jameson_JSC_2010}. The solver reconstructs the local solution inside each element as the tensor product of polynomials up to the user-specified order $p = N_p-1$, where $N_p$ is the number of solution points per dimension inside the element. Inter-element discontinuities in the solution are handled utilizing the~\cite{roe1981approximate}'s flux with the entropy correction by~\cite{harten1983self}. The numerical dissipation at the element interfaces scales as $\Delta_h^{N_p+1}$ where $\Delta_h$ is the characteristic length scale of neighbouring elements~\cite{lodato:14b,ChapelierLJ_CF_2016}. Sub-cell shock capturing is enabled through a Laplacian artificial diffusion term applied in regions of steep gradients, which are detected by means of a modal sensor based on a Legendre polynomial expansion~\cite{persson:06,lodato:16}. The time integration is carried out explicitly with a 3rd order Runge-Kutta scheme and discretization order $p=3$ during the transient, whereas a 5th order Runge-Kutta scheme is adopted at the shock-dominated limit cycle. The same solver has been used and validated in a wide variety of flow configurations, including turbulent channel flow~\cite{lodato2013discrete, lodato2014structural} and unsteady shock-wavy wall interaction problems~\cite{lodato:16,lodato:17}. 

Results shown at shock-dominated limit cycle hereafter correspond to the finest computational grid simulated. The reader is referred to~\cite{GuptaLS_JFM_2017} for the detailed grid-sensitivity analysis. 

\section{Regimes of Thermoacoustic Amplification}
\label{sec: Regimes}

Three different regimes of thermoacoustic wave amplification can be identified by visual inspection of the pressure time series in figure~\ref{fig: Intro_signal}. We attempt a rigorous classification here based on the dimensionless collapse of the nonlinear growth regime of the spectral energy density (figure~\ref{fig:intro_spectral}), derived in more detail in Section ~\ref{sec: Spectral energy distribution}.

\begin{figure}[!t]
\centering
  \centerline{\includegraphics[width=\textwidth]{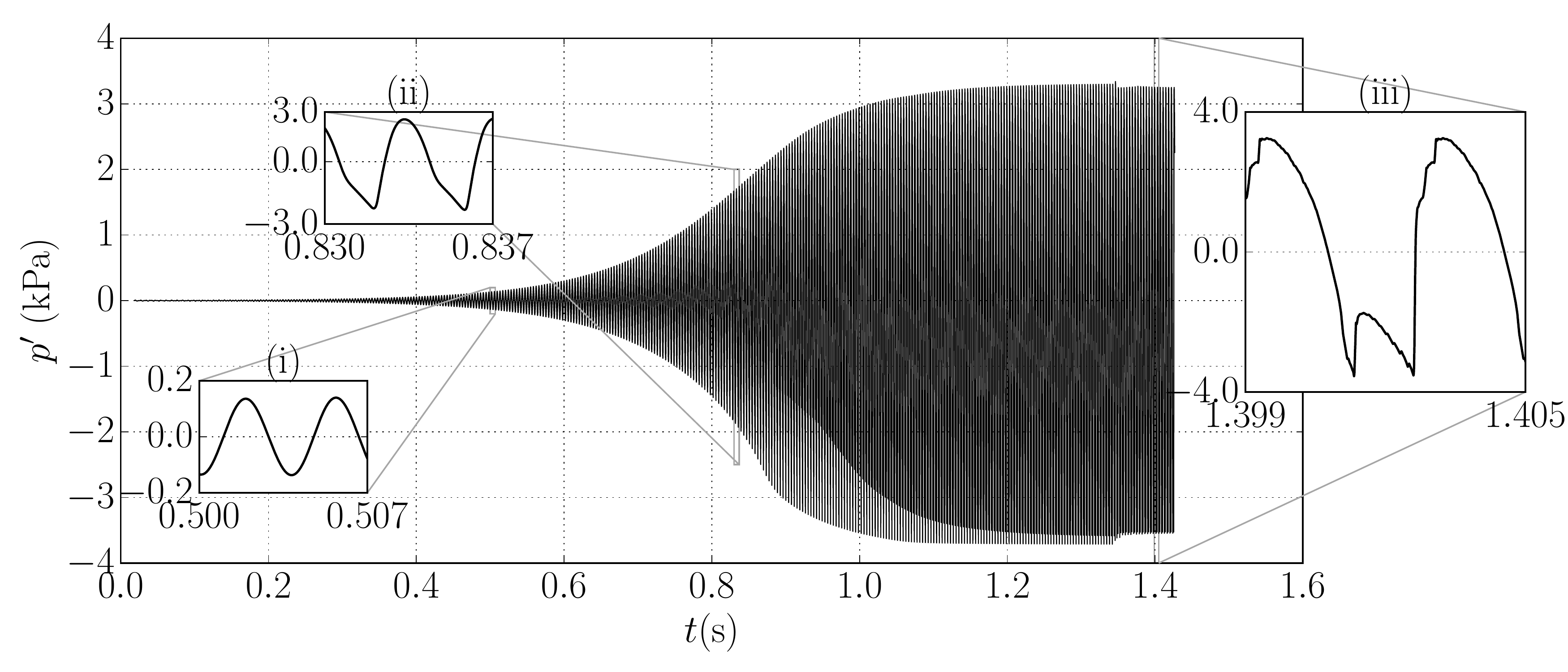}}
  \caption{Time series of pressure fluctuations at $x=1.54$ m (figure~\ref{fig:computational_setup}) for $T_H=$450 K with insets showing regimes ($i$) modal growth, ($ii$) hierarchical spectral broadening, and ($iii$) the shock-dominated limit cycle.}
\label{fig: Intro_signal}
\end{figure}
\subsection{Spectral energy density}
Any acoustic dynamical model (linear or nonlinear) can be written as 
\begin{equation}
\frac{\partial \mathbf{X}}{\partial t} = \mathbf{f}(\mathbf{X}),~~\mathrm{with}~~\mathbf{X} = \left(\frac{u'}{a_0}, \frac{p'}{\rho_0a^2_0}\right)^{\mathrm{T}},
\label{eq: AcousticDynamicalSystem}
\end{equation}
where $\mathbf{X}$ is the state vector containing the dimensionless perturbation variables. Throughout, we utilize the isentropic acoustic energy norm (see Section~\ref{sec: EnergyCorollary}), which is the squared $L_2$ norm, as the perturbation energy density~\cite{Naugol1975spectrum}:
\begin{equation}
E = \frac{1}{2}\rho_0a^2_0 \mathbf{X}^{\mathrm{T}}\mathbf{X} = \frac{1}{2}\rho_0u'^2 + \frac{p'^2}{2\rho_0a^2_0}.
\label{eq:energy_time}
\end{equation}
%The energy density of an isentropic one-dimensional perturbation is defined as 
%\begin{equation} \label{eq:energy_time}
% E = \frac{1}{2}\rho_0\,u'^2 + \frac{p'^2}{2\rho_0a^2_0}.
%\end{equation}
In the nonlinear growth and limit cycle regimes, velocity and pressure fluctuations, $u'$ and $p'$, respectively, are composed of higher harmonics of the linearly unstable mode. Substituting velocity and pressure fluctuations, expressed as complex Fourier expansions, 
\begin{align}
u'(x,t) = \sum_{\substack{k=-\infty\\k\ne0}}^{+\infty}\hat{u}_k(x,\epsilon t)\,e^{i\frac{k}{2}\omega_0t} ,& \quad p'(x,t) = \sum_{\substack{k=-\infty \\k\ne0}}^{+\infty} \hat{p}_k(x,\epsilon t)e^{i\frac{k}{2}\omega_0t},\\
\hat{u}_{-k} = \hat{u}^{*}_{k},& \quad \hat{p}_{-k} = \hat{p}^{*}_{k}, \nonumber
\end{align}
into Eq.~\eqref{eq:energy_time} and cycle averaging yields: 
\begin{equation}
\overline{E} = 2\sum_{k=1}^{\infty} E_k, \quad E_k = \frac{1}{2}\rho_0|\hat{u}_k|^2 + \frac{|\hat{p}_k|^2}{2\rho_0a^2_0}, 
\label{eq:energy_spectral}
\end{equation}
where $\epsilon\sim \alpha/\omega \ll 1$ is the smallness parameter such that $t$ and $\epsilon t$ correspond to fast and slow time scales, respectively, $\overline{(\cdot)}$ denotes the cycle averaging operator defined as
\begin{equation}
\overline{(\cdot)} = \frac{1}{T_0}\int_{\epsilon t}^{\epsilon t+T_0} (\cdot)dt, \quad T_0=\frac{2\pi}{\omega_0},
\end{equation}
and $(\cdot)^*$ denotes the complex conjugate. Here $\omega_0$ is the angular frequency of the unstable second harmonic, and $E_k$ is the spectral energy density of the $k^{\mathrm{th}}$ mode. The pressure and velocity amplitudes of the $k^{\mathrm{th}}$ harmonic ($|\hat{p}_k|$ and $|\hat{u}_k|$, respectively) are functions of the $x$ coordinate and the slow time $\epsilon t$, and are extracted via a short time-windowed Fourier transform (over 8 cycles of time period $T_0$) of the time series shown in figure~\ref{fig: Intro_signal}. In the nonlinear growth regime, the energy cascades from the unstable second mode ($k=2$) into its overtones only ($k=4, 6, 8, \dots$) with no energy content in the odd-numbered harmonics. 

\subsection{Regime classification}
\begin{figure}[!t]
\centering
  \centerline{\includegraphics[width=0.9\textwidth]{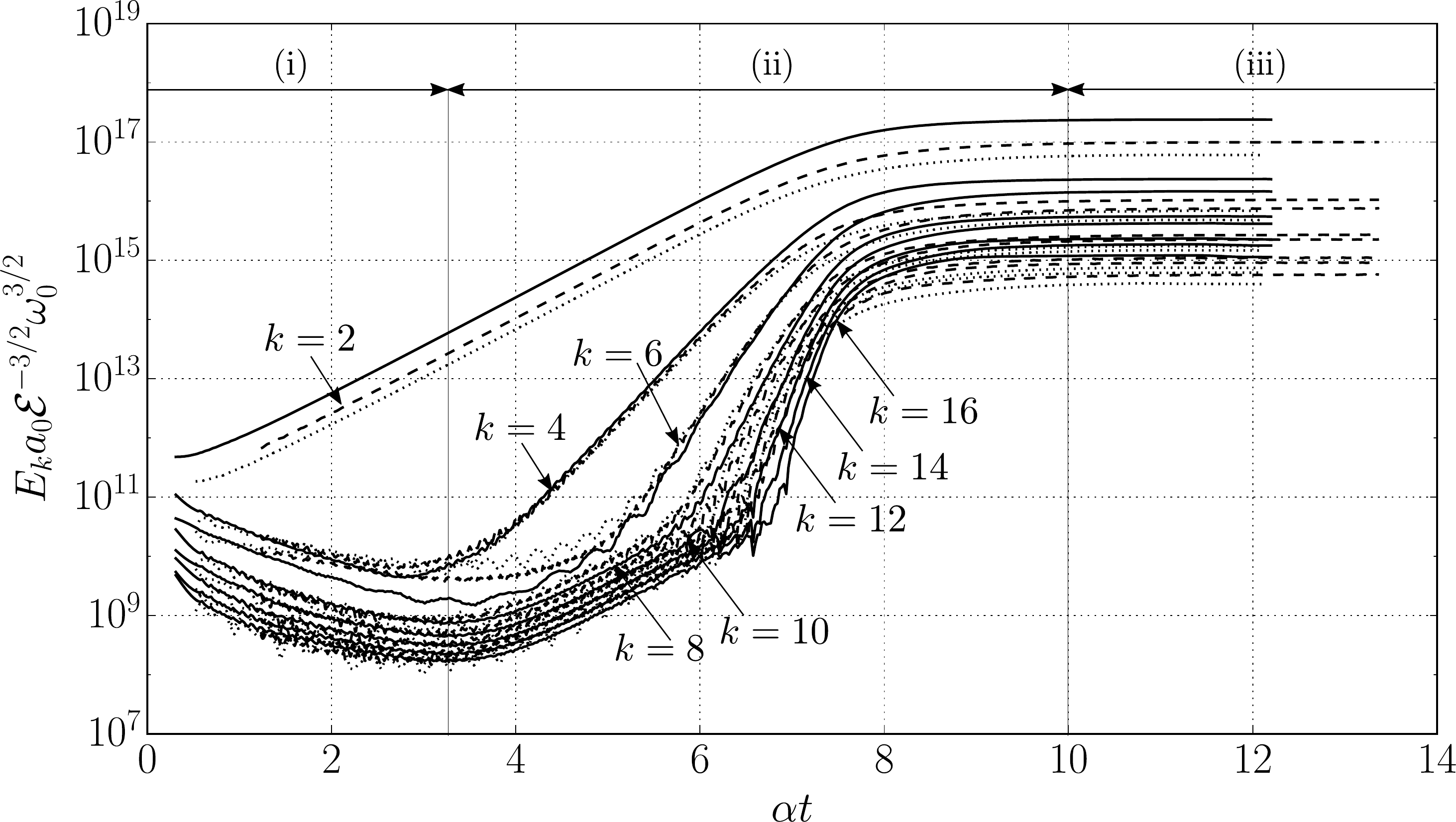}}
  \caption{Evolution of dimensionless spectral energy density of the unstable mode $k=2$ and its first seven overtones, scaled by the second mode angular frequency $\omega_0$, base speed of sound $a_0$, and rate of spectral energy transfer  $\mathcal{E}$ Eq.~\eqref{eq:dissipation_rate}. (--), $T_H=450~\mathrm{K}$; (- -), $T_H=500~\mathrm{K}$; ($\cdot \cdot \cdot$), $T_H = 550~\mathrm{K}$.}
\label{fig:intro_spectral}
\end{figure}
Based on a scale-by-scale analysis of the growth of spectral energy density (figure~\ref{fig:intro_spectral}), the aforementioned three regimes of thermoacoustic wave amplification are identified as:
\begin{enumerate}
 \item \textbf{Modal growth}: Only the thermoacoustically unstable mode amplifies exponentially and all the other modes excited by the initial perturbation field decay. Higher harmonics begin to grow after $\alpha t \approx 3.2$ (figure~\ref{fig:intro_spectral}), setting the end of a purely harmonic growth. In the modal growth regime, the system is well approximated by the linear system of equations.
 \item \textbf{Hierarchical spectral broadening}: Energy cascades down to higher harmonics hierarchically: the $k^{\mathrm{th}}$ harmonic ($k>2$) grows at a rate equal to $k/2$-times the modal growth rate of the second harmonic (see Section ~\ref{sec: NonlinearRegime_modeling} and Section ~\ref{sec: nonlinear_cascade}), $\alpha$, that is
 \begin{equation}
 \alpha_k = \alpha k/2, \quad\, k \in \{4,6,8, \dots \}.
 \end{equation}
 The saturation of the spectral energy density $E_k$ occurs at $\alpha t \approx 10$  followed by the formation of resonating shock waves at limit cycle.
 \item \textbf{Shock dominated limit cycle}: In this regime ($\alpha t>10$), the continued injection of energy in the second-mode harmonic is balanced by the cascade of the spectral energy density into the overtones of the second mode and terminates by viscous dissipation at very high overtones ($k\approx 300$). The maximum number of overtones generated is a function of the acoustic phasing of the unstable mode. Moreover, at the limit cycle, the spectral energy density scales with instability growth rate $\alpha$ approximately as $\alpha^3$ (Section ~\ref{sec: Spectral energy distribution}).
 %Maximum number of overtones generated is a function of both phasing and growth rate(s) of thermoacoustically unstable mode(s).
\end{enumerate}

The harmonic growth analysis corresponds to the linear stability analysis of the resonator by formulation of a system-wide differential eigenvalue problem. The reader is referred to~\cite{GuptaLS_JFM_2017} for further details on eigenvalue analysis and thermoacoustic energy budget analysis.

\section{Harmonic growth analysis} \label{sec: Linear}

In this section, the amplification of acoustic waves in the linear regime in the proposed minimal-unit model is discussed. A system-wide differential eigenvalue problem is formulated and solved numerically utilizing the strategy adopted by~\cite{LinSH_JFM_2016} in (Section ~\ref{sec:planar_thermoviscous_equations}). Utilizing the eigenvalue analysis, the computational setup has been optimized (Section ~\ref{sec: optimization}) and the acoustic energy budgets are derived (Section ~\ref{sec:Budgets}). This  provides analytical expressions for the cycle-averaged thermoacoustic production and dissipation, and elucidates the role of the acoustic phasing on thermoacoustic instability. Finally, the effects of varying the  hot-side temperature and the geometry on the thermoacoustic growth rates are discussed in~Section ~\ref{sec: GrowthrateEffect}. While the eigenvalue analysis is restricted to the minimal-unit model, the physical conclusions and analysis presented in this section hold for any thermoacoustically unstable device operating in the low-acoustic amplitude regime.

\subsection{Linear thermoviscous quasi-planar wave equations} \label{sec:planar_thermoviscous_equations} \label{sec: LinearRegime}

The time-domain linear thermoviscous governing equations for a two-dimensional perturbation are:
\begin{subequations}
\begin{eqnarray}
\frac{\partial \rho'}{\partial t} + \rho_0 \frac{\partial u'}{\partial x} + u'\frac{d\rho_0}{dx} + \rho_0\frac{\partial v'}{\partial y}=0,\label{eq: 1_mass}\\
\frac{\partial u'}{\partial t} + \frac{1}{\rho_0}\frac{\partial p'}{\partial x} - \nu_0 \frac{\partial^2 u'}{\partial y^2}=0,\label{eq: 1_mom}\\
\frac{\partial s'}{\partial t} + u'\frac{ds_0}{dx} - \frac{k_0}{\rho_0T_0}\frac{\partial^2 T'}{\partial y^2}= 0, \label{eq: 1_ent}
\end{eqnarray}
\label{eq: GoverningEqsLinear}
\end{subequations}
where primed variables $\left(\cdot\right)'$ represent the fluctuations in the corresponding quantities and the subscript $0$ denotes the base state.
Axial diffusion terms and the $y$-momentum equation in Eq.~\eqref{eq: GoverningEqsLinear} have been neglected based on the scaling analysis reported in Appendix~\ref{sec: appendixD}. Combining the cross-sectionally averaged Eq.~\eqref{eq: 1_mass} and Eq.~\eqref{eq: 1_ent} and cross-sectionally averaging Eq.~\eqref{eq: 1_mom}, accounting for isothermal and no-slip boundary conditions, yields 
\begin{subequations}
\begin{gather}
\frac{\partial p'}{\partial t} + \frac{\rho_0 a^2_0}{h}\frac{\partial U'}{\partial x} = \frac{\rho_0a^2_0}{h}\frac{q'}{\rho_0C_pT_0},\label{eq: thermoacoustic_time_a}\\
\frac{1}{h}\frac{\partial U'}{\partial t} + \frac{1}{\rho_0}\frac{\partial p'}{\partial x} = \frac{1}{h}\frac{\tau'_w}{\rho_0},\label{eq: thermoacoustic_time_b}
\end{gather}
\label{eq: thermoacoustic_time}
\end{subequations}
respectively, where $h$ denotes the cross-sectional width of the duct, $U'$ denotes the fluctuations in the flow rate,
\begin{equation}
U' = \int^{+h/2}_{-h/2}u'(x,y,t)dy,
\end{equation}
whereas $q'$ and $\tau'_w$ are the wall-heat flux and the wall-shear, respectively:
\begin{equation}
q' = 2k_0\left.\frac{\partial T'}{\partial y}\right|_{y=+h/2},\quad \tau'_w = 2\mu_0\left.\frac{\partial u'}{\partial y}\right|_{y=+h/2}.
\end{equation}
In sections $a, c,$ and $d$, linear wave propagation is assumed to be inviscid and adiabatic, hence isentropic, resulting in $q'=0$ and $\tau'_w=0$. Applying the normal mode assumption to Eq.~\eqref{eq: thermoacoustic_time}, namely,
\begin{equation}
p'(x,t) = \hat{p}(x)e^{\sigma t},\quad U'(x,t) = \hat{U}(x)e^{\sigma t},
\label{eq:normal_modes}
\end{equation}
where $\sigma = \alpha + i\omega$ is the complex eigenvalue of the system with growth rate $\alpha$ and angular frequency $\omega$, leads to the thermoviscous set of quasi-planar wave equations in the frequency domain~\cite{LinSH_JFM_2016}:
\begin{subequations}
\begin{gather}
\sigma\hat{p} = \frac{\rho_0 a^2_0}{h_b}\left[\frac{1}{1+\left(\gamma -1\right)f_k}\left(\frac{\Theta(f_k - f_{\nu})}{(1-f_{\nu})(1-Pr)} - \frac{d}{dx}\right)\right]\hat{U},~\Theta = \frac{1}{T_0}\frac{dT_0}{dx},\label{eq: eig_p}\\
\sigma\hat{U} = -\frac{h_b}{\rho_0}\left(1-f_{\nu}\right)\frac{d\hat{p}}{dx},\label{eq: eig_u}
\end{gather}
\label{eq: thermoacoustic_eig}
\end{subequations}
where the thermoviscous functions $f_\nu$ and $f_k$ are given by
\begin{equation}
f_{\nu} = \frac{\tanh(\eta h_b/2)}{\eta h_b/2},\quad f_k = \frac{\tanh(\eta h_b\sqrt{Pr}/2)}{\eta h_b\sqrt{Pr}/2},~~\mathrm{with}~~ \eta = \sqrt{i\omega/\nu_0}.
\label{eq: thermoviscous}
\end{equation}

\subsection{Acoustic energy budgets for quasi-planar wave perturbations} \label{sec:Budgets}
Multiplying Eq.~\eqref{eq: thermoacoustic_time_a} by $p'/(\rho_0a^2_0)$ and Eq.~\eqref{eq: thermoacoustic_time_b} by $\rho_0 U'/h_b$ and adding them, yields the conservation equation
\begin{equation}
\frac{\partial E}{\partial t} + \frac{\partial \mathcal{I}}{\partial x} = \mathcal{P}-\mathcal{D}, \label{eq: energy}
\end{equation}
for the one-dimensional acoustic energy density
\begin{equation} \label{eq:energy_planar}
E = \frac{1}{2}\frac{p'^2}{\rho_0a^2_0} + \frac{1}{2}\rho_0\left(\frac{U'}{h_b}\right)^2,
\end{equation}
consistent with the definition Eq.~\eqref{eq:energy_time}. The instantaneous acoustic flux $\mathcal{I}$ and the net energy production $\mathcal{P}-\mathcal{D}$ therein are given by
\begin{equation}
\mathcal{I} = \frac{p'U'}{h_b},\quad\mathcal{P}-\mathcal{D} = \frac{p'q'}{h_b\rho_0C_p T_0} + \frac{\tau'_w U'}{h_b^2}. 
\label{eq: energy_density_production}
\end{equation}
Averaging Eq.~\eqref{eq: energy} over one acoustic cycle and integrating axially over the periodic domain, $L$, yields:
\begin{equation}
\frac{d}{d(\epsilon t)}\int_L \overline{E}(x,\epsilon t) dx = \int_L \left(\overline{\mathcal{P}}-\overline{\mathcal{D}}\right) dx= \mathcal{R} ,
\label{eq: ModifiedRayleigh}
\end{equation}
where $\mathcal{R}$ is the Rayleigh index and $\epsilon t$ is the slow time scale (cf. Eq.~\eqref{eq:energy_spectral}). Relation Eq.~\eqref{eq: ModifiedRayleigh} allows to unambiguously identify the onset of an instability via the criterion $\mathcal{R}>0$. This expression also accounts for wall-shear and wall-heat flux losses outside the regenerator (if present), which attenuate the thermoacoustic instability. Such thermoviscous losses are captured in the heat flux $q'$ and shear stress $\tau'_w$ terms (cf. Eq.~\eqref{eq: energy_density_production}) in the respective duct sections. Utilizing the frequency domain linear equations Eq.~\eqref{eq: thermoacoustic_eig}, the wall-heat flux $\hat{q}$ and wall-shear $\hat{\tau}_w$ in the frequency domain are given by 
\begin{subequations}
\begin{gather}
\hat{\tau}_w = h_b\frac{\partial \hat{p}}{\partial x}f_\nu,\\
\hat{q} = h_b(i\omega)C_p T_0 \left(\frac{\Theta}{\left(1-Pr\right)\omega^2}\frac{\partial \hat{p}}{\partial x}\left(f_k - f_\nu\right) - \frac{\gamma - 1}{a^2_0}\hat{p}f_k\right).
\end{gather}
\label{eq: av_heat}
\end{subequations}
\begin{figure}[!t]
\centering
\includegraphics[width=\textwidth]{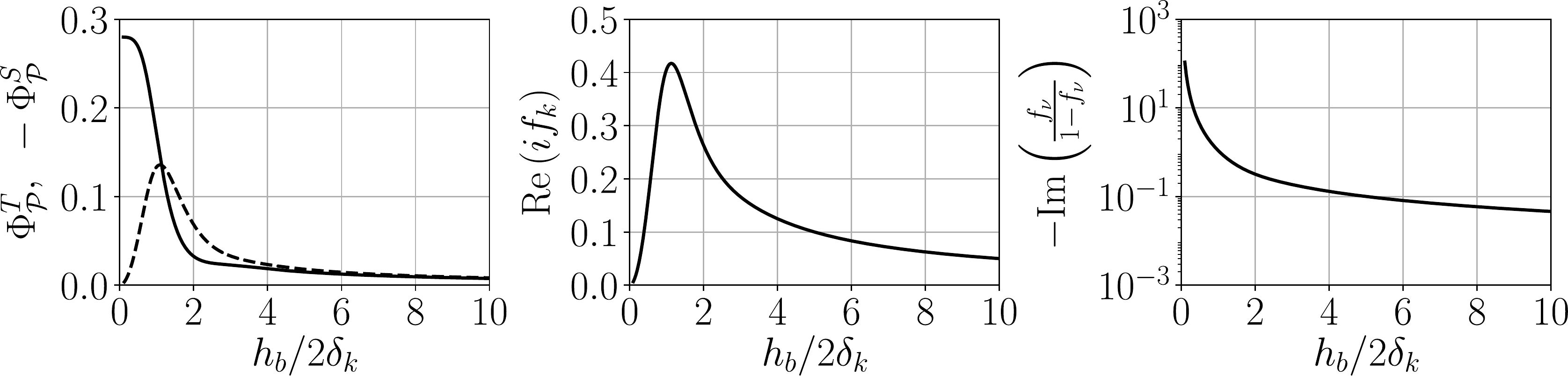}
\put(-433,105){$(a)$}
\put(-292,105){$(b)$}
\put(-140,105){$(c)$}
 \caption{Variation of thermoviscous functionals affecting the cycle averaged thermoacoustic production  $\overline{\mathcal{P}}$ ($a$), and dissipation $\overline{\mathcal{D}}$ ($b$ and $c$) of acoustic energy density versus the ratio of the regenerator half-width $h_b/2$ to the Stokes boundary layer thickness $\delta_k$ (cf. Eq.~\eqref{eq: energy_production}). ($a$): (--), $\Phi^T_{\mathcal{P}}$; (- -), $-\Phi^S_{\mathcal{P}}$.}
\label{fig: ProductionTerms}
\end{figure}
Combining Eq.~\eqref{eq: energy_density_production}, Eq.~\eqref{eq: av_heat}, and Eq.~\eqref{eq: eig_u}, an analytical expression  is obtained for the cycle-averaged production $\overline{\mathcal{P}}$ and dissipation $\overline{\mathcal{D}}$ of the acoustic energy density Eq.~\eqref{eq:energy_planar}: 
\begin{subequations}
\begin{gather}
\overline{\mathcal{P}} = \frac{\Theta}{2\left(1-Pr\right)h_b}\left[\Phi^T_{\mathcal{P}}\;\mathrm{Re}\left(\hat{p}^*\hat{U}\right) -\Phi^S_{\mathcal{P}}\;\mathrm{Im}\left(\hat{p}^*\;\hat{U}\right)\right], \label{eq: budgets_production}\\ 
\overline{\mathcal{D}} = \mathrm{Re}\left(if_k\right) \frac{\omega\left(\gamma -1\right)}{2\rho_0a^2_0}|\hat{p}|^2 -\mathrm{Im}\left(\frac{f_\nu}{1-f_\nu}\right)\,\frac{\rho_0\omega}{2h^2_b}|\hat{U}|^2.\label{eq: budgets_dissipation}
\end{gather}
\label{eq: energy_production}
\end{subequations}
In the above relations, $\Phi^{T}_\mathcal{P}$ and $-\Phi^{S}_\mathcal{P}$ weigh the contributions to the thermoacoustic energy production by the travelling-wave, $\mathrm{Re}(\hat{p}^*\hat{U})$, and the standing-wave, $\mathrm{Im}(\hat{p}^*\,\hat{U})$, components respectively. Their expressions read:
\begin{equation} \label{eq:travelling_standing_weights}
\Phi^T_{\mathcal{P}} = \mathrm{Re}\left(\frac{f_k - f_\nu}{1-f_\nu}\right), \quad 
\Phi^S_{\mathcal{P}} = \mathrm{Im}\left(\frac{f_k - f_\nu}{1-f_\nu}\right).
\end{equation}
\begin{figure}[!t]
\centering
\includegraphics[width=0.9\textwidth]{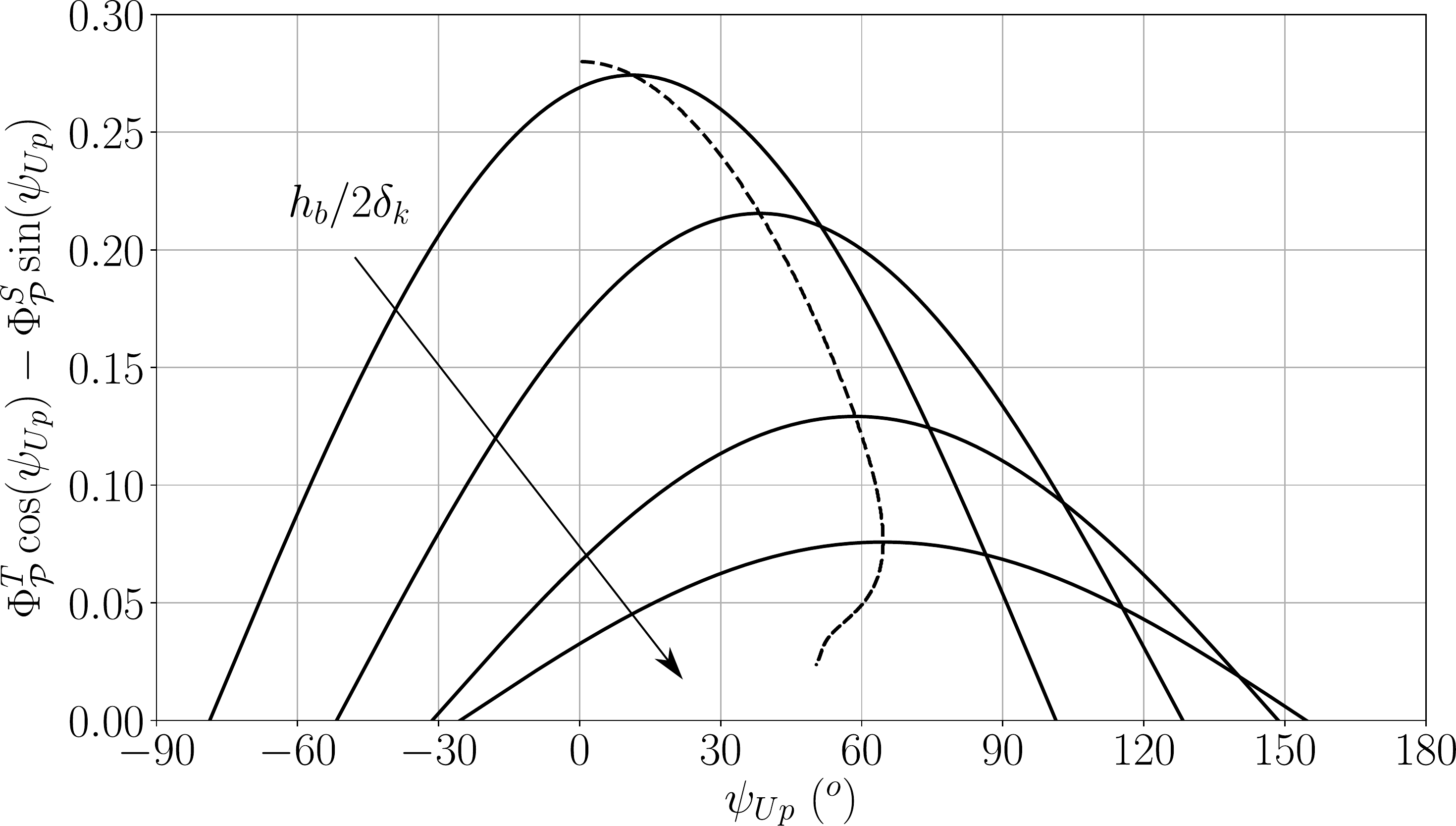}
\caption{Normalized thermoacoustic production $\Phi^T_{\mathcal{P}}\cos(\psi_{Up}) - \Phi^S_{\mathcal{P}}\sin(\psi_{Up})$ ~(cf. Eq.~\eqref{eq: budgets_production}) versus the phase angle difference between $\hat{p}$ and $\hat{U}$, $\psi_{Up}=\angle{\hat{U}}-\angle{\hat{p}}$, for increasing values of $h_b/2\delta_{k}=~$0.5, 1.0, 1.5, 2.0. (--); (- -), Optimum phasing maximizing the thermoacoustic production for continually varying $h_b/2\delta_{k}$.}
\label{fig: OptPhase}
\end{figure}
For $h_b/2\delta_k\leq1.13$, $|\Phi^T_{\mathcal{P}}|>|\Phi^S_{\mathcal{P}}|$, which implies that the regenerator half-width $h_b$ must remain comparable to or smaller than the Stokes boundary layer thickness $\delta_k$ to achieve higher thermoacoustic amplification of travelling waves ($\psi_{Up}\simeq 0^{\circ}$) (figures~\ref{fig: ProductionTerms}$a$ and~\ref{fig: OptPhase}). However, to maximize thermoacoustic energy production for standing waves ($\psi_{Up} \simeq \pm 90^{\circ}$), a larger regenerator half-width ($h_b/2\delta_k > 1.13$) is required. While production alone for a purely travelling wave ($\psi_{Up}=0$) is maximized in the limit $h_b/2\delta_k\rightarrow0$ (figure~\ref{fig: OptPhase}), for fixed temperature settings, dissipation also diverges (figure~\ref{fig: ProductionTerms}$c$). Therefore, \emph{pure travelling wave phasing, if at all achieved, always results in smaller net production of acoustic energy density compared to an optimal combination of standing and travelling waves}. Moreover, varying the temperature inside the regenerator results in a local variation of the ratio $h_b/2\delta_k$ which, in turn,  causes the optimum phase to vary along the regenerator (figure~\ref{fig: OptPhase}). For the temperature settings considered here (table~\ref{tbl:base_state}), the optimum phasing angle, averaged over the regenerator length changes from $43.17^{\circ}$ to $36.41^{\circ}$ as the temperature $T_{H}$ is increased.

\subsection{Effects of temperature gradient and geometry on growth rates}
\label{sec: GrowthrateEffect}
\begin{figure}[!t]
\centering
\includegraphics[width=0.8\textwidth]{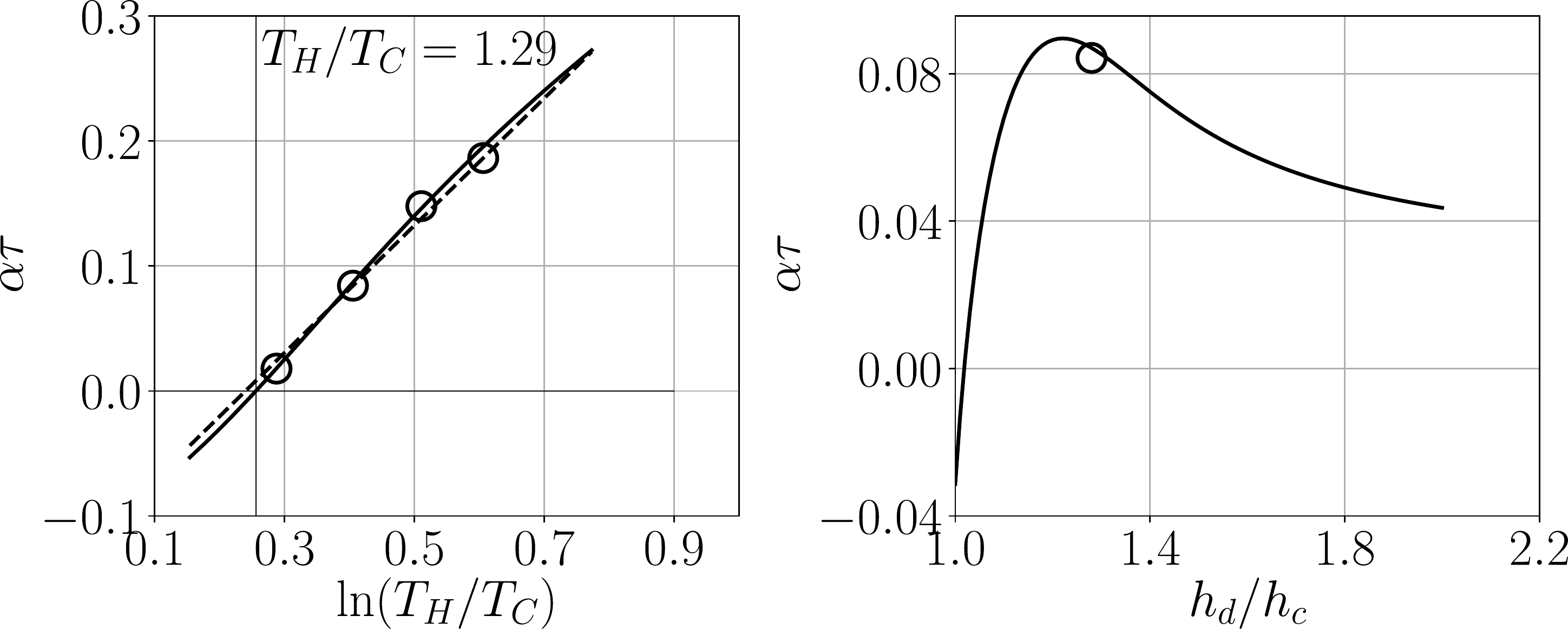}
\put(-350,135){$(a)$}
\put(-175,135){$(b)$}
 \caption{Dimensionless thermoacoustic growth rate, $\alpha\tau$ (cf. Eq.~\eqref{eq:tau}), versus the natural logarithm of the temperature ratio $T_H/T_C$ for $T_C=300~$\,K ($a$) and resonator area ratio $h_d/h_c$ ($b$). ($a$): (--), Linear stability analysis; (- -), logarithmic estimate of $\alpha\tau$ Eq.~\eqref{eq: functional_approx} fitted using values at $T_H = 450\,\mathrm{K}$ and $T_H = 550\,\mathrm{K}$; ($\circ$), Navier-Stokes simulations.}
\label{fig: GrowthRate}
\end{figure}
The variation of the instability growth rates with the temperature and the geometry is further analysed utilizing the acoustic energy budget formulation developed in the previous section. To this end, the production, dissipation, and Rayleigh index (normalized by the pressure amplitude) are plotted in the convenient dimensionless forms 
\begin{equation}
\overline{\mathcal{P}}_* = \frac{\overline{\mathcal{P}}\tau}{\rho_0a^2_0},\quad \overline{\mathcal{D}}_{*} = \frac{\overline{\mathcal{D}}\tau}{\rho_0a^2_0},\quad\mathcal{R}_* = \frac{\mathcal{R}\tau}{h_b\rho_0a^2_0}.\label{eq: dimensionless_PDR}
\end{equation}
Cycle averaged production of acoustic energy density due to travelling wave and standing wave components given by
\begin{subequations}
\begin{eqnarray}
\mathcal{R}_{T*} = \frac{\tau}{2(1-Pr)\rho_0a^2_0h^2_b}\int_b\Theta\,\mathrm{Re}\left(\hat{p}^*\hat{U}\right)\Phi^T_{\mathcal{P}}dx ,\label{eq: dimensionless_RT}\\ \mathcal{R}_{S*} = -\frac{\tau}{2(1-Pr)\rho_0a^2_0h^2_b}\int_b\Theta\,\mathrm{Im}\left(\hat{p}^*\hat{U}\right)\Phi^S_{\mathcal{P}}dx, \label{eq: dimensionaless_RS}
\end{eqnarray}
\end{subequations}
respectively, are also analysed, where
\begin{equation} \label{eq:tau}
\tau = h^2_b/\nu_0
\end{equation}
is a reference viscous time scale in the regenerator with $\nu_0$ evaluated at $T_C=300\,$K.

Increasing the hot side temperature $T_H$, the thermoacoustic production $\overline{\mathcal{P}}_*$ increases monotonically, approximately as
\begin{equation}
\overline{\mathcal{P}}_* \sim \Theta = \frac{d}{dx}\ln[T_0(x)],
\label{eq: ApproxProdDiss}
\end{equation}
and more rapidly than the dissipation $\overline{\mathcal{D}}_*$, yielding positive values of the Rayleigh index $\mathcal{R}_*$ (figure~\ref{fig: RayleighIndex}$a$) for $T_H/T_C>1.29$ (figure~\ref{fig: GrowthRate}$a$). The Rayleigh index can thus be used to quantify the thermoacoustic growth rate (figure~\ref{fig: GrowthRate}$a$,~\ref{fig: RayleighIndex}$a$) as
\begin{equation}
\mathcal{R}_*\sim \alpha\tau \approx A \ln(T_H/T_C) - B,
\label{eq: functional_approx}
\end{equation}
where $A$ and $B$ are geometry dependent fitting coefficients. 

\begin{figure}[!b]
\centering
\includegraphics[width=0.8\textwidth]{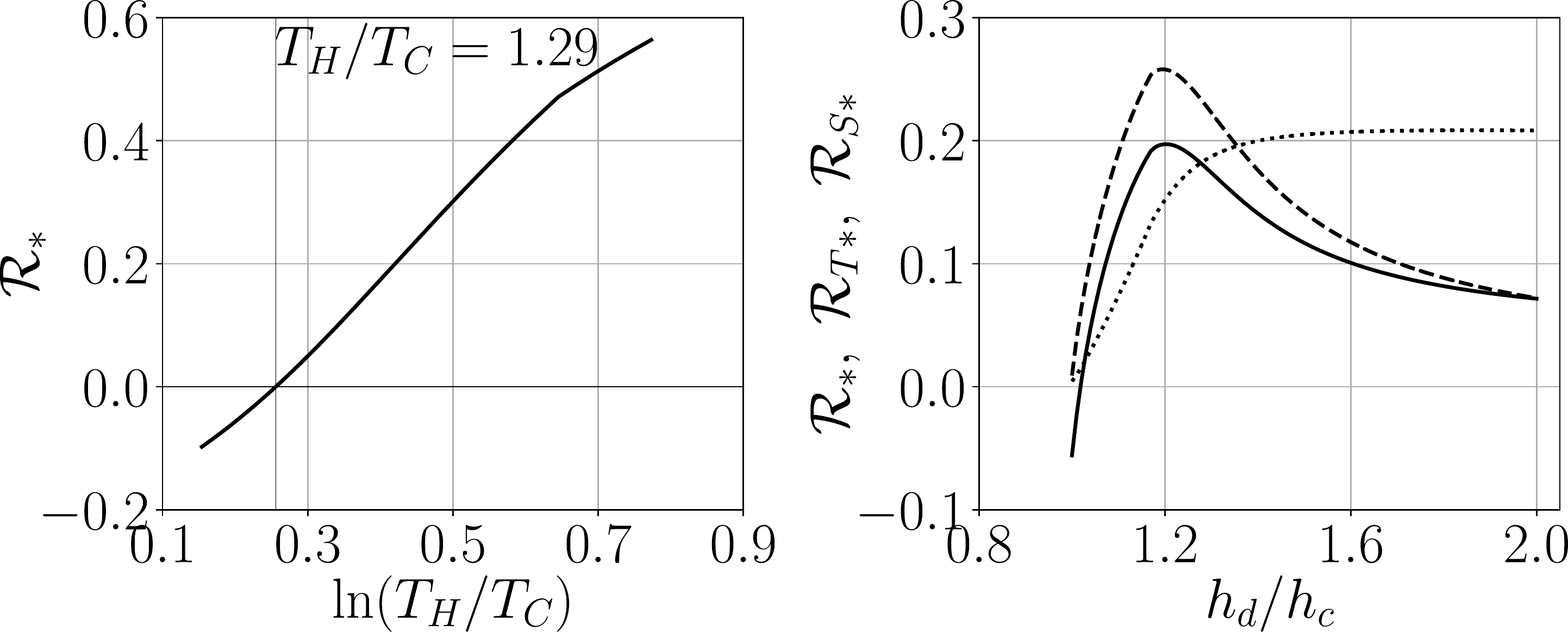}
\put(-350,135){$(a)$}
\put(-175,135){$(b)$}
\caption{Dimensionless Rayleigh index $\mathcal{R}_*$ versus the natural logarithm of the temperature ratio $T_H/T_C$ for $T_C=300\,$K ($a$) and the resonator area ratio $h_d/h_c$ ($b$). ($b$): (--), $\mathcal{R}_*$ ; (- -), $\mathcal{R}_{T*}$ ; ($\cdot \cdot \cdot$), $\mathcal{R}_{S*}$.}
\label{fig: RayleighIndex}
\end{figure}

With increasing resonator area ratio $h_d/h_c$, keeping $h_c$ fixed, the frequency decreases monotonically by approximately $4\%$ in the range of $h_d/h_c$ considered. However, the growth rates vary non-monotonically, reaching a local maximum  at $h_d/h_c \approx 1.28$ (figure~\ref{fig: GrowthRate}$b$). 
Also, the net production of acoustic energy density due to the travelling wave component $\mathcal{R}_{T*}$ (figure~\ref{fig: RayleighIndex}$b$) peaks at $h_d/h_c\approx1.28$. Moreover, the high mechanical impedance of the section $d$ ($\rho_0a_0h_d$) for high values of $h_d$ increases the standing wave component of the acoustic power $\mathcal{R}_{S*}$ and decreases the travelling wave  component $\mathcal{R}_{T*}$. The variation of the transverse geometrical parameters does not significantly alter the frequency. Therefore, changes in the ratio $(f_k - f_\nu)/(1-f_{\nu})$ are also negligible and do not influence the growth rates significantly (cf. Eq.~\eqref{eq:travelling_standing_weights}).

\section{Formulation of a Nonlinear Thermoacoustic Model}
\label{sec: NonlinearRegime_modeling}
As a result of the modal thermoacoustic instability, large pressure amplitudes ($\sim 160~\mathrm{dB}$) are generated, which result in the nonlinear steepening of the waveform. In the spectral space, the nonlinear steepening can be viewed as the cascade of energy from the unstable mode into higher harmonics with correspondingly shorter wavelengths. Moreover, inside the regenerator, large amplitude perturbations in thermodynamic quantities are responsible for thermoacoustic nonlinearities. As a result of nonlinear wave propagation, thermoacoustically sustained shock waves propagate in the system. While the analysis above highlights that quasi-travelling wave phasing is essential for high thermoacoustic growth rates, nonlinear steepening is also favoured by such phasing~\cite{BiwaEtAl_JASA_2014}.

In this section, a first-principles-based theoretical framework accounting for acoustic and thermoacoustic nonlinearities up to second order is developed, and a quasi one-dimensional evolution equation is obtained for nonlinear thermoacoustic waves Eq.~\eqref{eq: nonlinear_P}. In Section ~\ref{sec: governing_equations} mass, momentum, and energy (combined with the second law of thermodynamics) equations correct up to second order are introduced. Furthermore, in Section ~\ref{sec: nonlinear_wave} and Section ~\ref{sec: nonlinear_thermoacoustic}, cross-sectionally averaged nonlinear spatio-temporal evolution model equations are derived with the time-domain approximations of wall-shear and wall-heat flux outlined in~Section ~\ref{sec: shear_heat}. 
\subsection{Governing equations for nonlinear thermoviscous perturbations}
\label{sec: governing_equations}
The nonlinear governing equations, correct up to second order, for a two-dimensional perturbation read:
\begin{subequations}
\begin{align}
&\frac{\partial \rho'}{\partial t} + \rho_0 \frac{\partial u'}{\partial x} + u'\frac{d\rho_0}{dx} + \rho_0\frac{\partial v'}{\partial y}= \left[-\rho'\frac{\partial u'}{\partial x} - u'\frac{\partial \rho'}{\partial x}\right],\label{eq: 2_mass}\\
&\frac{\partial u'}{\partial t} + \frac{1}{\rho_0}\frac{\partial p'}{\partial x} - \nu_0 \frac{\partial^2 u'}{\partial y^2} - \frac{1}{\rho_0}\frac{\partial }{\partial x}\left[\mu_0\left(\xi_B + \frac{4}{3}\right)\frac{\partial u'}{\partial x}\right]=  \left[-\frac{\rho'}{\rho_0}\frac{\partial u'}{\partial t} - \frac{1}{2}\frac{\partial u'^2}{\partial x}\right],\label{eq: 2_mom}
\end{align}
\begin{align}
&\frac{\partial s'}{\partial t} + u'\frac{ds_0}{dx} - \frac{Rk_0}{p_0}\frac{\partial^2 T'}{\partial y^2} - \frac{R}{p_0}\frac{\partial}{\partial x}\left(k_0\frac{\partial T'}{\partial x}\right)= \left[-\frac{p'}{p_0}\left(\frac{\partial s'}{\partial t} + u'\frac{ds_0}{dx}\right) - u'\frac{\partial s'}{\partial x}\right], \label{eq: 2_ent}
\end{align}
\end{subequations}
where, primed variables $\left(\cdot\right)'$ represent the fluctuations in corresponding quantities, the subscript $0$ denotes the base state, whereas $\xi_B=2/3$ is the ratio of the bulk viscosity coefficient $\mu_B$ to the shear viscosity coefficient $\mu$. The terms on the left-hand side are linear in the perturbation variables while those on the right-hand side are nonlinear. The entropy generation due to viscous dissipation is neglected, as well are the pressure gradients and velocity in $y$ direction (boundary layer assumption), and the fluctuations in the  diffusivity coefficients $\mu$ and $k$. Higher harmonics with correspondingly shorter wavelengths are generated due to the nonlinear spectral energy cascade. Consequently, the axial diffusion terms in Eq.~\eqref{eq: 2_mom} and Eq.~\eqref{eq: 2_ent}, which have been neglected in the linear regime, become significant and act as the primary sink of energy at large harmonic scales in the spectral space. 

We seek to collapse Eq.~\eqref{eq: 2_mass}--\eqref{eq: 2_ent} to obtain a set of equations similar to Eq.~\eqref{eq: AcousticDynamicalSystem}. To this end, the following quadratic thermodynamic constitutive equation relating the density fluctuations $\rho'$ with the pressure and entropy fluctuations ($p'$ and $s'$, respectively) is considered: 
\begin{equation}
\rho' = \alpha_s p' + \alpha_ps' + \frac{1}{2}\left(\beta_s p'^2 + \beta_p s'^2 + 2\beta_{sp} s'p'\right), \label{eq: constitutive_eqn}
\end{equation}
where the thermodynamic coefficients $\alpha $ and $\beta$ are given by
\begin{subequations}
\begin{gather}
 \alpha_s = \left(\frac{\partial \rho}{\partial p}\right)_s = \frac{1}{a^2_0},\quad \alpha_p = \left(\frac{\partial \rho}{\partial s}\right)_p = -\frac{\rho_0}{C_p},\\
\beta_s = \left(\frac{\partial^2 \rho}{\partial p^2}\right)_s = -\frac{\gamma -1}{\rho_0 a^4_0},\quad \beta_p = \left(\frac{\partial^2 \rho}{\partial s^2}\right)_p = \frac{\rho_0}{C^2_p},\\
 \beta_{sp} = \left[\frac{\partial }{\partial s}\left(\frac{\partial \rho}{\partial p}\right)_s\right]_p = \left[\frac{\partial }{\partial s}\left(\frac{\rho}{\gamma p}\right)\right]_p = -\frac{1}{C_p a^2_0}.\label{eq: const_coeff}
\end{gather}
\end{subequations}
\begin{figure}[!t]
\centering
\includegraphics[width=0.7\textwidth]{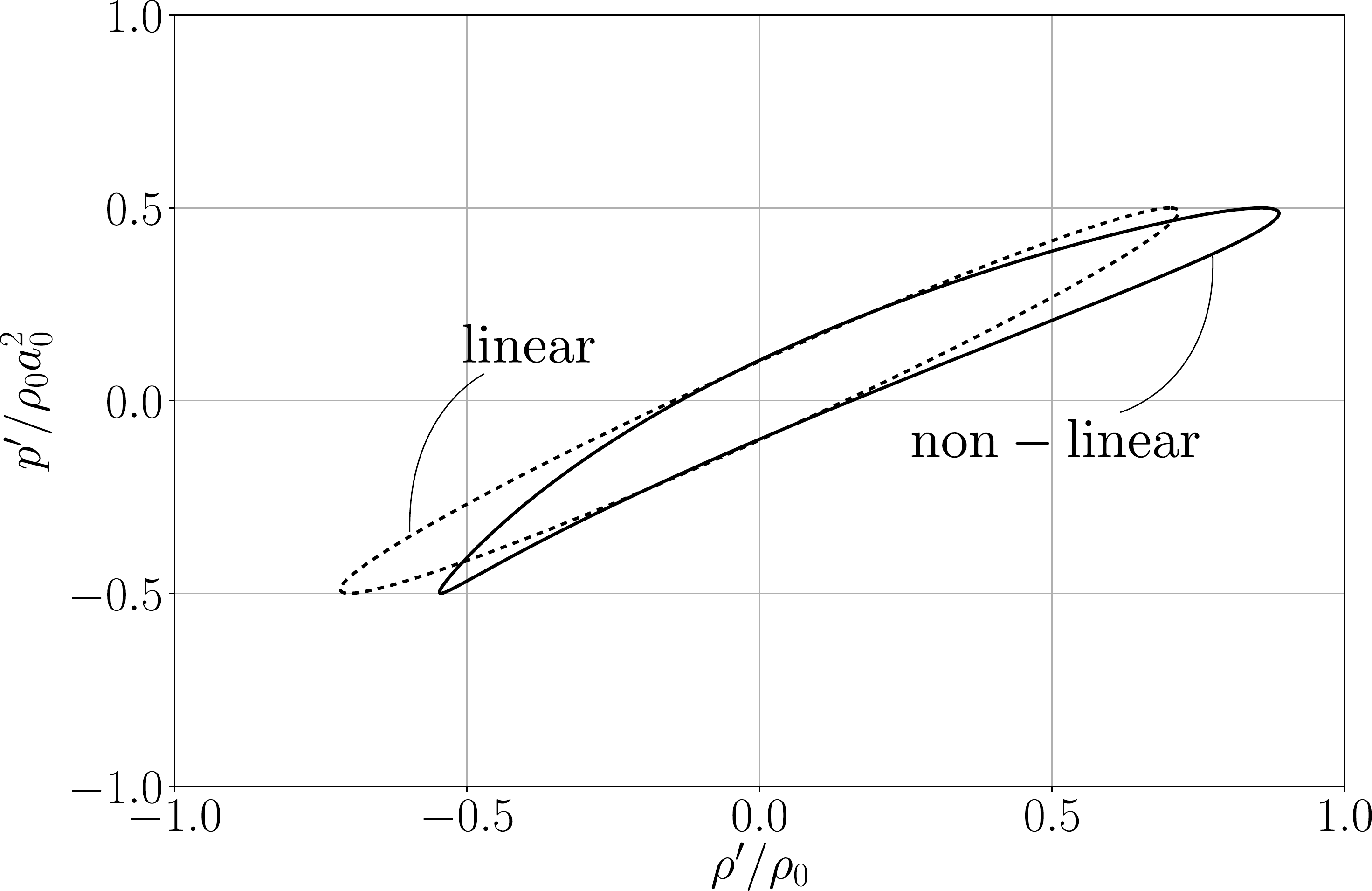}
\caption{Thermodynamic cycle in $p'-\rho'$ plane for non-isentropic thermoacoustic wave amplification. (--), Nonlinear; (- -), Linear. Sample perturbation fields are: $p' = 0.5p_0\cos\omega t$ and $s' = -Rp'/p_0 + 250\sin\left(\omega(t + \tau)\right)$ for $\tau = h^2_b/\nu = 0.0107~\mathrm{s}$.}
\label{fig: constitutive}
\end{figure}
The coefficients $\alpha_p$ and $\alpha_s$ contribute to first order wave propagation and thermoacoustic effects while the second-order coefficients $\beta_s,~\beta_p,$ and $\beta_{sp}$ in Eq.~\eqref{eq: constitutive_eqn} account for the corresponding nonlinear effects.~\cite{HedbergR_JAP_2011} have demonstrated the hysteretic effects of nonlinear wave propagation retaining only the $\beta_s$ nonlinear term and $\alpha_p$ term accounting for irreversible entropy changes. Figure~\ref{fig: constitutive} shows the hysteresis cycle for the second order constitutive relation in Eq.~\eqref{eq: constitutive_eqn} in comparison with the first order approximation.
Nonlinear wave propagation outside the regenerator (sections $a$, $c$, and $d$) is not affected by the no-slip and isothermal boundary conditions. Only higher order irreversible entropy fluctuations are generated due to the axial conduction terms~\cite{Hamilton_NLA_1998}. Consequently, the constitutive relation Eq.~\eqref{eq: constitutive_eqn} can be approximated with the terms corresponding to $\alpha_s, \alpha_p$, and $\beta_s$ retained. However, inside the regenerator, wall-shear and wall-heat flux from the no-slip isothermal boundaries generate first order reversible entropy fluctuations; hence, in order to capture the nonlinear thermoacoustic wave amplification inside the regenerator, nonlinear terms in entropy perturbations need to be included, as shown in Eq.~\eqref{eq: constitutive_eqn}. Starting from the general second order governing equations discussed above, the spatio-temporal evolution equations for the fluctuations in pressure $p'$ and flow rate $U'$ can be derived and the final results are presented in the following sections.

\subsection{Free-shear nonlinear wave propagation}
\label{sec: nonlinear_wave}
Waves outside the regenerator (sections $a$, $c$, and $d$) in the minimal unit setup (figure~\ref{fig:computational_setup}) propagate in the absence of wall-shear and wall-heat flux. As a result, terms involving the transverse gradient of $u'$ in Eq.~\eqref{eq: 2_mom} and $T'$ in Eq.~\eqref{eq: 2_ent} can be neglected and entropy fluctuations remain second order in the nonlinear regime as well~\cite{Hamilton_NLA_1998}. Up to second order, the wave propagation in the duct sections $a$, $c$, and $d$ is governed by
\begin{gather}
   \frac{\partial p'}{\partial t} = -\frac{\gamma p_0}{h}\left(1 + \frac{1+\gamma}{\gamma}\frac{p'}{p_0}\right)\frac{\partial U'}{\partial x} + \frac{k}{\rho_0}\left(\frac{1}{C_v} - \frac{1}{C_p}\right)\frac{\partial^2 p'}{\partial x^2},\label{eq: inviscid_p}\\
   \frac{\partial U'}{\partial t} = -\frac{h}{\rho_0}\frac{\partial p'}{\partial x} + \nu_0\left(\xi_B + \frac{4}{3}\right)\frac{\partial^2 U'}{\partial x^2},\label{eq: inviscid_U}
  \end{gather}
where $h = h_a,~h_c,$ or $h_d$. Second-order nonlinearities in Eq.~\eqref{eq: inviscid_p} cause the waveform distortion and steepening.

\subsection{Wall-shear and wall-heat flux}
\label{sec: shear_heat}
By design, the regenerator width is comparable to the local viscous and thermal Stokes layer thickness ($h_b/2\delta_\nu\sim1$, $h_b/2\delta_k\sim1$). Due to the wall-shear, the velocity fluctuations inside the regenerator vary in the $y$ direction as well. Moreover, the wall-heat flux contributes to the first order entropy fluctuations. Noting that nonlinear acoustic waves can be decomposed into acoustic, viscous, and entropic modes~\cite{ChuK_JFM_1958, Pierce_1989}, the following decomposition for the entropy and velocity fluctuations inside the regenerator is considered:
\begin{equation}
 u'(x,y,t) = \tilde{u}(x,t) + u'_{\nu}(x,y,t), \quad s' = \tilde{s}(x,t)+ s'_{q}(x,y,t),
\end{equation}
where $u'_{\nu}$ is the viscous velocity fluctuation and $\tilde{u}$ is the nonlinear acoustic wave field. The former is diffused by viscosity and is governed by the unsteady diffusion equation 
\begin{equation}
 \frac{\partial u'_{\nu}}{\partial t} = \nu_0\frac{\partial^2 u'_{\nu}}{\partial y^2},\quad u'_{\nu}(x,y=\pm h_b/2,t) = -\tilde{u}(x,t). \label{eq: shear}
\end{equation}
Similarly, $\tilde{s}$ accounts for the entropy changes due to the nonlinear acoustic wave propagation and $s'_q$ corresponds to the first order entropy changes due to the wall-heat flux inside the regenerator and is governed by the equation
\begin{equation}
 \frac{\partial s'_q}{\partial t} + u'\frac{ds_0}{dx} = \frac{\nu_0}{Pr}\frac{\partial^2 s'_q}{\partial y^2},\quad s'_q\left(x,y=\pm h_b/2, t\right) = -\tilde{s} + s'_\mathrm{w}, \label{eq: heat_transfer}
\end{equation}
where $s'_\mathrm{w}$ corresponds to the entropy fluctuations at the isothermal walls driven by pressure fluctuations
\begin{equation}
 s'_\mathrm{w} = -\frac{1}{\rho_0 T_0}p^{\prime} = -\frac{R}{p_0}p^{\prime}. \label{eq: entropy_wall}
\end{equation}
Equations~\eqref{eq: shear} and~\eqref{eq: heat_transfer} suggest the following infinite series solution forms for the viscous and the entropic fields: 
\begin{subequations}
\begin{gather}
 u'_{\nu} = -\tilde{u} + \sum_{j=0}^{\infty}\check{u}_j(x,t) \cos(\zeta_j y),\quad s'_q = -\tilde{s} + s_\mathrm{w} + \sum_{j=0}^{\infty}\check{s}_j(x,t)\cos \left(\zeta_j y\right),\label{eq: eigenfunctions_y}\\
\quad\text{with}\quad \zeta_j = \left(2j+1\right)\frac{\pi}{h_b}.\label{eq: wavenumber_y}
\end{gather}
\end{subequations}
Performing eigenfunction expansions along the $y$ direction Eq.~\eqref{eq: eigenfunctions_y} yields the following evolution equations for the Fourier coefficients corresponding to the viscous and entropic modes:
\begin{eqnarray}
 \frac{\partial \check{u}_j}{\partial t} + \nu_0 \zeta^2_j \check{u}_j = (-1)^{j+1}\frac{2}{\zeta_j h_b}\left(\frac{1}{\rho_0}\frac{\partial p'}{\partial x}\right),\label{eq: u_modes}\\
 \frac{\partial \check{s}_j}{\partial t} + \check{u}_j\frac{d s_0}{dx} + \frac{\nu_0}{Pr}\zeta^2_j \check{s}_j = (-1)^{j}\frac{2 R}{\zeta_jh_b p_0}\frac{\partial p'}{\partial t}.
 \label{eq: s_modes}
\end{eqnarray}
Equations~\eqref{eq: u_modes} and~\eqref{eq: s_modes} determine the evolution of the transverse modes of the longitudinal velocity $\check{u}_j$ and entropy $\check{s}_j$ fluctuations.

\subsection{Nonlinear thermoviscous wave equations}
\label{sec: nonlinear_thermoacoustic}

The axial velocity fluctuations $u'$ are governed by Eq.~\eqref{eq: 2_mom} up to second order accuracy. However, nonlinearities in Eq.~\eqref{eq: 2_mom} result in acoustic streaming, which is suppressed in the current analysis, and is therefore neglected~\cite{Hamilton_NLA_1998}. Integrating the resulting momentum equation in $y$ and substituting Eq.~\eqref{eq: eigenfunctions_y} yields 
\begin{eqnarray}
\frac{\partial U'}{\partial t} + \frac{h_b}{\rho_0}\frac{\partial p'}{\partial x} =  \tau'_w + \frac{1}{\rho_0}\frac{\partial }{\partial x}\left[\mu_0\left(\xi_B + \frac{4}{3}\right)\frac{\partial U'}{\partial x}\right],\label{eq: nonlinear_U}
\end{eqnarray}
where
\begin{eqnarray}
\tau'_w =  2\nu_0\sum^{\infty}_{j=0}(-1)^{j+1}\check{u}_j(x,t)\zeta_j.\label{eq: shear_stress}
\end{eqnarray}
Equation~\eqref{eq: inviscid_p} governs the evolution of the pressure fluctuations up to second order in the free-shear/adiabatic ducts. In order to derive an analogous governing equation for the regenerator, Eqs. ~\eqref{eq: 2_mass},~\eqref{eq: 2_ent}, and~\eqref{eq: constitutive_eqn} are combined to obtain 
\begin{eqnarray}
\underbrace{\frac{\partial p'}{\partial t} + \frac{\rho_0a^2_0}{h_b}\frac{\partial U'}{\partial x}}_\text{wave propagation}=\frac{\rho_0a^2_0}{h_b}\Bigg(\frac{q'}{C_p\rho_0T_0}+\underbrace{q_2  + \mathbb{T} -\mathbb{Q}}_{\substack{\text{thermodynamic}\\ \text{nonlinearities}}} +  \mathbb{D}_s\Bigg) - \mathbb{C}, \label{eq: nonlinear_P}
\end{eqnarray}
where 
\begin{equation}
 q' = \frac{2\nu_0\rho_0 T_0}{Pr}\sum^{\infty}_{j=0}(-1)^{j+1}\check{s}_j(x,t)\zeta_j,
 \label{eq: heat_flux_fluct}
\end{equation}
defines the fluctuating wall-heat flux and couples the pressure evolution Eq.~\eqref{eq: nonlinear_P} with the entropic mode evolution Eq.~\eqref{eq: s_modes}, whereas $\mathbb{Q}$ denotes the nonlinear interaction of pressure and wall-heat flux fluctuations, 
\begin{equation}
 \mathbb{Q} = \frac{\gamma p'q'}{C_p p_0\rho_0T_0},
 \label{eq: MacrosonicThermoacousticInteraction}
\end{equation}
 hereafter referred as \emph{macrosonic thermoacoustic interaction}. The term denoted by $q_2$ corresponds to the second order heat flux which is a quadratic function of the entropy gradient in $y$. The nonlinear terms denoted by the double faced $\mathbb{T}$ correspond to the constitutive (thermodynamic) nonlinearities which account for the second order density fluctuation due to first order entropic modes. The terms denoted by the double faced $\mathbb{C}$ correspond to the convective nonlinearities in Eq.~\eqref{eq: 2_mom} and Eq.~\eqref{eq: 2_ent} and those denoted by $\mathbb{D}_s$ account for the axial diffusion of gradients in highly nonlinear regimes of thermoacoustic wave amplification. A detailed derivation of~\eqref{eq: nonlinear_P} and expressions for the terms $q_2$, $\mathbb{D}_s$, $\mathbb{T}$, and $\mathbb{C}$ are given in~\cite{GuptaLS_JFM_2017}. Equations~\eqref{eq: u_modes},~\eqref{eq: s_modes},~\eqref{eq: nonlinear_U}, and~\eqref{eq: nonlinear_P} constitute the governing equations for the spatio-temporal evolution of large amplitude acoustic perturbations inside the regenerator. The macrosonic thermoacoustic interaction Eq.~\eqref{eq: MacrosonicThermoacousticInteraction} breaks the thermodynamic symmetry between the interactions of compressions and dilatations with the wall-heat flux inside the regenerator, thus highlighting that the entropy of a Lagrangian parcel of fluid changes by a small amount under high amplitude compressions ($\rho'>0$), compared to dilatations ($\rho'<0$), for the same amount of heat input or output. 

In general, for thermoacoustic devices in looped configuration, the length of the regenerator is very short compared to the total length of the device. As a result, higher order terms affecting only the propagation of the acoustic perturbations, such as convective nonlinearities, can be neglected inside the regenerator. Under such hypotheses,  the following approximate nonlinear governing equation for the pressure fluctuations $p'$ inside a short regenerator is obtained:
\begin{equation}
 \frac{\partial p'}{\partial t} + \frac{\rho_0a^2_0}{h_b}\frac{\partial U'}{\partial x}\approx \frac{\rho_0a^2_0}{h_b}\left\{\frac{1}{C_p}\left[\left(1-\frac{\gamma p'}{p_0}\right)\frac{q'}{\rho_0T_0}\right]\right\}. \label{eq: nonlinear_P_reduced}
\end{equation}
In the above equation, terms $\mathbb{T}$, $\mathbb{D}_s$, and $q_2$ are neglected for simplicity. 
Equations~\eqref{eq: u_modes} and~\eqref{eq: s_modes} can also be integrated in time analytically to express the wall-shear $\tau'_w$ and wall-heat flux $q'$ in terms of acoustic variables. 
The time integration of Eqs.~\eqref{eq: u_modes} and~\eqref{eq: s_modes} yields: 
\begin{gather}
\check{u}_j = (-1)^{j+1}\frac{2}{\zeta_j h_b\rho_0}\int^{t}_{-\infty}e^{-\frac{t - \eta}{\tau_j}}\frac{\partial p'}{\partial x}(x,\eta)d\eta, \label{eq: analytical_un}\\
%\end{equation}
%where $\tau_j = 1/\nu_0\zeta^2_j$ defines the viscous relaxation time for the $j^{\mathrm{th}}$ viscous mode. Similarly,~\eqref{eq: s_modes} for $\check{s}_j$ yields,
%\begin{equation}
\check{s}_j = -\frac{ds_0}{dx}\int^{t}_{-\infty}e^{-\frac{t - \eta}{Pr\tau_j}}\check{u}_j(x,\eta)d\eta + (-1)^j\frac{2R}{\zeta_jh p_0}\int^{t}_{-\infty}e^{-\frac{t - \eta}{Pr\tau_j}}\frac{\partial p'}{\partial \eta}(x,\eta) d\eta. \label{eq: analytical_sn}
\end{gather}
where $\tau_j = 1/\nu_0\zeta^2_j$ defines the viscous relaxation time for the $j^{\mathrm{th}}$ viscous mode.
Hence, writing the relaxation functional for some function $\phi(x,t)$, namely,
\begin{equation}
\mathcal{G}_j(\phi, \tau_j) = \int^{t}_{-\infty}e^{-\frac{t-\eta}{\tau_j}}\phi(x,\eta)d\eta, \label{eq: relaxation_functional} 
\end{equation}
and summing~\eqref{eq: analytical_un} and~\eqref{eq: analytical_sn} over $j$, the following expressions for the wall-shear and wall-heat flux are obtained:
\begin{gather}
\tau'_w = \frac{4\nu_0}{\rho_0 h_b}\sum^{\infty}_{j=0}\mathcal{G}_j\left(\frac{\partial p'}{\partial x}, \tau_j\right),\label{eq: analytical_tau}\\
 q' = \frac{2\rho_0\nu_0T_0}{Pr}\sum^{\infty}_{j=0}\left[(-1)^{j}\frac{ds_0}{dx}\mathcal{G}_j\left(\zeta_j\check{u}_j, \tau_jPr\right)-\frac{2R}{h_bp_0}\mathcal{G}_j\left(\frac{\partial p'}{\partial t}, \tau_j Pr\right)\right]. \label{eq: analytical_q}
\end{gather}
Equations~\eqref{eq: analytical_tau} and~\eqref{eq: analytical_q} provide first order expressions for the wall-shear and the wall-heat flux as a function of a generic acoustic field near the walls and, together with~\eqref{eq: nonlinear_U} and~\eqref{eq: nonlinear_P_reduced}, complete the nonlinear wave propagation model equations. However, in the present work,Eqs.~\eqref{eq: u_modes} and~\eqref{eq: s_modes} have been considered for time integration for simplicity.
%
%Sugimoto~\cite{Sugimoto_JFM_2010} systematically derived the functional form of $\mathcal{G}_j$ for extremely \emph{thick} and \emph{thin} diffusion layers using the linear acoustic field approximation. The functional $\mathcal{G}_j$ in equation~\eqref{eq: relaxation_functional} approximates the wall-shear and the wall-heat flux in terms of any acoustic field, linear or nonlinear, though within the restriction of linear decomposition of the field into viscous and entropic modes. Recently,Sugimoto~\cite{Sugimoto_JFM_2016} also developed a theoretical framework to elucidate high amplitude nonlinear wave propagation in a shear dominated duct with the restriction of very \emph{thick} diffusion layers and focusing on streaming. However, the wall-shear and wall-heat flux expressed in terms of the acoustic field variables by the relaxation functional $\mathcal{G}_j$ in~\eqref{eq: analytical_tau} and~\eqref{eq: analytical_tau} hold true irrespective of the relative thickness between the diffusion layers and the channel width. 

%\begin{figure}
%\centering{\includegraphics[width=0.6\textwidth]{Sol_technique.pdf}}
%\caption{Solution technique for integrating quasi one-dimensional governing equations in time across abrupt area changes. Interface bulk velocity at $n$ time step, $U^{n}_i$ is calculated such that difference of left and right extrapolations of pressure ($p^{-}_i$ and $p^{+}_i$ respectively) equals minor loss $\Delta p_{\mathrm{ml}}$ at $n+1$ time step which is a function of $U^{n}_i$.}
%\label{fig: solution_technique}
%\end{figure}

The solution technique for model Eqs. ~\eqref{eq: inviscid_p},~\eqref{eq: inviscid_U},~\eqref{eq: u_modes},~\eqref{eq: s_modes},~\eqref{eq: nonlinear_U}, and~\eqref{eq: nonlinear_P} is discussed in detail in~\cite{GuptaLS_JFM_2017}.

\section{Nonlinear Spectral Energy Dynamics} 
\label{sec: nonlinear_cascade}

\begin{figure}[!h]
\includegraphics[width=1.0\textwidth]{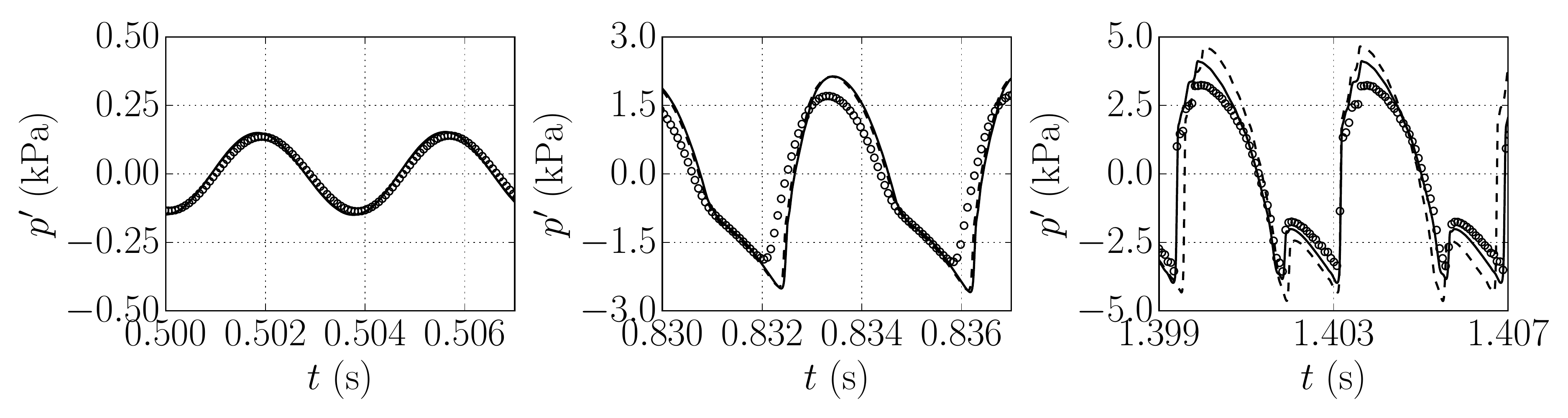}
\put(-433,105){$(a)$}
\put(-290,105){$(b)$}
\put(-150,105){$(c)$}
\caption{Comparison of filtered thermoacoustic signal as obtained from time integration of nonlinear model Eqs.~\eqref{eq: inviscid_p},~\eqref{eq: inviscid_U},~\eqref{eq: u_modes},~\eqref{eq: s_modes},~\eqref{eq: nonlinear_U}, and~\eqref{eq: nonlinear_P} probed at $x=1.54~\mathrm{m}$ in harmonic growth regime ($a$), hierarchical spectral broadening regime ($b$), and limit cycle ($c$). (--), Model with macrosonic thermoacoustic interaction $\mathbb{Q}$; (- -), Model without $\mathbb{Q}$, ($\circ$), Navier-Stokes simulations.}
\label{fig: model_signal}
\end{figure}
In order to elucidate the physics of the hierarchical spectral broadening regime, the results from the nonlinear model derived in the previous section are discussed here and compared to the high-order Navier-Stokes calculations. Pressure time series from the time integration of~\eqref{eq: nonlinear_P_reduced} are in fairly good agreement with the fully compressible Navier-Stokes simulations (see figure~\ref{fig: model_signal}). Since several nonlinearities are neglected in calculations via~\eqref{eq: nonlinear_P_reduced}, time integration results in spurious temporal variations of the time averaged pressure fluctuations which are removed in further discussions. The time integration of the nonlinear model up to the limit cycle offers a significant reduction in computational cost (about $500$ times faster than the fully compressible Navier-Stokes simulations) and predicts the captured limit cycle amplitudes within $80\%$ accuracy. Additionally, upon excluding the macrosonic thermoacoustic interaction term from the model~\eqref{eq: MacrosonicThermoacousticInteraction}, the accuracy of the predicted limit cycle amplitude gets reduced to $60\%$. 

Figure~\ref{fig: model_spectra} shows the time evolution of the spectral energy density $E_k$  of the unstable mode (cf. Eq.~\eqref{eq:energy_spectral}) and its first seven overtones, as obtained from the nonlinear model discussed in Section ~\ref{sec: NonlinearRegime_modeling}, and compares it to the results obtained from the fully compressible Navier-Stokes simulations for the signal shown in figure~\ref{fig: model_signal}. In the spectral broadening regime of thermoacoustic wave amplification, the growth of spectral energy density of the $k^{\mathrm{th}}$ harmonic obtained from linear interpolation is approximately $k\alpha_2/2$, where $\alpha_2=2\alpha$ (since $E\propto p'^2$) is the growth rate of the  spectral energy density of the unstable mode, i.e., $E_k\sim e^{k\alpha_2t/2}$. For instance, figure~\ref{fig: model_spectra} corresponds to the case $T_H = 450$\,K, for which the growth rate is $\alpha = 8.64\,\mathrm{s}^{-1}$; therefore, the growth rate of the spectral energy density is $\alpha_2 = 2\alpha=17.28\,\mathrm{s}^{-1}$ for the unstable mode.

Nonlinear energy cascade in the spectral space can be further explained using reduced order modeling. Assuming propagation of purely travelling waves in the system and eliminating $U'$ from Eqs.~\eqref{eq: inviscid_p} and~\eqref{eq: inviscid_U}, the following  Burgers equation for the pressure fluctuations $p'$ is obtained:
\begin{equation}
\frac{\partial p'}{\partial t} - \frac{(\gamma+1)}{4\rho_0a_0}\frac{\partial p'^2}{\partial \xi} = \frac{\delta}{2}\frac{\partial^2 p'}{\partial \xi^2}, \label{eq: Burgers}
\end{equation}
where $\xi = a_0t - x$ is the travelling wave coordinate and $\delta$ is the axial dissipation coefficient given by
\begin{equation}
 \delta = \nu_0\left(\frac{4}{3} + \xi_B\right) + \frac{k}{\rho_0}\left(\frac{1}{C_v} - \frac{1}{C_p}\right). \label{eq: dissipation}
\end{equation}
In general, the wall-shear and the wall-heat flux expressions can be used as forcing functions in the above Burgers equation. However, the abrupt area changes present in the setup under study make the generalization and further time domain simplification seemingly challenging. Hence, we seek to only qualitatively explain the temporal evolution of the nonlinear cascade utilizing the Burgers equation. Introducing thermoacoustic amplification by adding a simple linear forcing term in Eq.~\eqref{eq: Burgers} yields: 
\begin{equation}
 \frac{\partial p'}{\partial t} - \frac{(\gamma + 1)}{4\rho_0a_0}\frac{\partial p'^2}{\partial \xi} = \frac{\delta}{2}\frac{\partial^2 p' }{\partial \xi^2} + \alpha_{\mathrm{th}}p', \label{eq: Burgers_thermoacoustic}
\end{equation}
where $\alpha_{\mathrm{th}}$ accounts for the thermoacoustic growth rate. Substituting a Fourier expansion for acoustic pressure $p'$, namely,
\begin{equation}
 p' = \sum_{k}p_k(t)\sin\left(\frac{k\omega_0}{2a_0}\xi\right),~~\mathrm{where}~~k=2,4,6,8,\cdots
\end{equation}
the following modal evolution equation is obtained:
\begin{equation}
 \frac{dp_k}{dt} = \alpha_{\mathrm{th}}p_k + Q(p_{k}) - \frac{\delta}{16} \left(\frac{\omega_0}{a_0}\right)^2k^2p_k, \label{eq: CascadeFunction1}
\end{equation}
where 
\begin{equation}
 Q(p_k) = \frac{(\gamma+1)\omega_0}{8\rho_0a^2_0}\Bigg(\sum^{n\leq k-2}(k-n)p_np_{k-n} - k\sum_{n\geq k+2}p_np_{n-k}\Bigg), \label{eq: CascadeFunction}
\end{equation}
$\alpha_{\mathrm{th}}$ determines the rate of thermoacoustic amplification of the $k^{\mathrm{th}}$ mode ($\alpha_{\mathrm{th}} = 8.64\,\mathrm{s}^{-1}$ for $k=2$), and $\omega_0$ is the angular frequency of the unstable mode. 
 \begin{figure}[!h]
\centering{\includegraphics[width=1.0\textwidth]{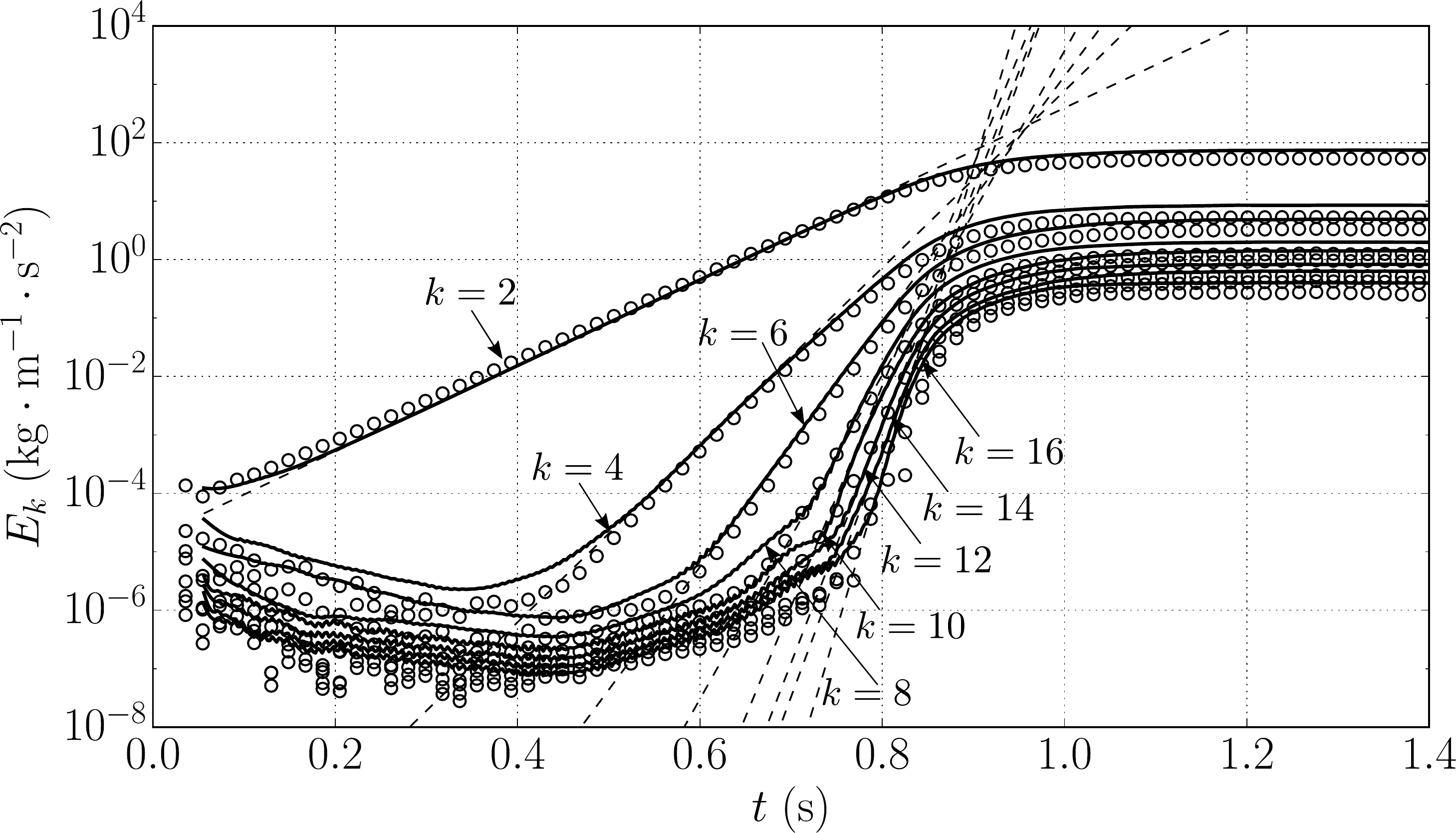}}
\caption{Comparison of evolution of the spectral energy density $E_k$ for the unstable mode and its first seven overtones as obtained from the signal shown in figure~\ref{fig: model_signal}. (--), Nonlinear model; ($\circ$), Navier-Stokes simulations; (- -), Linear interpolation of spectral energy density evolution in hierarchical spectral broadening regime.}
\label{fig: model_spectra}
\end{figure}
$Q(p_k)$ is the nonlinear cascade function quantifying the scale-by-scale flux of energy in the spectral space from the unstable mode to the harmonics which are dissipated by the molecular diffusion effects (momentum and thermal diffusivity). The third term on the right-hand side of Eq.~\eqref{eq: CascadeFunction1} signifies dissipation of the $k^{\mathrm{th}}$ mode through molecular diffusion. Assuming that the subsequent overtones of the unstable mode are characterized by pressure amplitudes which are an order of magnitude smaller ($p_{k+2}/p_{k} \ll 1$), the modal growth rate Eq.~\eqref{eq: CascadeFunction1} can be approximated such that 
 \begin{equation}
  \frac{dp_2}{dt}\approx\alpha p_2,\quad\frac{dp_4}{dt}\approx\frac{(\gamma+1)}{4\rho_0a^2_0}\omega_0p^2_2,\quad\frac{dp_6}{dt}\approx\frac{3(\gamma + 1)}{4\rho_0a^2_0}\omega_0p_2p_4, \label{eq: overtones}
 \end{equation}
  and so on for higher harmonics. Equation~\eqref{eq: overtones} shows that the growth rates of the overtones of the unstable mode due to the energy cascade are proportional to the ratio of the oscillation frequencies, that is to say, $p_k \sim e^{k\alpha t/2}$ in the hierarchical spectral broadening regime. It is important to note that the growth of higher overtones is not strictly exponential since they are not subject to thermoacoustic instability. Energy is cascaded into the overtones of the thermoacoustically unstable mode via nonlinear energy cascade due to the high acoustic wave amplitude. Nonetheless, it is possible to assume modal growth for a small time interval in the spectral broadening regime to quantify the exponential growth rate for each harmonic.

\section{Scales of thermoacoustically sustained spectral energy cascade}
\label{sec: Spectral energy distribution}
At the limit cycle, the energy of the unstable mode continues to increase and further cascades into higher harmonics on account of the nonlinear wave propagation. Higher harmonics have correspondingly shorter wavelengths due to which gradients in the longitudinal direction $x$ become large. Through the bulk viscosity and the thermal conductivity, the energy density is dissipated at higher harmonics, thus establishing a steady flow of energy from the unstable mode to the higher harmonics. The distribution of the spectral energy density in the harmonics can be derived utilizing an energy cascade modeling, analogous to the turbulent energy cascade~\cite{pope2000turbulent,Nazarenko_2011_WT}. Assuming travelling wave propagation, the total energy density of the $k^{\mathrm{th}}$ harmonic associated to planar wave propagation, $E^{(1\mathrm{D})}_k$, can be defined as 
\begin{equation}
 E^{(1\mathrm{D})} = \int^{\frac{2\pi a_0}{\omega_0}}_0E\,d\xi = \sum_k E^{(1\mathrm{D})}_k, \quad E^{(1\mathrm{D})}_k = \frac{\pi a_0}{\omega_0}E_k, \label{eq: TotalEnergy}
\end{equation}
where $E$ is the squared $L_2$ norm defined in Eq.~\eqref{eq:energy_time} where $p'$ and $u'$ are defined as a function of the travelling wave coordinate $\xi$. At the limit cycle, the rate of cascade of energy $\mathcal{E}$ in the spectral space balances the thermoacoustic wave amplification. As a result, higher harmonics (namely, the overtones of the unstable mode) are generated and later  dissipated by molecular dissipation, as indicated by the presence of  the molecular dissipation factor $\delta$ in Eq.~\eqref{eq: dissipation}. Thus, $\mathcal{E}$ is purely governed by thermoacoustic wave amplification at smaller harmonics and viscous dissipation at higher harmonics and scales as 
\begin{equation} \label{eq:dissipation_rate}
 \frac{\mathcal{E}}{\alpha^2_{\mathrm{eff}}\delta} = \mathrm{const.},
\end{equation}
where $\alpha_{\mathrm{eff}}$ is the effective energy amplification rate at the limit cycle.
\begin{figure}[!t]
\centering
  \centerline{\includegraphics[width=1.0\textwidth]{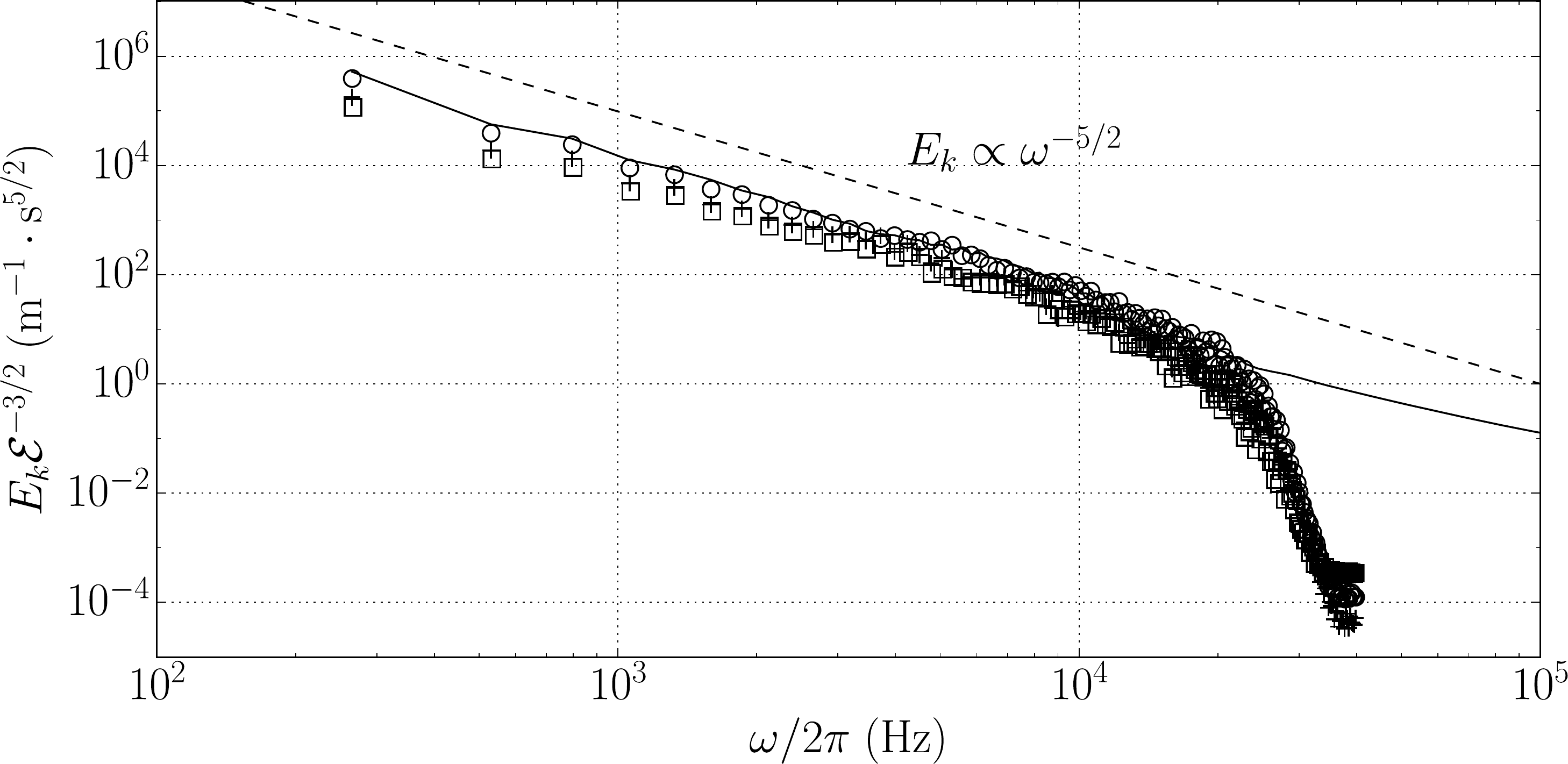}}
  \caption{Scaled spectral energy density $E_k\mathcal{E}^{-3/2}$ at the limit cycle against frequency of the harmonics. (--), Time domain nonlinear model at $T_H=450~$; Navier-Stokes simulations at ($\circ$), $T_H=450~$K; ($+$), $T_H = 500$\,K; ($\square$),  $T_H = 550$\,K; (- -) compares the variation of the energy with harmonic frequency with the power law derived in Eq.~\eqref{eq: SpectralScaling}.}
\label{fig: pressure_spectralDist}
\end{figure}
Based on the macrosonic thermoacoustic interaction, $\alpha_{\mathrm{eff}}$ can be estimated as
\begin{equation}
\alpha_{\mathrm{eff}}\approx(1-\gamma |p'|/p_0)\alpha,
\end{equation}
 where $\alpha$ is the growth rate of the unstable mode in the linear regime, and $|p'|$ is the amplitude of the pressure fluctuation at the limit cycle. Assuming that the total energy per unit mass $E^{(1\mathrm{D})}_k/\rho_0$ only depends on the rate of energy cascade $\mathcal{E}$ and on the angular frequency of the harmonic $\omega_k$, the following scaling is obtained: 
\begin{equation}
 \frac{E^{(1\mathrm{D})}_k \omega^{5/2}_k}{\rho_0\mathcal{E}^{3/2}} = \mathrm{const.}. \label{eq: SpectralScaling0}
\end{equation}
Substituting into Eq.~\eqref{eq: TotalEnergy} and eliminating $\rho_0$, the scaling for the spectral energy density $E_k$ remains the same such that 
\begin{equation}
 \frac{E_k\omega^{5/2}_k}{\mathcal{E}^{3/2}} = \mathrm{const.}. \label{eq: SpectralScaling}
\end{equation}
The scaling derived in Eq.~\eqref{eq: SpectralScaling} shows that the distribution of the energy density at limit cycle in the spectral space decays as $\omega^{-5/2}_k$---where $\omega_k$ is the frequency of the harmonic---as shown in figure~\ref{fig: pressure_spectralDist}. Moreover, in the spectral broadening regime, the energy density in the spectral space varies with the effective growth rate as $\alpha^{3}_{\mathrm{eff}}$ (figure~\ref{fig:intro_spectral}) which, in turn, allows the comparison among cases with varying hot side temperature $T_H$. Such a scaling arises purely from the mechanism of the nonlinear saturation resulting from the spectral energy cascade. It is however challenging to accurately estimate the acoustic energy production (quantified by $\alpha^{3}_{\mathrm{eff}}$) at the limit cycle utilizing the time domain nonlinear model, or even the fully compressible Navier-Stokes simulations, due to the shock capturing artificial viscosity that has to be added to ensure numerical stability in the presence of shocks. Overall, the spectral cascade of the energy density is balanced by the thermoacoustic wave amplification and the viscous dissipation, i.e., the shock waves are thermoacoustically sustained. However,  as shown previously~\cite{GuptaLS_SCITECH_2017}, at the location of the steepest gradient in pressure, the wall-heat flux from the acoustic field inside the isothermal walls and wall shear stress are maximum inside the regenerator. Thus, the propagation of sharp gradients in the acoustic field $p'$ and $U'$ inside the regenerator are \emph{counteracted} by the wall-shear stresses and the wall-heat flux resulting in smoothing of shock waves. Consequently, the scaling argument Eq.~\eqref{eq: SpectralScaling} can be further improved accounting for propagation of very large harmonics inside the regenerator.

\section{Summary}

% General summary: immediately point out turbulent analogy
The linear and nonlinear regimes of thermoacoustic wave amplification have been modelled up to the formation of shock waves in a minimal unit looped resonator with the support of high-fidelity fully compressible Navier-Stokes simulations. The computational setup is inspired by the experimental investigations conducted by Yazaki\emph{et al.}~\cite{Yazaki_PhysRevLet_1998} and geometrically optimized to maximize growth rates for the quasi-travelling wave mode. Three regimes of thermoacoustic wave amplification have been identified: ($i$) a monochromatic or modal growth regime, ($ii$) a hierarchical spectral broadening or nonlinear growth regime and ($iii$) a shock-dominated limit cycle. The modal growth regime is characterized by exponential amplification of thermoacoustically unstable modes. 
An acoustic energy budget formulation yielding a closed form expression of the Rayleigh index has been developed and the effect of variations in geometry and hot-to-cold temperature ratios on the thermoacoustic growth rates have been elucidated. The limit cycle regime exhibits many features of Kolmogorov's equilibrium turbulence, where energy, steadily injected at the integral length scale (wavelength of the second-mode harmonic), cascades towards higher wave numbers via inviscid mechanisms (wave steepening) and is finally dissipated at the Kolmogorov's length scale (of the order of the shock thickness). A grid sensitivity analysis has been carried out at the limit cycle to ensure that the entropy jump across the captured shock waves is grid convergent, hence assuring the same fidelity typically attributed to direct numerical simulations of turbulent flows, with the caveat that shocks in the present study are not fully resolved by the computational mesh.

The existence of an equilibrium \emph{thermoacoustic energy cascade} has thus been shown. The spectral energy density at the limit cycle, in particular, has been found to decay as $\omega^{-5/2}$ in  spectral space, the relevant intensity scaling with growth rate as $\alpha^3$. Such findings are confirmed by a novel theoretical framework to model  thermoacoustic nonlinearities, which has lead to the formulation of a one-dimensional time-domain nonlinear acoustic model. The model is correct up to second order in the perturbation variables and addresses the fundamental problem of high amplitude wave propagation in the presence of wall-shear and wall-heat flux, accounting for thermodynamic nonlinearities such as the second-order interactions between the pressure fluctuations and the wall-heat flux, namely the macrosonic thermoacoustic interaction term. 
%The model also confirms the dynamics of energy transfer across scales in the nonlinear growth regime: the growth of higher harmonics is hierarchical in nature and higher harmonics are amplified at faster rates. in particular the $k^{\mathrm{th}}$ harmonic grows at the rate of $\alpha k/2$, where $\alpha$ is the rate of growth of energy in the unstable mode. 

% \include{ThermoacousticShocks}
% %%%%%%%%%%%%%%%%%%%%%%%%%%%%%%

% %%%%%%%%%%%%%%%%%%%%%%%%%%%%%%
\chapter{Spark-plasma generated shock waves}
The contents of this chapter were presented in the AIAA-Scitech presentation in January 2019~\cite{BsinghEtAl_2019_AIAA} and have been reported here in abridged form. The work presented here was motivated by the experimental study of wave propagation due to spark-plasma heat deposition by Singh~\emph{et al.}~\cite{Bsingh2018_EinF}

\section{Introduction}
Plasma discharges are a growing field of interest with numerous applications across a wide range of areas in science and engineering including flow and combustion control, biomedicine, materials processing, nanotechnology, and environmental engineering~\cite{Corke_ARFM_2010}. In the area of aerodynamic flow control, considerable work over the past couple of decades has focused on using plasma actuators as active flow control devices on aerodynamic bodies. Nanosecond pulsed plasmas have recently been attracting great interest due to their extremely efficient generation of excited and ionized species and relatively low power consumption. Recent studies have investigated the effect of various parameters of the nanosecond pulse driven plasma discharges on the nature of the plasma~\cite{Pai_PSST_2010}. The aim of this study is to analyze the flow field generated as a result of high-frequency repeated plasma between two electrodes with the help of high-fidelity simulations, low order modeling, and experiments.

\section{Experimental study and motivation (Credit: Ms. Bhavini Singh, Mr. Lalit Rajendran, and Dr. Sally Bane)}
Figure~\ref{fig:SparkElectrodes} shows the snapshots of repeated spark discharge between two electrodes as seen in PIV measurements. Multiple bursts of pulses were sent to the pulse-generator which creates the repeated spark discharge. The instantaneous heat deposition due to the spark discharge generates a heat-induced cylindrical shock wave which propagates outward and is deflected by the electrode geometry. 
\begin{figure}[!t]
	\centering
	\includegraphics[width=\textwidth]{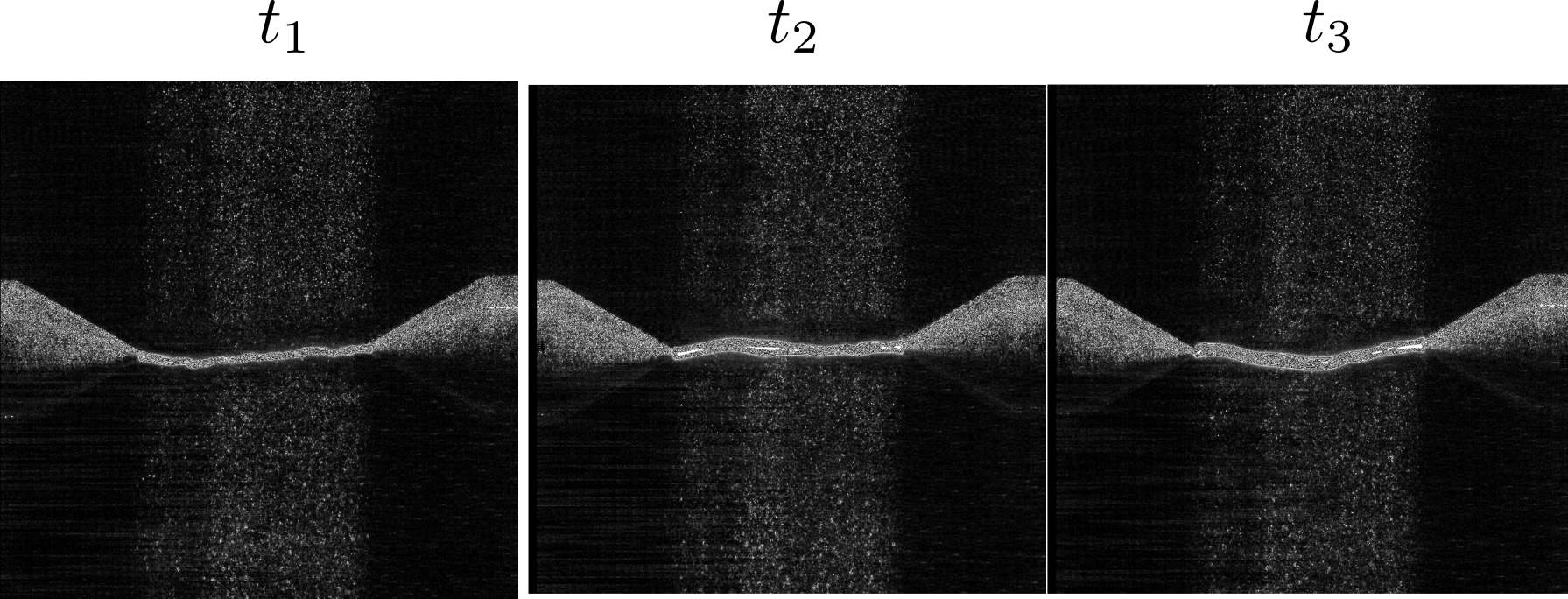}
	\caption{PIV measurements of repeated spark discharges at 1 kHz of frequency between the electrodes 5 mm apart (Courtesy of Ms. Bhavini Singh, Mr. Lalit Rajendran, and Dr. Sally Bane).}
	\label{fig:SparkElectrodes}
\end{figure}
\begin{figure}[!b]
	\centering
	\includegraphics[width=\textwidth]{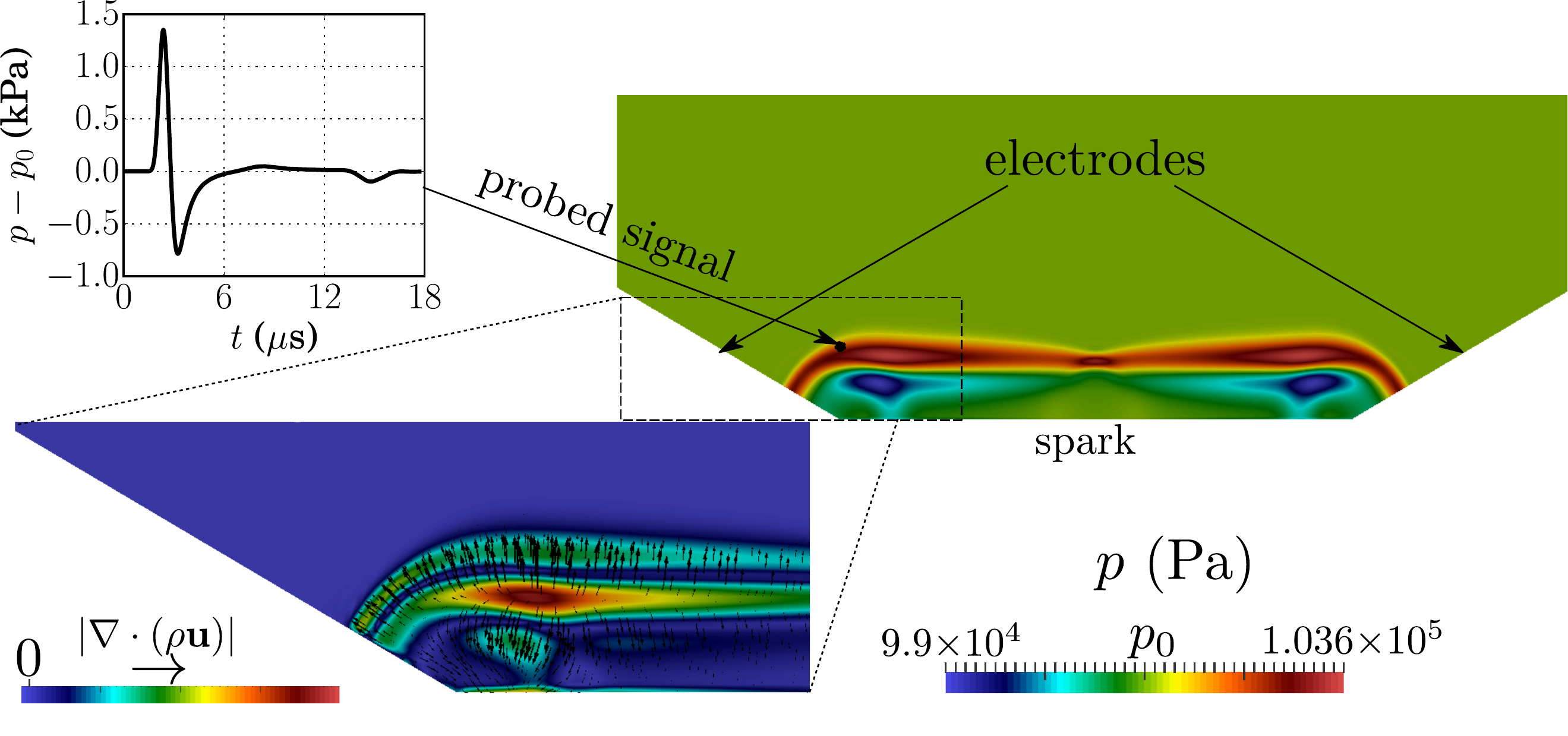}
	\caption{Shock wave induced due to the spark plasma heat deposition.}
	\label{fig:SparkElectrodesShocks}
\end{figure}
Behind the shock wave, vortical structures are generated due to grazing of the shock wave over hard no-slip walls of electrodes. Such vortical structures are utilized in the boundary layer re-attachment over aerodynamic objects to improve lift characteristics. Figure~\ref{fig:SparkElectrodesShocks} shows the preliminary simulation results obtained via modeling the heat deposited by a single spark event by a simplistic isothermal boundary condition.

\section{Computational setup}
A high-order unstructured, fully compressible Navier-Stokes solver (\textsc{sd3DvisP}) is used to simulate the shocks induced by the heat deposition by the spark. The \textsc{sd3DvisP} solver is an MPI parallelized fortran 90 code for compressible flows based on the high-order Spectral Difference (SD) scheme for unstructured hexahedral elements.~\cite{kopriva:96,sun:07} The solver is capable of running with arbitrary preselected orders of accuracy and provides minimal numerical dissipation.~\cite{lodato:14b} For the computation of high-speed compressible flows with shocks and discontinuities, a self-calibrating shock capturing methodology is available in the solver.
Originally developed for the Discontinuous Galerkin scheme,~\cite{persson:06} this method, enables sub-cell resolution of discontinuities which are detected through a modal sensor. The \textsc{sd3DvisP} solver has been successfully applied to the computation of flows with shocks, including shock/vortex interaction and shock/wavy-wall interactions.~\cite{lodato:15,lodato:16}
\begin{figure}[!b]
 \centerline{\includegraphics[width=\textwidth]{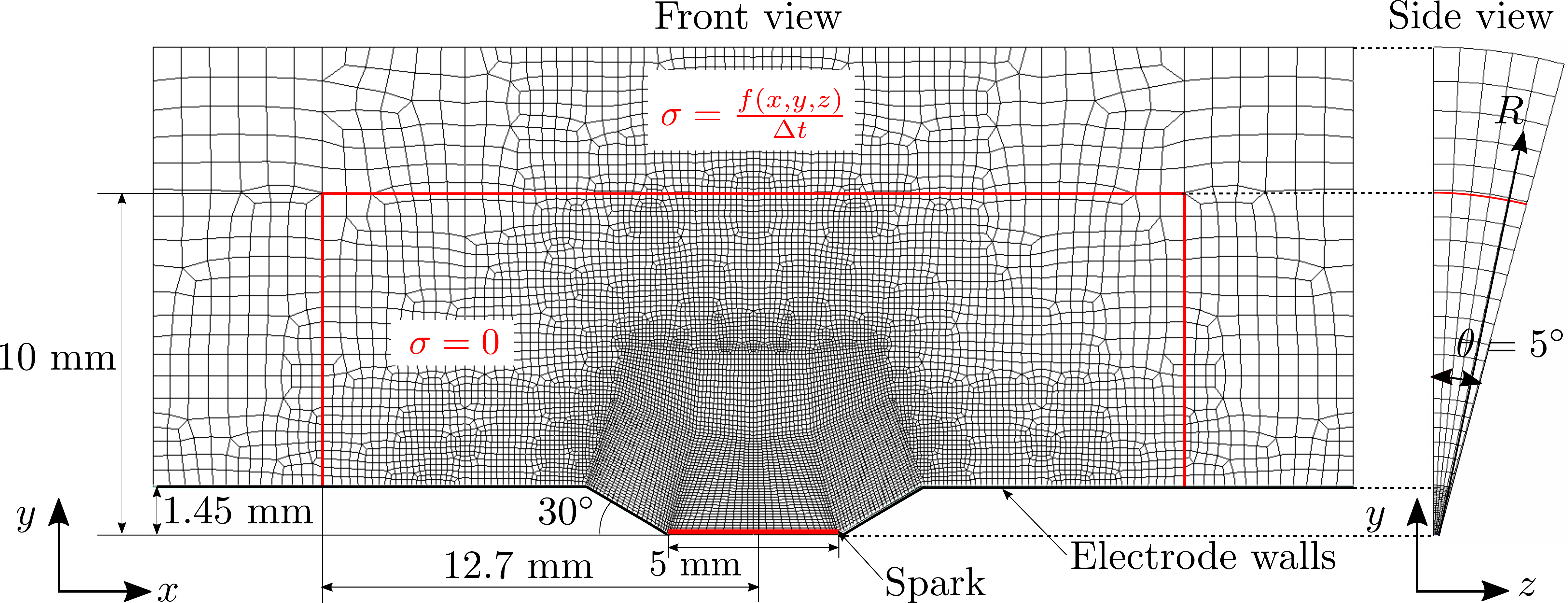}}
 \caption{Computational setup of the electrodes. The setup models axisymmetric flow conditions.}
\label{fig: Setup}
\end{figure}
In the scope of the current work, we investigate the flow field generated in an axisymmetric computational setup~(figure~\ref{fig: Setup}). The governing equations solved are conservation laws for mass, momentum, and total energy in three dimensions, given by,
\begin{subequations}
	\label{eq:navierstokes_spark_plasma}
	\begin{align}
		\frac{\partial}{\partial t} \left(\rho\right) &+ \frac{\partial}{\partial x_j} \left(\rho u_j \right)  = \sigma\left(\rho_{\infty}-\rho\right),
		\label{subeq:ns1_sp}
		\\
		\frac{\partial}{\partial t} \left(\rho u_i\right) &+ \frac{\partial}{\partial x_j} \left(\rho u_i u_j\right)  =  -\frac{\partial}{\partial x_i} p  +
		\frac{\partial}{\partial x_j} \tau_{ij} + \sigma\left(\left(\rho u\right)_{\infty}-\left(\rho u\right)\right),
		\label{subeq:ns2_sp}
		\\
		\frac{\partial}{\partial t} \left(\rho \, E\right) &+ \frac{\partial}{\partial x_j} \left[ u_j \left(\rho \, E + p \right) \right] =
		\frac{\partial}{\partial x_j } \left(u_i \tau_{ij} - q_j\right) + \dot{\mathcal{Q}}_{\mathrm{spark}} + \sigma\left(\left(\rho E\right)_{\infty} - \rho E\right),
		\label{subeq:ns3_sp}
	\end{align}
\end{subequations}
respectively, are solved where $x_i$($x_1$, $x_2$,  $x_3$ or equivalently, $x$, $y$, $z$) are the cartesian coordinates, $u_i$ are the velocity components in each of those directions, and $p$, $\rho$, $T$, and $E$ are the instantaneous pressure, density, temperature, and total energy per unit mass, respectively. We model the energy deposition due to the spark plasma utilizing the $\dot{\mathcal{Q}}_{\mathrm{spark}}$ term in the energy equation given by, 
\begin{equation}
\dot{\mathcal{Q}}_{\mathrm{spark}} = \mathcal{Q}_0\exp\left(-\alpha\left({t-\tau}\right)^2\right),
\end{equation}
where, 
\begin{equation}
\mathcal{Q}_0 =\rho_0 C_p{\alpha \tau T_s}g(x,y,z),
\end{equation}
$T_s$ is the characteristic spark temperature, and $\alpha$ and $\tau$ are the spark energy deposition rate parameters. Function $g(x,y,z)$ determines the shape of the spark. For simplicity, we model the spark as a straight channel with heat deposition such that, 
\begin{equation}
 g(x,y,z) = \exp\left(-8\times 10^8\left(R-R_{cut}\right)^2\right),
\end{equation}
where $R_{cut} = 0.6~\mathrm{mm}$. 
The exponential relaxation terms (characterized by $\sigma$ in Eqs.~\ref{subeq:ns1}-\ref{subeq:ns3}) are used for non-reflective sponge layers (borders distinguished by red line in figure~\ref{fig: Setup}) which model the outflow boundaries. Within the domain $|x|<x_s$ and $R<R_s$, $\sigma=0$ is used and outside the domain, 
\begin{align}
 \sigma = \frac{f(x,y,z)}{\Delta t} &= \frac{1}{4\Delta t}\left(1+\tanh\left(\beta|x - x_s|\right)\right)\left(1+\tanh\left(\beta\left(R(y,z)  - R_s\right)\right)\right),\\ ~~&\mathrm{for}~~|x|>x_s,~~R>R_s, \nonumber
\end{align}
is used. Parameter $\beta$ determines the spatial variation of the sponge layers. In the current simulations $x_s = 12.7~\mathrm{mm}$ and $R_s = 10~\mathrm{mm}$ are chosen. 

Below we discuss two simulation results, one corresponding to no-slip adiabatic walls of the electrodes (viscous), and the other corresponding to hard-slip walls (inviscid). These simulations enable comparison of viscous and inviscid vorticity generation mechanisms. The spark parameters used in the current simulations are, 
\begin{equation}
 T_s = 2000~\mathrm{K},~\tau = 20~\mathrm{ns}~\mathrm{and}~ \alpha = 1\times 10^{17}~\mathrm{s}^{-2}.
\end{equation}

\section{Results}
Figure~\ref{fig: FlowModeling} illustrates the basic flow features observed behind the curved shock induced by the heat deposition. As the temperature of the discharge region drops after reaching a maximum, an expansion wave is generated behind the shock wave which creates a backward flow (flow towards the electrode center) in the domain. Moreover, due to the shock curvature, the vorticity generated also enhances the flow towards the domain. 
Below, we quantitatively discuss the effect of baroclinic torque on the vorticity.
\begin{figure}[!t]
 \centerline{\includegraphics[width=\textwidth]{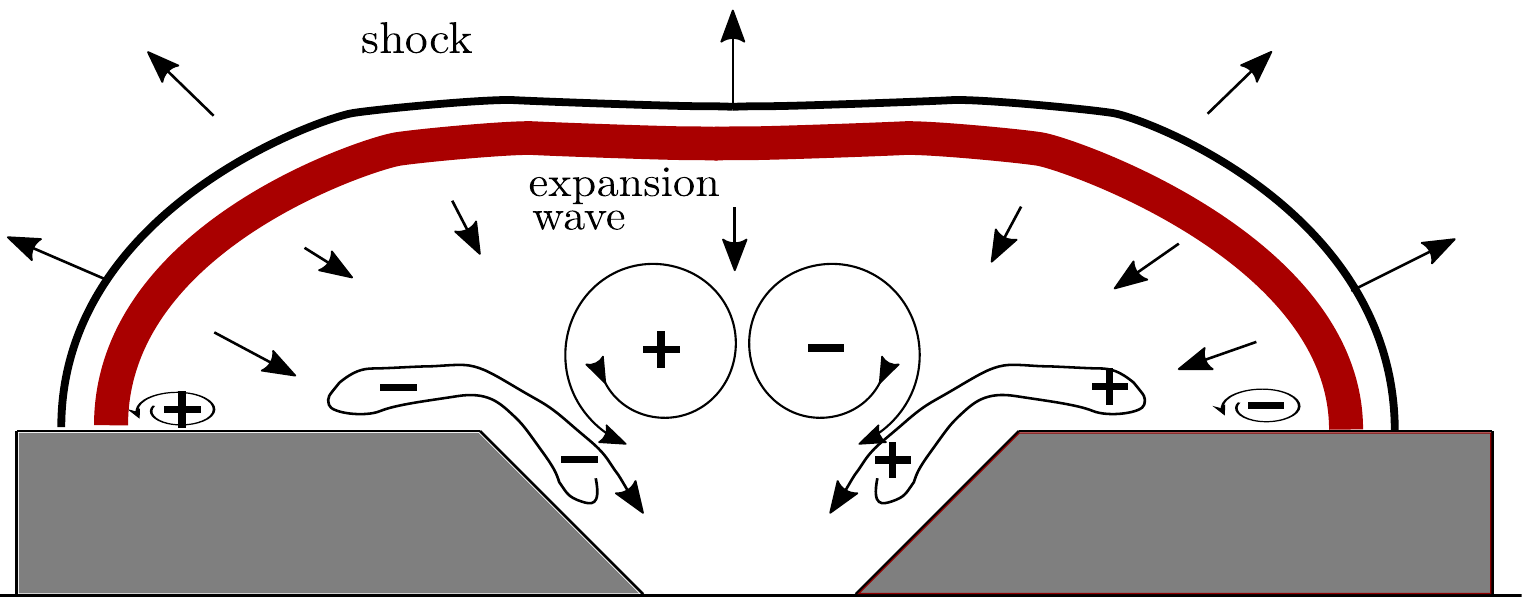}}
 \caption{Schematic illustrating the flow features generated by the heat deposition induced shock wave and the curvature in the shock wave caused by the electrode geometry. }
\label{fig: FlowModeling}
\end{figure}

Figure~\ref{fig: InitialStagesVorticity_noslip} shows the initial stages of vorticity $\omega_n$ normal to $x-y$ plane. As the heat deposition peaks, a cylindrical shock wave emanates from the spark. Due to the slant boundaries, the shock wave diffracts and develops curvature. At the initial stages, the shock wave diffraction from the walls causes high localized baroclinic torque which results in production of vorticity. Moreover, behind the curved shocks, boundary layer develops in case of no-slip electrode walls (figure~\ref{fig: InitialStagesVorticity_noslip}). The vorticity field near the electrode walls is modified considerably due to no-slip walls (figure~\ref{fig: InitialStagesVorticity_noslip}) compared to slip walls (figure~\ref{fig: InitialStagesVorticity_slip}).
 
Behind the shock wave, due to sudden decrease in heat rate (after reaching maximum), an expansion wave is generated. As a result, the flow decelerates and is directed inward (towards the center of the spark) by the electrode walls. Since, the current simulations model spark boundary as a symmetric boundary, the flow is re-directed upwards, hence driving the large scale vortices generated. While the magnitude of vorticity is higher, the sense of vorticity is similar to the one observed in experiments. Due to the simulations being axisymmetric, the vortices do not breakdown. Thus, no smaller scale vortices are observed and the large scale vortices simply decay due to fluid viscosity.

\begin{figure}[!b]
 \centerline{\includegraphics[width=\textwidth]{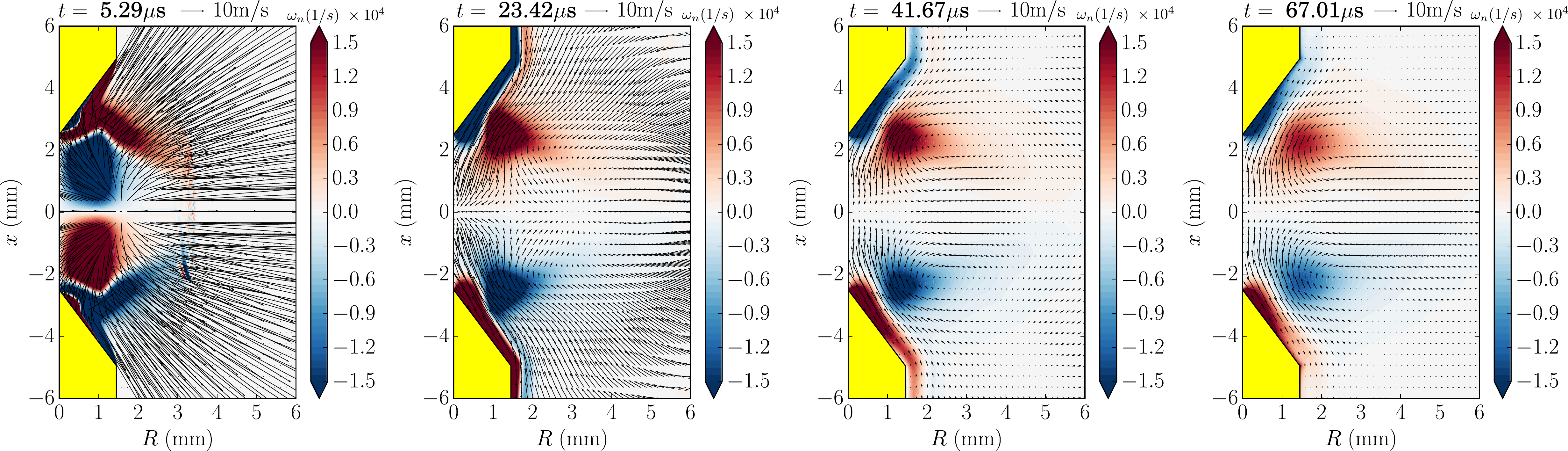}}
 \caption{Evolution of the vorticity $\omega_n$ normal to the $x-y$ plane in the initial stages of the shock wave propagation for no-slip electrode walls.}
\label{fig: InitialStagesVorticity_noslip}
\end{figure}

\begin{figure}[!b]
 \centerline{\includegraphics[width=\textwidth]{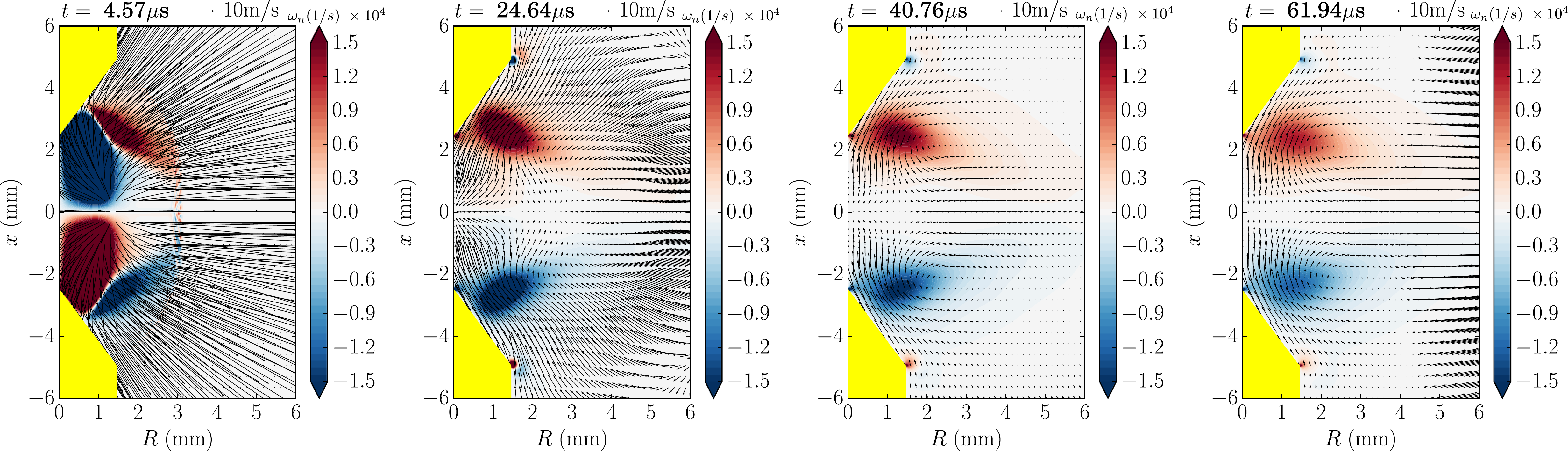}}
 \caption{Evolution of the vorticity $\omega_n$ normal to the $x-y$ plane in the initial stages of the shock wave propagation for slip electrode walls.}
\label{fig: InitialStagesVorticity_slip}
\end{figure}

Figure~\ref{fig: InitialStagesVorticity_slip} shows the the vorticity evolution for a simulation with identical $\alpha$, $\tau$, and $T_s$ but only hard-wall symmetric condition imposed on the electrode walls. Due to fluid slip, the walls do not act as a source of vorticity. However, the large scale vortices can still be observed with similar magnitude of $\omega_n$ thus suggesting that inviscid baroclinic terms result in initial generation of vorticity.

\begin{figure}[!t]
 \centerline{\includegraphics[width=\textwidth]{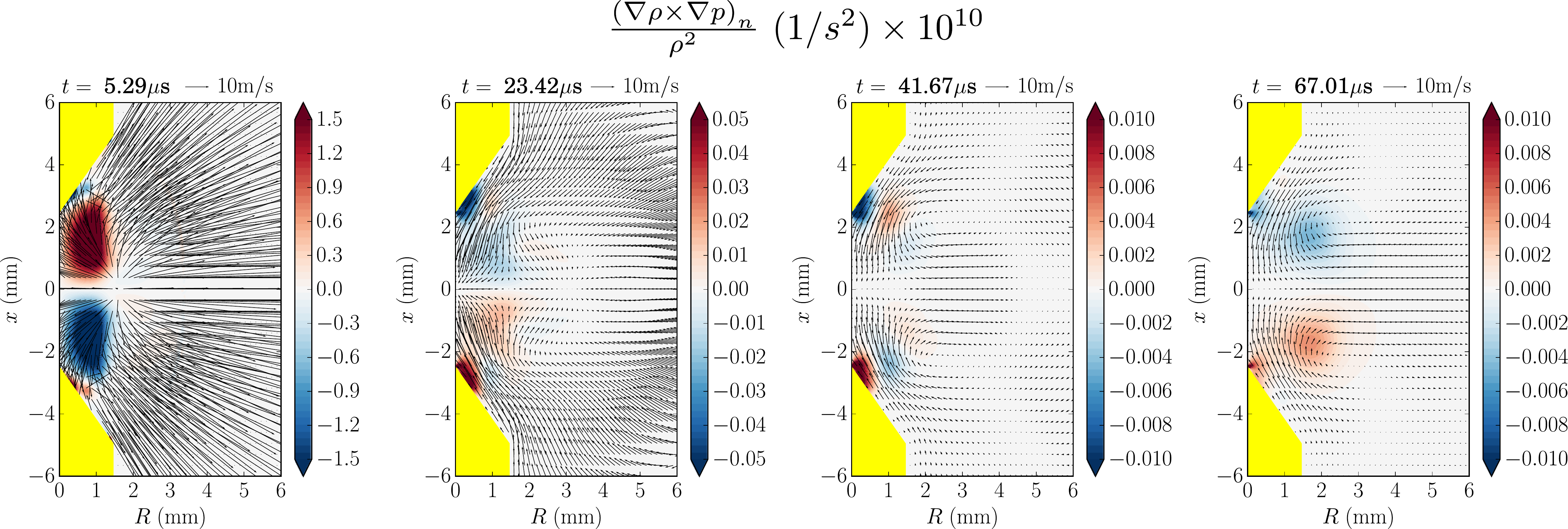}}
 \caption{Evolution of the normalized baroclinic torque  $\Gamma_n$ normal to the $x-y$ plane in the initial stages of the shock wave propagation for no-slip electrode walls.}
\label{fig: InitialStagesBaroclinic_NoSlip}
\end{figure}

Figures~\ref{fig: InitialStagesBaroclinic_NoSlip} and~\ref{fig: InitialStagesBaroclinic_Slip} show the evolution of baroclinic torque $\Gamma_n$ normal to the x-y plane, given by,
\begin{equation}
 \Gamma_n = \frac{\left(\nabla \rho \times \nabla p\right)_n}{\rho^2}\left(\frac{\rho_0^2 l^2}{\Delta p \Delta \rho}\right),
 \label{eq: baroclinic}
\end{equation}
where $l$ is the characteristic length scale of pressure and density gradients ($l\sim1~\mathrm{mm}$), $\Delta p$ and $\Delta \rho$ are the pressure and density jumps across the shock wave. In the current simulations (both no-slip wall and slip wall), the pressure jump varies from $\Delta p\sim 110~\mathrm{kPa}$ in the center to $\Delta p\sim 50~\mathrm{kPa}$
near the electrode walls. Similarly, the density jump varies from $\Delta \rho\sim 0.8~\mathrm{kg/m}^3$ in the center to $\Delta \rho\sim 0.5\sim \mathrm{kg/m}^3$ near the electrode walls. Values near the electrode walls have been used for normalization in Eq.~\ref{eq: baroclinic}. 

\begin{figure}[!t]
 \centerline{\includegraphics[width=\textwidth]{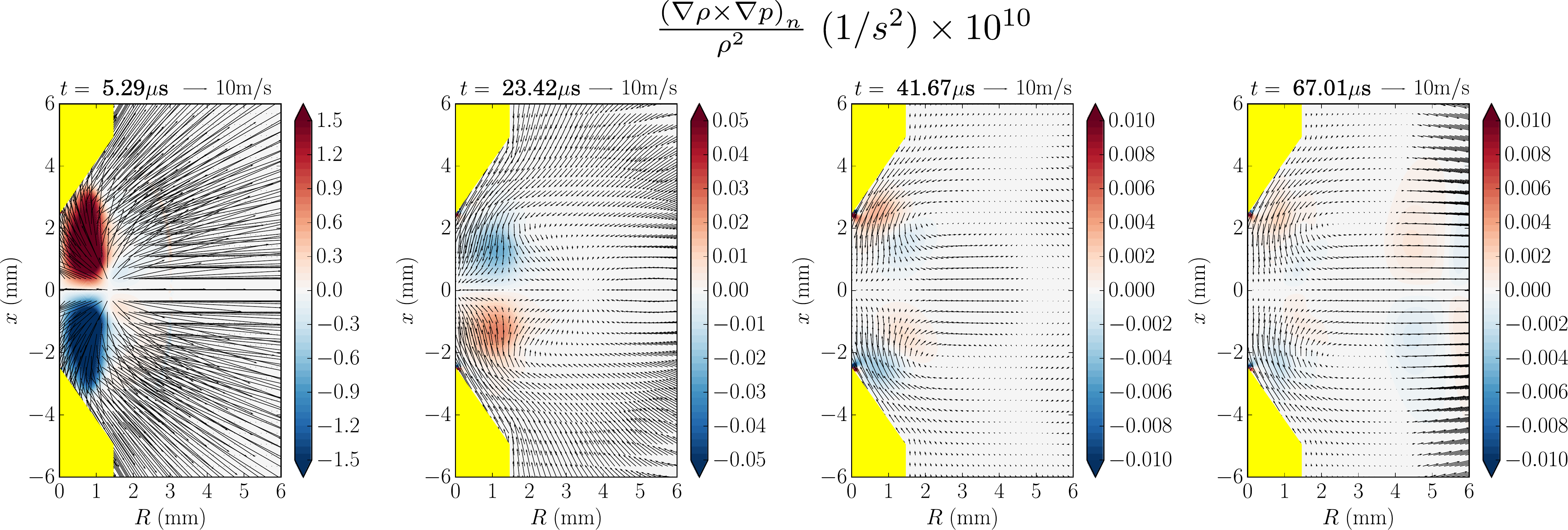}}
 \caption{Evolution of the normalized baroclinic torque  $\Gamma_n$ normal to the $x-y$ plane in the initial stages of the shock wave propagation for slip electrode walls.}
\label{fig: InitialStagesBaroclinic_Slip}
\end{figure}
In figures~\ref{fig: InitialStagesBaroclinic_NoSlip} and~\ref{fig: InitialStagesBaroclinic_Slip}, we note that the baroclinic torque $\Gamma_n$ in both the slip wall and no-slip wall cases evolves similarly. Moreover, the sense of the torque matches the vorticity alignment shown in figures~\ref{fig: InitialStagesVorticity_noslip} and ~\ref{fig: InitialStagesVorticity_slip} thus indicating that baroclinic torque indeed plays an important role in initial generation of the large scale vortices observed in both the simulations and the experiments.

% \include{SparkPlasmaShockWaves}
% %%%%%%%%%%%%%%%%%%%%%%%%%%%%%%

% %%%%%%%%%%%%%%%%%%%%%%%%%%%%%%
\chapter{Numerical modeling of detonation waves}
The contents of this chapter were presented in the AIAA-Aviation meeting in June 2018~\cite{GuptaSLSS_Aviation_2018} and have been reported here in abridged form. The work presented here was motivated by the experimental study by Schwinn~\emph{et al.}~\cite{Schwinn_CandF_2018} at Maurice J. Zucrow Laboratories at Purdue. 
\section{Introduction}

Combustion waves are primarily classified as deflagration or detonation waves, the latter being high amplitude pressure waves propagating at supersonic speeds. Heating caused by a propagating adiabatic shock-compression results in ignition of fuel-oxidizer mixture. The fluid containing reaction products expands behind the shock wave and depending on downstream boundary conditions, further accelerates the wave front sustaining its propagation \cite{JHlee_2008}. Classical Zel’dovich, von Neumann and D\"oring (ZND) model postulates equilibrium one-dimensional detonation waves. In experiments and detailed theoretical studies, unstable detonation wave propagation with complex reactive chemistry and compressible flow physics interactions are often observed~\cite{JHlee_2008, SHEPHERD2009}. Consequently, unsteady dynamics of unstable one-dimensional pulsating detonation waves have received wide-spread attention \cite{Erpenbeck_PoF_1964, Lee_JFM_1990, Sharpe_PRSA_1999}. 

Detonation waves result in very high pressure gains, which exhibit higher thermodynamic efficiencies when utilized for mechanical work~\cite{Wolanski_RDE_2015,WintenbergerShepherd_JPP_2006}. Rotating Detonation Engines (RDEs) are propulsion devices, which utilize such pressure gains through continuously spinning detonation waves for generating thrust~\cite{Bykovskii_JPP_2006}. Fuel-oxidizer mixture is injected axially into an annular shaped device which is undergoing combustion due to the rotating detonation waves. High pressure combustion expels products axially from the opposite end thus generating thrust. However, the complex combustion wave propagation dynamics in such devices require further attention and careful analysis for effective design~\cite{Schwinn_CandF_2018}. In this work, we perform numerical investigations of such sustained detonation dynamics in a periodic domain, inspired by the recent experimental study of Schwinn \emph{et al.}~\cite{Schwinn_CandF_2018} in which a straight-line detonation chamber was analyzed experimentally exhibiting sustained resonance of detonation waves.

Numerical studies of RDEs usually focus on the general flow features exhibited by sustained detonation waves~\cite{Schwer_AIAA_2010,Schwer_CandF_2011,Lietz_AIAA_2018}. In this work, we numerically investigate the effect of fuel-oxidizer injection rates on the dynamics of detonation waves and sustenance in a periodic one-dimensional domain. We model fuel-oxidizer injection rates utilizing a one-dimensional model reduction which represents solving the governing equations close to the injector plate in an experimental setup. Moreover, we adopt a device-scale dynamical system perspective in the current study to elucidate the wave propagation dynamics.

 \section{Experimental study and motivation (Credit: Mr. Kyle Schwinn and Dr. Carson Slabaugh)}
\label{sec: ExperimentalMotivation}

Schwinn \emph{et al.}~\cite{Schwinn_CandF_2018} conducted an experimental study on sustained detonation wave resonance in a linear semi-bounded channel with CH$_4$+O$_2$ mixture. Figure~\ref{fig:DRONE_Instrumentation_Diagram} shows a schematic of the experimental setup of Schwinn \emph{et al.}~\cite{Schwinn_CandF_2018} with the instrumentation points and the two possible window locations for obtaining optical measurements. The experimental setup provided modular capability between making single-point measurements with pressure transducers and ion probes, and obtaining high frame rate images of the chamber dynamics through large, fused quartz windows. Pressure fluctuations in the chamber were measured using piezoelectric pressure transducers (PCB 113B26). 
\begin{figure}[!t]
	\centering
	\includegraphics[width=0.9\textwidth]{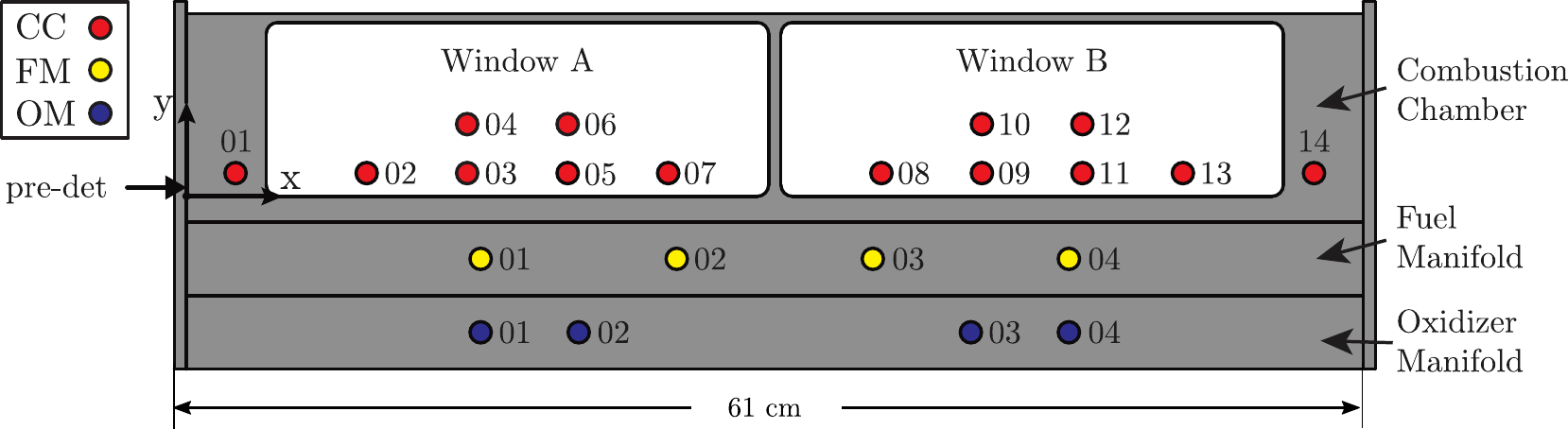}
	\caption{Schematic diagram of instrumentation port nomenclature for the combustion chamber (CC), the fuel manifold (FM), and the oxidizer manifold (OM). White panels in the combustion chamber indicate regions of optical access.(Courtesy of Schwinn \emph{et al.}~\cite{Schwinn_CandF_2018})}
	\label{fig:DRONE_Instrumentation_Diagram}
\end{figure}

Using different diagnostics on opposites sides of the narrow chamber, simultaneous pressure and chemiluminescence measurements of the chamber combustion dynamics were performed. Figure~\ref{fig:Test40_ChemiluminescencePCB}$(a)$ shows a detonation wave (\ref{fig:Test40_ChemiluminescencePCB}A) on the left side of the image, propagating in the positive x-direction and approaching the instrumentation port location CC-09 (see figure~\ref{fig:DRONE_Instrumentation_Diagram}). The average intensity at probe location CC-09, which is proportional to the local combustion heat release, is shown in figure~\ref{fig:Test40_ChemiluminescencePCB}$(b)$ along with the in-phase local pressure fluctuation. 

\begin{figure}[!t]
	\centering
	\includegraphics[width=0.9\textwidth]{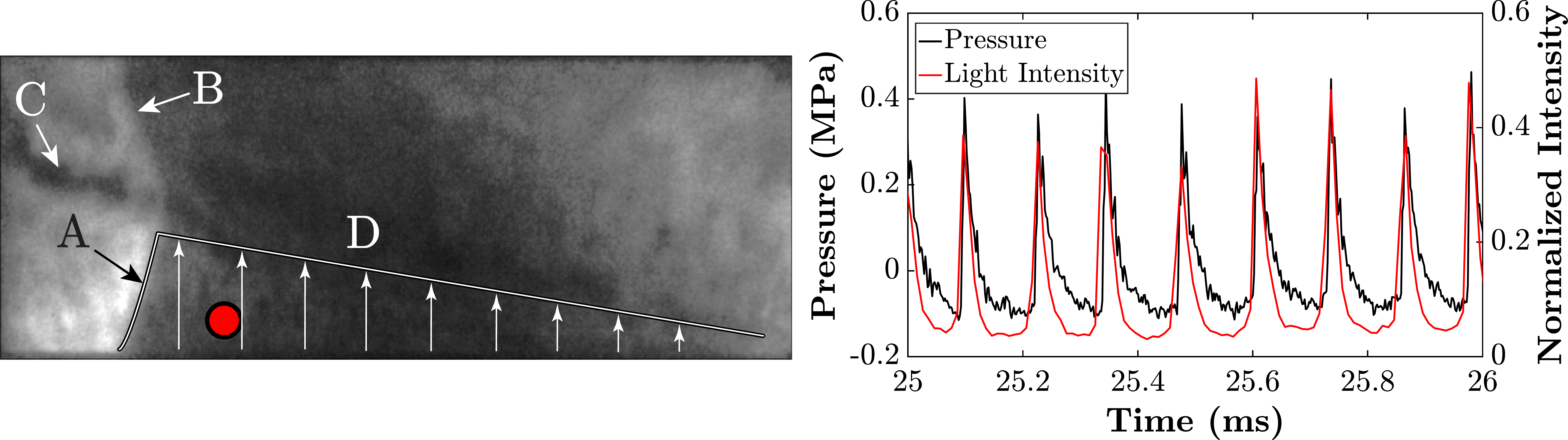}
	\put(-400,110){$(a)$}
    \put(-190,110){$(b)$}
	\caption{a) Detonation imaged using OH*-chemiluminescence (Window B), with instrumentation port CC-09 marked with red circle. b) Overlay of intensity data with pressure data from instrument at location CC-09. (Courtesy of Schwinn \emph{et al.}~\cite{Schwinn_CandF_2018})}
	\label{fig:Test40_ChemiluminescencePCB}
\end{figure}

Schwinn \emph{et al.}~\cite{Schwinn_CandF_2018} observed wave steepening and amplification caused by the fuel-oxidizer injection through a time series of OH*-chemiluminescence, as shown in figure~\ref{fig:Chemiluminescence}. In the first frame (with arbitrary time assignment $+0 \, \mu s$), a combustion front (figure~\ref{fig:Chemiluminescence}A) propagates from an auto-ignition source at the interface between the reactants and product gases. In this same frame, a high-intensity compact combustion wave (\ref{fig:Chemiluminescence}B) from the previous cycle is traveling in the positive x-direction. The recently-ignited combustion front propagates upstream into the reactant jets and reaches the injector face, and then travels along the injector face in the positive x-direction (\ref{fig:Chemiluminescence}C). Between two successive waves, deflagrative burning in the reactant fill is observed as a region of low intensity (\ref{fig:Chemiluminescence}D). As the combustion front continues to propagate in the positive x-direction into the reactants, a distinct contact surface forms between the reactants and product gases (\ref{fig:Chemiluminescence}E). The combustion wave accelerates due to the heat release behind the shock, intensifying and becoming more spatially compact (\ref{fig:Chemiluminescence}F). As the steep-fronted combustion wave continues to propagate through the channel, the expanding combustion front from an auto-ignition event appears (\ref{fig:Chemiluminescence}G), beginning the process for the next cycle.

\begin{figure}[!t]
	\centering
	\includegraphics[width=0.99\textwidth]{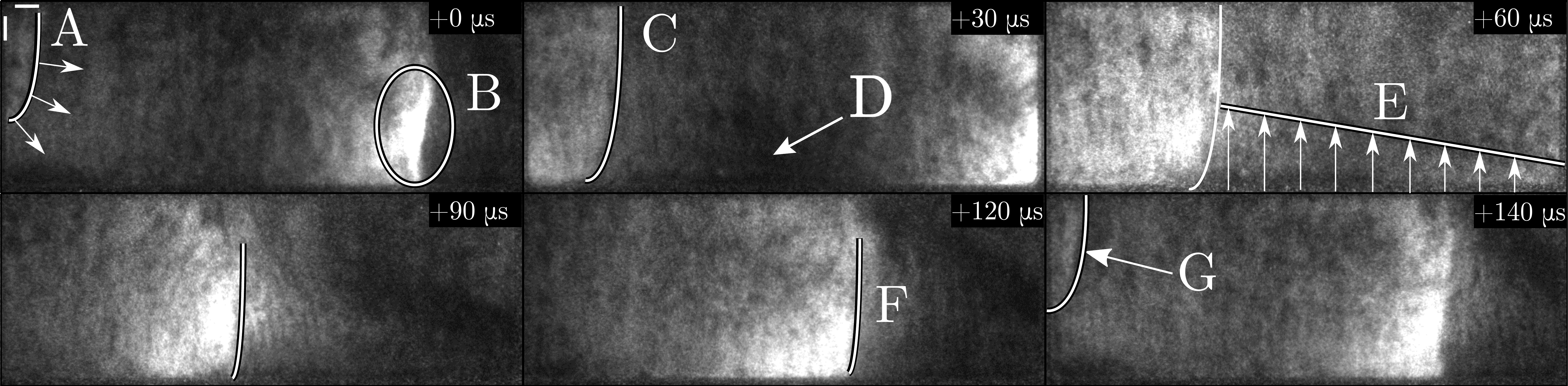}
	\caption{Time series of OH*-chemiluminescence images (imaged in Window A) detailing the combustion zones in the development region of the combustion chamber (Scale is $10 \, mm$; see first frame).(Courtesy of Schwinn \emph{et al.}~\cite{Schwinn_CandF_2018})}
	\label{fig:Chemiluminescence}
\end{figure}

By inspection of images in figure~\ref{fig:Chemiluminescence}, the combustion waves are determined to be supersonic and the nature of the limit-cycle behavior of the chamber dynamics was linked to periodic injection, ignition, amplification, and acceleration processes. The high-speed chemiluminescence images obtained in Window B, as seen in figure~\ref{fig:Test40_ChemiluminescencePCB} enable the identification of structures in the reaction zone that confirm the presence of self-excited, periodic detonation waves.

Motivated by the experimental study conducted by Schwinn \emph{et al.}~\cite{Schwinn_CandF_2018}, we study the possibility of a reduced order one-dimensional detonation wave resonance model in this work to elucidate the nature of sustained detonation wave resonance from the perspective of dynamical systems. We note that the fuel-oxidizer injection and coupling of injector nozzles is crucial to allow the spatial variation of fuel injection coupled with the local pressure variations caused by the detonation waves. In our study, we assume premixed fuel-oxidizer injec human being tion of H$_2$+O$_2$+Ar (due to ease of achieving detonations) mixture through sonic (converging) nozzles in the chamber. Detailed model for hydrogen-oxygen-argon reaction is a 9-species and 19-step reaction scheme based on Jachimowski's mechanisms for hydrogen-air combustion~\cite{Jachimowski1988_Hydrogen}. In the following sub-section, we discuss the validation of detailed chemistry model in a high-order spectral difference solver, \textsc{sd3DvisP}~\cite{lodato2013discrete,lodato2014structural,lodato:14b,GuptaLS_JFM_2017} used in the previous chapter.

\section{Numerical validation}
\label{sec: numerics}
 Detonation systems are usually modeled with single step reactive chemistry models based on induction times in which progress variable of reaction replaces the species concentrations in dynamical equations. The dynamics of resonating detonations are elucidated in this work based on detailed chemistry models utilizing the state-of-the-art reaction mechanisms and spectral difference solver utilized in previous chapters.

\subsection{Validation with Shock-to-detonation transition}
 \begin{figure}[!t]
\centering
\includegraphics[width=0.9\textwidth]{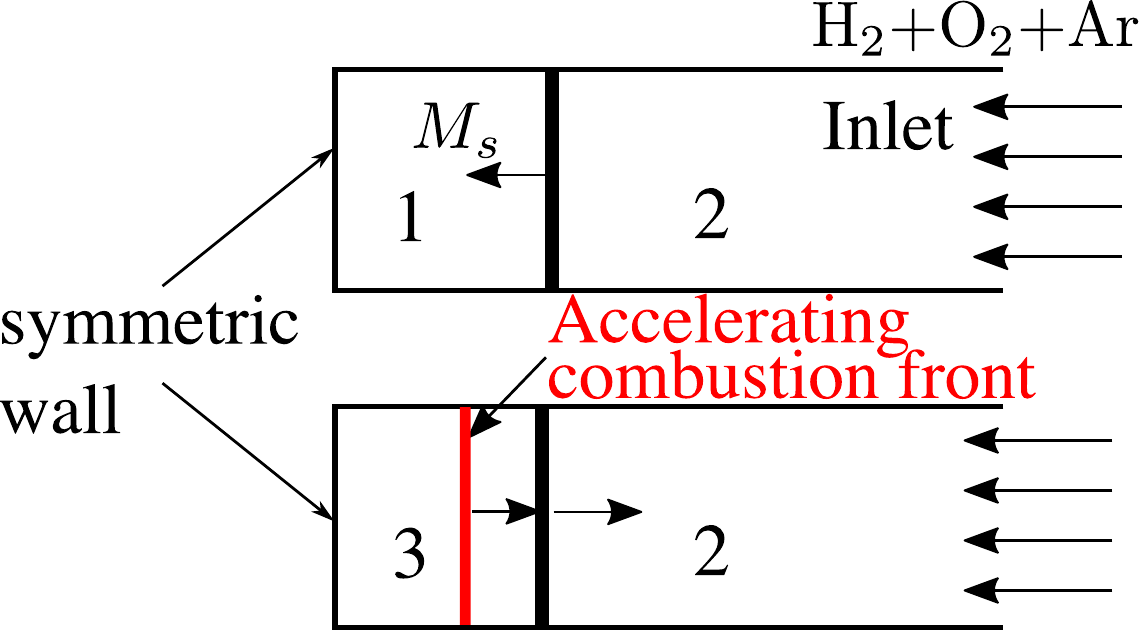}
\put(-390,300){$(a)$}
\put(-190,300){$(b)$}
\put(-390,150){$(c)$}
  \caption{Validation test setup $(a)$ identical to the one studied by Oran \emph{et al.}~\cite{Oran_CandF_1982} and Yungster and Radhakrishnan~\cite{YungsterRadha_AIAA_1996}. The combustion front due to shock-heating at the adiabatic (symmetric) wall accelerates $(b)$ eventually merging with the reflected shock wave. The resulting steady state detonation wave propagates steadily at $v_D\approx 1600~\mathrm{m/s}$ $(c)$. The pressure peak in the detonation wave corresponds to the Von-Neumann state $p_{vn} = 560156.5~\mathrm{Pa}$.}
\label{fig: Validation1DDet}
\end{figure}
The numerical model is validated against one-dimensional steady detonation wave propagation, as shown in figure~\ref{fig: Validation1DDet}. The validation setup is the one defined by Yungster and Radhakrishnan~\cite{YungsterRadha_AIAA_1996}. In the validation study, the simulation is initialized with a shock wave of mach number $M_s = 3$, propagating towards an adiabatic (symmetric) wall and fluid mixture at $p_1 = 6666.12~\mathrm{Pa}, T_1 = 300 K$. On the right hand side, an inlet is defined driving the shock wave in a mixture of H$_2$+O$_2$+$Ar$ with molar ratio $2:1:7$. The reflection of the shock wave results in instantaneous shock-heating at the adiabatic wall, which ignites the mixture. The combustion wave generated due to the instantaneous heating initially lags behind the reflected shock wave~(figure.~\ref{fig: Validation1DDet}$(a)$). The combustion front accelerates (figure.~\ref{fig: Validation1DDet}$(b)$) eventually merging with the reflected shock wave and steadily propagates with the velocity $v_D$ thus resulting in a steady propagating detonation wave. The steady state velocity $v_D$ saturates to $1600~\mathrm{m}/\mathrm{s}$ as approximated by the steady state ZND theory.
The merging of combustion wave front and the shock wave increases the peak pressure (figure~\ref{fig: Validation1DDet}$(c)$) as well. The peak pressure corresponds to the Von-Neumann state in the steady state detonation propagation theory~\cite{JHlee_2008} and matches the value $p_{vn} = 560156.5~\mathrm{Pa}$. The temperature of the  Von-Neumann state is $T_{vn}=2459.33~\mathrm{K}$ and the maximum heat release rate behind the Von-Neumann state $\omega_T = 2.58\times 10^{11}~\mathrm{W}/\mathrm{m}^3$.

The shock-to-detonation transition validation study (figure~\ref{fig: Validation1DDet}) validates the numerical model for detailed chemistry implemented. We note that the shock-capturing viscosity is approximately $100$ times higher than the maximum physical diffusivity in the system and hence only supports stable numerical computations. The inviscid and chemical reaction based characteristics of the detonation waves are captured accurately. In the section below, we outline the one-dimensional problem formulation based on previous numerical studies~\cite{Schwer_CandF_2011, Lietz_AIAA_2018} and experimental studies~\cite{Fotia_AIAA_2014,Schwinn_CandF_2018} for sustained detonation waves.

\subsection{Pre-mixed laminar flame}

The pre-mixed laminar CH$_4$-air flame was simulated at various equivalence ratios $\phi$ for validating the combined implementation of transport and combustion terms in the governing equations. Since the formulation in the current work is fully compressible, the results differ from the usual low-Mach number implementation results of the energy conservation equation. The computation is initialized with a fully converged steady state solution obtained from CANTERA~\cite{goodwin2009cantera}. Due to fully compressible formulation and fixed Lewis numbers for all the species, the spatial distribution of velocity and temperature adjusts to the steady state solution shown in figure~\ref{fig: LaminarFlame} with a corresponding change in pressure. Relaxation sponges are used at the extremes of the domain to eliminate thermoacoustic interactions and convergence to steady state solution. The $\mathrm{CH}_4 - \mathrm{air}$ mixture (left) flows towards a stationary flame. Flow speed before the flame corresponds to the flame velocity $u_f$ in an unsteady case.
\begin{figure}[!t]
\centering
\includegraphics[width=\textwidth]{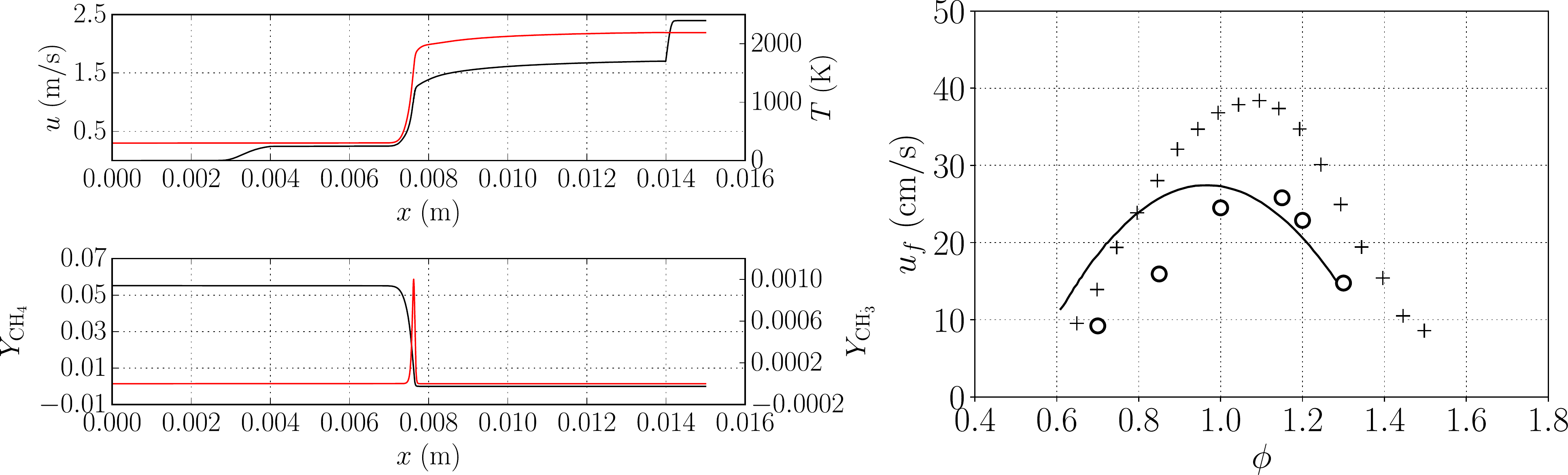}
  \put(-440,120){$(a)$}
\put(-195,120){$(b)$}
  \caption{Spatial distribution of fluid velocity and temperature (top) and mass fractions of $\mathrm{CH}_4$ and $\mathrm{CH}^{.}_3$ (bottom) for a laminar flame testcase for equivalence ratio $\phi=1$ $(a)$. Comparison of CH$_4$-air laminar flame speeds for various equivalence ratios with experimental data \cite{vagelopoulos1994}, low-mach number unity lewis number calculation ($-$)~\cite{PoinsotV_numComb_2011}, and current calculations ($\circ$) $(b)$}
\label{fig: LaminarFlame}
\end{figure}
Smaller intermediate species like H$\cdot$ and H$_2$ have higher diffusivity coefficients which would tend to increase the flame speed ($u_f\propto \sqrt{K \mathcal{D}}$ where $K$ is some measure of rate of reaction and $\mathcal{D}$ is the species diffusivity). Hence, the simplified assumption of $Le_k = 1$ for all species results in under-prediction of flame speeds compared to the experiments.

\section{One-dimensional problem formulation}
\label{sec: 1DModel}

Detonation propagation can be sustained via fuel injection and exhaust purge as shown in various experiments~\cite{Fotia_AIAA_2014,Schwinn_CandF_2018}. Various numerical studies of such sustained detonations in two dimensions have also been conducted~\cite{Schwer_CandF_2011, Lietz_AIAA_2018}. However, two-dimensional studies focus more on the flow characteristics induced by sustained detonations and are computationally expensive. In this study, particular focus is given to establishing which parameters affect the sustenance of detonation waves. To the best of our knowledge, no previous attempt has been made to derive first-principle based 1D model capable of simulating sustained detonation waves. 
\begin{figure}[!t]
\centering
\includegraphics[width=0.65\textwidth]{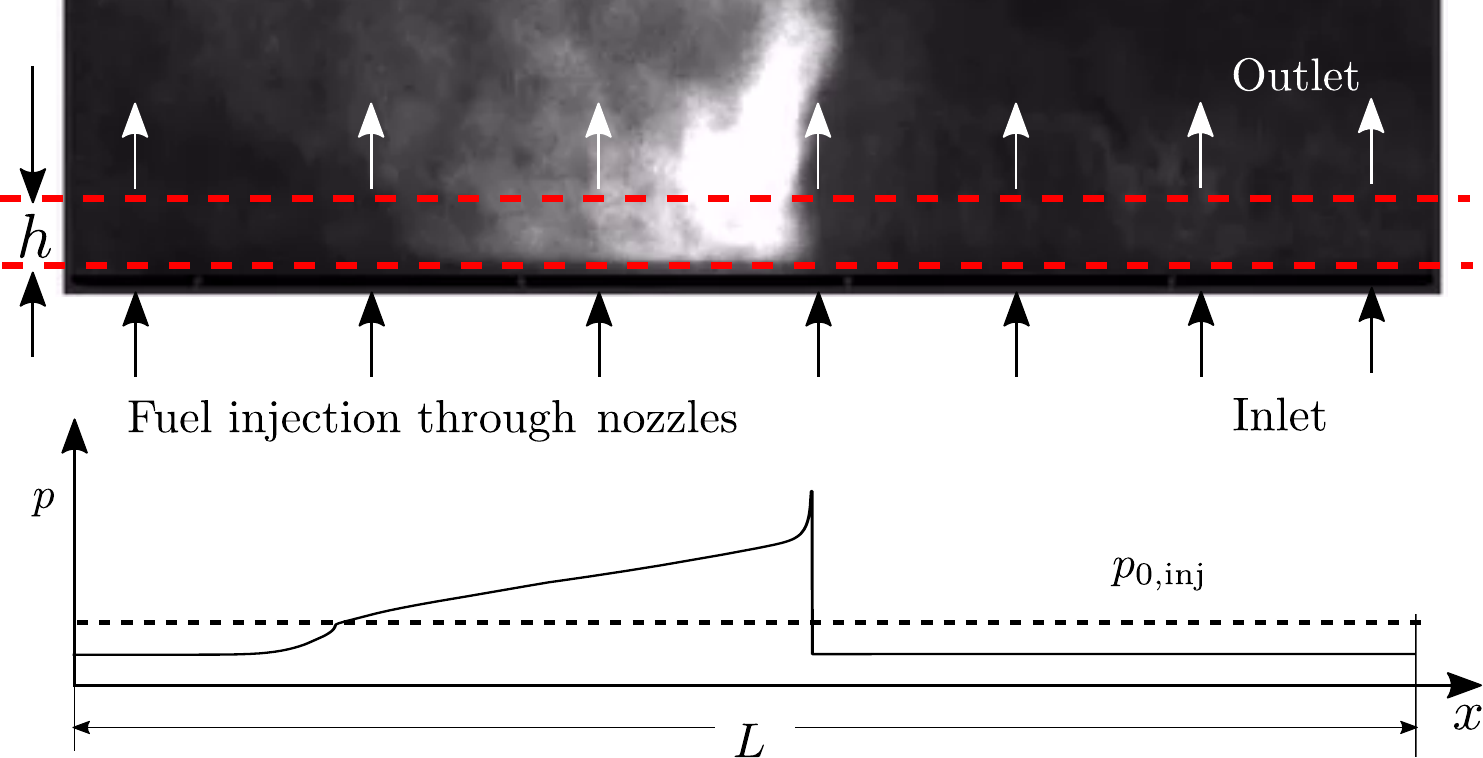}
\put(-320,140){$(a)$}
\put(-320,70){$(b)$}
  \caption{Schematic illustrating the one-dimensional model reduction based on the OH$^*$ chemiluminescence visualization of experimental study by Schwinn \emph{et al.}~\cite{Schwinn_CandF_2018} $(a)$ and sketch of the spatial profile of pressure corresponding to the visualized detonation wave $(b)$. Dashed line (- -) shows the stagnation injection pressure.}
\label{fig: ConceptProblem}
\end{figure}

We reduce the problem to a one-dimensional problem as shown in figure~\ref{fig: ConceptProblem}$(a)$ by considering a small region of height $h$ in which gradients in $y$ direction can be neglected. Two dimensional governing equations can be written as, 
\begin{align}
&\frac{\partial \rho}{\partial t} + \frac{\partial \left(\rho u\right)}{\partial x} + \frac{\partial \left(\rho v\right)}{\partial y} = 0,\label{eq: 2D_governingEqs1}\\
&\frac{\partial \rho u}{\partial t} + \frac{\partial \rho u^2}{\partial x} + \frac{\partial \rho v u}{\partial y} = -\frac{\partial p}{\partial x} + \frac{\partial }{\partial x}\left(\frac{4\mu}{3}\frac{\partial u}{\partial x}\right) + \frac{\partial }{\partial y}\left(\mu\frac{\partial u}{\partial y} + \mu\frac{\partial v}{\partial x}\right),\label{eq: 2D_governingEqs2}
\end{align}
\begin{align}
&\frac{\partial \rho E}{\partial t} + \frac{\partial}{\partial x}\left(u(\rho E +p)\right) + \frac{\partial}{\partial y}\left(v(\rho E +p)\right) = \frac{\partial }{\partial x}\left(k\frac{\partial T}{\partial x}\right) + \frac{\partial }{\partial y}\left(k\frac{\partial T}{\partial y}\right) \nonumber \\
&+ \frac{\partial}{\partial x}\left(u\frac{4\mu}{3}\frac{\partial u}{\partial x} + v\mu\left(\frac{\partial u}{\partial y} + \frac{\partial v}{\partial x}\right)\right) + \frac{\partial }{\partial y}\left(u\mu\left(\frac{\partial u}{\partial y} + \frac{\partial v}{\partial x}\right) + v\frac{4\mu}{3}\frac{\partial v}{\partial y}\right) +\nonumber \\
&\omega_T -\frac{\partial}{\partial x}\left(\rho\sum^{N}_{k=1}h_{s,k}Y_kV_{k,x}\right) - \frac{\partial}{\partial y}\left(\rho\sum^{N}_{k=1}h_{s,k}Y_kV_{k,y}\right),\label{eq: 2D_governingEqs3}
\\
&\frac{\partial \rho Y_k}{\partial t} + \frac{\partial}{\partial x}\left(\rho (u + V_{k,x}) Y_k\right) + \frac{\partial}{\partial y}\left(\rho (v + V_{k,y} + V^{C}_{k,y}) Y_k\right) = \omega_k.
\label{eq: 2D_governingEqs4}
\end{align}
Integrating the Eqs.~\eqref{eq: 2D_governingEqs1}-\ref{eq: 2D_governingEqs4} in $y$ within the region of height $h$ shown in figure~\ref{fig: ConceptProblem}$(a)$ and neglecting stresses, diffusion velocities, and temperature gradients at inlet and outlet, we obtain the following system of equations in 1D, 
\begin{align}
&\frac{\partial \rho}{\partial t} + \frac{\partial \left(\rho u\right)}{\partial x}  = \frac{1}{h}\left(\left(\rho v\right)_{\mathrm{inlet}} - \left(\rho v\right)_{\mathrm{outlet}}\right),\label{eq: 1D_ReducedEqs1}\\
&\frac{\partial (\rho u)}{\partial t} + \frac{\partial (\rho u^2)}{\partial x}  = -\frac{\partial p}{\partial x} + \frac{\partial }{\partial x}\left(\frac{4\mu}{3}\frac{\partial u}{\partial x}\right) + \frac{1}{h}\left(\left(\rho u v\right)_{\mathrm{inlet}} - \left(\rho u v\right)_{\mathrm{outlet}}\right),\label{eq: 1D_ReducedEqs2}
\end{align}
\begin{align}
&\frac{\partial (\rho E)}{\partial t} + \frac{\partial}{\partial x}\left(u(\rho E +p)\right)  = \frac{\partial }{\partial x}\left(k\frac{\partial T}{\partial x}\right) + \frac{\partial}{\partial x}\left(u\frac{4\mu}{3}\frac{\partial u}{\partial x} + v\mu\left(\frac{\partial u}{\partial y} + \frac{\partial v}{\partial x}\right)\right)\nonumber \\
&+\omega_T -\frac{\partial}{\partial x}\left(\rho\sum^{N}_{k=1}h_{s,k}Y_kV_{k,x}\right) + \frac{1}{h}\left(\left(v(\rho E +p)\right)_{\mathrm{inlet}} - \left(v(\rho E +p)\right)_{\mathrm{outlet}}\right),\label{eq: 1D_ReducedEqs3}\\
&\frac{\partial (\rho Y_k)}{\partial t} + \frac{\partial}{\partial x}\left(\rho (u + V_{k,x}) Y_k\right)  = \omega_k + \frac{1}{h}\left(\left(\rho v Y_k\right)_{\mathrm{inlet}} - \left(\rho v Y_k\right)_{\mathrm{outlet}}\right).
\label{eq: 1D_ReducedEqs4}
\end{align}

Equations~\eqref{eq: 1D_ReducedEqs1}-\ref{eq: 1D_ReducedEqs4} represent the one-dimensional model equations accounting for fuel/oxidizer injection and outlet close to the injector plate in the full experimental setup. We assume $h =  1~\mathrm{mm}$ in the current study based on experimental visualizations. We note that values of $h$ should be much smaller than the overall transverse dimension of the rig to account for negligible $y$ gradients in the domain. Furthermore, we assume that $v_{\mathrm{outlet}}=v_{\mathrm{inlet}} = v_{\mathrm{inj}}$ where $v_{\mathrm{inj}}$ is evaluated assuming isentropic expansion through injector nozzles, as introduced by Schwer and Kailasnath~\cite{Schwer_AIAA_2010}. The assumption of $v_{\mathrm{outlet}}=v_{\mathrm{inlet}}$ will be corrected in future studies to account for the flux of reaction products out of the domain correctly. The outlet quantities $\rho$, $p$, $\rho u$, $\rho E$, and $\rho Y_k$ correspond to the quantities in the domain. Consequently, the model equations become, 
\begin{align}
&\frac{\partial \rho}{\partial t} + \frac{\partial \left(\rho u\right)}{\partial x}  = \frac{v_{\mathrm{inj}}}{h}\left(\rho_{\mathrm{inlet}} - \rho\right),\label{eq: 1D_ModelEqs1}\\
&\frac{\partial (\rho u)}{\partial t} + \frac{\partial (\rho u^2)}{\partial x}  = -\frac{\partial p}{\partial x} + \frac{\partial }{\partial x}\left(\frac{4\mu}{3}\frac{\partial u}{\partial x}\right) + \frac{v_{\mathrm{inj}}}{h}\left(- \left(\rho u \right)\right),\label{eq: 1D_ModelEqs2}\\
&\frac{\partial (\rho E)}{\partial t} + \frac{\partial}{\partial x}\left(u(\rho E +p)\right)  = \frac{\partial }{\partial x}\left(k\frac{\partial T}{\partial x}\right) + \frac{\partial}{\partial x}\left(u\frac{4\mu}{3}\frac{\partial u}{\partial x} + v\mu\left(\frac{\partial u}{\partial y} + \frac{\partial v}{\partial x}\right)\right)\nonumber \\
&+\omega_T -\frac{\partial}{\partial x}\left(\rho\sum^{N}_{k=1}h_{s,k}Y_kV_{k,x}\right) + \frac{v_{\mathrm{inj}}}{h}\left(\left(\rho E +p\right)_{\mathrm{inlet}} - (\rho E +p)\right),\label{eq: 1D_ModelEqs3}\\
&\frac{\partial (\rho Y_k)}{\partial t} + \frac{\partial}{\partial x}\left(\rho (u + V_{k,x}) Y_k\right)  = \omega_k + \frac{v_{\mathrm{inj}}}{h}\left(\left(\rho Y_k\right)_{\mathrm{inlet}} - \rho Y_k\right).
\label{eq: 1D_ModelEqs4}
\end{align}
To calculate $v_{\mathrm{inj}}$ and $\rho_{\mathrm{inlet}}$ in the model Eqs.~\eqref{eq: 1D_ModelEqs1}-\ref{eq: 1D_ModelEqs4}, we specify the stagnation conditions at the injectors $p_{0,\mathrm{inj}}, T_{0,\mathrm{inj}}$. Through isentropic nozzle expansion relations, the mass flow rate of the injected fluid is given by, 
\begin{equation}
\rho_{\mathrm{inlet}}v_{\mathrm{inj}} = p_{0,\mathrm{inj}}\sqrt{\frac{\gamma}{RT_{0,\mathrm{inj}}}}M\left(1 + \frac{\gamma - 1}{2} M^2\right)^{-\frac{1}{2}\left(\frac{\gamma + 1}{\gamma - 1}\right)}, 
\label{eq: flowrate_nozzle}
\end{equation}
where $M$ is the exit Mach number of the premixed fuel/oxidiser mixture governed by the pressure in the domain as, 
\begin{equation}
M = \sqrt{\frac{2}{\gamma - 1}\left(\left(\frac{p_{0,\mathrm{inj}}}{p}\right)^{\frac{\gamma - 1}{\gamma}}- 1\right)},
\label{eq: M_nozzle}
\end{equation}
and $\gamma$ is the adiabatic exponent of the mixture. Where $p>p_{0,\mathrm{inj}}$, the mixture injection does not take place. 
We note that, isentropic expansion relations also yield the density injected $\rho_{\mathrm{inj}}$ inside the domain as, 
\begin{equation}
\rho_{\mathrm{inj}} = \rho_{0,\mathrm{inj}}\left(1+\frac{\gamma - 1}{2}M^2\right)^{\frac{1}{1-\gamma}}.
\label{eq: rho_nozzle}
\end{equation}
For $p>p_{0,\mathrm{inj}}$, there is no fuel injection. For $p_{s,\mathrm{inj}}<p<p_{0,\mathrm{inj}}$ the fuel injection rates are determined by Eqs.~\eqref{eq: flowrate_nozzle}-\ref{eq: rho_nozzle}, where $p_{s,\mathrm{inj}}$ is the sonic pressure corresponding to $M=1$ in equation~\eqref{eq: M_nozzle}. For $p<p_{s,\mathrm{inj}}$ in the domain, all the quantities are calculated at the sonic state, $M=1$.
\begin{figure}[!t]
\centering
\includegraphics[width=0.8\textwidth]{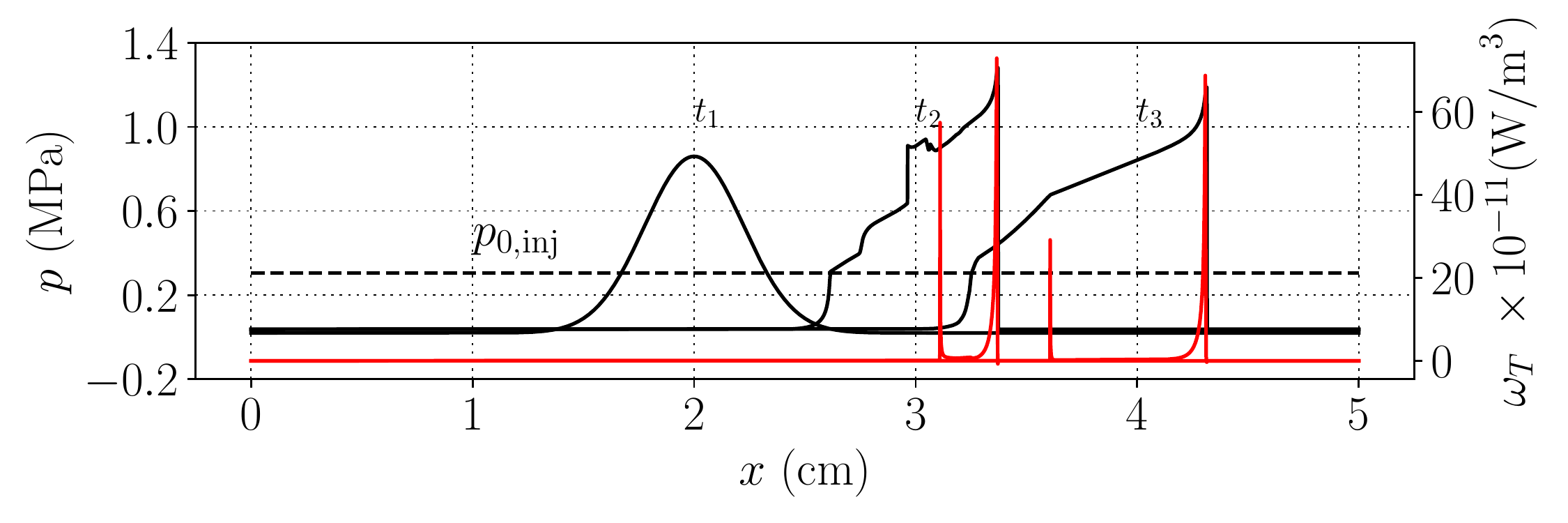}
 \caption{Initial propagation of detonation waves in a periodic domain at $t_1 = 0$, $t_2 = 10~\mu\mathrm{s} $, and $t_3 = 20~\mu\mathrm{s}$ for $p_{0,\mathrm{inj}} = 0.3~\mathrm{MPa}$ and $w_i = 0.2$.}
\label{fig: DetonationInit}
\end{figure}

\section{Results}
\label{sec: results}
\begin{table}[!b]
\centering
\caption{Injection stagnation pressures $p_{0,\mathrm{inj}}$ and the weighting factors $w_i$ considered.}
\begin{tabular}{|c|c|c|c|}
\hline
$p_{0,\mathrm{inj}}$ & 0.2 MPa & 0.3 MPa & 0.4 MPa \\
\hline
& 0 & 0 & 0 \\
& 0.1 & 0.1 & 0.1 \\
 & 0.2 & 0.2 & 0.2 \\
$w_i$ & - & 0.4 & - \\
& - & 0.6 & - \\
& - & 0.8 & - \\
& - & 1.0 & - \\
\hline
\end{tabular}\quad
\label{tab: TestCases}
\end{table}
In the current study, we restrict the analysis to stoichiometric H$_2$+O$_2$+Ar mixture with molar fractions $2:1:3$ and consider three injection pressures $p_{0,\mathrm{inj}}$ to elucidate the dynamics of detonation wave propagation with premixed fuel injection. We note that increasing stagnation pressure $p_{0,\mathrm{inj}}$ causes higher density of the mixture injected in the domain which results in higher heat release rates due to combustion. Since high heat release rates cause very large gradients in species concentrations, the numerical simulations become unstable. In this study, to achieve stable computations, we assume the inlet density $\rho_{\mathrm{inlet}}$ as an average between the injected density $\rho_{\mathrm{inj}}$ and the initialized density $\rho_{\infty}$ as, 
\begin{equation}
\rho_{\mathrm{inlet}} = w_i\rho_{\mathrm{inj}} + (1-w_i)\rho_{\infty}.
\end{equation}
\begin{figure}[!b]
\centering
\includegraphics[width=\textwidth]{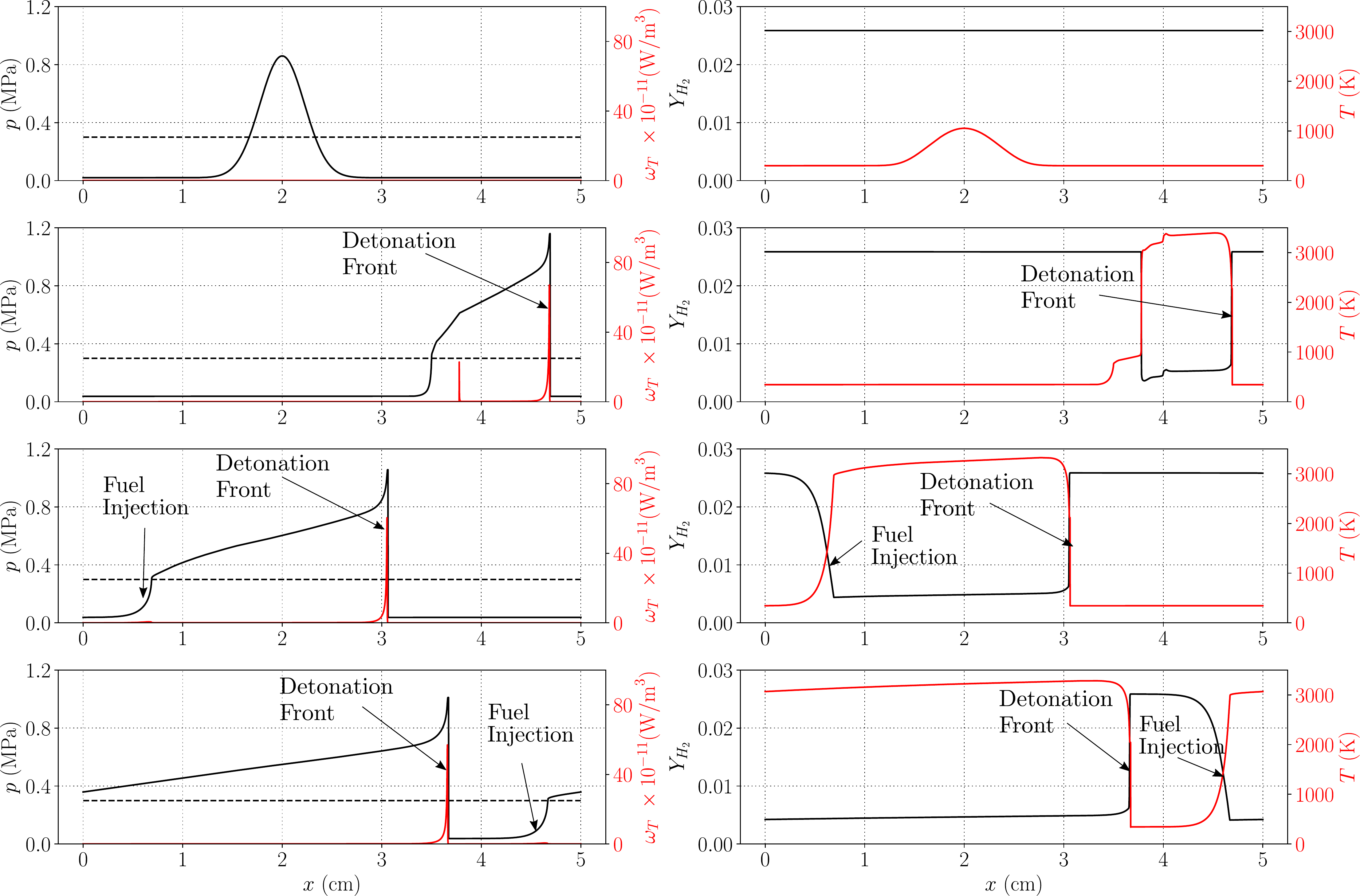}
\put(-440,285){$(a)$}
\put(-440,210){$(b)$}
\put(-440,145){$(c)$}
\put(-440,75){$(d)$}
 \caption{Spatio-temporal evolution of pressure $p$ (--) and heat release $\omega_T$ ({\color{red}--}) (left) ; mass fraction of Hydrogen $Y_{H_2}$ (--) and temperature $T$ ({\color{red}--}) (right) for times $t = 0$ $(a)$, $12~\mu\mathrm{s}$ $(b)$, $30~\mu\mathrm{s}$ $(c)$, and $58~\mu\mathrm{s}$ $(d)$ for $p_{0,\mathrm{inj}}=0.3~\mathrm{MPa}$ and $w_i = 0.2$. The initialized pulse propagates to the right and steepens thus forming a shock wave of $M\approx 6$. The rapid temperature rise behind the shock wave causes the ignition of the fuel thus resulting in detonation wave formation.}
\label{fig: Detonation3P0}
\end{figure}
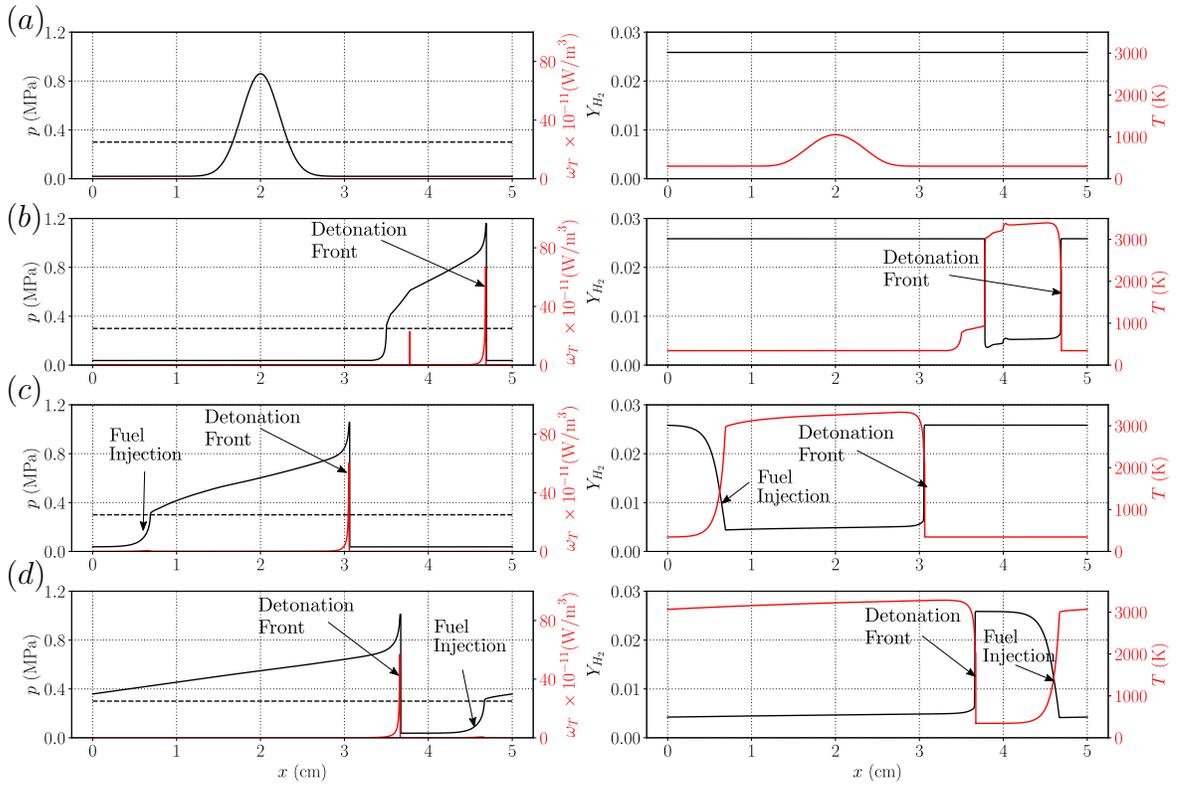
We note that such choice of averaged inlet density is akin to a numerical under-relaxation and is un-physical. However, it yields preliminary physical insights into the detonation waves reproducible by one-dimensional model reduction. We consider a periodic domain of length $L=5~\mathrm{cm}$ (see figure~\ref{fig: ConceptProblem}) and initialize the system with a stoichiometric mixture of H$_2$+O$_2$+Ar with molar fractions $2:1:3$. The pressure $p$, density $\rho$, and velocity $u$ in the domain are initialized with an isentropic Gaussian pulse (see figure~\ref{fig: DetonationInit}) propagating in $+x$ direction. The initial field is given as, 
\begin{align}
p(x, t=0) &= p_{\infty} + \Delta p f_x,\\ \rho(x, t=0) &= \rho_{\infty}\left(\frac{p}{p_{\infty}}\right)^{1/\gamma_{\infty}},\\ u(x,t=0) &= \frac{2a_{\infty}}{\gamma_{\infty} - 1}\left(\left(\frac{\rho}{\rho_{\infty}}\right)^{\frac{\gamma_{\infty} - 1}{2}} - 1\right),
\end{align} 
where $p_{\infty} = 0.02~\mathrm{MPa}$,  $\Delta p$ is the initial pressure jump and $f_x$ is the Gaussian given by,
\begin{equation}
\Delta p = 0.82~\mathrm{MPa},~ f_x = \exp\left(-10^5\left(x - 0.02\right)^2\right),
\end{equation}
respectively. The pulse propagates and steepens thus forming a shock wave of $M \approx 6$. The  temperature jump thus caused ignites the mixture of H$_2$+O$_2$+Ar such that a combustion front is created behind the shock wave. Unlike the validation case shown in figure~\ref{fig: Validation1DDet}, the detonation wave is not driven by an external flow as a result of which, the pressure falls behind the wave instead of saturating to a constant value. As the pressure falls below $p_{0,\mathrm{inj}}$ the fuel injection is activated thus reducing the temperature and increasing the fuel/oxidizer fractions in the mixture. Table~\ref{tab: TestCases} lists the different values of $p_{0,\mathrm{inj}}$ and $w_i$ considered in the current study. Throughout the computations, we assume constant injection stagnation temperature $T_{0,\mathrm{inj}} = 700~\mathrm{K}$.

Figure~\ref{fig: Detonation3P0} below shows the spatio-temporal evolution of pressure $p$, heat release rate $\omega_T$, mass fraction of Hydrogen $Y_{H_2}$, and the temperature $T$ in the domain for $p_{0,\mathrm{inj}} = 0.3~\mathrm{MPa}$ and $w_i=0.2$. The shock wave formation due to steepening of the initialized pulse causes rapid heating of the mixture behind the shock wave up to temperature $T\approx 1000~\mathrm{K}$. As a result, the mixture in the domain ignites resulting in detonation waves. Due to no driving support (for instance, a moving piston), the pressure downstream of the detonation wave decreases generating an expansion wave. As the pressure decreases below the injection stagnation pressure $p_{0,\mathrm{inj}}$, the fuel injection is activated locally. As the detonation wave continues to propagate, the fuel injection regime follows the wave. However, the expansion wave behind the detonation front flattens in space thus causing the fuel injection zone to lag behind the detonation front. 
\begin{figure}[!t]
\centering
\includegraphics[width=\textwidth]{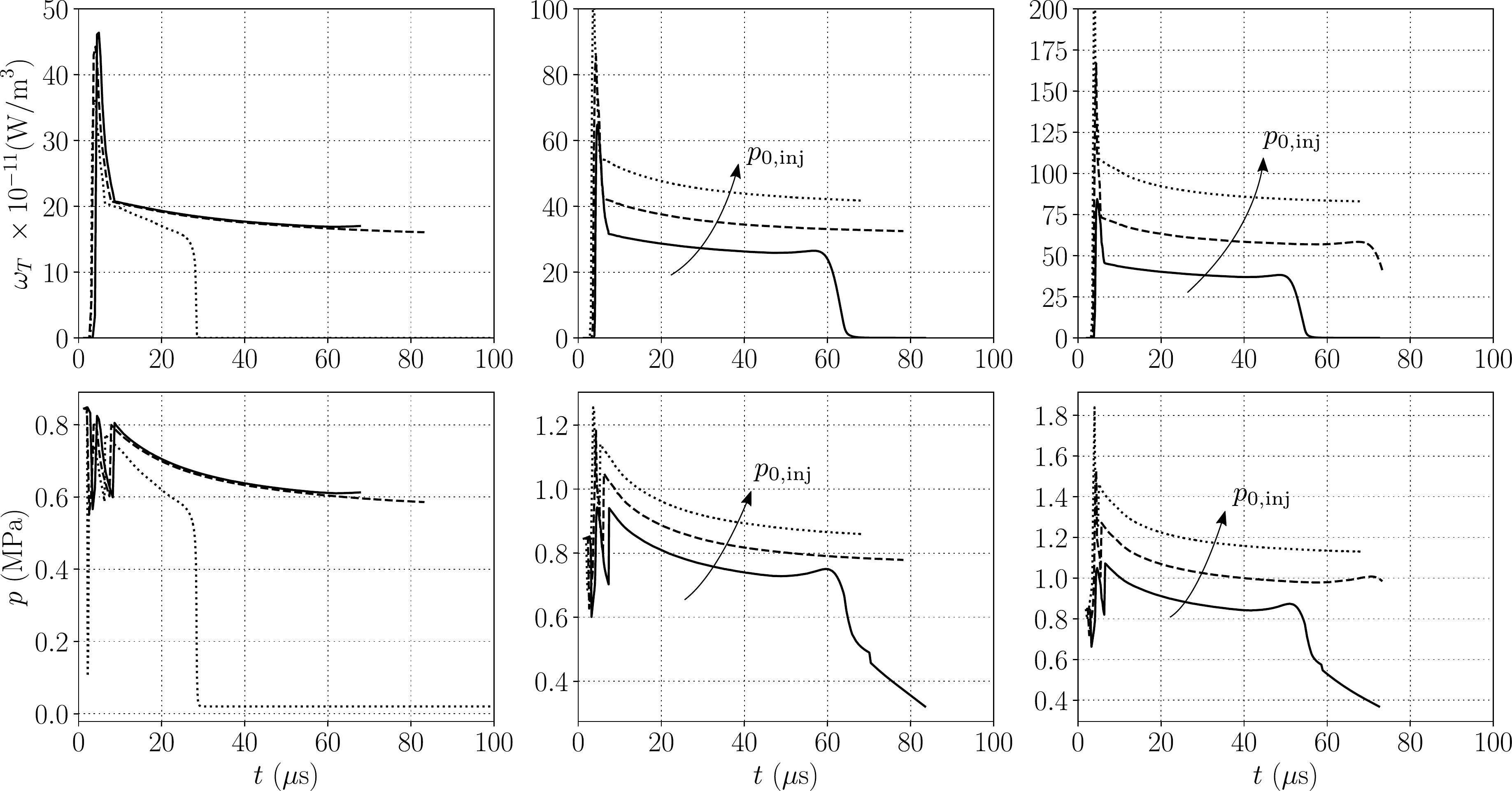}
\put(-431,230){$(a)$}
\put(-291,230){$(b)$}
\put(-146,230){$(c)$}
\put(-431,119){$(d)$}
\put(-291,119){$(e)$}
\put(-146,119){$(f)$}
\put(-380,230){$w_i = 0$}
\put(-255,230){$w_i = 0.1$}
\put(-90,230){$w_i = 0.2$}
 \caption{Evolution of the maximum heat release $\omega_T$ behind the detonation wave in the system for $w_i = 0$ $(a)$, $w_i = 0.1$ $(b)$, and $w_i = 0.2$ $(c)$ for injection stagnation pressures $p_{0,\mathrm{inj}} = 0.2~\mathrm{MPa}$ (--), 0.3 MPa ($--$), and 0.4 MPa ($\cdots$).}
\label{fig: DetonationTrackHeatRelease}
\end{figure}
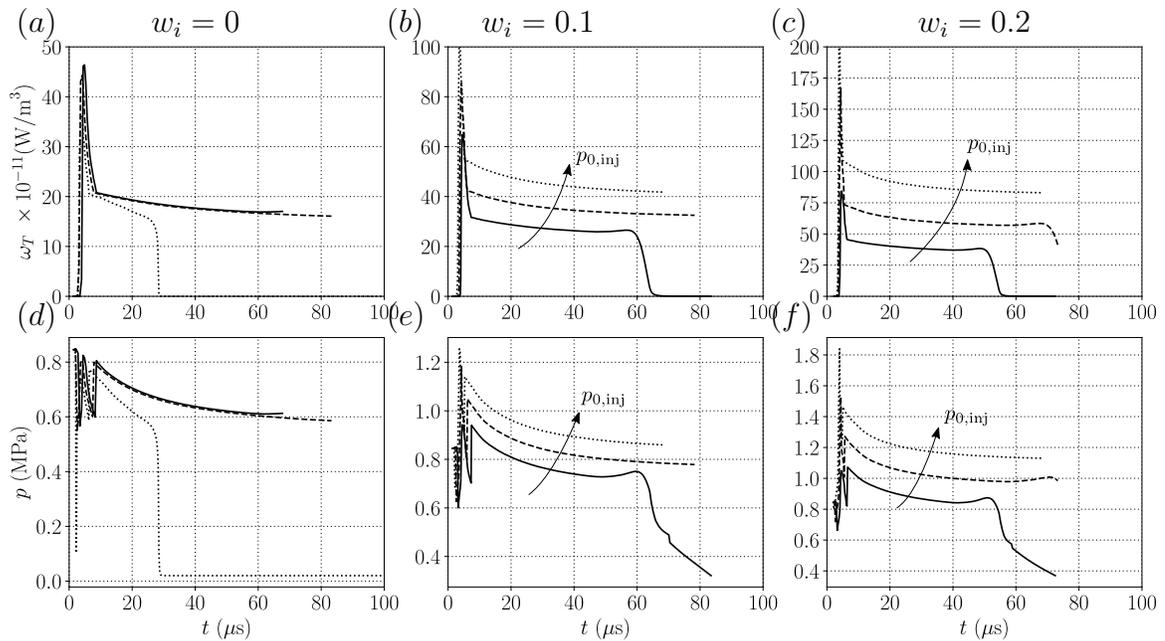
The flattening of the expansion wave downstream of the detonation wave results in widening of the gap between the fuel injection zone and the detonation wave. Consequently, detonation wave eventually merges with the fuel injection zone resulting in quenching of the reaction and failure of the detonation wave sustenance. Figure~\ref{fig: DetonationTrackHeatRelease} shows the evolution of maximum heat release $\omega_T$ behind the detonation wave as the wave propagates in the periodic domain for $w_i = 0,~0.1,~0.2$ and increasing values of injection stagnation pressures $p_{0,\mathrm{inj}}=0.2~\mathrm{MPa},~0.3~\mathrm{MPa},~0.4~\mathrm{MPa}$.  With increasing stagnation pressures, the density of the injected fuel increases. Since the chemical composition of the mixture is fixed, the total amount of fuel and oxidizer increases in the system thus causing higher heat release behind the detonation wave.

As shown in figure~\ref{fig: DetonationTrackHeatRelease}, detonations exhibiting higher heat release sustain for more time compared to those which exhibit lower heat release. However, for $w_i = 0$, the heat release is independent of the injection stagnation pressure since the density of the fluid remains constant in the domain. For such conditions, lower injection stagnation pressures increase the separation between the detonation front and the fuel injection zone. Hence, the waves are quenched by the fuel injection at later times for lower injection stagnation pressures sustaining waves for longer. Figure~\ref{fig: DetonationTrajectory} shows the trajectory of the combustion front in time compared with the corresponding C-J velocities, obtained by CEA~\cite{CEAcite}. The combustion front velocity is very close to the C-J velocity initially. However, the detonation wave decelerates to velocity slightly below C-J velocity slowly and eventually gets quenched. 

\begin{figure}[!t]
\centering
\includegraphics[width=\textwidth]{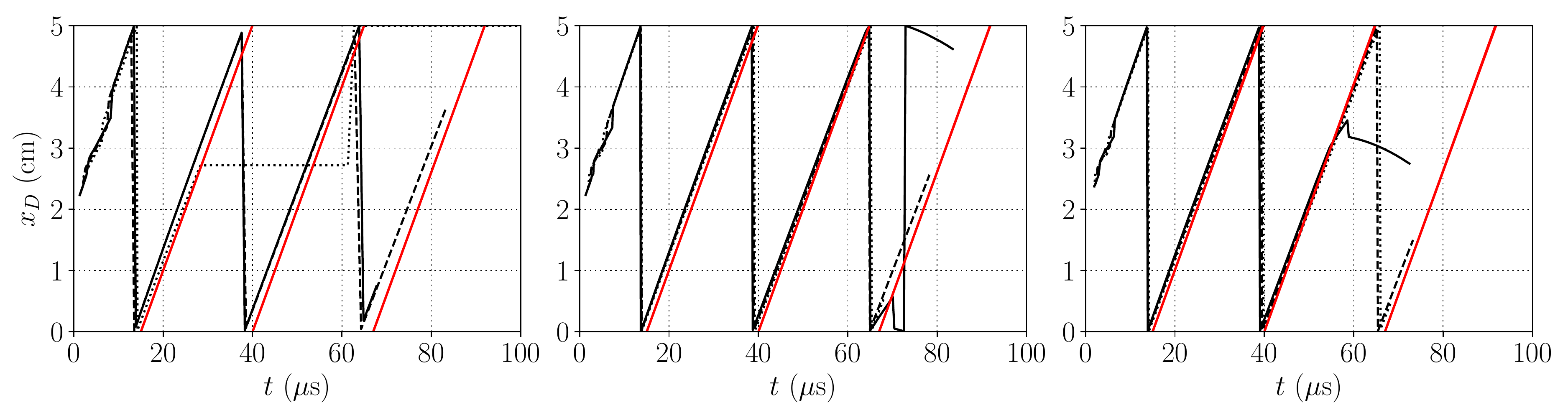}
\put(-420,115){$(a)$}
\put(-280,115){$(b)$}
\put(-140,115){$(c)$}
\put(-380,115){$w_i = 0$}
\put(-255,115){$w_i = 0.1$}
\put(-90,115){$w_i = 0.2$}
 \caption{Evolution of the detonation wave trajectory $x_D$ in time for $w_i = 0$ $(a)$, $w_i = 0.1$ $(b)$, and $w_i = 0.2$ $(c)$ for injection stagnation pressures $p_{0,\mathrm{inj}} = 0.2~\mathrm{MPa}$ (--), 0.3 MPa ($--$), and 0.4 MPa ($\cdots$). ({\color{red}--}) are the C-J velocities for $p_{0,\mathrm{inj}} = 0.4$ for various $w_i$.}
\label{fig: DetonationTrajectory}
\end{figure}

\begin{figure}[!b]
\centering
\includegraphics[width=0.5\textwidth]{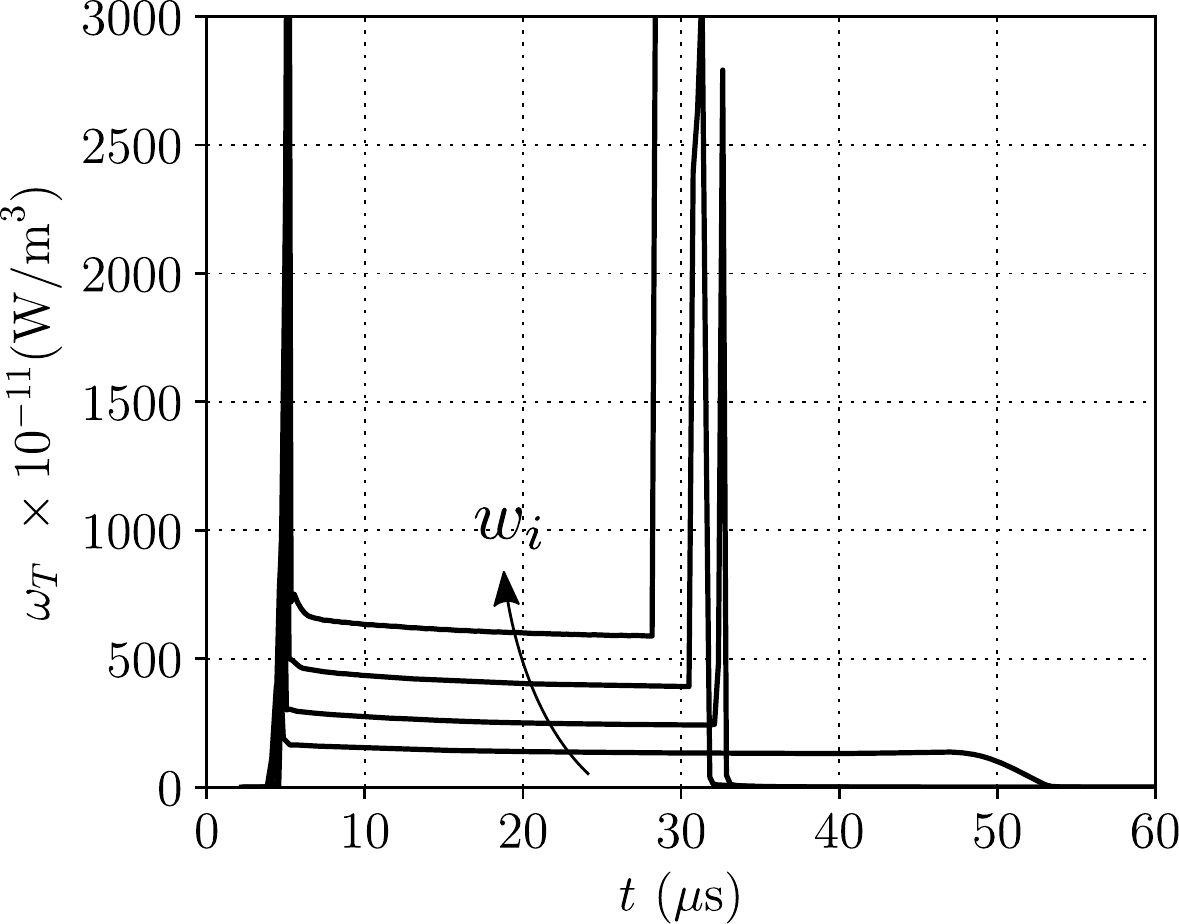}
 \caption{Evolution of the maximum heat release $\omega_T$ behind the detonation wave in the system for $w_i = 0.4,~0.6,~0.8$ and $w_i = 1.0$ for injection stagnation pressure $p_{0,\mathrm{inj}} = $ 0.3 MPa ($--$).}
\label{fig: DetonationTrackHeatRelease_ws}
\end{figure}
Similar to increasing the injection stagnation pressure $p_{0,\mathrm{inj}}$, with increasing the weighting factor $w_i$ also, the density of the injected mixture increases thus causing high heat release rates, as shown in figure~\ref{fig: DetonationTrackHeatRelease_ws}. With increasing release rates, the detonation velocity becomes higher. This results in the merging of detonation front and the fuel injection zone. Consequently, the detonation wave fails and the pressure becomes greater than the injection stagnation pressure everywhere in the domain. Such over-pressure in the domain highlights the unphysical nature of the assumption $v_{\mathrm{outlet}}=v_{\mathrm{inlet}}$. In reality, high pressure generated due to combustion of reactants expels the products at very high speeds compared to the inlet speeds. Such increase in outlet momentum, in-fact, results in generation of useful thrust. 

\section{Conclusions}
A numerical investigation of planar detonation wave sustenance due to fuel injection in a periodic domain, inspired by the experimental investigation of Schwinn \emph{et al.}~\cite{Schwinn_CandF_2018}, was performed utilizing spectral difference scheme. The detonation waves were modeled using detailed chemical reaction scheme and the results were analyzed for H$_2$+O$_2$+Ar mixture in a periodic domain of $5~\mathrm{cm}$ length. The fuel injection was modeled with one-dimensional model reduction representative of a region close to the injector plates in an original experimental setup. The assumptions in the fuel-injection model result in failure of detonation wave sustenance.

The initiation of detonation waves was done by a high pressure pulse which steepens and forms shock waves. Rapid heating due to the shock wave causes the fuel/oxidizer mixture in the domain to ignite. The combustion front thus formed follows the shock wave resulting in a detonation wave. Downstream of the detonation wave, an expansion wave is generated. The pressure profile becomes flattened as the expansion and the detonation waves propagate. The fuel injection is activated as the pressure falls below the injection stagnation pressure. The fuel injection zone travels with the expansion wave, lagging behind the detonation wave. As a result, the detonation wave merges with the fuel-injection zone, thus causing quenching of the wave. We reported the wave propagation and failure for various densities and stagnation pressure of injected fuel.

We note that, the model assumptions are highly restrictive in representing the fuel-injection and exhaust phenomenon observed in experiments~\cite{Schwinn_CandF_2018}, or multi-dimensional simulations~\cite{Lietz_AIAA_2018,Schwer_AIAA_2010} thus causing failure of detonation wave sustenance. Moreover, the length of the domain, restricted by the computational cost in the current study, also plays an important role in wave sustenance due to the time taken by the detonation wave to merge with the fuel-injection zone. 

% \include{DetonationWaves}
% %%%%%%%%%%%%%%%%%%%%%%%%%%%%%%

% %%%%%%%%%%%%%%%%%%%%%%%%%%%%%%
\chapter{Development of an order adaptive Navier Stokes solver}
The contents of this chapter summarize the development of a mesh adaptive compressible flow solver, undertaken as a part of summer internship in 2018 at HySonic Technologies, LLC. The numerical method discussed here-in is central to the numerical results presented previously in this dissertation and all the computations showed were performed utilizing the presented numerical method. However, the particular mesh adaptive solver discussed here is developed as a future tool to conduct more in-depth and efficient analysis of the problems presented in previous chapters, in particular, resonating detonation waves.

\section{Introduction}
High order discontinuous finite element methods such as spectral element methods~\cite{karniadakis2005spectral}, spectral volume methods~\cite{Wang_2002_JCP}, and multidomain spectral difference methods~\cite{kopriva1996conservative} provide flexibility of obtain highly accurate solutions to partial differential equations in geometrically complex domains~\cite{GuptaLS_JFM_2017, LodatoJ_2019_LES, Zhou_2010_JCP}. In particular, spectral difference methods are characterized by very less numerical diffusion for increasing orders of accuracy~\cite{lodato:14b, ChapelierLJ_CF_2016} thus providing highly accurate solutions to nonlinear wave propagation problems ~\cite{GuptaScalo_PRE_2018,GuptaLS_JFM_2017, lodato:16} and simulation of compressible turbulent flows~\cite{lodato2013discrete, LodatoJ_2019_LES}. In such methods, each subdomain is solved utilizing high order polynomial interpolations local to the subdomain. Consequently, the methodology exhibits compactness of the discretization stencil and can be optimally parallelized. 

In this work, we develop a methodology to adapt the local orders of the subdomains to obtain a high degree of polynomial reconstruction near the flow features exhibiting high gradients utilizing the spectral difference methods. 

\section{Spectral difference methods : Overview}
In this section, an overview of spectral difference methods for fully compressible Navier Stokes equations for an ideal gas is given. We begin with presenting the governing equations, followed by discussing the discretized solution evaluation methodology specific to unstructured hexahedral meshes with straight-edged elements. However, we note that, such methods can be easily extended to generic shaped meshes containing tetrahedra and prisms~\cite{Taylor_2000_SIAM, Luo_2006_Luo, karniadakis2005spectral}. In further sections, we discuss the adaptive order refinement strategy for multidomain spectral difference methods. 

\subsection{Governing equations}
Fully compressible three dimensional Navier-Stokes equations can be written in the following conservative form, 
\begin{align}
\frac{\partial \mathbf{Q}}{\partial t} + \frac{\partial \mathbf{F}_{i}}{\partial x_i} = \frac{\partial \mathbf{F^v}_{i}}{\partial x_i},
\end{align}
where $t$ is time, $x_i$ represent spatial directions for $i=1,2,3$, $\mathbf{Q}$ is the vector of conservative variables, namely, mass, momentum in $1,2,3$ directions, and total energy, with $\mathbf{F}_{i}$ and $\mathbf{F^v}_{i}$ being the inviscid and viscous flux vectors in $i^{th}$ direction respectively, given by, 
\begin{equation}
 \mathbf{Q} = \left(\begin{matrix}
               \rho \\
               \rho u_1\\
               \rho u_2\\
               \rho u_3\\
               \rho e\\
              \end{matrix}\right), \quad \mathbf{F}_{i} = \left(\begin{matrix}
               \rho u_i\\
               \rho u_1 u_i + p\delta_{i1}\\
               \rho u_2 u_i + p\delta_{i2}\\
               \rho u_3 u_i + p\delta_{i3}\\
               \left(\rho e + p\right)u_i\\
              \end{matrix}\right), \quad \mathrm{and} \quad \mathbf{F^v}_i = \left(\begin{matrix}
               0\\
               \tau_{1i}\\
               \tau_{2i}\\
               \tau_{3i}\\
               \tau_{ij}u_j - q_i\\
              \end{matrix}\right).
\label{eq: NS_EQ_3D}
\end{equation}
For an ideal gas, the total specific energy $e$ is given by, 
\begin{equation}
e = \frac{p}{\rho\left(\gamma - 1\right) } + \frac{u_i u_i}{2}.
\end{equation}
The viscous stress tensors $\tau_{ij}$, assuming a Newtonian fluid, are given as, 
\begin{equation}
 \tau_{ij} = 2\mu\left(\frac{\partial u_i}{\partial x_j} + \frac{\partial u_j}{\partial x_i}+ \left(\beta - \frac{2\mu}{3}\right)\frac{\partial u_k}{\partial x_k}\delta_{ij}\right), 
 \label{eq: stress_tensor}
\end{equation}
where $\mu$ and $\beta$ are the coefficients of dynamic (shear) viscosity and bulk viscosity, respectively. The heat flux $q_i$ is given by, 
\begin{equation}
 q_i = -k\frac{\partial T}{\partial x_i}, 
 \label{eq: heat_flux}
\end{equation}
where $k$ is the coefficient of conduction, and related to the dynamic shear viscosity via the Prandtl number, 
\begin{equation}
 Pr = \frac{\mu C_p}{k},
\end{equation}
where $C_p$ is the heat capacity at constant pressure. 

\subsection{Mapping and spatial discretization}
In the spectral difference method, the domain is assumed to be divided into non-overlapping subdomains or elements. The elements can be hexahedral, prismatic, or even tetrahedral. However, in this work, we focus on development for hexahedral elements only. Each hexahedral element is mapped on a standard computational domain, with the mapping function given by, 
\begin{equation}
x_i  = \sum^{N} _{n=1}\chi^n\left(\xi_j\right) \xi^n_i,
\label{eq: mapping}
\end{equation}
where $\chi^n$ is the mapping function corresponding to the $n^{th}$ vertex used for defining the physical element, $x_i$ are the physical coordinates, and $\xi_i$ are the computational coordinates. The standard computational domain or element is assumed to extend from $\left[-1, 1\right]$ such that, 
\begin{equation}
 \left(\xi_1, \xi_2, \xi_3\right) \in [-1,1]\times[-1,1]\times[-1,1].
\end{equation}
We note that the total number of vertices $N$ depends on the physical shape of the element. For quadratic or higher order elements, isoparametric mapping functions of same order can be utilized for mapping. However, in this work, we focus only on trilinear elements, i.e., linear in all directions, which correspond to $N=8$ for hexahedra.

\begin{figure}[!t]
\centering
\includegraphics[width=0.8\linewidth]{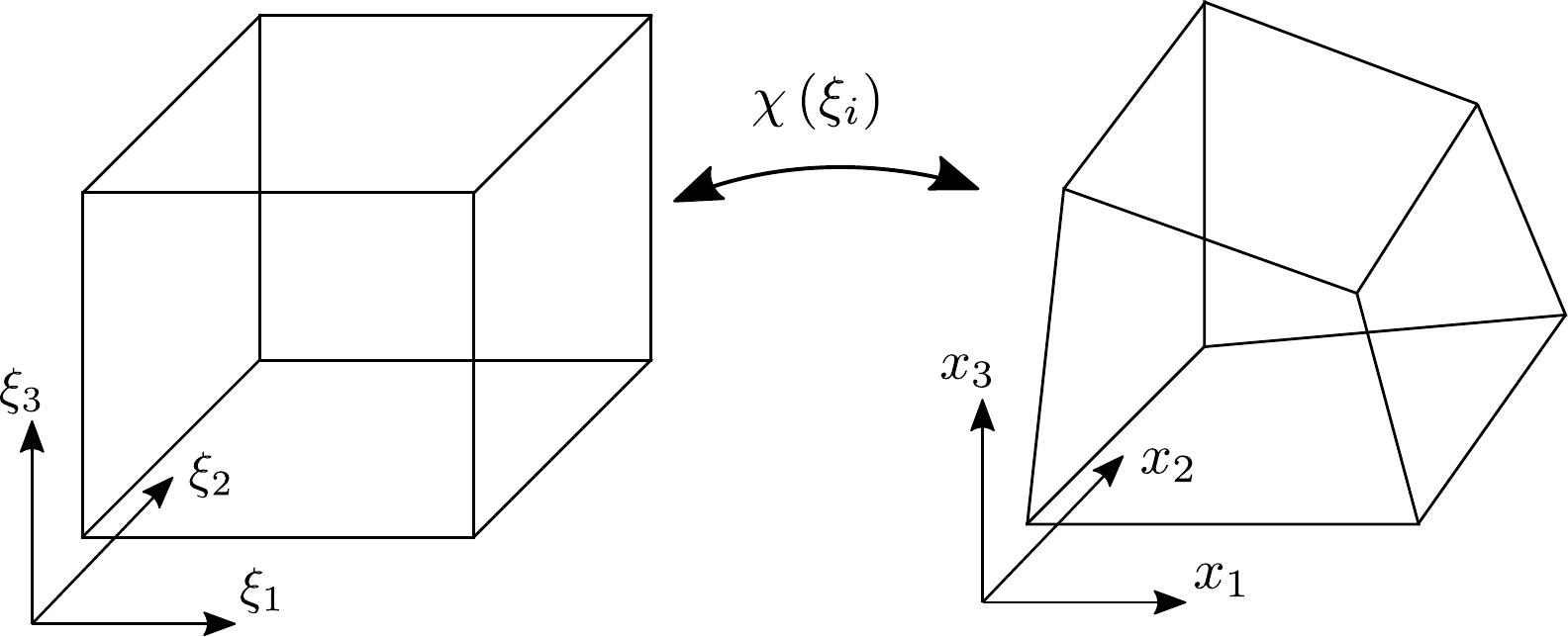}
\caption{Mapping $\chi$ between a standard hexahedral domain (left) defined in computational space ($\xi_1,~\xi_2,~\xi_3$) and a skewed trilinear hexahedral defined in physical space ($x_1,~x_2, ~x_3$).}
\label{fig: mapping}
\end{figure}

For the mapping function $\chi$ in Eq.~\ref{eq: mapping}, the jacobian of transformation is given by the determinant, 
\begin{equation}
 J = \left|\frac{\partial x_i}{\partial \xi_j}\right|,
\end{equation}
which plays an important role in transformation of the conservation Eqs.~\ref{eq: NS_EQ_3D}, which take the following form in the mapped computational coordinate system~\cite{Landmann_2010_AIAA}, 
\begin{equation}
 \frac{\partial\tilde{\mathbf{Q}}}{\partial t} + \frac{\partial \tilde{\mathbf{F}}_i}{\partial \xi_i} = \frac{\partial \tilde{\mathbf{F^v}}_i}{\partial \xi_i},
\end{equation}
where $\tilde{\cdot}$ denotes quantities in the computational coordinate system and are given by, 
\begin{equation}
 \tilde{\mathbf{Q}} = J\mathbf{Q}, ~\mathrm{and} ~\tilde{\mathbf{F}}_i = J \frac{\partial \xi_i}{\partial x_j}\mathbf{F}_j, ~\mathrm{and}~ \tilde{\mathbf{F^v}}_i = J \frac{\partial \xi_i}{\partial x_j}\mathbf{F^v}_j.
\end{equation}

\begin{figure}[!t]
\centering
\includegraphics[width=0.8\linewidth]{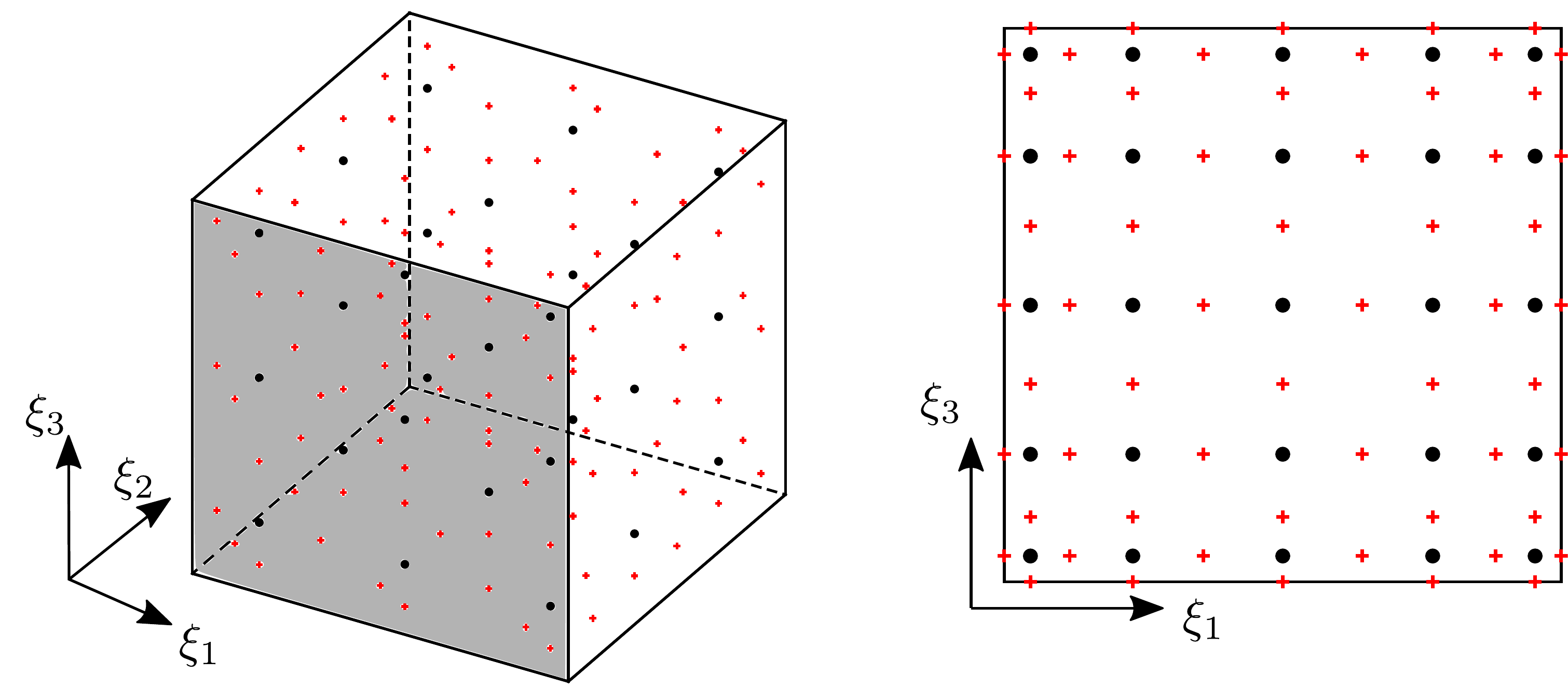}
\put(-360, 150){$(a)$}
\put(-160, 150){$(b)$}
\caption{Solution and flux point distribution in 3D ($a$) and 2D projection on the $\xi_1 - \xi_3$ plane ($b$) for $P=5$. Solution points are shown by black circles and red crosses denote flux points.}
\label{fig: HEXA_FLUX}
\end{figure}
Inside the standard hexahedron, two sets of points are defined namely, solution points (figure~\ref{fig: HEXA_FLUX}) and flux points (figure~\ref{fig: HEXA_FLUX}), which are staggered w.r.t. each other. The conservative variables $\mathbf{Q}$ are stored on the solution points inside the hexahedron and the fluxes are computed on the flux points. The solution points and flux points are the tensor product of one-dimensional Gauss-Legendre and Gauss-Lobatto-Legendre points, respectively~\cite{kopriva1996conservative}. Here onwards, we denote the set of solution points as, 
\begin{equation}
 {X_i},~\mathrm{for}~ i = 1,2,.. P, ~\mathrm{and}~{X_{i+1/2}},~\mathrm{for}~ i = 0, 1,2,.. P,
\end{equation}
respectively. With $P$ solution points and $P+1$ in each direction inside an element, following Lagrange basis functions are used for interpolations and differentiations, 
\begin{equation}
 h_i(X) = \prod^{P}_{j=1, j\neq i}\left(\frac{X - X_j}{X_i - X_j}\right),~\mathrm{and}~l_{i+1/2}(X) = \prod^{P}_{j=0, j\neq i}\left(\frac{X - X_{j+1/2}}{X_{i+1/2} - X_{j+1/2}}\right).
 \label{eq: Lagrange}
\end{equation}
Polynomials in Eq.~\ref{eq: Lagrange} yield the following reconstructions of the vectors $\tilde{\mathbf{Q}}$ and $\tilde{\mathbf{F}}_i, \tilde{\mathbf{F^v}}_i$ in the computational space, 
\begin{align}
 \tilde{\mathbf{Q}} &= \sum^{P}_{k=1}\sum^{P}_{j=1}\sum^{P}_{i=1}\tilde{\mathbf{Q}}_{ijk}h_i(\xi_1)h_j(\xi_2)h_k(\xi_3), \\
 \tilde{\mathbf{F}}_1, \tilde{\mathbf{F^v}}_1 &= \sum^{P}_{k=1}\sum^{P}_{j=1}\sum^{P}_{i=0}\tilde{\mathbf{F}}_{1,ijk},\tilde{\mathbf{F^v}}_{1,ijk} l_{i+1/2}(\xi_1)h_j(\xi_2)h_k(\xi_3),\\
 \tilde{\mathbf{F}}_2, \tilde{\mathbf{F^v}}_2 &= \sum^{P}_{k=1}\sum^{P}_{j=0}\sum^{P}_{i=1}\tilde{\mathbf{F}}_{2,ijk},\tilde{\mathbf{F^v}}_{2,ijk}h_i(\xi_1)l_{j+1/2}(\xi_2)h_k(\xi_3),\\
  \tilde{\mathbf{F}}_3 , \tilde{\mathbf{F^v}}_3 &= \sum^{P}_{k=0}\sum^{P}_{j=1}\sum^{P}_{i=1}\tilde{\mathbf{F}}_{3,ijk} , \tilde{\mathbf{F^v}}_{3,ijk}h_i(\xi_1)h_j(\xi_2)l_{k+1/2}(\xi_3),
  \label{eq: LagrangeReconstruction}
 \end{align}
which are used to advance the conservative variable vector $\tilde{\mathbf{Q}}$ in time at solution points. Normal flux at the element faces is calculated utilizing approximate Riemann solver (HLLC)~\cite{Toro_2013_Riemann} and averaged diffusion flux. For time integration, we utilize the strong stability preserving, five step RK3~\cite{ruuth2006global} scheme, however, we note that any order time integration can be allowed by the spatial discretization methodology described above. 

\section{Adaptive order refinement}
Local order of each computational subdomain can be adapted according to the order of accuracy required. For steady state problems, Kopriva~\cite{Kopriva_1996_Mortar} deviced a way utilizing mortar elements to calculate the interface fluxes for non-conforming set of points for steady state problems. In this work, we extend the mortar element method to the unsteady problems which require high spatial resolution locally. 

Due to local order refinement, interface of various computational subdomains (called mortars hereafter) might be connected across non-conformal set of flux points, as shown in figure~\ref{fig: Mortar}. As a result, local interpolation along the face of computational elements is required.  
\begin{figure}[!t]
\centering
\includegraphics[width=0.7\linewidth]{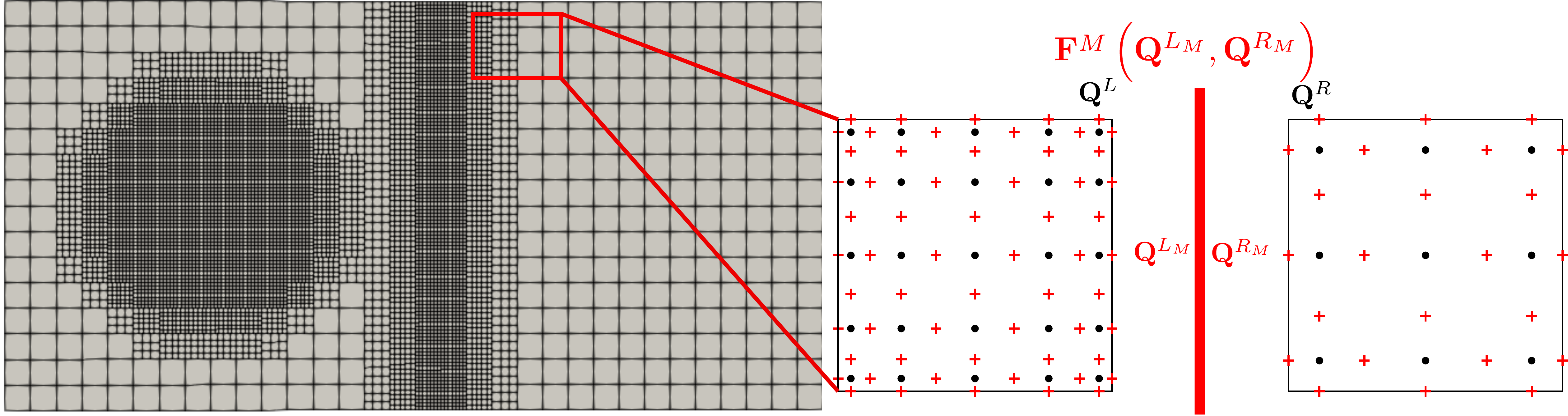}
\caption{Mortar element connecting elements with different local orders with non-conforming flux points.}
\label{fig: Mortar}
\end{figure}
To this end, we consider the reconstruction of left and right states along the interface, $\tilde{\mathbf{Q}}^L$, $\tilde{\mathbf{Q}}^R$ respectively, 
\begin{align}
 \tilde{\mathbf{Q}}^L(X) = \sum^{P_L}_{i=1}\sum^{P_L}_{j=1}\tilde{\mathbf{Q}}^L_{ij}h^L_i(\xi_1)h^L_j(\xi_2), \quad
 \tilde{\mathbf{Q}}^R(X) = \sum^{P_R}_{i=1}\sum^{P_R}_{j=1}\tilde{\mathbf{Q}}^R_{ij}h^R_i(\xi_1)h^R_j(\xi_2),
\end{align}
where $P_L$ and $P_R$ denote the orders of computational elements on the left and right of the mortar and $\xi_1, \xi_2$ are the local coordinates on the mapped face. On the mortar, we consider total number of points to be $P_M = \mathrm{max}\left(P_L, P_R\right)$ in each direction. To interpolate, we utilize the least squares approximation, as done by Kopriva~\cite{Kopriva_1996_Mortar}. To this end, we assume left and right interpolated states on the mortar to have the following reconstruction, 
\begin{equation}
 \tilde{\mathbf{Q}}^{L_M,R_M} = \sum^{P_M}_{i=1}\sum^{P_M}_{j=1}\tilde{\mathbf{Q}}^{L_M,R_M}_{ij}h^M_i(\xi_1)h^M_j(\xi_2).
\end{equation}
The least squares approximation yields the following relation for the interpolation from element to mortar on the left side~\cite{Kopriva_1996_Mortar},

\begin{align}
 \sum^{P_M}_{i=1}\sum^{P_M}_{j=1}\tilde{\mathbf{Q}}^{L_M}_{ij}\int_{\mathrm{mortar}} h^M_i(\xi_1)h^M_j(\xi_2)h^M_p(\xi_1)h^M_q(\xi_2)d\xi_1 d\xi_2 =  \nonumber \\ \sum^{P_L}_{i=1}\sum^{P_L}_{j=1}\tilde{\mathbf{Q}}^{L}_{ij}\int_{\mathrm{mortar}} h^L_i(\xi_1)h^L_j(\xi_2)h^M_p(\xi_1)h^M_q(\xi_2)d\xi_1 d\xi_2,
 \label{eq: MortarLeastSquare}
\end{align}

and similar for the right side. Solving Eq.~\ref{eq: MortarLeastSquare} for every mortar yields the interpolated values $\tilde{\mathbf{Q}}^{L_M}_{ij}$ and $\tilde{\mathbf{Q}}^{R_M}_{ij}$  from the left and right sides of the mortar on the mortar points. Utilizing these, fluxes $\tilde{\mathbf{F}}_{i, pq}\left(\tilde{\mathbf{Q}}^{L_M}_{ij}, \tilde{\mathbf{Q}}^{R_M}_{ij}\right)$ on the mortar points are evaluated and interpolated back on the left and right elements utilizing the Eq.~\ref{eq: MortarLeastSquare}. 

To adapt the local element orders based on the flow features, such as vortices and shock waves, we detect the local flow velocity gradients. For shock waves, we also add a localized shock capturing bulk viscosity~\cite{KawaiLele_2008_JCP, Premasuthan_2014_I, Premasuthan_2014_II} based on the local magnitude of divergence of the velocity field, which indicates the degree of compression/rarefaction in the flow. 

To detect the shocks, we calculate the local normalized divergence of the velocity field
\begin{equation}
 \epsilon_s = \frac{1}{2}\left(\tanh\left( c_1 - c_2 \frac{\Delta x}{a}\left|\sum^3_{j=1}\frac{\partial u_i}{\partial \xi_j}\frac{\partial \xi_j}{\partial x_i}\right| \right) - \tanh\left(c_1\right)\right),
 \label{eq: ShockSensor}
\end{equation}
where $c_1>0$ and $c_2>0$ are the tuning parameters, $a$ is the local speed of sound, and $\Delta x$ is the characteristic element thickness. The error is bounded by $0$ and $1$ $0 < \epsilon_s < 1$, and based on predefined tolerances, the local order of the elements is increased. Furthermore, a shock capturing bulk viscosity and corresponding increase in thermal conductivity are also added in the stress tensor calculation (see Eq.~\ref{eq: stress_tensor} ) and heat flux calculation (see Eq.~\ref{eq: heat_flux}), 
\begin{align}
 \beta^* = c_\beta \rho \epsilon_s\left(\Delta x\right)^2 \left|\sum^3_{j=1}\frac{\partial u_i}{\partial \xi_j}\frac{\partial \xi_j}{\partial x_i}\right|,  \\
 k^* = C_p\beta^*/Pr.
 \label{eq: ArtificialViscosity}
\end{align}

\begin{figure}[!b]
\centering
\includegraphics[width=0.9\linewidth]{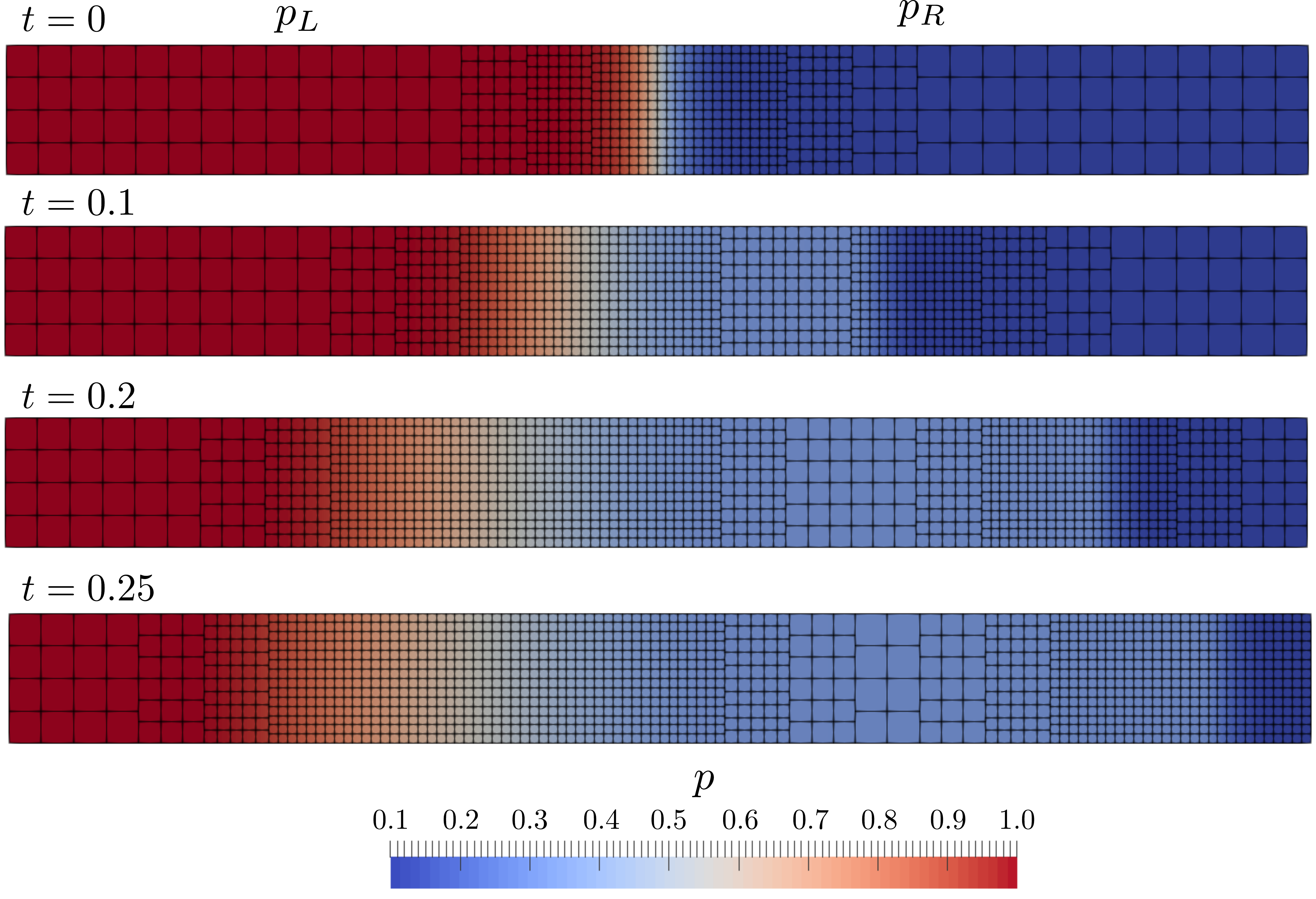}
\caption{Spatio-temporal evolution of the one dimensional Sod shock tube problem and the adaptive order refinement.}
\label{fig: SOD_p}
\end{figure}

To detect the local vorticity, we utilize the ratio of vorticity to velocity magnitude as the sensing parameter, and define the vorticity sensor as, 
\begin{equation}
 \epsilon_v = \frac{1}{2}\left(\tanh\left( c_1 - c_2 \Delta x\frac{\omega_i\omega_i}{u_i u_i} \right) - \tanh\left(c_1\right)\right).
 \label{eq: VorticitySensor}
\end{equation}
Local resolution error is defined as the $\epsilon = \mathrm{max}\left(\epsilon_s, \epsilon_v\right)$ and refinement or coarsening is done comparing $\epsilon$ with predefined tolerances.

\section{Test cases}

In this section, we discuss the three test cases which test the shock and vorticity detecting sensors for adaptive order refinement.

\subsection{1D Shock tube}

The first test case that we consider is the 1D Sod shock tube problem~\cite{SOD_1978_JCP} with the dimensionless left and right states defined as, $\left(\rho_L, u_{1, L}, p_L\right)$ = $\left(1, 0, 1\right)$ and $\left(\rho_R, u_{1,R}, p_R\right) = \left(0.125, 0, 0.1\right)$ in a 1 unit length domain with 10$\times$2 elements. Order refinement is also donw corresponding to the initial conditions so as to capture the initial conditions as accurately as possible. Figure~\ref{fig: SOD_p} shows the spatio-temporal evolution of the dimensionless pressure and adaptive order refinement of the mesh. The test case was initialized with $P=2$ throughout and a maximum order of 7. 
\begin{figure}[!t]
\centering 
\includegraphics[width=0.9\linewidth]{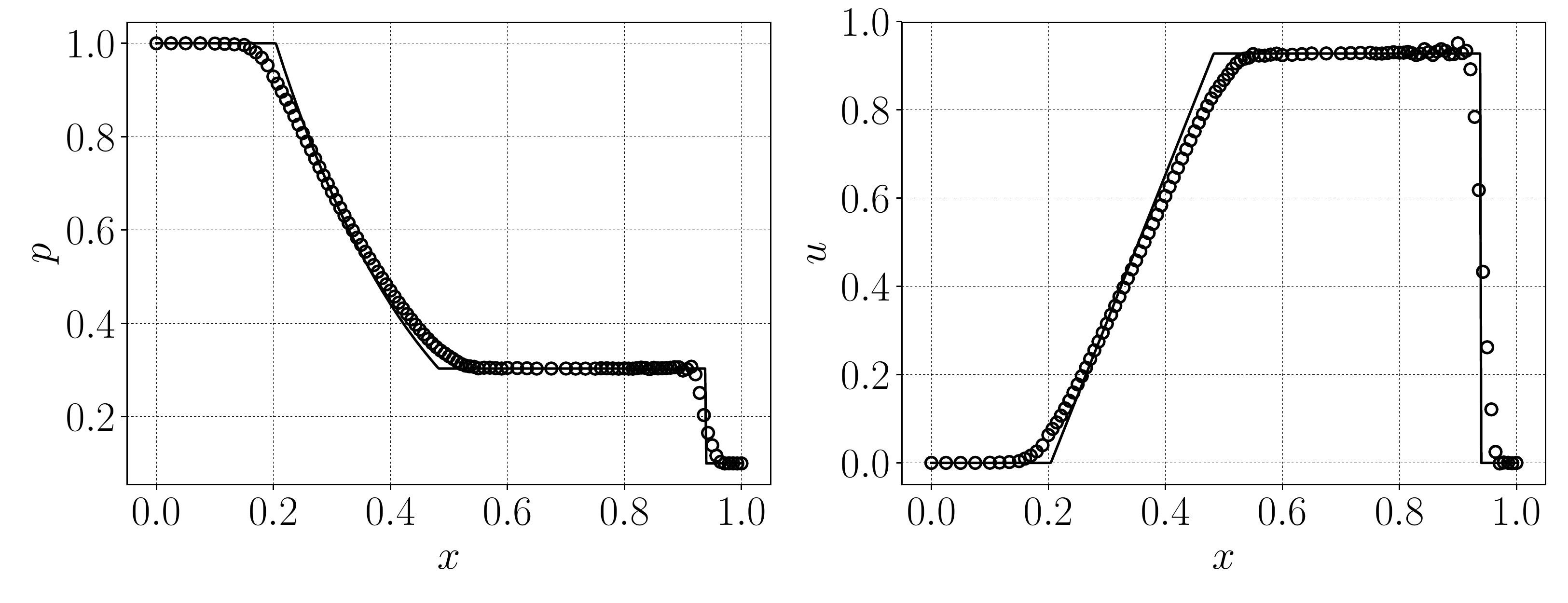}
\caption{Comparison of the solution obtained from the adaptive $p$-refinement strategy with the analytical solution of the Sod shock tube case at dimensionless time $t = 0.25$.}
\label{fig: SOD_p_u}
\end{figure}We note that the sensor tracks the right propagating shock and the left propagating expansion wave very effectively. Figure~\ref{fig: SOD_p_u} shows the comparison of the solution at $t=0.25$ units with the analytical solution of the inviscid problem.

\subsection{Convecting isentropic vortex }

The second test case considered is a convecting isentropic vortex in a base flow of 0.5 Mach number in domain with $19\times10$ elements. The vortex is defined through the following dimensionless velocity and temperature values, 
\begin{align}
 u_1 &= U_{\infty} + \varepsilon\frac{r}{r_0}e^{\alpha\left(1 - \left(r/r_0\right)^2\right)}\sin\theta,\\
 u_2 &= -\varepsilon\frac{r}{r_0}e^{\alpha\left(1 - \left(r/r_0\right)^2\right)}\cos\theta, \\
 T &= T_{\infty} - \frac{\left(\gamma - 1\right)\varepsilon^2 e^{2\alpha\left( 1 - \left(r/r_0\right)^2\right)}}{4\alpha \gamma},
 \label{eq: isentropicVortex}
\end{align}
where $U_{\infty} = 0.5$ and $T_{\infty} = 1$. Initial pressure and density are evaluated through the isentropic relation $p/\rho^\gamma = \mathrm{const.}$ and the ideal gas equation of state. The vortex related parameters considered were $\varepsilon = 0.3, r_0 = 0.05, \alpha = 0.204$.~\cite{Jiang_1996_WENO}

The vortex core propagates at the base flow velocity. Moreover, acoustic waves are emanated from the vortex initialization which propagate due to temperature initialization and the corresponding pressure and density changes. 
\begin{figure}[!t]
\centering 
\includegraphics[width=\linewidth]{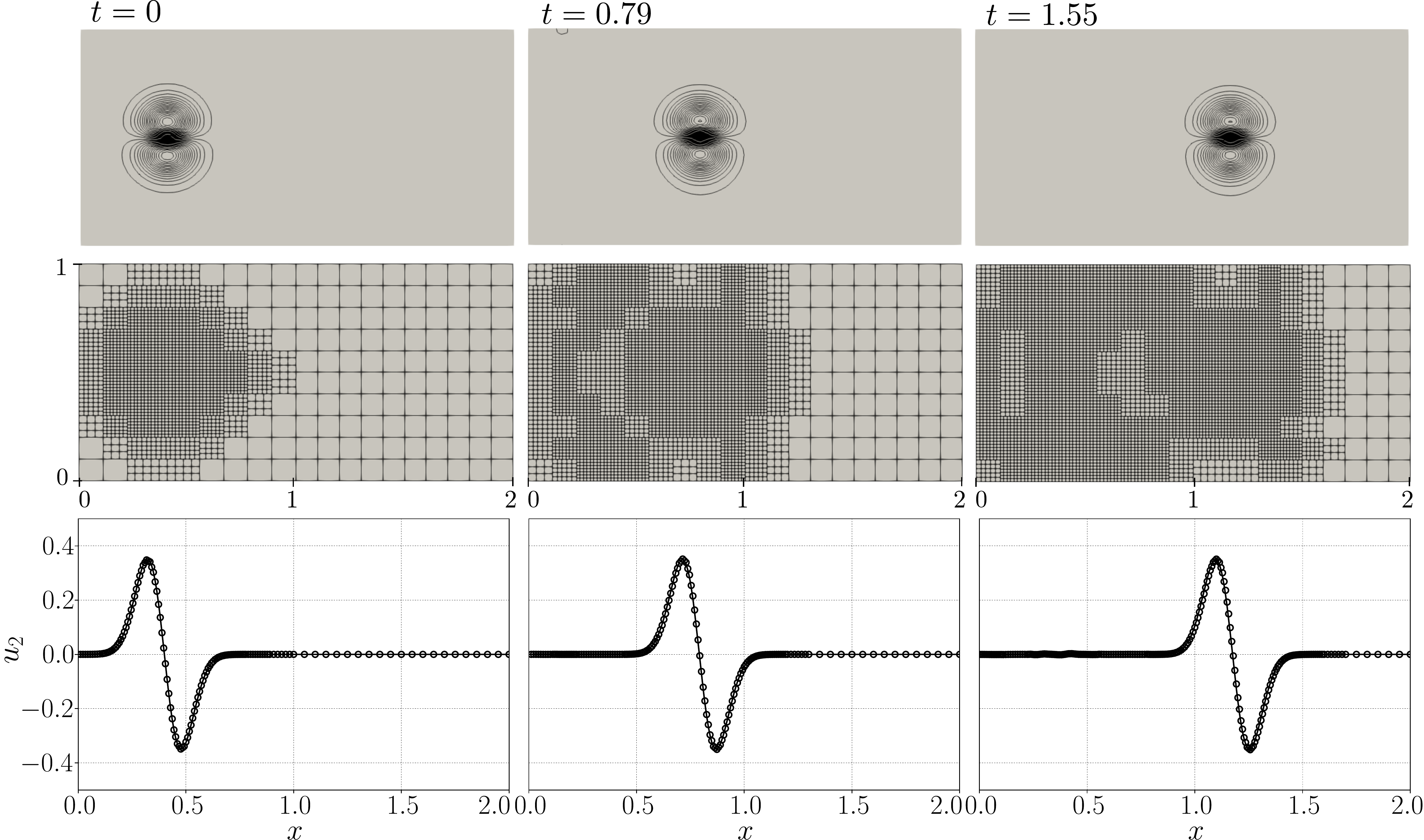}
\caption{Spatio temporal evolution of the isocontours of $u_1$ of the vortex defined by velocity and temperature field in Eq.~\ref{eq: isentropicVortex} (top), adaptive order refinement as the vortex core moves (mid), and comparison of the vertical velocity $u_2$ at $x_2=0.5$ with the analytical solution. }
\label{fig: IsentropicVortex}
\end{figure}
The adaptive order refinement follows the vortex core as well as the acoustic waves propagating away from the vortex, as shown in figure~\ref{fig: IsentropicVortex}. The domain is initialized with $P=1$ in all the cells and then pre-refined before the first time integration step based on the initial conditions. The maximum allowed order is $P=7$. In figure~\ref{fig: IsentropicVortex}, the comparison of $u_2$ with the analytical result of pure advection with a velocity of $U_{\infty}$ is also shown. 
\begin{figure}[!t]
\centering 
\includegraphics[width=\textwidth]{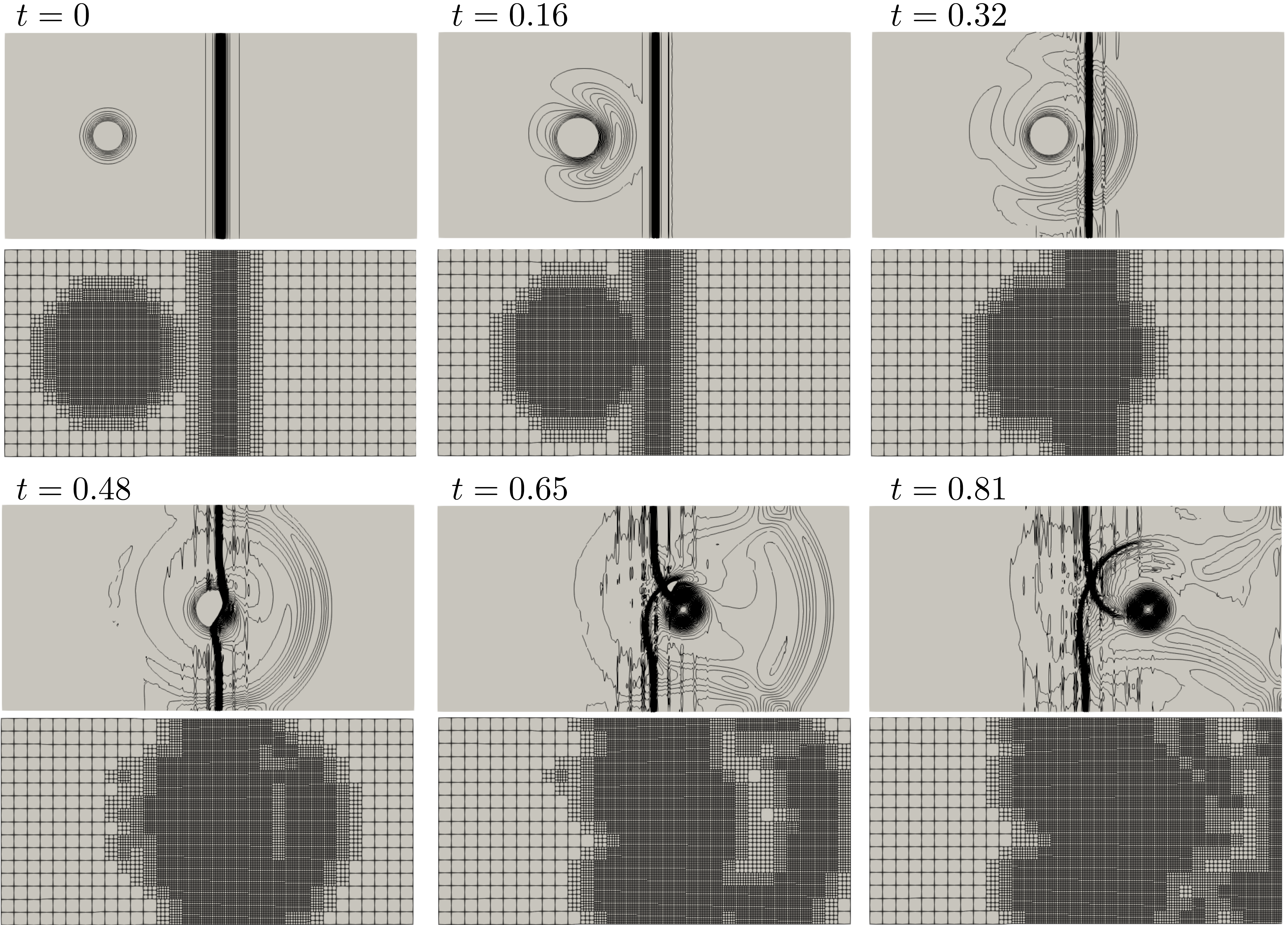}
\caption{Spatio temporal evolution of pressure isocontours for the shock vortex interaction test case and the corresponding dynamic mesh refinement based on both the shock sensor (see Eq.~\ref{eq: ShockSensor}) and the vorticity sensor (see Eq.~\ref{eq: VorticitySensor}).}
\label{fig: VortexShockInteraction}
\end{figure}
\subsection{Shock Vortex interaction}
The third and final test case considered in this work is the shock-vortex interaction case with a shock of Mach number 1.1. Figure~\ref{fig: VortexShockInteraction} shows the adaptive order refinement and iso-contours of pressure field as the vortex passes through the shock in a 2D domain with $32\times16$ elements. The domain is initialized with $P=1$ in all the cells and then pre-refined before the first time integration step based on the initial conditions. The maximum allowed order is $P=7$. 
We note that when the vortex passes through the shock, the shock bifurcates thus creating triple points in the domain. Moreover, due to the shock-vortex interaction, intense pressure oscillations at the element boundaries are generated due to high order polynomial interpolation. However, the hydrodynamic features are captured very well by the high-order refinement.

\section{Summary and future work}
In this work, we developed and tested an unstructured fully compressible Navier Stokes solver which adaptively refines the mesh by increasing the local order of refinement of the interpolation and derivative calculation. The solver is based on the spectral difference technique which allows definition of order of interpolation local to the mesh elements. We tested the solver on three test cases. Results for one-dimensionl shock wave and a vortex advection show very good agreement with analytical results. However, for the complex test case of shock vortex interaction, we note that element boundaries exhibit large oscillations of pressure. 
Future research from the developed solver includes further development of adaptive element refinement ($h-$refinement) which breaks a given element into subelements thus allowing higher resolution with low order interpolation as well. 

% \include{NumericalMethods}
% %%%%%%%%%%%%%%%%%%%%%%%%%%%%%%

% %%%%%%%%%%%%%%%%%%%%%%%%%%%%%%
\chapter{Contributions, conclusions, and future research}
In this chapter, major contributions and conclusions of this dissertation are summarized. We also present recommendations for future research.

\section{Major contributions}
\subsection{Spectral energy dynamics in nonlinear acoustics}
\begin{itemize}
	\item Second order scaling of maximum entropy generated within weak shock waves with pressure jump for an ideal gas was shown.
	\item Utilizing the second order scaling of maximum entropy, the correct second order equations for planar nonlinear acoustics were derived which hold for any general nonlinear acoustic system (within second order approximation), unlike Burgers equation which holds only for a planar propagating wave.
	\item The second order equations yielded the correct perturbation energy corollary for nonlinear acoustics and the analytical expression of the correct perturbation energy norm for nonlinear acoustics. 
	\item From the correct perturbation energy corollary, the spectral energy, spectral energy flux, and spectral energy dissipation expressions were obtained and analyzed utilizing DNS of fully compressible one-dimensional Navier-Stokes equations.
\end{itemize}

\subsection{Spectral energy cascade in thermoacoustic shock waves}
\begin{itemize}
	\item First high fidelity numerical and analytical modeling of thermoacoustic shock waves was done utilizing the fully compressible Navier-Stokes simulations, pushing the thermoacoustic amplification of acoustic waves to the formation of shock waves.
	\item In the linear regime of thermoacoustic amplification of waves, thermoacoustic energy budgets were analyzed yielding the optimum phasing required for thermoacoustic growth. 
	\item In the nonlinear regime, novel framework of nonlinear acoustic waves propagating near solid no-slip isothermal walls was developed, yielding the macrosonic interaction of wall-heat flux and wall-shear on the nonlinear acoustic waves. 
	\item Scaling of spectral acoustic energy was shown ($\widehat{E}_k \propto \omega^{-5/2}_k$), utilizing the computational models and theoretical dimensionless scaling arguments. 
\end{itemize}

\subsection{Development of an order adaptive Navier-Stokes solver}
\begin{itemize}
	\item Motivated by the study of spark plasma shock wave induced flow field and sustained planar detonation wave dynamics, a high fidelity order adaptive unstructured Navier-Stokes solver was developed, which can be further utilized for simulating even more complex flow phenomena.
\end{itemize}
\section{Conclusions and recommendations for future research}
\subsection{Spectral energy cascade and decay in planar nonlinear acoustic waves}
\begin{itemize}
	\item The maximum entropy perturbations $s'$ in an ideal gas scale as $p'^2$ for normalized pressure perturbation $\pFluct \sim \mathcal{O}\left(10^{-3} - 10^{-1}\right)$.
	\item In traveling waves (TW) and standing waves (SW), spatial average of perturbation energy decays as $\left\langle E^{(2)}\right\rangle\sim t^{-2}$ and dissipation rate as $\epsilon \sim t^{-3}$ in time. Moreover, the spectral energy for both traveling and standing waves assumes the self-similar form: $\widehat{E}_k k^2 \epsilon^{-2/3}\ell^{1/3}\sim 0.075 f(k\eta)$, where $\ell$ is the integral length scale and $\eta$ is the Kolmogorov length scale. 
	\item In acoustic wave turbulence (AWT), due to gradual increase of the integral length scale $\ell$ caused by the shock coalescence, the approximate decay laws are $\left\langle E^{(2)}\right\rangle \sim t^{-2/3}$ and $\epsilon \sim t^{-5/3}$.
	\item The correct energy corollary can be utilized to evaluate nonlinear interactions of high amplitude sound and heat release due to combustion or thermal gradients. 
	\item Spectral energy cascade due to the nonlinear acoustic effects plays an important role in the study of high Reynolds number compressible turbulence~\cite{Lele_AnnRev_1994}. Hence, importance of nonlinear compressible effects can be quantified via spectral energy transfer terms derived in this work. 
\end{itemize}
\subsection{Spectral energy cascade in thermoacoustic shock waves}
\begin{itemize}
	\item Three regimes of thermoacoustic wave amplification were identified: ($i$) a monochromatic or modal growth regime, ($ii$) a hierarchical spectral broadening or nonlinear growth regime and ($iii$) a shock-dominated limit cycle. The modal growth regime is characterized by exponential amplification of thermoacoustically unstable modes. 
	\item The existence of an equilibrium \emph{thermoacoustic energy cascade} was shown. The spectral energy density at the limit cycle, in particular, was found to decay as $\omega^{-5/2}$ in  spectral space, the relevant intensity scaling with growth rate as $\alpha^3$.
	\item Nonlinear wave propagation equations in the presence of wall-shear and wall-heat transfer can be extended to higher dimensions establishing a generalized framework for understanding nonlinear wave propagation and effects of wall-shear and wall-heat transfer on the acoustic variables.
\end{itemize}

\subsection{Mesh adaptive Navier-Stokes solver}
\begin{itemize}
	\item Axisymmetric simulations of flow field behind a spark induced shock wave were carried out, revealing the importance of baroclinic torque in the generation of vorticity field behind the shock wave. 
	\item Utilizing the mesh adaptive solver, full three-dimensional simulations of the flow field behind the spark-plasma induced shock wave can be carried out, thus enabling the study of low Reynolds number turbulence observed in the experiments~\cite{BsinghEtAl_2019_AIAA}.
	\item Further development of the mesh adaptive solver (implementation of detailed chemistry models) would enable the study of detonation wave propagation in three dimensions and resonant dynamics sustained by fuel injection and exhaust.
\end{itemize}

% \include{Conclusions}
% %%%%%%%%%%%%%%%%%%%%%%%%%%%%%%

% %%%%%%%%%%%%%%%%%%%%%%%%%%%%%%
\chapter{Status of publications}
\section{Refereed journal articles}
\begin{itemize}
\item \textbf{Prateek Gupta}, Guido Lodato, Carlo Scalo, \emph{Spectral energy cascade in thermoacoustic shock waves}, \emph{Journal of Fluid Mechanics}, 2017, \textbf{831}; 358-393
\item Danish Patel, \textbf{Prateek Gupta}, Carlo Scalo, \emph{Acoustic Impedance Calculation via Numerical Solution of the Inverse Helmholtz Problem}, In review \emph{Journal of Sound and Vibration}, 2018
\item \textbf{Prateek Gupta}, Carlo Scalo, \emph{Spectral energy cascade and decay in nonlinear acoustic waves}, Submitted to \emph{Physical Review E}, 2018
\end{itemize}

\section{Presentations and non-refereed conference publications}
\begin{itemize}
\item \textbf{Prateek Gupta}, Guido Lodato, Carlo Scalo, \emph{Nonlinear modeling of thermoacoustically driven energy cascade}, 69th APS DFD (Portland, Oregon, USA), Nov. 2016
\item \textbf{Prateek Gupta}, Guido Lodato, Carlo Scalo, \emph{Numerical investigation of thermoacoustic shock waves}, 55th AIAA Scitech (Grapevine, Texas, USA), Jan. 2017
\item  Danish Patel, \textbf{Prateek Gupta}, Thomas Rothermel, Markus Kuhn, Carlo Scalo, \emph{mpedance Characterization of Ultrasonically Absorptive Coatings via Numerical Solution of the Inverse Helmholtz Problem}, 55th AIAA Scitech (Grapevine, Texas, USA), Jan. 2017
\item  Mario Tindaro Migliorino, \textbf{Prateek Gupta}, Carlo Scalo, \emph{Real fluid effects on thermoacoustic standing-wave resonance in supercritical $CO_2$}, 55th AIAA Aviation (Denver, Colorado, USA), June 2017
\item \textbf{Prateek Gupta}, Carlo Scalo, \emph{Glimpses of Kolmogorov energy cascade in nonlinear acoustic waves}, 70th APS DFD (Denver, Colorado, USA), Nov. 2017
\item \textbf{Prateek Gupta}, Kyle Schwinn, Guido Lodato, Carson Slabaugh, Carlo Scalo, \emph{Numerical investigation of sustained planar detonation waves in a periodic domain}, 56th AIAA Aviation (Atlanta, Georgia, USA), June 2018
\end{itemize}

\chapter{Derivation of second order equations}
\label{sec: appedixA}
{Decomposing the variables in base state and perturbation fields, we obtain, }
\begin{subequations}
\begin{align}
& \rho^* = \rho^{*}_0 + {\rho^*}',\quad p^* = p^*_0 + {p^*}', \\
& u^* = {u^*}', \quad s^* = {s^*}', \quad T^* = T^*_0 + {T^*}',
 \label{eq: PertDecom_APP}
 \end{align}
\end{subequations}
where the superscript $( )_0$ denotes the base state, the superscript $({ })'$ denotes the perturbation values, no mean flow $u^*_0 = 0$ is considered, and $s^*_0$ is arbitrarily set to zero.
We neglect the fluctuations in the dynamic viscosity as well, i.e.,
\begin{equation}
\mu^* = \mu^*_0.
\end{equation}
{\color{black}{While in classic gas dynamics, pre-shock values are used to normalize fluctuations or jumps across the shock (e.g. see Eq.~\eqref{eq: PressureMach}), here we choose base state values to non-dimensionalize the variables,
\begin{subequations}
\begin{align}
& \rho = \frac{{\rho^*}}{\rho^*_0} = 1 + \rhoFluct ,~~p = \frac{p^*}{\gamma p^*_0} = \frac{1}{\gamma} + \pFluct, \\
& u = \frac{u^*}{a^*_0} = \uFluct, ~~s = \frac{s^*}{R^*} = \sFluct,~~ T = \frac{T^*}{T^*_0} = 1 + \TFluct,\\
&x = \frac{x^*}{L^*}, ~~ t = \frac{a^*_0 t^*}{L^*}.
\end{align}
\label{eq: per_norm_APP}
\end{subequations}
where $L^*$ is the length of the one-dimensional periodic domain.
For a chemically inert generic gas, infinitesimal changes in dimensionless density $\rho (p,s)$ in terms of pressure $p$ and entropy $s$ are given by,
\begin{align}
d\rho &= \left(\frac{\partial \rho}{\partial p}\right)_{s} dp + \left(\frac{\partial \rho}{\partial s}\right)_{p} ds,\nonumber\\
& = \frac{\rho}{\gamma p } dp -\left(\frac{\rho^{*}_0T^*_0R^*}{\gamma p^*_0}\right)\frac{\rho^2 T}{p}\left(\frac{\gamma - 1}{\gamma}\right)ds.
\label{eq: ConstitutiveEquationAPP}
\end{align}
Substituting the above relation in the dimensionless continuity Eq.~\eqref{eq: PressureStep1}, we obtain, 
\begin{align}
\frac{\partial p}{\partial t} &+ u\frac{\partial p}{\partial x} + \gamma p\frac{\partial u}{\partial x} \nonumber \\
&= \left(\frac{\rho^{*}_0T^*_0R^*}{p^*_0}\right)\left(\frac{\gamma - 1}{\gamma}\right){\rho T}\left(\frac{\partial s}{\partial t} + u\frac{\partial s}{\partial x}\right).
\label{eq: PressureStepAPP1}
\end{align}
Non-dimensionalizing the entropy Eq.~\eqref{eq: NS_eqns3} utilizing the Eq.~\eqref{eq: per_norm_APP}, we obtain, 
\begin{equation}
\rho T \left(\frac{\partial s}{\partial t} + u\frac{\partial s}{\partial x}\right) = \frac{\nu_0}{Pr}\frac{C^*_p}{R^*}\frac{\partial^2 T}{\partial x^2} + \frac{4\nu_0}{3}\frac{a^{*2}_0}{R^*T^*_0}\left(\frac{\partial u}{\partial x}\right)^2.
\label{eq: NonDimEnt}
\end{equation}
Substituting the above equation in Eq.~\eqref{eq: PressureStepAPP1}, we obtain, 
\begin{align}
&\frac{\partial p}{\partial t} + u\frac{\partial p}{\partial x} + \gamma p\frac{\partial u}{\partial x} \nonumber \\
&= \left(\frac{\rho^{*}_0T^*_0R^*}{ p^*_0}\right)\frac{\gamma - 1}{\gamma }\left(\frac{\nu_0}{Pr}\frac{C^*_p}{R^*}\frac{\partial^2 T}{\partial x^2} + \frac{4\nu_0}{3}\frac{a^{*2}_0}{R^*T^*_0}\left(\frac{\partial u}{\partial x}\right)^2\right).
\label{eq: PressureStepAPP2}
\end{align}
Substituting the decomposition of variables (cf. Eq.~\eqref{eq: per_norm_APP}) in the above Eq.~\eqref{eq: PressureStepAPP2}, we obtain the pressure perturbation equation for a generic gas, 
\begin{align}
&\frac{\partial \pFluct}{\partial t} + \frac{\partial \pFluct}{\partial x} + \uFluct\frac{\partial \pFluct}{\partial x} + \gamma \pFluct\frac{\partial \uFluct}{\partial x} \nonumber \\
&= \left(\frac{\rho^{*}_0T^*_0R^*}{ p^*_0}\right)\frac{\gamma - 1}{\gamma }\left(\frac{\nu_0}{Pr}\frac{C^*_p}{R^*}\frac{\partial^2 T'}{\partial x^2} + \frac{4\nu_0}{3}\frac{a^{*2}_0}{R^*T^*_0}\left(\frac{\partial \uFluct}{\partial x}\right)^2\right).
\label{eq: PressureStepAPP3}
\end{align}
As shown in Section~\ref{sec: 2ndOrderEntropy}, the entropy perturbations are atmost $2^{\mathrm{nd}}$ order in pressure, independent of viscosity. Consequently, the first and second term on right hand side of Eq.~\eqref{eq: NonDimEnt} are second and third order in pressure perturbations, respectively. Truncating the Eq.~\eqref{eq: PressureStepAPP3} up to second order, we obtain the second order equation for pressure perturbations for a generic fluid as, 
\begin{align}
&\frac{\partial \pFluct}{\partial t} + \uFluct\frac{\partial \pFluct}{\partial x} + \frac{\partial \uFluct}{\partial x} + \gamma \pFluct\frac{\partial \uFluct}{\partial x} = \nonumber \\
&\frac{\nu_0}{Pr}\left(\frac{\rho^{*}_0T^*_0R^*}{ p^*_0}\right)\left(\frac{\gamma - 1}{\gamma }\right)\left(\frac{\partial T}{\partial p}\right)_{s,0}\frac{C^*_p}{R^*}\frac{\partial^2 \pFluct}{\partial x^2} \nonumber \\
&+ \mathcal{O}\left(\pFluct\sFluct, \sFluct^2, \pFluct^3, \left(\frac{\partial \uFluct}{\partial x}\right)^2\right),
\label{eq: PressureStepAPP4}
\end{align}
Substituting the decomposition of variables (cf. Eq.~\eqref{eq: per_norm_APP}) in dimensionless Eq.~\eqref{eq: NS_eqns2} and neglecting changes in kinematic viscosity, we obtain, 
\begin{equation}
 \frac{\partial \uFluct}{\partial t} + \uFluct\frac{\partial \uFluct}{\partial x} + \frac{1}{1 + \rho'}\frac{\partial \pFluct}{\partial x} = \frac{4}{3}\nu_0\frac{\partial^2 \uFluct}{\partial x^2}.
 \label{eq: velocityStep1}
\end{equation}
Equations~\eqref{eq: PressureStepAPP4} and~\eqref{eq: velocityStep1} do not involve any assumption regarding the thermal equation of state of the gas and hold for any chemically inert generic gas.\\

Assuming a thermal equation of state for an ideal gas in Eq.~\eqref{eq: PressureStepAPP4} and utilizing binomial expansion in Eq.~\eqref{eq: velocityStep1}, we obtain Eqs.~\eqref{eq: pressure} and~\eqref{eq: velocity} as, 
\begin{align}
\frac{\partial \pFluct}{\partial t} + \frac{\partial \uFluct}{\partial x} + \gamma \pFluct \frac{\partial \uFluct}{\partial x} &+ \uFluct\frac{\partial \pFluct}{\partial x} = \nu_0\left(\frac{\gamma - 1}{Pr}\right)\frac{\partial^2 \pFluct}{\partial x^2} \nonumber \\
& + \mathcal{O}\left(\pFluct\sFluct, \sFluct^2, \pFluct^3, \left(\frac{\partial \uFluct}{\partial x}\right)^2\right),\label{eq: pressureAPP}
\end{align}
\begin{align}
\frac{\partial \uFluct}{\partial t} + \frac{\partial \pFluct}{\partial x} + \frac{\partial}{\partial x}\left(\frac{\uFluct^2}{2} - \frac{\pFluct^2}{2}\right) &= \frac{4}{3}\nu_0\frac{\partial^2 \uFluct}{\partial x^2} \nonumber \\
&+\mathcal{O}\left(\rho'^2\pFluct,\rho'^3\pFluct \right). \label{eq: velocityAPP}
\end{align}
We note that the LHS of Eqs~\eqref{eq: PressureStepAPP4} and \eqref{eq: velocityStep1} (up to second order) are identical to those of Eqs.~\eqref{eq: pressureAPP} and \eqref{eq: velocityAPP}, respectively, hence independent from the thermal equation of state. As shown in section~\ref{sec: EnergyCorollary}, the functional form of the second order perturbation energy norm $E^{(2)}$ (Eq \eqref{eq: Energy_norm}) is exclusively dictated by such terms, and hence is also independent from the thermal equation of state. The results shown in this work focus on ideal-gas simulations merely for the sake of simplicity, with no loss of generality pertaining to inviscid nonlinear (up to second order) spectral energy transfer dynamics.}

\chapter{Derivation of Spectral Energy Transfer}
\label{sec: appedixB}
Equation~\eqref{eq: spectral_conservation} can be obtained from the conservation of perturbation energy upon considering the second order governing relations (Eqs.~\eqref{eq: pressure} and~\eqref{eq: velocity}) and substituting the Fourier expansions of $\pFluct$ and $\uFluct$,
\begin{equation}
\pFluct = \sum^{\infty}_{k=-\infty}\widehat{p}_k e^{ikx}, ~~ \uFluct=\sum^{\infty}_{k=-\infty}\widehat{u}_k e^{ikx},
\end{equation}
yielding,
\begin{align}
\frac{d\widehat{p}_k}{dt} + a_0ik\widehat{u}_k + &ia_0\gamma\sum^{\infty}_{k'=-\infty}k'\widehat{p}_{k-k'}\widehat{u}_{k'} + ia_0\sum^{\infty}_{k'=-\infty}k'\widehat{p}_{k'}\widehat{u}_{k-k'} = -\nu_0\left(\frac{\gamma - 1}{Pr}\right) k^2 \widehat{p}_k,\label{eq: p_spectral}\\
\frac{d\widehat{u}_k}{dt} + a_0ik\widehat{p}_k &+ a_0i\sum^{\infty}_{k'=-\infty}k'\widehat{u}_{k-k'}\widehat{u}_{k'} - a_0i\sum^{\infty}_{k'=-\infty}k'\widehat{p}_{k-k'}\widehat{p}_{k'} = -2\nu_0k^2\widehat{u}_k.\label{eq: u_spectral}
\end{align} 
Multiplying eqs.~\eqref{eq: p_spectral} and~\eqref{eq: u_spectral} by $\widehat{p}_{-k}$ and $\widehat{u}_{-k}$ and adding the complex conjugate, we obtain,
\begin{align}
&\frac{d}{dt}\left(\frac{|\widehat{p}_k|^2}{2} + \frac{|\widehat{u}_k|^2}{2}\right) + a_0\gamma \Re\left(\widehat{p}_{-k}\sum^{\infty}_{k'=-\infty}ik'\widehat{u}_{k'}\widehat{p}_{k-k'}\right) \nonumber \\
&+ a_0\Re\left(\widehat{p}_{-k}\widehat{\left(u\frac{\partial p}{\partial x}\right)}_k + \frac{\widehat{u}_{-k}}{2}\reallywidehat{\left(\frac{\partial }{\partial x}\left(u^2-p^2\right)\right)}_k \right) \nonumber \\
& = -\nu_0\frac{\gamma - 1}{Pr} k^2|\widehat{p}_k|^2 - 2\nu_0k^2|\widehat{u}_k|^2.
\label{eq: LinEq1}
\end{align}
The second term in the above equation can be evaluated recursively utilizing the Eq.~\eqref{eq: pressure} yielding,
\begin{align}
&a_0\gamma \Re\left(\widehat{p}_{-k}\sum^{\infty}_{k'=-\infty}ik'\widehat{u}_{k'}\widehat{p}_{k-k'}\right) = \Re\left(\widehat{p}_{-k}\frac{d\widehat{g}_k}{dt}\right) +\nonumber\\
&a_0\Re\left(\widehat{p}_{-k}\widehat{\left(\frac{\partial ug}{\partial x}\right)}_k\right) - \nu_0\left(\frac{\gamma - 1}{Pr}\right)\Re\left(\widehat{p}_{-k}\widehat{\left(\frac{\partial g}{\partial p}\frac{\partial^2 p}{\partial x^2}\right)}_k\right),
\end{align}
which, upon substitution in Eq.~\eqref{eq: LinEq1} yields, 
\begin{equation}
\frac{d \widehat{E}_k}{dt} + \widehat{T}_k = \widehat{\mathcal{D}}_k,
\end{equation}
where, the spectral energy transfer function $T_k$ is given by, 
\begin{align}
\widehat{T}_k = a_0\Re\Bigg(\widehat{p}_{-k}\widehat{\left(\frac{\partial ug}{\partial x}\right)}_k + \widehat{p}_{-k}\widehat{\left(u\frac{\partial p}{\partial x}\right)}_k +\frac{\widehat{u_{-k}}}{2}\reallywidehat{\left(\frac{\partial }{\partial x}\left(u^2-p^2\right)\right)_k\Bigg)}, 
\end{align}
and the spectral dissipation term $\mathcal{D}_k$ is given by,
\begin{align}
\mathcal{D}_k &= -\nu_0\left(\frac{\gamma - 1}{Pr}\right) \left(k^2|\widehat{p}_k|^2 - \Re\left(p_{-k}\reallywidehat{\left(\frac{\partial g}{\partial p}\frac{\partial^2 p}{\partial x^2}\right)_k} \right)\right)-2\nu_0k^2|\widehat{u}_k|^2.
\end{align}
Summation of Eq.~\eqref{eq: LinEq1} for $k'<k$ yields Eq.~\eqref{eq: spectral_conservation} and the expressions thereafter.
%%%%%%%%%%%%%%%%%%%%%%%%%%%%%%%%%%%%%%%%%%%%%%%%%%%%%%%%%%%%%%%%%%%%%%%%%%%%%%%%%%%%%%%%%%%

\chapter{Experimental Validation of linear model}
\label{sec: appedixC}
\begin{figure}[!b]
\centering
\includegraphics[width=0.95\textwidth]{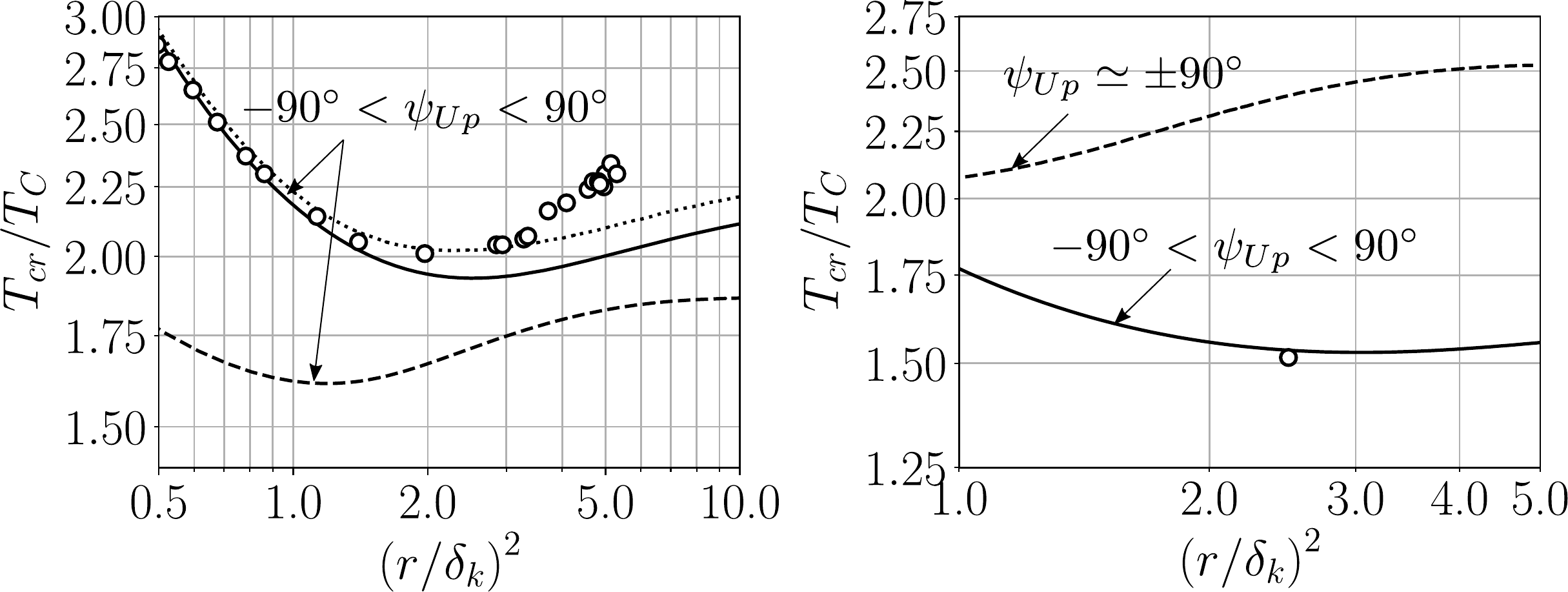}
\put(-414,150){$(a)$}
\put(-207,150){$(b)$}
\caption{Neutral stability curves obtained from linear analysis (\S~\ref{sec: Linear}), ~\cite{Yazaki_PhysRevLet_1998}'s data ($a$), and~\cite{BiwaTY_JAcousSocAm_2011}'s reported instability limit ($b$). $r$ is the hydraulic radius of regenerator pore (square cross-section). ($a$): (--), Y-1998 in 3D ; (- -), 3D minimal unit corresponding to Y-1998; ($\cdot \cdot\cdot$),~\cite{guedra_use_2012}'s results; ($\circ$),~\cite{Yazaki_PhysRevLet_1998}'s reported data. ($b$): (--), B-2011 in 3D (with $\ell_{buffer}=0.4~$m); (- -), 3D minimal unit corresponding to B-2011 without buffer length ; ($\circ$), reported instability limit.}
\label{fig: ExpValid}
\end{figure}
Figure~\ref{fig: ExpValid}$a$ shows the comparison between the neutral stability curve evaluated from the linear eigenvalue analysis discussed in \S~\ref{sec: Linear}, experimental data obtained by~\cite{Yazaki_PhysRevLet_1998} (setup referenced as Y-1998), and the numerical predictions obtained by~\cite{guedra_use_2012}. The lengths of smaller cross-section ducts ($\ell_a$ and $\ell_c$ in figure~\ref{fig:computational_setup}) have been estimated from the laser Doppler velocimetry (LDV) data reported by~\cite{Yazaki_PhysRevLet_1998} ($\ell_a=0.153$ m and $\ell_c=0.5$ m). The rest of the missing geometrical details have been taken from~\cite{guedra_use_2012}. 
The looped geometry of ~\cite{BiwaTY_JAcousSocAm_2011} (setup referenced as B-2011), is shown to be thermoacoustically unstable at the reported critical $T_{H}$ value, ($T_{cr}/T_{C}=1.52$) only if a mean negative temperature gradient from $\ell_b<x<\ell_{\mathrm{buffer}}+\ell_b$, is assumed in the linear analysis. Moreover, the unstable mode is a quasi-travelling wave ($-90^{\circ}<\psi_{Up}<90^{\circ}$). However, by considering an abrupt temperature change from $T_{H}$ to ambient temperature $T_{C}$, the ratio $T_{cr}/T_{C}$ is very large ($\sim$2.3) and the unstable mode is a standing wave ($\psi_{Up}\simeq \pm 90^{\circ}$). Since the nonlinear cascade leading to shock formation is inhibited by the standing wave phasing compared to the quasi-travelling wave phasing~\cite{BiwaEtAl_JASA_2014}, a geometry similar to Y-1998 has been chosen in the current study which exhibits quasi-travelling wave unstable mode due to a wider duct section.

\chapter{Scaling of governing equations}
\label{sec: appendixD}
The governing equations for compressible flows in two dimensions read: 
\begin{subequations}
 \begin{gather}
  \frac{\partial \rho}{\partial t} + \frac{\partial (\rho u)}{\partial x} + \frac{\partial (\rho v)}{\partial y} = 0,\\
  \rho\left(\frac{\partial u}{\partial t} + u\frac{\partial u}{\partial x} + v\frac{\partial u}{\partial y}\right) = -\frac{\partial p}{\partial x} + \frac{\partial}{\partial x}\left[\mu\left(\xi_B + \frac{4}{3}\right)\frac{\partial u}{\partial x}\right] + \frac{\partial }{\partial y}\left(\mu\frac{\partial u}{\partial y}\right),\\
  \rho\left(\frac{\partial v}{\partial t} + u\frac{\partial v}{\partial x} + v\frac{\partial v}{\partial y}\right) = -\frac{\partial p}{\partial y} + \frac{\partial}{\partial y}\left[\mu\left(\xi_B + \frac{4}{3}\right)\frac{\partial v}{\partial y}\right] + \frac{\partial}{\partial x}\left(\mu\frac{\partial v}{\partial x}\right),\\
  \rho T \left(\frac{\partial s}{\partial t} + u\frac{\partial s}{\partial x} + v\frac{\partial s}{\partial y}\right) = \frac{\partial }{\partial y}\left(k\frac{\partial T}{\partial y}\right) + \frac{\partial }{\partial x}\left(k\frac{\partial T}{\partial x}\right) + \Phi,
 \end{gather}
\end{subequations}
where $\Phi$ accounts for entropy generation due to viscous gradients. These equations are written in terms of perturbations, denoted by primed symbols, in the relevant variables $u$, $v$, $\rho$, $p$, $T$, and $~s$.
%, and drop the $\Phi$ terms (following~\citealt{Chester_JFM_1964}). 
It shall be noted that, for a nonlinear acoustic field of velocity amplitude scale $\mathcal{U}$, the acoustic Mach number $\mathcal{M}=\mathcal{U}/a_0$ is $\mathcal{O}(10^{-1})$ and the aspect ratio of the regenerator $h_b/l_b$ is $\mathcal{O}(10^{-2})$. As a result, the perturbations in the $y$ component of velocity (namely, $v'/u' \sim h_b/l_b$) from the momentum and entropy equations can be neglected and the following equations, including second order terms  in $\mathcal{M}$, are obtained:
\begin{subequations}
\begin{gather}
\frac{\partial \rho'}{\partial t} + \rho_0 \frac{\partial u'}{\partial x} + u'\frac{d\rho_0}{dx}  + \rho_0\frac{\partial v'}{\partial y} = \left[-\rho'\frac{\partial u'}{\partial x} - u'\frac{\partial \rho'}{\partial x}\right],\label{eq: app_mass}\\
\frac{\partial u'}{\partial t} + \frac{1}{\rho_0}\frac{\partial p'}{\partial x} - \frac{1}{\rho_0}\frac{\partial}{\partial y}\left(\mu\frac{\partial u'}{\partial y}\right) - \frac{1}{\rho_0}\frac{\partial }{\partial x}\left[\mu\left(\xi_B + \frac{4}{3}\right)\frac{\partial u'}{\partial x}\right] =   \left[-\frac{\rho'}{\rho_0}\frac{\partial u'}{\partial t} - \frac{1}{2}\frac{\partial u'^2}{\partial x}\right],\label{eq: app_mom}\\
\frac{\partial s'}{\partial t} + u'\frac{ds_0}{dx} - \frac{R}{p_0}\frac{\partial}{\partial y}\left(k\frac{\partial T'}{\partial y}\right) - \frac{R}{p_0}\frac{\partial}{\partial x}\left(k\frac{\partial T'}{\partial x}\right) = \left[-\frac{p'}{p_0}\left(\frac{\partial s'}{\partial t} + u'\frac{ds_0}{dx}\right) - u'\frac{\partial s'}{\partial x}+\Phi\right].\label{eq: app_ent}
\end{gather}
\end{subequations}
In the above equations, the subscript $0$ denotes the base state. Also, the terms on the left-hand side of the equations are linear in the perturbation variables, whereas the terms on the right-hand side are non-linear. Assuming $\omega^{-1}$ as the characteristic time scale of the acoustic field,  the viscous dissipation terms on the left-hand side of~\eqref{eq: app_mom} and~\eqref{eq: app_ent}, and the entropy generation due to viscous gradients $\Phi$ scale relatively, as
\begin{eqnarray}
 \frac{\left\lvert \frac{1}{\rho_0}\frac{\partial}{\partial y}\left(\mu\frac{\partial u'}{\partial y}\right) \right\lvert}{\left\lvert \frac{\partial u'}{\partial t} \right\lvert} \sim \left(\frac{\delta_\nu}{h_b}\right)^2 \left(1 + \mathcal{M}\right),\quad \frac{\left\lvert \mu \left(\frac{\partial u'}{\partial y}\right)^2\right\lvert}{\rho T \frac{\partial s}{\partial t}} \sim \left(\frac{\delta_\nu}{h_b}\right)^2 \mathcal{M},\\ \frac{\left\lvert \frac{1}{\rho_0}\frac{\partial }{\partial x}\left[\mu\left(\xi_B + \frac{4}{3}\right)\frac{\partial u'}{\partial x}\right] \right\lvert}{\left\lvert \frac{\partial u'}{\partial t} \right\lvert} \sim \left(\frac{\delta_\nu}{h_b}\right)^2 \left(\frac{h_b\omega}{a_0}\right)^2 \left(1 + \mathcal{M}\right),
\end{eqnarray}
where the following scaling relations have been used:
\begin{equation}
 \mu \sim \mu_0(1 + \mathcal{M}),\quad \delta_{\nu} \sim \sqrt{\frac{\nu_0}{\omega}}. 
\end{equation}
%{\color{blue}{
%Furthermore, the entropy generation term $\Phi$ on the right hand side of \eqref{eq: app_ent} scales as, 
%\begin{equation}
%\frac{\left\lvert \mu \left(\frac{\partial u'}{\partial y}\right)^2\right\lvert}{\rho T \frac{\partial s}{\partial t}} \sim \left(\frac{\delta_\nu}{h_b}\right)^2 \mathcal{M}
%\end{equation}}}
Due to the nonlinear energy cascade, higher harmonics are generated, as a result of which, the characteristic time scale of the acoustic field $\omega^{-1}$ decreases. Consequently, the characteristic Stokes layer thickness, $\delta_\nu$, decreases and the ratio $h_b\omega/a_0$ increases. Accordingly, the terms which scale as $\mathcal{O}((\delta_\nu/h_b)^2\mathcal{M})$ can be dropped and terms corresponding to the local perturbations in viscosity and thermal conductivity, as well as the entropy generation term due to viscous gradients, are neglected. Finally, the equations governing the spatio-temporal evolution of the acoustic field, correct up to the second order, are given as 
\begin{subequations}
\begin{gather}
\frac{\partial \rho'}{\partial t} + \rho_0 \frac{\partial u'}{\partial x} + u'\frac{d\rho_0}{dx} + \rho_0\frac{\partial v'}{\partial y} = \left[-\rho'\frac{\partial u'}{\partial x} - u'\frac{\partial \rho'}{\partial x}\right],\label{eq: 2_massApp}\\
\frac{\partial u'}{\partial t} + \frac{1}{\rho_0}\frac{\partial p'}{\partial x} - \nu_0 \frac{\partial^2 u'}{\partial y^2} - \frac{1}{\rho_0}\frac{\partial }{\partial x}\left[\mu_0\left(\xi_B + \frac{4}{3}\right)\frac{\partial u'}{\partial x}\right] =  \left[-\frac{\rho'}{\rho_0}\frac{\partial u'}{\partial t} - \frac{1}{2}\frac{\partial u'^2}{\partial x}\right],\label{eq: 2_momApp}\\
\frac{\partial s'}{\partial t} + u'\frac{ds_0}{dx} - \frac{Rk_0}{p_0}\frac{\partial^2 T'}{\partial y^2} - \frac{R}{p_0}\frac{\partial}{\partial x}\left(k_0\frac{\partial T'}{\partial x}\right) = \left[-\frac{p'}{p_0}\left(\frac{\partial s'}{\partial t} + u'\frac{ds_0}{dx}\right) - u'\frac{\partial s'}{\partial x}\right], \label{eq: 2_entApp}
\end{gather}
\end{subequations}

\chapter{Time domain model and convergence}
\label{sec: appedixE}
Integrating~\eqref{eq: 2_massApp} and~\eqref{eq: 2_entApp} in $y$, the following equations are obtained:
\begin{equation}
\frac{\partial \tilde{\rho}^{(2)}}{\partial t} + \rho_0\frac{\partial U'^{(1)}}{\partial x} + U'^{(1)}\frac{d\rho_0}{dx} = \int^{h_b/2}_{-h_b/2}\left[-\rho'^{(1)}\frac{\partial u'^{(1)}}{\partial x} - u'^{(1)}\frac{\partial \rho'^{(1)}}{\partial x}\right]dy,\label{eq: ContApp}
\end{equation}
\begin{equation}
  \frac{\partial S'^{(2)}}{\partial t} + U'^{(1)}\frac{ds_0}{dx} = \left[1-\frac{p'^{(1)}}{p_0}\right]\frac{q'^{(1)}}{\rho_0T_0} + q_2 + \frac{R}{p_0}\frac{\partial }{\partial x}\left[\frac{k_0T_0}{C_p}\frac{\partial S'^{(1)}}{\partial x}\right] -
\int^{h_b/2}_{-h_b/2}u'^{(1)}\frac{\partial s'^{(1)}}{\partial x}dy\label{eq: EntApp},
\end{equation}
where,
\begin{gather}
 \tilde{\rho}^{(2)} = \int^{h_b/2}_{-h_b/2}\rho^{(2)} dy,\quad S'^{(2)}= \int^{h_b/2}_{-h_b/2}s'^{(2)} dy, \\
 q'^{(1)} = \frac{2\nu_0\rho_0 T_0}{Pr}\left.\frac{\partial s'^{(1)}}{\partial y}\right|_{y=h_b/2}.
\end{gather}
Also, the density constitutive equation, up to second order, is given by 
\begin{equation}
 \rho'^{(2)} = \alpha_s p'^{(2)} + \alpha_ps'^{(2)} + \frac{1}{2}\left(\beta_s p'^{(1)2} + \beta_p s'^{(1)2} + 2\beta_{sp} s'^{(1)}p'^{(1)}\right), \label{eq: ConstitutiveApp}
\end{equation}
and the entropy perturbations $s'^{(1)}$, up to first order, are 
\begin{equation}
 s'^{(1)} = -\frac{R}{p_0}p'^{(1)} + \sum_{j=0}^{\infty}\check{s}_j(x,t)\cos \left(\zeta_j y\right).\label{eq: EntropyModesApp}
\end{equation}
Notice that the superscripts $(1)$ and $(2)$ are dropped hereafter for convenience of notation. Integrating equation~\eqref{eq: ConstitutiveApp} along $y$, differentiating in time, and combining with~\eqref{eq: ContApp} and~\eqref{eq: EntApp} to eliminate $\tilde{\rho}$, the following equation is obtained: 
\begin{equation}
\frac{\partial p'}{\partial t} + \frac{\rho_0a^2_0}{h_b}\frac{\partial U'}{\partial x}=\frac{\rho_0a^2_0}{h_b}\left(\frac{q'}{C_p\rho_0T_0}+q_2  + \mathbb{T} -\mathbb{Q} +  \mathbb{D}_s\right) - \mathbb{C}, \label{eq: nonlinear_P_App}
\end{equation}
where 
\begin{equation}
 q' = \frac{2\nu_0\rho_0 T_0}{Pr}\sum^{\infty}_{j=0}(-1)^{j+1}\check{s}_j(x,t)\zeta_j,\quad\mathbb{Q} = \frac{\gamma p'q'}{C_p p_0\rho_0T_0},
 \label{eq: heat_flux_fluct_app}
\end{equation}
and,
\begin{subequations}
\begin{gather}
q_2 = \frac{h_b\nu_0}{C_p Pr}\sum^{\infty}_{j=0}\left(\zeta_j \check{s}_j\right)^2,\\
\mathbb{T} = \frac{\left(\gamma-1\right)}{p_0}p'\frac{\partial U'}{\partial x} + \frac{1}{C_pa^2_0}\frac{\partial}{\partial t}\Bigg[\frac{p'}{p_0}\sum_{j=0}^{\infty}(-1)^{j+1}\frac{2\check{s}_j}{\zeta_j}\Bigg]-\frac{h_b}{4C^2_pa^2_0}\frac{\partial}{\partial t}\sum^{\infty}_{j=0}s^2_j,\\
\mathbb{C} = -\frac{\gamma - 1}{a^2_0}U'\frac{\partial p'}{\partial x} -\gamma\frac{\partial}{\partial x}\left(\frac{p'U'}{a^2_0}\right)- \frac{\rho_0 h_b}{2C_p}\sum^{\infty}_{j=0}\check{u}_j\frac{\partial \check{s}_j}{\partial x} + \frac{h_b}{2}\frac{\partial}{\partial x}\Bigg(\frac{\rho_0}{C_p}\sum^{\infty}_{j=0}\check{u}_j\check{s}_j\Bigg),\\
\mathbb{D}_s = \frac{1}{Pr}\frac{\partial}{\partial x}\Bigg[\nu_0\Bigg(\sum^{\infty}_{j=0}(-1)^{j}\frac{2}{\zeta_j}\frac{\partial \check{s}_j}{\partial x} + \frac{h_b R}{p_0}\frac{\partial p'}{\partial x}\Bigg)\Bigg].
\end{gather}
\label{eq: nonlinearities_App}
\end{subequations}
The above equations account for thermodynamic, as well as, convective nonlinearities, with $\mathbb{D}_s$ denoting the axial conduction term. The treatment of quadratic nonlinearities gets significantly simplified  due to the Fourier expansions of the viscous and entropic modes. For simplicity, only the nonlinear macrosonic thermoacoustic interaction $\mathbb{Q}$ is retained  in the present one-dimensional computations. However, the complete model equations~\eqref{eq: nonlinear_P_App}--\eqref{eq: nonlinearities_App} should be considered in the case of relatively large regenerators. 

In order to show the convergence of the viscous and entropic modes, $\check{u}_j$ and  $\check{s}_j$, respectively, \eqref{eq: u_modes} and~\eqref{eq: s_modes} are pre-multiplied by $(-1)^{j+1}\zeta_j$ and summed over $j$ to obtain: 
\begin{gather}
 \frac{\partial \tau'_w}{\partial t} + 2\nu^2_0\sum^{N}_{j=0}(-1)^{j+1}\check{u}_j\zeta^3_j = N\frac{4\nu_0}{h\rho_0}\frac{\partial p'}{\partial x},\label{eq: shear_conv1}\\
 \frac{1}{\rho_0T_0}\frac{\partial q'}{\partial t} + \frac{\tau'_w}{Pr}\frac{ds_0}{dx} + 2\left(\frac{\nu_0}{Pr}\right)^2\sum^{N}_{j=0}(-1)^{j+1}\check{s}_j\zeta^3_j = -N\frac{4\nu_0}{h_bPr\rho_0 T_0}\left(\frac{\partial p'}{\partial t}\right). \label{eq: heat_conv1}
\end{gather}
For the wall-shear $\tau'_w$ and the wall-heat flux $q'$ to converge, \eqref{eq: shear_conv1} and~\eqref{eq: heat_conv1} must yield the same values of $q'$ and $\tau'_w$ for $N$ and $N+1$ in the limit of $N\rightarrow \infty$. Hence, the following conditions are obtained:  
\begin{equation}
 \lim_{N\to\infty} |u_N| = \frac{2}{\zeta^3_N}\frac{1}{h_b\rho_0\nu_0}\frac{\partial p'}{\partial x}, \quad \lim_{N\to\infty} |s_N| = \frac{2}{\zeta^3_N}\frac{Pr}{h_b\rho_0\nu_0 T_0}\frac{\partial p'}{\partial t}. \label{eq: modes_convergence}
\end{equation}
The above relations show that, in order for convergence to be ensured, the magnitudes of the viscous modes $\check{u}_j$ and $\check{s}_j$ must decay as $\zeta^{-3}_j$ for large values of the index $j$. The relaxation functional used in~\eqref{eq: analytical_un} and~\eqref{eq: analytical_sn} for very large $j$ yields: 
\begin{equation}
 \int^{t}_{-\infty}e^{-\frac{t-\eta}{\tau_j}}\phi(x,\eta)d\eta \approx \tau_j\phi(x,t), \label{eq: approx_kernel}
\end{equation}
for some function $\phi(x,t)$. After substituting the above approximation in~\eqref{eq: analytical_un} and~\eqref{eq: analytical_sn}, it is  shown that the magnitudes of the viscous modes $\check{u}_j$ and $\check{s}_j$ decay as $\zeta^{-3}_j$ in the limit of $j$ such that the approximation in~\eqref{eq: approx_kernel} holds valid, which, in turn,  guarantees the convergence of the infinite series approximation for the  viscous and entropic modes. 

\chapter{Windowed shock capturing}
\label{sec: appedixF}
At the limit cycle, the perturbation fields exhibit shock wave propagation causing the formation of large gradients at the limit of grid resolution. Spatially windowed Legendre polynomial expansions are therefore used in order to add an artificial viscosity term to ensure a proper resolution of longitudinal gradients in the model equations. The implemented  strategy, which is briefly summarized below, is similar to the one proposed by Persson and Peraire~\cite{persson:06} for discontinuous Galerkin methods.
For a spatial window of size $N_p$, the local pressure field is reconstructed using  Legendre polynomial basis expansions up to orders $N_p$ and $N_p-1$, such that 
\begin{equation}
 p'=\sum^{N_p}_{n=0}p_n\psi_n,\quad \tilde{p}' = \sum^{N_p-1}_{n=0}p_n\psi_n,
\end{equation}
where $\psi_n$ denotes the Legendre polynomial of order $n$ and $p_n$ is the relevant $n^{\mathrm{th}}$ mode. 
A modal smoothness indicator $S_e$ is hence constructed from the ratio of the inner products of  $p'-\tilde{p}'$ and $p'$, such as to detect the onset of excessively high high-frequency modes, which is typical of insufficiently unresolved signals:
\begin{equation}
 S_e = \frac{\langle p'-\tilde{p}',p'-\tilde{p}'\rangle}{\langle p',p' \rangle}.
\end{equation}
The sensor is then used to trigger the shock capturing artificial diffusivity $\epsilon_e$, which is evaluated as 
\[
\epsilon_e=
\begin{cases}
0 &\text{if } s_e < s_0  ,\\
\epsilon_0\sin\left(\frac{\pi(s_e-s_0)}{2\kappa}\right) &\text{if } s_0<s_e<s_0+\kappa,\\
\epsilon_0 &\text{if } s_e>s_0.
\end{cases}
\]
where $s_e = \log S_e$ and the corresponding parameters are  $s_0 = 3.2$, $\kappa = 0.9(\mathrm{max}(s_e) - s_0)$, and $\epsilon_0 = 100\Delta x/N_p$, $\Delta x$ being the grid spacing of the discretized one-dimensional domain.

Finally, the artificial viscosity augmented equations to capture the nonlinear acoustic fields become: 
\begin{gather}
   \frac{\partial p'}{\partial t} = -\frac{\gamma p_0}{h}\left(1 + \frac{1+\gamma}{\gamma}\frac{p'}{p_0}\right)\frac{\partial U'}{\partial x} + \left[\epsilon_{e}+\frac{k}{\rho_0}\left(\frac{1}{C_v} - \frac{1}{C_p}\right)\right]\frac{\partial^2 p'}{\partial x^2},\label{eq: inviscid_pApp}\\
   \frac{\partial U'}{\partial t} = -\frac{h}{\rho_0}\frac{\partial p'}{\partial x} + \left[\epsilon_e Pr+\nu_0\left(\xi_B + \frac{4}{3}\right)\right]\frac{\partial^2 U'}{\partial x^2}.\label{eq: inviscid_UApp}
\end{gather}

% \include{appendix}
% %%%%%%%%%%%%%%%%%%%%%%%%%%%%%%

% Recommendations are optional.
% You may include recommendations as a major division if your
% subject matter and research dictate.
% Reference: TM2006 page 32.
% CHANGE NEXT LINE?
%\include{recommendations}

% Bibliography is required if you consulted any outside references.
% Reference: TM2006 page 32.
\addcontentsline{toc}{chapter}{REFERENCES}
\bibliographystyle{unsrt}
\bibliography{./references}

% Appendices are optional.
% Appendices are not necessarily a part of every thesis.
% An appendix is used for supplementary illustrative material,
% original data, computer programs, and other material that
% is not necessarily appropriate for inclusion within the
% text of your thesis.
% Reference: TM2006 page 33.
% Use "\appendix" for one appendix or "\appendices" for more than one
% appendix.
% CHANGE NEXT 7 LINES?
%\appendices
%\include{demo-citations}
%\include{demo-figures}
%\include{demo-mathematics}
%\include{demo-multicols}
%\include{demo-tables}
%\include{demo-text}

% Notes and footnotes are optional.
% Reference: TM2006 page 34.
% I have not implemented this yet.  Mark Senn 2002-06-03
%%\include{notes}

% A vita is optional for masters theses
% and required for doctoral dissertations.
% Reference: TM2006 page 13.
% CHANGE NEXT LINE?
% \include{vita}

\end{document}